


%
%
%
\newbox\leftpage \newdimen\fullhsize \newdimen\hstitle \newdimen\hsbody
\tolerance=1000\hfuzz=2pt
\def\printertype{}
\def\qms{\def\printertype{qms: }
\ifx\answ\bigans\else\voffset=-.4truein\hoffset=.125truein\fi}
\def\bigans{b }
\let\answ\bigans
%
\ifx\answ\bigans\message{(This will come out unreduced.}
\magnification=1200\baselineskip=16pt plus 2pt minus 1pt
\hsbody=\hsize \hstitle=\hsize 
%
%
\else\message{(This will be reduced.} \let\lr=L
\magnification=1000\baselineskip=16pt plus 2pt minus 1pt
\voffset=-.31truein\vsize=7truein\hoffset=-.59truein
\hstitle=8truein\hsbody=4.75truein\fullhsize=10truein\hsize=\hsbody
\output={\ifnum\pageno=0 
  \shipout\vbox{\special{\printertype landscape}\makeheadline
    \hbox to \fullhsize{\hfill\pagebody\hfill}}\advancepageno
  \else
  \almostshipout{\leftline{\vbox{\pagebody\makefootline}}}\advancepageno
  \fi}
\def\almostshipout#1{\if L\lr \count1=1 \message{[\the\count0.\the\count1]}
      \global\setbox\leftpage=#1 \global\let\lr=R
  \else \count1=2
    \shipout\vbox{\special{\printertype landscape}
      \hbox to\fullhsize{\box\leftpage\hfil#1}}  \global\let\lr=L\fi}
\fi
%
\catcode`\@=11 
\def\noblackbox{\overfullrule=0pt}
\hyphenation{anom-aly anom-alies coun-ter-term coun-ter-terms}
\def\@Roman#1{\uppercase{\romannumeral #1}}
\def\Roman{\@Roman}
\newdimen\shrinkdim
\ifx\ans\bigans\shrinkdim=-2in\else\shrinkdim=-1.5in\fi
\def\mytopinsert#1{\insert\topins{\penalty100
\splittopskip=0pt\splitmaxdepth=\maxdimen\floatingpenalty=0
\vbox{#1}\nobreak\bigskip}}
%
\newcount\yearltd\yearltd=\year\advance\yearltd by -1900

\def\Title#1#2{\nopagenumbers\abstractfont\hsize=\hstitle\rightline{#1}%
\vskip 1in\centerline{\titlefont #2}\abstractfont\vskip .5in\pageno=0}
\def\Date#1{\vfill\leftline{#1}\tenpoint\supereject\global\hsize=\hsbody%
\footline={\hss\tenrm\folio\hss}}
%

\def\draftmode{\message{ DRAFTMODE }\def\draftdate{{\rm preliminary draft:
\number\month/\number\day/\number\yearltd\ \ \hourmin}}%
\headline={\hfil\draftdate}\writelabels\baselineskip=20pt plus 2pt minus 2pt
 {\count255=\time\divide\count255 by 60 \xdef\hourmin{\number\count255}
  \multiply\count255 by-60\advance\count255 by\time
  \xdef\hourmin{\hourmin:\ifnum\count255<10 0\fi\the\count255}}}
\def\nolabels{\def\wrlabel##1{}\def\eqlabel##1{}\def\reflabel##1{}}
\def\writelabels{\def\wrlabel##1{\leavevmode\vadjust{\rlap{\smash%
{\line{{\escapechar=` \hfill\rlap{\sevenrm\hskip.03in\string##1}}}}}}}%
\def\eqlabel##1{{\escapechar-1\rlap{\sevenrm\hskip.05in\string##1}}}%
\def\reflabel##1{\noexpand\llap{\noexpand\sevenrm\string\string\string##1}}}
\nolabels
\newwrite\ffile  
\newif\iffclosed \fclosedtrue 
\def\fwrite#1{\iffclosed\immediate\openout\ffile=\jobname.fwd
	      \global\fclosedfalse\fi
              \immediate\write\ffile{#1}}
%
%
\def\freference#1{\fwrite{\noexpand\xdef\noexpand#1{#1}}}
\newread\tfile
\def\testinput#1{
                 \immediate\openin\tfile=#1
                 \immediate\ifeof\tfile\else\closein\tfile\input#1\fi
                 \closein\tfile
                  }
\testinput{\jobname.fwd}
\newif\if@afterindent \@afterindenttrue
\newdimen\@tempdima
\def\contentdepth{3}	
\def\@pnumwidth{20pt}
\def\@tocrmarg{80pt}
\def\@dotsep{1.7}

%
%
\def\dottedtocline#1#2#3#4#5{\ifnum #1>\contentdepth \else
  \vskip \z@ plus .2pt
  {\leftskip #2\relax \rightskip \@tocrmarg \parfillskip -\rightskip
    \parindent #2\relax\@afterindenttrue
   \interlinepenalty\@M
   \leavevmode
   \@tempdima #3\relax \advance\leftskip \@tempdima \hbox{}\hskip -\leftskip
   #4\nobreak\leaders\hbox{$\m@th \mkern \@dotsep mu.\mkern \@dotsep
       mu$}\hfill \nobreak \hbox to\@pnumwidth{\hfil\rm #5}\par}\fi}
\def\numberline#1{\hbox to\@tempdima{#1\hfil}}
\def\tocline#1#2#3{\ifnum #1>\contentdepth \else
  \vskip \z@ plus .2pt
  {\leftskip #2\relax \rightskip \@tocrmarg
\noindent #3\hfill\par}\fi}
%
\def\ourfolio{\ifnum\pageno<0 \romannumeral-\pageno \else\number\pageno\fi}
\def\leaderfill{\leaders\hbox to 1em{\hss . \hss}\hfill}
%
%
%
%
\newwrite\cfile			
\newif\ifcclosed \cclosedtrue	
\newwrite\pfile			
\newif\ifpclosed \pclosedtrue
\newif\iftoclisted \toclistedfalse	
\def\writetoc{
\def\cwrite##1{\iftoclisted
	      \ifcclosed\immediate\openout\cfile=\jobname.toc
	      \global\cclosedfalse\fi
              \write\cfile{##1}		
	      \else\pwrite{##1}\fi}
%
\def\pwrite##1{\ifpclosed\immediate\openout\pfile=\jobname.pre
	      \global\pclosedfalse\fi
              \write\pfile{##1}}	
}
\def\cwrite#1{}\def\pwrite#1{}
%
%
\def\listtoc{\vfill\supereject
             \immediate\closeout\pfile
             \centerline{{\extrafont Table of Contents}}\bigskip
             {\baselineskip=12pt
	     \catcode`\@=11		
             \testinput{\jobname.pre}
             \testinput{\jobname.toc}
             \vfill\supereject
	     \catcode`\@=12}
	     \global\toclistedtrue
            }
\def\extrahead#1{\vfil\supereject
\centerline{\extrafont #1}\bigskip\bigskip}
\def\extra#1{\extrahead{#1}
  \cwrite{\noexpand \dottedtocline{0}{0pt}{20pt}
  {#1}{\ourfolio} } }
\def\extraearly#1{\extrahead{#1}
  \pwrite{\noexpand \dottedtocline{0}{0pt}{20pt}
  {#1}{\ourfolio}} }
%
%
\newif\if@meqn\@meqntrue
\def\eqnbychapter{\global\@meqnfalse}	
\xdef\chapsym{}\xdef\secsym{}\xdef\eqnsym{}
\global\newcount\chapno \global\chapno=0
\global\newcount\secno \global\secno=0
\global\newcount\subsecno \global\subsecno=0
\global\newcount\meqno \global\meqno=0
\global\newcount\thmno \global\thmno=1
\global\newcount\figno \global\figno=1
\global\newcount\tblno \global\tblno=1
%
%
\def\tchapter#1#2#3{
\xdef#1{#2}
\xdef\chapsym{#2.}\xdef\secsym{#2.}\xdef\eqnsym{#2.}
\global\secno=0\global\subsecno=0
\global\thmno=1\global\figno=1\global\meqno=0\global\tblno=1
\vfill\supereject
\message{(#2. #3)}
\ \bigskip\centerline{\chapterfont\chapsym\ #3}\bigskip\bigskip
{\edef\writet@c{
    \noexpand\cwrite{\noexpand\noexpand\noexpand\dottedtocline{0}{0pt}{15pt}
	{\noexpand\noexpand\noexpand\numberline{#1.}{#3}}{\noexpand\ourfolio}}
}\writet@c}
\nobreak\freference{#1}
}

\def\tnewchap#1#2{\global\advance\chapno by 1
                 \tchapter#1{\the\chapno}{#2}
                 }
\def\newchap#1{\tnewchap\tempt@g{#1}}
\def\chapter#1{\tnewchap\tempt@g{#1}}
%
%
\def\tsection#1#2#3{
\xdef#1{\chapsym#2}
\xdef\secsym{\chapsym#2.}
\if@meqn\xdef\eqnsym{\chapsym#2.}\global\meqno=0\fi
\global\subsecno=0
\message{(\chapsym#2. #3)}
\bigbreak\bigskip\noindent{\bf\chapsym#2. #3}\par\nobreak\medskip\nobreak
{\edef\writet@c{
    \noexpand\cwrite{\noexpand\noexpand\noexpand\dottedtocline{1}{15pt}{25pt}
	{\noexpand\noexpand\noexpand\numberline{#1.}{#3}}{\noexpand\ourfolio}}
}\writet@c}
\nobreak\freference{#1}
}
\def\tnewsec#1#2{\global\advance\secno by 1
                 \tsection#1{\the\secno}{#2}
                 }
\def\newsec#1{\tnewsec\tempt@g{#1}}

%
\def\tsubsection#1#2#3{
\xdef#1{\secsym#2}
\message{(\secsym#2. #3)}
\ifnum\lastpenalty>9000\else\bigbreak\fi
\noindent{\it\secsym#2. #3}\par\nobreak\medskip\nobreak
{\edef\writet@c{
    \noexpand\cwrite{\noexpand\noexpand\noexpand\dottedtocline{2}{40pt}{30pt}
	{\noexpand\noexpand\noexpand\numberline{#1.}{#3}}{\noexpand\ourfolio}}
}\writet@c}
\nobreak\freference{#1}
}
\def\tnewsubsec#1#2{\global\advance\subsecno by 1
                   \tsubsection#1{\the\subsecno}{#2}}
\def\subsec#1{\tnewsubsec\tempt@g{#1}}

\def\newsubsec#1{\tnewsubsec\tempt@g{#1}}
%
%
\def\subsubsec#1{\medbreak\vskip0pt plus3pt
\centerline{\it #1}\nobreak\smallskip\nobreak
\cwrite{\noexpand\tocline{3}{85pt}{\noexpand\it #1}}
\nobreak
}
%

%
%
\def\tappendix#1#2#3{
\xdef#1{{\hbox{\chapsym#2}}}
\xdef\secsym{\hbox{\chapsym#2.}}\xdef\eqnsym{\hbox{\chapsym#2.}}
\global\meqno=0\global\subsecno=0
\message{(\chapsym#2. #3)}
\bigbreak\bigskip\noindent{\bf Appendix \chapsym#2. #3}\par%
\nobreak\medskip\nobreak
{\edef\writet@c{
    \noexpand\cwrite{\noexpand\noexpand\noexpand\dottedtocline{1}{15pt}{25pt}
	{\noexpand\noexpand\noexpand\numberline{#1.}{#3}}{\noexpand\ourfolio}}
}\writet@c}
\nobreak\freference{#1}
}
\def\appendix#1#2{\tappendix\tempt@g{#1}{#2}}
%
%
\def\tCappendix#1#2#3{
\xdef#1{{\hbox{#2}}}
\xdef\secsym{\hbox{#2.}}\xdef\eqnsym{\hbox{#2.}}
\global\meqno=0\global\subsecno=0
\message{(#2. #3)}
\vfill\supereject\ %
\bigskip\centerline{\chapterfont Appendix #2. #3}\bigskip\bigskip
{\edef\writet@c{
    \noexpand\cwrite{\noexpand\noexpand\noexpand\dottedtocline{0}{0pt}{60pt}
	{\noexpand\noexpand\noexpand\numberline{Appendix #1.}{#3}}
	{\noexpand\ourfolio}}
}\writet@c}
\freference{#1}
}
\def\Cappendix#1#2{\tappendix\tempt@g{#1}{#2}}
%
%

%
%
\def\seclab#1{\xdef #1{\the\secno}\writedef{#1\leftbracket#1}\wrlabel{#1=#1}}
\def\subseclab#1{\xdef #1{\secsym\the\subsecno}%
\writedef{#1\leftbracket#1}\wrlabel{#1=#1}}
\def\newthm#1{\xdef#1{{\chapsym\the\thmno}}#1%
\freference{#1}
\global\advance\thmno by 1
}
%
%
\def\eqnn#1{\global\advance\meqno by1\xdef #1{(\eqnsym\the\meqno)}%
\writedef{#1\leftbracket#1}\wrlabel#1\global\advance\foo by1}
\def\eqna#1{\global\advance\meqno by1%
\xdef #1##1{\hbox{$(\eqnsym\the\meqno##1)$}}%
\writedef{#1\numbersign1\leftbracket#1{\numbersign1}}\wrlabel{#1$\{\}$}%
\global\advance\foo by1}
\def\eqn#1#2{\global\advance\meqno by1\xdef #1{(\eqnsym\the\meqno)}%
\writedef{#1\leftbracket#1}$$#2\eqno#1\eqlabel#1$$\global\advance\foo by1}
\def\eqnt#1{\global\advance\meqno by1\xdef #1##1{\hbox{$(T\the\meqno##1)$}}%
\writedef{#1\numbersign1\leftbracket#1{\numbersign1}}\wrlabel{#1$\{\}$}}



%
\newskip\footskip\footskip14pt plus 1pt minus 1pt 
\def\f@@t{\baselineskip\footskip\bgroup\aftergroup\@foot\let\next}
\setbox\strutbox=\hbox{\vrule height9.5pt depth4.5pt width0pt}
\global\newcount\ftno \global\ftno=0
\def\foot{\global\advance\ftno by1\footnote{$^{\the\ftno}$}}
%
\newwrite\ftfile
\def\footend{\def\foot{\global\advance\ftno by1\chardef\wfile=\ftfile
$^{\the\ftno}$\ifnum\ftno=1\immediate\openout\ftfile=\jobname.fts\fi%
\immediate\write\ftfile{\noexpand\smallskip%
\noexpand\item{f\the\ftno:\ }\pctsign}\findarg}%
\def\footatend{\immediate\closeout\ftfile{\parindent=20pt
\extra{Footnotes}\input \jobname.fts }}}
\def\footatend{}
%
%
\global\newcount\refno \global\refno=1
\newwrite\rfile
\def\ref{[\the\refno]\nref}
\def\nref#1{\xdef#1{[\the\refno]}\writedef{#1\leftbracket#1}%
\ifnum\refno=1\immediate\openout\rfile=\jobname.ref\fi
\global\advance\refno by1\chardef\wfile=\rfile\immediate
\write\rfile{\noexpand\item{#1\ }\reflabel{#1\hskip.31in}\pctsign}\findarg}
\def\findarg#1#{\begingroup\obeylines\newlinechar=`\^^M\pass@rg}
{\obeylines\gdef\pass@rg#1{\writ@line\relax #1^^M\hbox{}^^M}%
\gdef\writ@line#1^^M{\expandafter\toks0\expandafter{\striprel@x #1}%
\edef\next{\the\toks0}\ifx\next\em@rk\let\next=\endgroup\else\ifx\next\empty%
\else\immediate\write\wfile{\the\toks0}\fi\let\next=\writ@line\fi\next\relax}}
\def\striprel@x#1{} \def\em@rk{\hbox{}}
\def\lref{\begingroup\obeylines\lr@f}
\def\lr@f#1#2{\gdef#1{\ref#1{#2}}\endgroup\unskip}

\def\addref#1{\immediate\write\rfile{\noexpand\item{}#1}} 
\def\listrefs{\footatend\immediate\closeout\rfile\writestoppt
\baselineskip=14pt\extra{References}{\frenchspacing%
\parindent=20pt\testinput{\jobname.ref}\vfill\eject}%
\nonfrenchspacing%
}
\def\startrefs#1{\immediate\openout\rfile=refs.tmp\refno=#1}
\def\xref{\expandafter\xr@f}\def\xr@f[#1]{#1}
\def\refs#1{\count255=1[\r@fs #1{\hbox{}}]}
\def\r@fs#1{\ifx\und@fined#1\message{reflabel \string#1 is undefined.}%
\nref#1{need to supply reference \string#1.}\fi%
\vphantom{\hphantom{#1}}\edef\next{#1}\ifx\next\em@rk\def\next{}%
\else\ifx\next#1\ifodd\count255\relax\xref#1\count255=0\fi%
\else#1\count255=1\fi\let\next=\r@fs\fi\next}
%
%
%

%
\newwrite\figfile\newif\iffigclosed\figclosedtrue
\def\fig{fig.~\chapsym\the\figno\nfig}
\def\nfig#1{\xdef#1{fig.~\chapsym\the\figno}%
\writedef{#1\leftbracket fig.\noexpand~\chapsym\the\figno}%
\iffigclosed\immediate\openout\figfile=\jobname.fgs\global\figclosedfalse\fi%
\chardef\wfile=\figfile%
\immediate\write\figfile{\noexpand\medskip\noexpand\item{Fig.\ \the\figno. }
\reflabel{#1\hskip.55in}\pctsign}\global\advance\figno by1\findarg}
\def\listfigs{\immediate\closeout\figfile\extra{Figure Captions}
{\catcode`\@=11\parindent40pt\baselineskip16pt
\testinput{\jobname.fgs}\vfill\supereject\catcode`\@=12}}
\def\listfigtoc{\immediate\closeout\figfile\extra{List of Figures}
{\catcode`\@=11			
\baselineskip=12pt plus2pt minus1pt\parskip=3pt plus2pt minus1pt
\testinput{\jobname.fgs}\vfill\supereject\catcode`\@=12}}
%
%
%
%
\ifx\answ\bigans
\else				

\fi
\def\writefigs{\def\writefigref##1##2{%
\iffigclosed\immediate\openout\figfile=\jobname.fgs\global\figclosedfalse\fi%
\write\figfile{\noexpand \dottedtocline{0}{0pt}{50pt}
{\noexpand\numberline{Fig.\ \xfig##1}{##2}}
{\ourfolio}}%
}}
\def\writefigref#1#2{}
%
%
\def\figPS{fig.~\chapsym\the\figno\nfigPS}
\def\nfigPS#1#2#3{\xdef#1{fig.~\chapsym\the\figno}%
\freference{#1}%
\writedef{#1\leftbracket fig.\noexpand~\chapsym\the\figno}%
\mytopinsert{\@PSfig{#2}{#3}\writefigref{#1}{#3}}%
\global\advance\figno by1%
}%
%
%
\def\figPSmid{fig.~\chapsym\the\figno\nfigPSmid}
\def\nfigPSmid#1#2#3{\xdef#1{fig.~\chapsym\the\figno}%
\freference{#1}%
\writedef{#1\leftbracket fig.\noexpand~\chapsym\the\figno}%
\bigbreak\@PSfig{#2}{#3}\writefigref{#1}{#3}\bigskip%
\global\advance\figno by1%
}%
%
%
\def\figuresPS{\extra{Figures}\global\figno=1%
\def\chapsym{}%
\def\figPS##1##2##3{
\xdef##1{fig.~\chapsym\the\figno}%
\freference{##1}%
\writedef{##1\leftbracket fig.\noexpand~\chapsym\the\figno}%
\ \vfill\@PSfig{##2}{##3}\writefigref{##1}{##3}%
\vskip1in\eject%
\global\advance\figno by1%
}}
%
%
\def\@PSfig#1#2{
\centerline{\epsffile{#1}}
\medskip
\centerline{\hfill
\hbox {\bf Figure \chapsym\the\figno:}
\vtop{\advance\hsize by \shrinkdim\baselineskip=14pt\noindent #2}
\hfill}}
%
%
%
\def\xfig{\expandafter\xf@g}\def\xf@g fig.\penalty\@M\ {}
\def\figs#1{figs.~\f@gs #1{\hbox{}}}
\def\f@gs#1{\edef\next{#1}\ifx\next\em@rk\def\next{}\else
\ifx\next#1\xfig #1\else#1\fi\let\next=\f@gs\fi\next}
%
\def\tbl{table~\chapsym\the\tblno\ntbl}
\def\ntbl#1#2#3{\xdef#1{table~\chapsym\the\tblno}%
\writedef{#1\leftbracket table\noexpand~\chapsym\the\tblno}%
\mytopinsert{\centerline{#2}
\medskip
\centerline{\hfill
\hbox {\bf Table \chapsym\the\tblno:}
\vtop{\advance\hsize by \shrinkdim\baselineskip=14pt\noindent #3}
\hfill}}\global\advance\tblno by1%
}%
%
%
\newwrite\lfile
{\escapechar-1\xdef\pctsign{\string\%}\xdef\leftbracket{\string\{}
\xdef\rightbracket{\string\}}\xdef\numbersign{\string\#}}

\def\writestop{\def\writestoppt{\immediate\write\lfile{\string\pageno%
\the\pageno\string\startrefs\leftbracket\the\refno\rightbracket%
\string\def\string\secsym\leftbracket\secsym\rightbracket%
\string\secno\the\secno\string\meqno\the\meqno}\immediate\closeout\lfile}}
\def\writestoppt{}\def\writedef#1{}
%
\catcode`\@=12 
%
%
\font\chapterfont=cmbx10 scaled \magstep2
\font\extrafont=cmbx10 scaled \magstep2
\font\thesistitlefont=cmr10 scaled \magstep3
%
\ifx\answ\bigans
\font\titlerm=cmr10 scaled\magstep3 \font\titlerms=cmr7 scaled\magstep3
\font\titlermss=cmr5 scaled\magstep3 \font\titlei=cmmi10 scaled\magstep3
\font\titleis=cmmi7 scaled\magstep3 \font\titleiss=cmmi5 scaled\magstep3
\font\titlesy=cmsy10 scaled\magstep3 \font\titlesys=cmsy7 scaled\magstep3
\font\titlesyss=cmsy5 scaled\magstep3 \font\titleit=cmti10 scaled\magstep3
\else
\font\titlerm=cmr10 scaled\magstep4 \font\titlerms=cmr7 scaled\magstep4
\font\titlermss=cmr5 scaled\magstep4 \font\titlei=cmmi10 scaled\magstep4
\font\titleis=cmmi7 scaled\magstep4 \font\titleiss=cmmi5 scaled\magstep4
\font\titlesy=cmsy10 scaled\magstep4 \font\titlesys=cmsy7 scaled\magstep4
\font\titlesyss=cmsy5 scaled\magstep4 \font\titleit=cmti10 scaled\magstep4
\font\absrm=cmr10 scaled\magstep1 \font\absrms=cmr7 scaled\magstep1
\font\absrmss=cmr5 scaled\magstep1 \font\absi=cmmi10 scaled\magstep1
\font\absis=cmmi7 scaled\magstep1 \font\absiss=cmmi5 scaled\magstep1
\font\abssy=cmsy10 scaled\magstep1 \font\abssys=cmsy7 scaled\magstep1
\font\abssyss=cmsy5 scaled\magstep1 \font\absbf=cmbx10 scaled\magstep1
\skewchar\absi='177 \skewchar\absis='177 \skewchar\absiss='177
\skewchar\abssy='60 \skewchar\abssys='60 \skewchar\abssyss='60
\fi
\skewchar\titlei='177 \skewchar\titleis='177 \skewchar\titleiss='177
\skewchar\titlesy='60 \skewchar\titlesys='60 \skewchar\titlesyss='60
\def\titlefont{\def\rm{\fam0\titlerm}
\textfont0=\titlerm \scriptfont0=\titlerms \scriptscriptfont0=\titlermss
\textfont1=\titlei \scriptfont1=\titleis \scriptscriptfont1=\titleiss
\textfont2=\titlesy \scriptfont2=\titlesys \scriptscriptfont2=\titlesyss
\textfont\itfam=\titleit \def\it{\fam\itfam\titleit} \rm}
\ifx\answ\bigans\def\abstractfont{\tenpoint}\else
\def\abstractfont{\def\rm{\fam0\absrm}
\textfont0=\absrm \scriptfont0=\absrms \scriptscriptfont0=\absrmss
\textfont1=\absi \scriptfont1=\absis \scriptscriptfont1=\absiss
\textfont2=\abssy \scriptfont2=\abssys \scriptscriptfont2=\abssyss
\textfont\itfam=\bigit \def\it{\fam\itfam\bigit}
\textfont\bffam=\absbf \def\bf{\fam\bffam\absbf} \rm} \fi
\def\tenpoint{\def\rm{\fam0\tenrm}
\textfont0=\tenrm \scriptfont0=\sevenrm \scriptscriptfont0=\fiverm
\textfont1=\teni  \scriptfont1=\seveni  \scriptscriptfont1=\fivei
\textfont2=\tensy \scriptfont2=\sevensy \scriptscriptfont2=\fivesy
\textfont\itfam=\tenit \def\it{\fam\itfam\tenit}
\textfont\bffam=\tenbf \def\bf{\fam\bffam\tenbf} \rm}
%
%
\def\inv{^{\raise.15ex\hbox{${\scriptscriptstyle -}$}\kern-.05em 1}}

\def\Dsl{\,\raise.15ex\hbox{/}\mkern-13.5mu D} 
\def\dsl{\raise.15ex\hbox{/}\kern-.57em\partial}
\def\del{\partial}

\font\bigit=cmti10 scaled \magstep1
\def\lspace{\ifx\answ\bigans{}\else\qquad\fi}
\def\lbspace{\ifx\answ\bigans{}\else\hskip-.2in\fi} 
\def\boxeqn#1{\vcenter{\vbox{\hrule\hbox{\vrule\kern3pt\vbox{\kern3pt
	\hbox{${\displaystyle #1}$}\kern3pt}\kern3pt\vrule}\hrule}}}
\def\mbox#1#2{\vcenter{\hrule \hbox{\vrule height#2in
		\kern#1in \vrule} \hrule}}  

\def\darr#1{\raise1.5ex\hbox{$\leftrightarrow$}\mkern-16.5mu #1}

\def\half{{\textstyle{1\over2}}} 
\def\roughly#1{\raise.3ex\hbox{$#1$\kern-.75em\lower1ex\hbox{$\sim$}}}

\def\Thesis#1#2#3{
%
%
\ifx\answ\bigans
\advance\hsize by -.5truein
\ifsided\else \onesided \fi
\fi
{
\nopagenumbers
\
\baselineskip=16pt
\vskip 0.5in
\centerline {\thesistitlefont #2}
\vskip 1.3in
\centerline {#1}
\vskip 1.5in
\centerline {A dissertation}
\centerline {presented to the faculty}
\centerline {of Princeton University}
\centerline {in candidacy for the degree}
\centerline {of Doctor of Philosophy}
\vskip 1.0in
\centerline {Recommended for acceptance}
\centerline {by the Department of Physics}
\vskip .25in
\centerline {#3}
\vfill\supereject
\footline={\hss\tenrm\folio\hss}
}
%
%
\ifx\answ\bigans
\baselineskip=18pt plus 1pt minus .5pt
\parskip=6pt plus 3pt minus 2pt
\tolerance=1000\hfuzz=.1pt\vfuzz=.1pt
\gdef\makefootline{\baselineskip=36pt \line{\the\footline}}
\fi
}
\newdimen\fixhoffset \fixhoffset=0truein
\newdimen\hstagger \hstagger=0truein
\newif\ifsided \sidedfalse
\def\centered{\ifx\answ\bigans \global\advance\hoffset by \fixhoffset
	\global\advance\hoffset by .25truein
	\global\fixhoffset=-.25truein
	\global\hstagger=0truein \fi
	\sidedtrue}
\def\offsetleft{\ifx\answ\bigans \global\advance\hoffset by \fixhoffset
	\global\advance\hoffset by .5truein
	\global\fixhoffset=-.5truein
	\global\hstagger=0truein \fi
	\sidedtrue}
\let\onesided=\offsetleft
\def\twosided{
\ifx\answ\bigans
\global\advance\hoffset by \fixhoffset
\global\advance\hoffset by .5truein
\global\fixhoffset=-.5truein
\global\hstagger=.5truein
\gdef\advancepageno{\ifnum\pageno<0 \global\advance\pageno by -1
  \else\global\advance\pageno by 1 \fi
  \ifodd\pageno \global\advance\hoffset by \hstagger
	\global\advance\fixhoffset by -\hstagger
  \else \global\advance\hoffset by -\hstagger
	\global\advance\fixhoffset by \hstagger \fi}
\gdef\startpreface{\ifodd\pageno\else
	\global\advance\hoffset by \hstagger
	\global\advance\fixhoffset by -\hstagger \fi
	\global\pageno=-1}
\gdef\startbody{\ifodd\pageno\else
	\global\advance\hoffset by \hstagger
	\global\advance\fixhoffset by -\hstagger \fi
	\global\pageno=1}
\fi
\sidedtrue}
\def\startpreface{\global\pageno=-1}
\def\startbody{\global\pageno=1}

\global\newcount\foo \global\foo=0
\def\Tchapter#1#2#3#4{
\xdef#1{#2}
\xdef\chapsym{#2.}\xdef\secsym{#2.}\xdef\eqnsym{#2.}
\global\secno=0\global\subsecno=0
\global\thmno=1\global\figno=1\global\meqno=0\global\tblno=1
\vfill\supereject
\message{(#2. #3 #4)}
\ \bigskip\centerline{\chapterfont\chapsym\
\vtop{\hbox{#3}\smallskip\hbox{#4}}}
\bigskip\bigskip
{\edef\writet@c{
    \noexpand\cwrite{\noexpand\noexpand\noexpand\dottedtocline{0}{0pt}{15pt}
	{\noexpand\noexpand\noexpand\numberline{#1.}{#3 #4}}
	{\noexpand\ourfolio}}
}\writet@c}
\freference{#1}
}

\def\Tnewchap#1#2#3{\global\advance\chapno by 1
                 \Tchapter#1{\the\chapno}{#2}{#3}
                 }

\eqnbychapter
\noblackbox
\writetoc		


\lref\fabjr{M. Fabbrichesi, R. Jengo and K. ROland, Nucl.
Phys. {\bf B402} (1993) 360.}

\lref\jxthesis{J. X. Lu, {\it Supersymmetric Extended
Objects}, Ph.D. Thesis, Texas A\&M University (1992),
UMI {\bf 53 08B}, Feb. 1993.}

\lref\coleman{S. Coleman, ``Classical Lumps and Their Quantum
Descendants'', in {\it New Phenomena in Subnuclear Physics},
ed A. Zichichi (Plenum, New York, 1976).}

\lref\jackiw{R. Jackiw, Rev. Mod. Phys. {\bf 49} (1977) 681.}

\lref\hink{M. B. Hindmarsh and T. W. B. Kibble, SUSX-TP-94-74,
IMPERIAL/TP/94-95/5, NI 94025, hepph/9411342.}

\lref\senmod{A. Sen, Mod. Phys. Lett. {\bf A8} (1993) 2023.}

\lref\tduality{K. Kikkawa and M. Yamasaki, Phys. Lett. {\bf B149}
(1984) 357; N. Sakai and I. Senda, Prog. Theor. Phys. {\bf 75}
(1986) 692; T. Busher, Phys. Lett. {\bf B159} (1985) 127;
V. Nair, A. Shapere, A. Strominger and F. Wilczek, Nucl. Phys.
{\bf B322} (1989) 167; M. J. Duff, Nucl. Phys. {\bf B335} (1990)
610; A. A. Tseytlin and C. Vafa, Nucl. Phys. {\bf B372} (1992)
443; A. A. Tseytlin, Class. Quantum Grav. {\bf 9} (19920 979,
A. Giveon, M. Porrati and E. Rabinovici, Phys. Rep. C (to appear).}

\lref\gibkal{G. W. Gibbons and R. Kallosh, NI-94003,
hepth/9407118.}

\lref\vafw{C. Vafa and E. Witten, HUTP-94-A017, hepth/9408074.}

\lref\seiw{N. Seiberg and E. Witten, Nucl. Phys. {\bf B426}
(1994) 19; RU-94-60, IASSNS-HEP 94/55, hepth/9408099.}

\lref\gauh{J. Gauntlett and J. H. Harvey, EFI-94-30,
hepth/9407111.}

\lref\frak{P. H. Frampton and T. W. Kephart, IFP-708-UNC,
VAND-TH-94-15.}

\lref\hawhr{S. W. Hawking, G. T. Horowitz and S. F. Ross,
NI-94-012, DAMTP/R 94-26, UCSBTH-94-25, gr-qc/9409013.}

\lref\ellmn{J. Ellis, N. E. Mavromatos and D. V. Nanopoulos,
Phys. Lett. {\bf B278} (1992) 246.}

\lref\kaln{S. Kalara and N. Nanopoulos, Phys. Lett. {\bf B267}
(1992) 343.}

\lref\glasgow{M. J. Duff, NI-94-016, CTP-TAMU-48/94,
hepth/9410210.}

\lref\fere{R. C. Ferrell and D. M. Eardley,
Phys. Rev. Lett. {\bf 59} (1987) 1617.}

\lref\reyt{S. J. Rey and T. R. Taylor, Phys. Rev. Lett. {\bf 71}
(1993) 1132.}

\lref\senzwione{A. Sen and B. Zwiebach, Nucl. Phys.
{\bf B414} (1994) 649.}

\lref\senzwitwo{A. Sen and B. Zwiebach, Nucl. Phys.
{\bf B423} (1994) 580.}

\lref\bddo{T. Banks, A. Dabholkar, M. R. Douglas and
M. O'Loughlin, Phys. Rev. {\bf D45} (1992) 3607.}

\lref\bos{T. Banks, M. O'Loughlin and A. Strominger,
Phys. Rev. {\bf D47} (1993) 4476.}

\lref\kir{E. Kiritsis, Nucl. Phys. {\bf B405} (1993) 109.}

\lref\dufr{M. J. Duff and J. Rahmfeld, CTP-TAMU-25/94,
hepth/9406105.}

\lref\duffkk{M. J. Duff, NI-94-015, CTP-TAMU-22/94,
hepth/9410046.}

\lref\bakone{I. Bakas, Nucl. Phys. {\bf B428} (1994) 374.}

\lref\baktwo{I. Bakas, CERN-TH.7472/94, hepth/9410104.}

\lref\back{C. Bachas and E. Kiritsis, Phys. Lett. {\bf B325}
(1994) 103.}

\lref\bko{E. Bergshoeff, R. Kallosh and T. Ortin,
UG-8/94, SU-ITP-94-19, QMW-PH-94-13, hepth/9410230.}

\lref\vilsh{A. Vilenkin and E. P. Shellard,
{\it Cosmic String and Other Topological Defects},
(Cambridge University Press, 1994).}

\lref\bfrm{M. Bianchi, F. Fucito, G. C. Rossi and M. Martellini,
hepth/9409037.}

\lref\hult{C. M. Hull and P. K. Townsend, QMW-94-30,
R/94/33, hepth/9410167.}

\lref\ght{G. W. Gibbons, G. T. Horowitz and P. K. Townsend,
R/94/28, UCSBTH-94-35, hepth/9410073.}

\lref\dufkmr{M. J. Duff, R. R. Khuri, R. Minasian and
J. Rahmfeld, Nucl. Phys. {\bf B418} (1994) 195.}

\lref\sentd{A. Sen, TIFR/TH/94-19, hepth/9408083.}

\lref\maha{J. Maharana, Newton Institute preprint, October 1994.}

\lref\gresss{M. B. Green and J. Schwarz, Phys. Lett. {\bf B136}
(1984) 367.}

\lref\sie{W. Siegel, Phys. Lett. {\bf B128} (1983) 397.}

\lref\dir {P. A. M. Dirac, Pro. R. Soc. {\bf A133} (1931) 60.}

\lref\tho {G. t'Hooft, Nucl. Phys. {\bf B79} (1974) 276.}

\lref\pol {A. M. Polyakov, Sov. Phys. JETP Lett. {\bf 20}
(1974) 194.}

\lref\mono {C. Montonen and D. Olive, Phys. Lett. {\bf B72}
(1977) 117.}

\lref\col {S. Coleman, Phys. Rev. {\bf D11} (1975) 2088.}

\lref\grohmr {D. J. Gross, J. A. Harvey, E. Martinec and
 R. Rohm,
Nucl. Phys. {\bf B256} (1985) 253.}

\lref\ginone {P. Ginsparg,
 {\it Conformal Field Theory}, Lectures given
at Trieste Summer School, Trieste, Italy, 1991.}

\lref\calhstwo {C. Callan, J. Harvey and A. Strominger,
 {\it Supersymmetric String Solitons}, Lectures given
at Trieste Summer School, Trieste, Italy, 1991.}

\lref\sch {J. H. Schwarz, {\it Supersymmetry and Its
 Applications}
ed G. W. Gibbons {\it et al} (Cambridge University Press,
1986).}

\lref\huglp {J. Hughes, J. Liu and J. Polchinski, Phys. Lett.
{\bf B180} (1986)
370.}

\lref\berst {E. Bergshoeff, E. Sezgin and P. K. Townsend,
 Phys. Lett. {\bf B189} (1987) 75.}

\lref\achetw {A. Achucarro, J. Evans, P. K.  Townsend and
D. Wiltshire,
 Phys.
Lett. {\bf B198} (1987) 441.}

\lref\belpst {A. A. Belavin, A. M. Polyakov, A. S. Schwartz and
Yu. S. Tyupkin,
Phys. Lett. {\bf B59} (1975) 85.}

\lref\oset {D. O'Se and D. H. Tchrakian, Lett. Math. Phys.
{\bf 13} (1987) 211.}

\lref\groks {B. Grossman, T. W. Kephart and J. D. Stasheff,
 Commun. Math. Phys. {\bf 96} (1984) 431;
Commun. Math. Phys. {\bf 100} (1985) 311.}

\lref\grokstwo {B. Grossman, T. W. Kephart and J. D. Stasheff,
 Phys. Lett. {\bf B220}
 (1989)  431.}

\lref\tch{D. H. Tchrakian, Phys. Lett. {\bf B150}  (1985)  360.}

\lref\fubn{S. Fubini and H. Nicolai,
 Phys. Lett. {\bf B155}  (1985)  369}

\lref\fain{D. B. Fairlie and J. Nuyts,
 J. Phys. {\bf A17}  (1984)  2867.}

\lref\str {A. Strominger,  Nucl. Phys. {\bf B343}
(1990) 167.}

\lref\duflhs {M. J. Duff and J. X. Lu,
Phys. Rev. Lett. {\bf 66} (1991) 1402.}

\lref\godo {P. Goddard and D. Olive, Rep. Prog. Phys. {\bf 41}
(1978) 1357.}

\lref\colone {S. Coleman, Proc. 1975 Int. School on Subnuclear
Physics,
Erice, ed A. Zichichi  (Plenum, New York, 1977); Proc. 1981
Int. School
on Subnuclear Physics, Erice, ed A. Zichichi (Plenum, New York,
1983).}

\lref\wuy {T. T. Wu and C. N. Yang, Nucl. Phys. {\bf B107}
(1976) 365.}

\lref\dufldl {M. J. Duff and J. X. Lu, Class. Quantum Grav.
{\bf 9} (1992) 1.}

\lref\tei {C. Teitelboim,
Phys. Lett. {\bf B167} (1986) 69.}

\lref\nep {R. I. Nepomechie,
Phys. Rev. {\bf D31} (1984) 1921.}

\lref\raj {R. Rajaraman, {\it Solitons and Instantons}
(North--Holland, Amsterdam, 1982).}

\lref\wito {E. Witten and D. Olive, Phys. Lett.
{\bf B78} (1978) 97.}

\lref\pras {M. K. Prasad and C. M. Sommerfield, Phys. Rev. Lett.
{\bf 35} (1975) 760.}

\lref\corg {E. Corrigan and P. Goddard, Commun. Math. Phys.
{\bf 80} (1981)
575.}

\lref\godno {P. Goddard, J. Nuyts and D. Olive, Nucl. Phys.
{\bf B125} (1977)
1.}

\lref\osb {H. Osborn, Phys. Lett. {\bf B83} (1979) 321.}

\lref\egugh {T. Eguchi, P. B. Gilkey and A. J. Hanson,
Phys. Rep. {\bf 66}
(1980) 213.}

\lref\corf {E. F. Corrigan and D. B. Fairlie, Phys. Lett.
{\bf B67} (1977) 69.
}

\lref\atidhm {M. F. Atiyah, V. G. Drinfeld, N. J. Hitchin and
Y. I. Manin,
Phys. Lett. {\bf A65} (1978) 185.}

\lref\dun {A. R. Dundarer,
 Mod. Phys. Lett. {\bf A5} (1991) 409.}

\lref\cha {A. H. Chamseddine, Phys. Rev. {\bf D24} (1981) 3065.}

\lref\berrwv {E. A. Bergshoeff, M. de Roo, B. de Wit and
 P. van Nieuwenhuizen,
Nucl. Phys. {\bf B195}  (1982)  97}

\lref\cham{G. F. Chapline and N. S. Manton, Phys. Lett.
{\bf B120} (1983) 105.}

\lref\gatn {S. J. Gates and H. Nishino, Phys. Lett. {\bf B173}
(1986) 52.}

\lref\sala{A. Salam and E. Sezgin, Physica Scripta {\bf 32}
(1985) 283.}

\lref\duf {M. J. Duff,  Class.
 Quantum  Grav. {\bf 5} (1988) 189.}

\lref\duflfb {M. J. Duff and J. X. Lu, Nucl. Phys. {\bf B354}
(1991) 141.}

\lref\dabghr {A. Dabholkar, G. W. Gibbons, J. A. Harvey and
F. Ruiz Ruiz,
 Nucl. Phys. {\bf B340} (1990) 33.}

\lref\berdps {E. Bergshoeff, M. J. Duff, C. N. Pope and
E. Sezgin,
 Phys. Lett.
{\bf B199} (1987) 69.}

\lref\tow {P. K. Townsend,
Phys. Lett. {\bf B202} (1988) 53.}

\lref\dufhis {M. J. Duff, P. S. Howe, T. Inami and K. Stelle,
 Phys. Lett.
{\bf B191} (1987) 70.}

\lref\dufs {M. J. Duff and K. Stelle, Phys. Lett. {\bf B253}
(1991) 113.}

\lref\berst {E. Bergshoeff, E. Sezgin and P. K. Townsend, Ann.
 Phys. {\bf 199}
(1990) 340.}

\lref\duflrsfd {M. J. Duff and J. X. Lu, Nucl. Phys. {\bf B354}
(1991) 129.}

\lref\gresone {M. Green and J. Schwarz, Phys. Lett. {\bf B151}
(1985) 21.}

\lref\bercgw {C. W. Bernard, N. H. Christ, A. H. Guth and
E. J. Weinberg,
 Phys. Rev.
 {\bf D16} (1977) 2967.}

\lref\hars {J. Harvey and A. Strominger,
 Phys. Rev. Lett. {\bf 66} (1991) 549.}

\lref\gres {M. Green and J. Schwarz, Phys. Lett. {\bf B149}
(1984) 117.}

\lref\elljm {J. Ellis, P. Jetzer and L. Mizrachi,
Nucl. Phys. {\bf B303} (1988)
1.}

\lref\dixds {J. Dixon, M. J. Duff and E. Sezgin, Phys. Lett.
{\bf B279} (1992) 265.}

\lref\berrs {E. Bergsheoff, M. Rakowski and E. Sezgin, Phys.
 Lett. {\bf
B185}  (1987)  371}

\lref\berd{E. Bergsheoff and M. de Roo, Nucl. Phys. {\bf B328}
(1989)
439}

\lref\dersw{M. de Roo, H. Suelmann and A. Wiedemann, preprint
UG--1/92  (1992).}

\lref\gresw {M. Green, J. Schwarz and E. Witten,
{\it Superstring
theory} (Cambridge University Press, 1987).}

\lref\duflloop {M. J. Duff and J. X. Lu, Nucl. Phys. {\bf B357}
  (1991)  534.}

\lref\ven {G. Veneziano,
Europhys. Lett. {\bf 2}  (1986)  199.}

\lref\cain {Y. Cai and C. A. Nunez, Nucl. Phys. {\bf B287}
(1987)  41}

\lref\gros{D. J. Gross and J. Sloan, Nucl. Phys. {\bf B291}
(1987)  41.}

\lref\ellm {J. Ellis and L. Mizrachi,
  Nucl. Phys. {\bf B327}  (1989)  595.}

\lref\calfmp {C. G. Callan, D. Friedan, E. J. Martinec and
M. J. Perry,
Nucl. Phys. {\bf B262}  (1985)  593.}

\lref\grestwo {M. Green and J. Schwarz, Phys. Lett. {\bf B173}
 (1986)  52.}

\lref\lin {U. Lindstrom, in Supermembranes and Physics in 2 + 1
Dimensions, ed. M. J. Duff, C. N. Pope and E. Sezgin
(World Scientific,
Singapore) (1990).}

\lref\dufone {M. J. Duff, Class. Quantum Grav. {\bf 6}  (1989)
1577.}

\lref\callny {C. Callan, C. Lovelace, C. Nappi and S. Yost,
Nucl. Phys.
{\bf B308}  (1988)  221.}

\lref\frat {E. Fradkin and A. Tseytlin, Phys. Lett. {\bf B158}
(1985)  316.}

\lref\duflselft {M. J. Duff and J. X. Lu, Phys. Lett.
{\bf B273}  (1991)  409.}

\lref\hors {G. Horowitz and A. Strominger, Nucl. Phys.
{\bf B360}  (1991) 197. }

\lref\witone {E. Witten,  Phys. Lett. {\bf B86}  (1979)
283.}

\lref\schone {J. Schwarz, Nucl. Phys. {\bf B226}  (1983)  269.}

\lref\zwa {D. Zwanziger, Phys. Rev. {\bf 176}  (1968)  1480,
1489.}

\lref\schwing {J. Schwinger, Phys. Rev. {\bf 144} (1966) 1087;
{\bf 173}
(1968) 1536.}

\lref\gibt{G.W. Gibbons and P.K. Townsend, Phys. Rev. Lett.
{\bf 71} (1993)
3754.}

\lref\duflblacks {M. J. Duff and J. X. Lu,
 Nucl. Phys. {\bf B416} (1994) 301.}

\lref\dufklsin {M. J. Duff, R. R. Khuri and J. X. Lu,
 Nucl. Phys. {\bf B377}
(1992) 281.}

\lref\calk {C.~G.~Callan and R.~ R.~Khuri,
Phys. Lett. {\bf B261} (1991) 363.}

\lref\gib {G. W. Gibbons, Nucl. Phys. {\bf B207} (1982) 337.}

\lref\gibm {G. W. Gibbons and K. Maeda, Nucl. Phys. {\bf B298}
(1988) 741.}

\lref\rey{S. J. Rey, in Proceedings of Tuscaloosa
Workshop on Particle Physics, (Tuscaloosa, Alabama, 1989).}

\lref\reyone {S. J.~Rey, Phys. Rev. {\bf D43} (1991) 526.}

\lref\antben {I.~Antoniadis, C.~Bachas, J.~Ellis and
 D.~V.~Nanopoulos,
Phys. Lett. {\bf B211} (1988) 393.}

\lref\antbenone {I.~Antoniadis, C.~Bachas, J.~Ellis and
D.~V.~Nanopoulos,
Nucl. Phys. {\bf B328} (1989) 117.}

\lref\mett {R.~R.~Metsaev and A.~A.~Tseytlin, Phys. Lett.
{\bf B191} (1987) 354.}

\lref\mettone {R.~R.~Metsaev and A.~A.~Tseytlin,
Nucl. Phys. {\bf B293} (1987) 385.}

\lref\calkp {C.~G.~Callan,
I.~R.~Klebanov and M.~J.~Perry, Nucl. Phys. {\bf B278} (1986)
78.}

\lref\lov {C.~Lovelace, Phys. Lett. {\bf B135} (1984) 75.}

\lref\friv {B.~E.~Fridling and A.~E.~M.~Van de Ven,
Nucl. Phys. {\bf B268} (1986) 719.}

\lref\gepw {D.~Gepner and E.~Witten, Nucl. Phys. {\bf B278}
(1986) 493.}

\lref\din {M.~Dine, Lectures delivered at
TASI 1988, Brown University (1988) 653.}

\lref\berdone {E.~A.~Bergshoeff and M.~de Roo, Phys. Lett.
{\bf B218} (1989)
210.}

\lref\calhs{C.~G.~Callan, J.~A.~Harvey and A.~Strominger,
Nucl. Phys.
{\bf B359} (1991) 611.}

\lref\calhsone{C.~G.~Callan, J.~A.~Harvey and A.~Strominger,
Nucl. Phys.
{\bf B367} (1991) 60.}

\lref\thoone{G.~'t~Hooft, Phys. Rev. Lett. {\bf 37} (1976) 8.}

\lref\wil{F.~Wilczek, in
{\it Quark confinement and field theory},
Eds. D.~Stump and D.~Weingarten, (John Wiley and Sons, New York,
1977).}

\lref\jacnr{R.~Jackiw, C.~Nohl and C.~Rebbi, Phys. Rev.
{\bf D15} (1977)
1642.}

\lref\khuinst{R.~R.~Khuri, Phys. Lett.
{\bf B259} (1991) 261.}

\lref\khumant{R.~R.~Khuri, Nucl. Phys.
 {\bf B376} (1992) 350.}

\lref\khumono{R.~R.~Khuri,
 Phys. Lett. {\bf B294} (1992) 325.}

\lref\khumonscat{R.~R.~Khuri,
Phys. Lett. {\bf B294} (1992) 331.}

\lref\khumonex{R.~R.~Khuri,
Nucl. Phys. {\bf B387} (1992) 315.}

\lref\khumonin{R.~R.~Khuri,
 Phys. Rev. {\bf D46} (1992) 4526.}

\lref\khugeo{R.~R.~Khuri,
Phys. Lett. {\bf 307} (1993) 302.}

\lref\khuscat{R.~R.~Khuri,
 Nucl. Phys. {\bf B403} (1993) 335.}

\lref\khuwind {R.~R.~Khuri, Phys. Rev. {\bf D48} (1993) 2823.}

\lref\gin{P.~Ginsparg, Lectures delivered at
Les Houches summer session, June 28--August 5, 1988.}

\lref\alljj{R. W. Allen, I. Jack and D. R. T. Jones,
 Z. Phys. {\bf C41}
(1988) 323.}

\lref\sev{A. Sevrin, W. Troost and A. van Proeyen,
Phys. Lett. {\bf B208} (1988) 447.}

\lref\schout{K. Schoutens, Nucl. Phys. {\bf B295} [FS21] (1988)
634.}

\lref\harl{J.~A.~Harvey and J.~Liu, Phys. Lett. {\bf B268}
(1991) 40.}

\lref\man{N.~S.~Manton, Nucl. Phys. {\bf B126} (1977) 525.}

\lref\manone{N.~S.~Manton, Phys. Lett. {\bf B110} (1982) 54.}

\lref\mantwo{N.~S.~Manton, Phys. Lett. {\bf B154} (1985) 397.}

\lref\atihone{M.~F.~Atiyah and N.~J.~Hitchin, Phys. Lett.
{\bf A107}
(1985) 21.}

\lref\atihtwo{M.~F.~Atiyah and N.~J.~Hitchin, {\it The Geometry
and
Dynamics of Magnetic Monopoles}, (Princeton University Press,
1988).}

\lref\polc{J.~Polchinski, Phys. Lett. {\bf B209} (1988) 252.}

\lref\gibhp{G.~W.~Gibbons and S.~W.~Hawking, Phys. Rev.
{\bf D15}
(1977) 2752.}

\lref\gibhpone{G.~W.~Gibbons, S.~W.~Hawking and M.~J.~Perry,
 Nucl. Phys.
{\bf B318} (1978) 141.}

\lref\brih{D.~Brill and G.~T.~Horowitz, Phys. Lett. {\bf B262}
(1991)
437.}

\lref\gids{S.~B.~Giddings and A.~Strominger, Nucl. Phys.
{\bf B306}
(1988) 890.}

\lref\gidsone{S.~B.~Giddings and A.~Strominger, Phys. Lett.
{\bf B230}
(1989) 46.}

\lref\canhsw{P.~Candelas, G.~T.~Horowitz, A.~Strominger and
E.~Witten,
Nucl. Phys. {\bf B258} (1984) 46.}

\lref\bog{E.~B.~Bogomolnyi, Sov. J. Nucl. Phys. {\bf 24} (1976)
449.}

\lref\war{R.~S.~Ward, Comm. Math. Phys. {\bf 79} (1981) 317.}

\lref\warone{R.~S.~Ward, Comm. Math. Phys. {\bf 80} (1981) 563.}

\lref\wartwo{R.~S.~Ward, Phys. Lett. {\bf B158} (1985) 424.}

\lref\grop{D.~J.~Gross and M.~J.~Perry, Nucl. Phys. {\bf B226}
(1983)
29.}

\lref\ash{{\it New Perspectives in Canonical Gravity}, ed.
A.~Ashtekar,
(Bibliopolis, 1988).}

\lref\lic{A.~Lichnerowicz, {\it Th\' eories Relativistes de la
Gravitation et de l'Electro-magnetisme}, (Masson, Paris 1955).}

\lref\gol{H.~Goldstein, {\it Classical Mechanics},
Addison-Wesley,
1981.}

\lref\ros{P.~Rossi, Physics Reports, 86(6) 317-362.}

\lref\dixdp{J.~A.~Dixon, M.~J.~Duff and J.~C.~Plefka,
 Phys. Rev.
Lett.
{\bf 69} (1992) 3009.}

\lref\chad{J.~M.~Charap and M.~J.~Duff, Phys. Lett. {\bf B69}
(1977) 445.}

\lref\dufkexst{M.~J.~Duff and R.~R.~Khuri,
Nucl. Phys. {\bf B411} (1994) 473.}

\lref\khubifb{R.~R.~Khuri,
Phys. Rev. {\bf D48} (1993) 2947.}

\lref\khustab{R.~R.~Khuri, Phys. Lett. {\bf B307} (1993) 298.}

\lref\sor{R.~D.~Sorkin, Phys. Rev. Lett. {\bf 51} (1983) 87.}

\lref\dabh{A.~Dabholkar and J.~A.~Harvey,
 Phys. Rev. Lett. {\bf 63} (1989) 478.}

\lref\fels{A.~G.~Felce and T.~M.~Samols, Phys. Lett.
{\bf B308} (1993) 30.}

\lref\dufipss{M.~J.~Duff, T.~Inami, C.~N.~Pope, E.~Sezgin and
K.~S.~Stelle,
Nucl. Phys. {\bf B297}
(1988) 515.}

\lref\fujku{K.~Fujikawa and J.~Kubo,
 Nucl. Phys. {\bf B356} (1991) 208.}

\lref\cvet{M.~Cveti\v c, Phys. Rev. Lett. {\bf 71} (1993) 815.}

\lref\cvegs{M.~Cveti\v c, S. Griffies and H. H. Soleng, Phys.
Rev. Lett. {\bf 71} (1993) 670; Phys. Rev. {\bf D48} (1993)
2613.}

\lref\la{H. S. La, Phys. Lett. {\bf B315} (1993) 51.}

\lref\gresvy{B.~R.~Greene, A.~Shapere, C.~Vafa and S.~T.~Yau,
 Nucl. Phys.
{\bf B337} (1990) 1.}

\lref\fonilq{A.~Font, L.~Ib\'a\~nez, D.~Lust and F.~Quevedo,
Phys. Lett.
{\bf B249} (1990) 35.}

\lref\bin{P.~Bin\'etruy, Phys. Lett. {\bf B315} (1993) 80.}

\lref\koun{C.~Kounnas, in {\it Proceedings of INFN Eloisatron
Project, 26th Workshop: ``From Superstrings to Supergravity",
 Erice, Italy,
Dec. 5-12, 1992}, Eds. M.~Duff, S.~Ferrara and R.~Khuri,
(World Scientific, 1994).}

\lref\duftv{M.J. Duff, P.K. Townsend and P. van Nieuwenhuizen,
Phys. Lett. {\bf B122} (1983) 232.}

\lref\dufgt{M.J. Duff, G.W. Gibbons and P.K. Townsend
Phys. Lett. {\bf B} (1994) }

\lref\guv{R. G\"uven, Phys. Lett. {\bf B276} (1992) 49.}

\lref\guven{R. G\"uven, Phys. Lett. {\bf B212} (1988) 277.}

\lref\dobm{P.~Dobiasch and D.~Maison, Gen. Rel. Grav.
{\bf 14} (1982) 231.}

\lref\chod{A.~Chodos and S.~Detweiler, Gen. Rel. Grav.
{\bf 14} (1982) 879.}

\lref\pol{D.~Pollard, J. Phys. {\bf A16} (1983) 565.}

\lref\duffkr{M.~J.~Duff, S.~Ferrara, R.~R.~Khuri and
J.~Rahmfeld,
 in preparation.}

\lref\lu{J. X. Lu,
 Phys. Lett. {\bf B313} (1993) 29.}

\lref\grel{R. Gregory and R. Laflamme, Phys. Rev. Lett.
{\bf 70} (1993) 2837.}

\lref\rom{L. Romans, Nucl. Phys. {\bf B276} (1986) 71.}

\lref\sala {A. Salam and E. Sezgin, {\it Supergravities in
Diverse Dimensions},
(North Holland/World Scientific, 1989).}

\lref\strath {J. Strathdee, Int. J. Mod. Phys. {\bf A2}
(1987) 273.}

\lref\dufliib{M.~J.~Duff and J.~X.~Lu,
 Nucl. Phys. {\bf B390} (1993) 276.}

\lref\nictv{H. Nicolai, P. K. Townsend and P. van Nieuwenhuizen,
 Lett. Nuovo
Cimento {\bf 30} (1981) 315.}

\lref\towspan{P. K. Townsend, in Proceedings of the 13th GIFT
Seminar
on Theoretical Physics: {\it Recent Problems in Mathematical
Physics} Salamanca, Spain, 15-27 June, 1992.}

\lref\dufm{M.~J.~Duff and R.~Minasian,
 CTP-TAMU-16/94, hepth/9406198.}

\lref\gropy{D. J. Gross, R. D. Pisarski and L. G. Yaffe,
Rev. Mod. Phys.
{\bf 53} (1981) 43.}

\lref\rohw{R. Rohm and E. Witten,
 Ann. Phys. {\bf 170} (1986) 454.}

\lref\banddf{T. Banks, M. Dine, H. Dijkstra and W. Fischler,
Phys. Lett. {\bf B212} (1988) 45.}

\lref\ferkp{S. Ferrara, C. Kounnas and M. Porrati, Phys. Lett.
{\bf B181} (1986) 263.}

\lref\ter{M. Terentev, Sov. J. Nucl. Phys. {\bf 49} (1989) 713.}

\lref\hass{S. F. Hassan and A. Sen, Nucl. Phys. {\bf B375}
(1992) 103.}

\lref\mahs{J. Maharana and J. Schwarz, Nucl. Phys. {\bf B390}
(1993) 3.}

\lref\senrev{A. Sen,
 Int. J. Mod. Phys. {\bf A9} (1994) 3707.}

\lref\senone{A.~Sen,
Nucl. Phys. {\bf B404} (1993) 109.}

\lref\sentwo{A.~Sen,
  Int. J. Mod. Phys. {\bf A8} (1993) 5079.}

\lref\schsen{J.~H.~Schwarz and A.~Sen,
Nucl. Phys. {\bf B411} (1994) 35.}

\lref\schsentwo{J.~H.~Schwarz and A.~Sen,
 Phys. Lett. {\bf B312} (1993) 105.}

\lref\schtwo{J.~Schwarz,
CALT-68-1815.}

\lref\senph{A.~Sen, Phys. Lett. {\bf B303} (1993) 22.}

\lref\schwarz{J.~H.~Schwarz, CALT-68-1879, hepth/9307121.}

\lref\dufldr{M. J. Duff and J. X. Lu, Nucl. Phys. {\bf B347}
(1990) 394.}

\lref\salstr{A. Salam and J. Strathdee, Phys. Lett. {\bf B61}
(1976) 375.}

\lref\jjj{J. Gauntlett, J. Harvey and J. T. Liu,
Nucl. Phys. {\bf B409} (1993) 363.}

\lref\dupo{M.~J.~Duff and C.~N.~Pope, Nucl. Phys {\bf B255}
(1985) 355.}

\lref\sgone{E.~Cremmer, S.~Ferrara, L.~Girardello and
 A.~Van Proeyen,
 Nucl. Phys. {\bf B212} (1983) 413.}

\lref\ghs{D.~Garfinkle, G.~T.~Horowitz and A.~Strominger,
Phys. Rev. {\bf D43}
(1991) 3140.}

\lref\hor{G.~T.~Horowitz, in Proceedings of Trieste '92,
{\it String theory and quantum gravity '92} p.55.}

\lref\gidps{S.~B.~Giddings, J.~Polchinski and A.~Strominger,
Phys. Rev. {\bf D48} (1993) 5784.}

\lref\shatw{A.~Shapere, S.~Trivedi and F.~Wilczek,
 Mod. Phys. Lett. {\bf A6}
(1991) 2677.}

\lref\klopv{R.~Kallosh, A.~Linde, T.~Ortin, A.~Peet and
A.~Van~Proeyen,
       Phys. Rev. {\bf D46} (1992) 5278.}

\lref\kal{R.~Kallosh, Phys. Lett. {\bf B282} (1992) 80.}

\lref\ko{R.~Kallosh and T.~Ortin, Phys. Rev. {\bf D48} (1993)
742.}

\lref\hw{C.~F.~E.~Holzhey and F.~Wilczek, Nucl. Phys.
{\bf B360} (1992) 447.}

\lref\dauria{R.~D'Auria, S.~Ferrara and M.~Villasante,
Class. Quant. Grav. {\bf 11} (1994) 481.}

\lref\gibp{G.~W.~Gibbons and M.~J.~Perry, Nucl. Phys.
{\bf B248} (1984) 629.}

\lref\salam{S. W. Hawking, Monthly Notices Roy. Astron. Soc.
 {\bf 152} (1971) 75; Abdus Salam in
   {\it Quantum Gravity: an Oxford Symposium} (Eds. Isham,
Penrose
and Sciama, O.U.P. 1975); G. 't Hooft, Nucl. Phys. {\bf B335}
(1990) 138.}

\lref\susskind{ L. Susskind, RU-93-44, hepth/9309145;
   J. G. Russo and L. Susskind, UTTG-9-94,
hepth/9405117.}

\lref\gibbons{ G. W. Gibbons, in {\it Supersymmetry,
Supergravity
and Related Topics}, Eds. F. del Aguila, J. A. Azcarraga and
L. E. Ibanez
(World Scientific, 1985).}

\lref\aichelburg{ P. Aichelburg and F. Embacher, Phys. Rev.
{\bf D37}
(1986) 3006.}

\lref\geroch{ R. Geroch, J. Math. Phys. {\bf 13} (1972) 394.}

\lref\hosoya{ A. Hosoya, K.
Ishikawa, Y. Ohkuwa and K. Yamagishi, Phys. Lett. {\bf B134}
(1984) 44.}

\lref\gibw{ G. W. Gibbons
and D. L. Wiltshire, Ann. of Phys. {\bf 167} (1986) 201.}

\lref\senprl{ A. Sen, Phys. Rev. Lett. {\bf 69}
(1992) 1006.}

\lref\schild{ G. C. Debney, R. P. Kerr and
   A. Schild, J. Math. Phys. {\bf 10} (1969) 1842.}

\lref\hort{G. T. Horowitz and A. A. Tseytlin, Phys. Rev.
{\bf D50} (1994) 5204.}

\lref\hortsey{G. T. Horowitz and A. A. Tseytlin,
Imperial/TP/93-94/51, UCSBTH-94-24, hepth/9408040;
Imperial/TP/93-94/54, UCSBTH-94-31, hepth/9409021.}

\lref\cvey{M. Cveti\v c and D. Youm, UPR-623-T, hepth/9409119.}

\lref\tseytlin{A. A. Tseytlin, Imperial-TP-93-94-46,
hepth/9407099.}

\lref\klim{C. Klimcik and A. A. Tseytlin, Nucl. Phys.
{\bf B424} (1994) 71.}

%
%

\Title{\vbox{\baselineskip12pt\hbox{NI-94-017}
\hbox{CTP/TAMU-67/92}
\hbox{McGill/94--53}
\hbox{CERN-TH.7542/94}
\hbox{hep-th/9412184}
}}
{String Solitons}
\centerline{M.~J.~Duff$^{1,2}$\footnote{$^\dagger$}
{Work supported in part by NSF grant PHY-9411543.},
Ramzi R.~ Khuri$^{3,4}$\footnote{$^*$}
{Work supported in part by a World Laboratory Fellowship.} and
J.~X.~Lu$^2$\footnote{$^{**}$}{Current address:
Department of Nuclear Engineering, Texas A\&M University,
College Station, TX 77843 USA}}
\bigskip\centerline{$^1${\it Isaac Newton Institute,
Cambridge University}}
\centerline{\it 20 Clarkson Road, Cambridge CB3 OEH U.K.}
\bigskip\centerline{$^2${\it Center for Theoretical Physics, Texas A\&M
University}}
\centerline{\it College Station, TX 77843 USA}
\bigskip\centerline{$^3${\it Physics Department,
McGill University}}
\centerline{\it 3600 University, St. Montr\'eal, PQ, H3A 2T8
Canada}
\bigskip\centerline{$^4${\it Theory Division, CERN}}
\centerline{\it CH-1211, Geneva 23, Switzerland}

\Date{12/94}

\pageno=-1
\extra{{\bf Abstract}}
{\baselineskip=16truept

We review the status of solitons in superstring theory, with a
view to
understanding the strong coupling regime. These {\it solitonic}
solutions are
non-singular field configurations which solve the empty-space
low-energy field
equations (generalized, whenever possible, to all orders in
$\alpha'$), carry a
non-vanishing topological ``magnetic" charge and are stabilized
by a topological
conservation law.  They are compared and contrasted with the
{\it elementary}
solutions which are singular solutions of the field equations
with a
$\sigma$-model source term and carry a non-vanishing Noether
``electric" charge.
In both cases, the solutions of most interest are those
which preserve half the
spacetime supersymmetries and saturate a Bogomol'nyi bound.
They typically arise
as the extreme mass=charge limit of more general two-parameter
solutions with
event horizons. We also describe the theory {\it dual} to the
fundamental string
for which the roles of elementary and soliton solutions are
interchanged.  In
ten spacetime dimensions, this dual theory is a superfivebrane
and this gives
rise to a string/fivebrane duality conjecture according to
which the fivebrane
may be regarded as fundamental in its own right, with the
strongly coupled
string corresponding to the weakly coupled fivebrane and
vice-versa.  After
compactification to four spacetime dimensions, the fivebrane
appears as a magnetic
monopole or a dual string according as it  wraps around five or
four of the
compactified dimensions. This gives rise to a four-dimensional
string/string
duality conjecture which subsumes a Montonen-Olive type duality
in that the
magnetic monopoles of the fundamental string correspond to the
electric winding
states of the dual string.
This leads to a {\it duality of dualities} whereby
under string/string duality
the strong/weak coupling $S$-duality trades places with the
minimum/maximum
length $T$-duality.  Since these magnetic monopoles are
extreme black holes,
a prediction of $S$-duality is that the
corresponding electric massive states of the fundamental string
are also extreme black holes. This is indeed the case.

\vfil\eject
}

\listtoc
\vfill\supereject
\pageno=1
\chapter{Introduction}

\newsec{Preliminaries}
{\it Solitons} \refs{\coleman,\jackiw,\raj}, sometimes called
{\it topological defects}
\vilsh,
 are important in
quantum field theory for a variety of reasons. Their existence
 means that the
full non-perturbative theory may have a much richer structure
than is apparent
in perturbation theory. For example, the electrically charged
{\it elementary}
particle spectrum may need to be augmented by magnetically
charged
{\it solitonic} particles. The former carry a Noether charge
 following from
the equations of motion while the latter carry a topological
charge associated
with the Bianchi identities. That these magnetic monopoles are
intrinsically
 non-perturbative is apparent from their mass formula which
depends inversely
on the coupling constant. The spectrum might also contain
{\it dyons},
particles carrying both electric and magnetic charge. In four
spacetime
dimensions, extended solitonic objects, strings and domain walls,
 are also
possible. In modern parlance, they might be known as
{\it ($d-1$)-branes} with
$d=1,2,3$
 where $d$ is the dimension of the worldvolume swept out by the
soliton.
In this report we shall use the word {\it solitons} to mean any
such
non-singular lumps of field energy which solve the field
equations, which have
finite mass per unit ($d-1$)-volume and which are prevented
from dissipating by
some topological conservation law. Their existence, which in
some grand
unified theories is actually mandatory, has far-reaching
implications both at
the microscopic and cosmic scales \hink.

However, it is now commonly believed that ordinary quantum
field theory is
inadequate for a unification of all the forces including
gravity and that it
must be supplanted by superstring theory. If this is true then
soliton
solutions of string theory must be even more important.
 Although the revival
of the string idea is now ten years old, it is only recently
that much
attention has been devoted to the subject of solitons. This
interest has in
 part been brought about by the realization that the really
crucial questions
 of string theory:
``How does the string choose a vacuum state?'';
``How does
the string break supersymmetry?''; ``How does the string cope
with the
cosmological constant problem?'' cannot be answered within the
framework of a
weak coupling perturbation expansion. Consequently, despite the
enormous
 number of reviews on superstring theory, comparatively little
has been
 written on bringing together this new wealth of information on
its
non-perturbative sector and trying to
make sense of it. The purpose of this Report is to do just that.

The current interest in string solitons has another, ostensibly
different,
origin. While all the activity in superstring theory was going
on a small but
dedicated group of theorists was asking a seemingly very
different question:
if we are going to replace $0$-dimensional point particles by
$1$-dimensional
strings, why not $2$-dimensional membranes or in general
$(d-1)$-dimensional
objects or ``$(d-1)$-branes''? Since superstrings can have
$D\leq 10$ spacetime
dimensions, there is plenty of room for higher dimensional
 extended objects
 with $d<D$. Progress in super ($d-1$)-branes was hampered by
the belief that
$\kappa$-symmetry, so crucial to Green-Schwarz superparticles
($d=1$) \sie\
and superstrings ($d=2$) \gresss, could not be generalized to
$d > 2$.
The breakthrough came when Hughes et al. \huglp\ showed that
$\kappa$-symmetry
could, in fact, be generalized and proceeded to construct a
threebrane
displaying an explicit $D=6, N=1$ spacetime supersymmetry and
$\kappa$-invariance on the worldvolume. Moreover, the
motivation for this
paper
was precisely to find the superthreebrane as a
{\it topological defect} of a
supersymmetric field theory in $D=6$. The discovery of the
other supermembranes
 proceeded in the opposite direction. First of all, Bergshoeff
et al.
\berst\ found corresponding Green-Schwarz actions for other
 values of $d$ and
$D$, in particular, the eleven-dimensional supermembrane. We
shall discuss
this theory in section 4 and show how to derive from it the
Type IIA string in
ten dimensions by a simultaneous dimensional reduction of the
worldvolume and
the spacetime. Their method was to show such Green-Schwarz
super $p$-brane actions
are possible whenever there is a closed $(p+2)$-form in
superspace. As
described in section 1.2, the four ``fundamental''
super $p$-branes are then given
 by $p=2$ in $D=11$, $p=5$ in $D=10$, $p=3$ in $D=6$ and $p=2$
in $D=4$ \achetw.
 Applying the above mentioned simultaneous reduction $k$
times, we find four
sequences of $(p-k)$-branes in $(D-k)$ dimensions, which
include the well
known Green-Schwarz superstrings in $D=10,6,4$ and $3$.
These four sequences, known as the octonionic (${\cal O}$),
quaternionic (${\cal H}$), complex (${\cal C}$) and real
(${\cal R}$)
 sequences, make up the brane-scan of section 1.2.

Of particular interest was the $D=10$ fivebrane, whose
Wess-Zumino term
coupled to a rank six antisymmetric tensor potential
$A_{MNPQRS}$ just as the
Wess-Zumino term of the string coupled to a rank two potential
$B_{MN}$.
Spacetime supersymmetry therefore demanded that the fivebrane
coupled to
the
$7$-form field strength formulation of $D=10$ supergravity
\cha\ just as the
string coupled to the $3$-form version \refs{\berrwv,\cham}.
These dual
formulations of $D=10$ supergravity have long been something of
an enigma from
the point of view of superstrings. As field theories, each
 seems equally valid.
 In particular, provided we couple them to $E_8 \times E_8$ or
$SO(32)$
super-Yang-Mills, then both are anomaly free
\refs{\gres,\sala,\gatn}.
Since the $3$-form version
corresponds to the field theory limit
of the heterotic string, Duff conjectured \duf\ that there
ought to exist a
{\it heterotic fivebrane} which could be viewed as a
 fundamental anomaly-free
theory in its own right and whose field theory limit
corresponds to the dual $7$-form version. We shall refer to
this as the
{\it string/fivebrane duality conjecture}. One of the purposes
of this Report
will be to summarize the evidence in its favor. At this stage,
however, the
solitonic element had not yet been introduced.

The next development came when Townsend \tow\ pointed out that
not merely the
$D=6$ threebrane but all the points on the ${\cal H,C,R}$
sequences correspond
to topological defects of some globally supersymmetric field
theory which
break half the spacetime supersymmetries. This partial breaking
of
supersymmetry, already discussed in \huglp, is a key idea and
is intimately
connected with the worldvolume $\kappa$-symmetry which allows
one to gauge
away half of the fermionic degrees of freedom. He conjectured
that the
$p$-branes in the ${\cal O}$ sequence would also admit such a
solitonic
interpretation within the context of supergravity. Another
purpose of this
Report will be to examine to what extent this conjecture is
true.

The first hint in this direction came from  Dabholkar
{\it et al.} \dabghr,
who presented a multi-string solution which in $D=10$ indeed
 breaks half the
supersymmetries. They obtained the solution by solving the
low-energy $3$-form supergravity equations of motion coupled to
a string
$\sigma$-model source and demonstrated that it saturated a
Bogomol'nyi bound
and satisfied an associated zero-force condition, these
properties being
intimately connected with the existence of unbroken spacetime
supersymmetry.
The authors of \dabghr\ went on to interpret these solutions as
macroscopic
fundamental string states and presented evidence in favour of
this conjecture.
Since that time, further evidence has been obtained
(e.g. \refs{\calk,\khugeo,\khuscat}) to support this
identification.
In section 2.1 we rederive this string solution and point out
the existence of
the Bogomol'nyi bound  between the ADM mass per unit length and
charge per
 unit length and discuss the zero-force condition which arises
 from the
preservation of half the spacetime supersymmetries.

However, this $D=10$ string was clearly not the soliton
anticipated by
Townsend because it described a singular configuration with a
$\delta$-function source
at the string location. Moreover, its charge per unit length
 $e_2$ was an
``electric'' Noether charge associated with the equation of
 motion of the
antisymmetric tensor field rather than a ``magnetic''
topological charge
associated with the Bianchi identities. Consequently, in the
current
literature on the subject, this solution is now referred to as
the
``fundamental'' or ``elementary'' string. Similarly, the
supermembrane
solution of $D=11$ supergravity found in \dufs\ did not
seem to be
solitonic either
 because it was also obtained by coupling to a membrane
$\sigma$-model source.
\footnote{$^\dagger$}{Curiously, however, as discussed in
 section 3.7, the
curvature computed from its $\sigma$-model metric is finite at
the
location of the source, in contrast to
the case of the elementary string.}

The next major breakthrough for $p$-branes as solitons came
 with the paper of
Strominger \str, who showed that $D=10$ supergravity coupled to
 super
Yang-Mills (without a $\sigma$-model source), which is the
field theory
limit
of the heterotic string \grohmr, admits as a solution the
heterotic fivebrane.
In contrast to the elementary string, this fivebrane is a
genuine soliton,
being everywhere nonsingular and carrying a topological
magnetic charge $g_6$.
A crucial part of the construction was a Yang-Mills instanton
in the four
directions transverse to the fivebrane. He went on to suggest a
complete
strong/weak coupling duality with the strongly coupled string
corresponding
to the weakly coupled fivebrane and vice-versa, thus providing
a solitonic
interpretation of the string/fivebrane duality conjecture. In
this form,
string/fivebrane duality is in a certain sense an analog of the
Montonen-Olive
conjecture discussed in section 1.3, according to which the
magnetic monopole
states of four-dimensional spontaeously broken supersymmetric
Yang-Mills
theories may be viewed from a dual perspective as fundamental
in their own
right and in which the roles of the elementary and solitonic
states are
interchanged.

This strong/weak coupling was subsequently confirmed from the
point of view of
Poincar\'e duality in \duflrsfd. There it was shown that the
just as the
string loop expansion parameter is given by
${\rm g}_2=e^{\phi_0}$, where
$\phi_0$ is the dilaton vev, so the analogous fivebrane
parameter is given
by
${\rm g}_6=e^{-\phi_0/3}$ and hence that
\eqn\first{ {\rm g}_6={\rm g}_2^{-1/3}.}
The same paper also established a Dirac quantization rule
\eqn\second{\kappa^2 T_2T_6=n\pi, \qquad\qquad n={\rm integer}}
relating the fivebrane tension $T_6$ to the string tension
$T_2$, which
 followed from the corresponding rule for the electric and
 magnetic charges
generalized to extended objects
\refs{\nep,\tei}\ $e_2 g_6=2n\pi$.

For the purposes of generalizing the Dirac quantization rule
for extended objects, it is instructive to
write the Maxwell's equations in terms of the electromagnetic
tensor $F_{MN}$ as
\eqn\maxfone{\partial_{[M} F_{NP]}= 0, }
\eqn\maxftwo{\partial_{[M} {^\ast}F_{NP]} =  {^\ast}J_{MNP}, }
where
\eqn\dualfj{^\ast F^{MN} = {1\over 2} \varepsilon^{MNOP} F_{OP},
 \quad  ^\ast J^{MNO}=
\varepsilon^{MNOP} J_P ,}
and
\eqn\jay{J^M = (\rho, {\bf j}), \quad F^{0m} = E^m,
\quad F^{mn} =
\epsilon^{mnl} B_l ,}
and in the absence of the monopole, the field strength
 $F_{MN}$ can be written
in terms of the four-vector potential $A^M = (\phi, {\bf A})$
as
\eqn\fda{F_{MN} \equiv 2\partial_{[M} A_{N]} = \partial_M A_N -
\partial_N A_M.}
The asymmetry between the equation for $F$ and that for
$^\ast F$ corresponds physically to the statement that there
 are no magnetic
monopoles. If we want to restore the symmetry of the
source-free Maxwell's
equations by introducing magnetic monopoles then we must
 replace \fda\ by
\eqn\fdaom{F_{MN} = \partial_M A_N - \partial_N A_M +
\omega_{MN} ,}
so that
\eqn\dfx{\partial_{[M} F_{NP]} =  X_{MNP} ,}
with
\eqn\xdom{X_{MNP} = \partial_{[M} \omega_{NP]} ,}
and hence
\eqn\dxzero{\partial_{[M} X_{NPQ]} \equiv 0.}
The monopole may be of the Dirac type where $X$ is singular
\eqn\xdelt{X_{123} = g \delta^3 (y),}
 in analogy with the elementary electric
charge
\eqn\jdelt{^\ast J_{123} = e \delta^3 (y),}
 or the source may be smeared out so as to
be regular at the origin as in the 't Hooft-Polyakov monopole.
 In both cases, we have
\eqn\eintj{e \equiv \int \limits_{S^2} {^\ast F} = \int
\limits_{M^3} {^\ast J},
}
\eqn\gintx{g \equiv \int \limits_{S^2} F = \int \limits_{M^3} X.
 }
In the language of \maxftwo, \fdaom\ and \dfx, the electric
charge
is conserved by virtue of the field equations and hence
corresponds to a
Noether charge, whereas the magnetic charge is identically
conserved and
corresponds to a topological charge.

Just as a charged particle couples to an Abelian vector
potential $A_M$
displays a gauge invariance
\eqn\abgi{A_M \rightarrow A_M + \partial_M \Lambda}
and has a gauge invariant field strength
\eqn\abf{F_{MN} = 2\partial_{[M} A_{N]} \equiv \partial_M A_N
- \partial_N A_M
,}
a string couples to a rank-2 antisymmetric tensor potential
$A_{MN} = - A_{NM}$
with a gauge invariance
\eqn\antitwo{A_{MN} \rightarrow A_{MN} + \partial_{[M}
\Lambda_{N]} ,}
and field strength
\eqn\fmnp{F_{MNP} = 3\partial_{[M} A_{NP]}.}
In general, a $(d-1)$-brane couples to a $d$-form $A_{M_1 M_2
\cdots M_d}$ with
\eqn\antid{A_{M_1 M_2 \cdots M_d} \rightarrow A_{M_1 M_2
 \cdots M_d} +
\partial_{[ M_1} \Lambda_{M_2 \cdots M_d ]},}
and
\eqn\fsd{F_{M_1 M_2 \cdots M_{d+1}} = (d + 1) \partial_{[ M_1}
 A_{M_2
\cdots M_{d+1} ]}.}
In the language of differential forms we may write for
 arbitrary $d$ and $D$
\eqn\adgi{A_d \rightarrow A_d + d \Lambda_{d-1},}
and
\eqn\adf{F_{d+1} = dA_d ,}
from which the Bianchi identity
\eqn\dbianchi{dF_{d+1} \equiv 0 }
follows immediately. In the absence of other interactions, the
equation of
motion for the $d$-form potential is
\eqn\eomdf{d {^\ast} F_{D-d-1} = {^\ast}J_{D-d},}
where the source $J$ is a $d$-form. Here we have introduced the
Hodge dual
operation $^\ast$ which converts a $d$-form into a $(D-d)$-form,
 e.g.
\eqn\hodge{(^\ast J)^{M_1 M_2 \cdots M_{D-d}}
\equiv {1\over d!}
\varepsilon^{M_1 M_2 \cdots M_D} J_{M_{D-d+1} \cdots M_D},}
where $\varepsilon^{M_1 \cdots M_D}$ is the $D$-dimensional
 alternating symbol
with $\varepsilon^{01 \cdots D-1} = 1$.

Just as the usual Maxwell's equations, \adf, \dbianchi\ and
\eomdf\
imply the presence of an ``electric'' charge, i.e. a
 $(d-1)$-brane, but no
``magnetic'' charge, i.e. no $(D-d-3)$-brane. To restore the
duality symmetry by
introducing a $(D-d-3)$-brane we must modify \adf\ to
\eqn\adfom{F_{d+1} = dA_d + \omega_{d+1},}
so that the Bianchi identity \dbianchi\ becomes
\eqn\dbiantwo{dF_{d+1} = X_{d+2},}
with
\eqn\xdomtwo{X_{d+2} = d\omega_{d+1}.}
Once again $X$ may be singular
\eqn\xdeltd{X_{123 \cdots d+2} = g_{D-d-2} \delta^{d+2} (y),}
or may be smeared out so as
to be regular at the origin. We then have
\eqn\edintj{e_d = \int \limits_{S^{D-d-1}} {^\ast}F_{D-d-1} =
\int \limits_{M^{D-d}} {^\ast} J_{D-d},}
\eqn\gdintx{g_{D-d-2} = \int \limits_{S^{d+1}} F_{d+1} =
\int \limits_{M^{d+2}}
X_{d+2}.}
The Dirac quantization condition is again obtained by using the
generalization
of either the Dirac string \tei\ or Wu-Yang construction
\nep\ as
\eqn\diquand{{e_d g_{D-d-2} \over 4\pi} ={1\over 2}(n =
{\rm integer}).}
Note that, $e_d$ and $g_{D- d -2}$ are not in general
dimensionless but rather
\eqn\dimeg{[e_d] = - {1\over 2} (D-2d -2),
\qquad [g_{D-2d -2}] = {1\over 2}
(D - 2d -2).}
They do become dimensionless when
\eqn\nodimeg{D= 2 (d + 1),}
of which the point particle ($d = 1$) in $D = 4$ is the most
familiar special
case.

In keeping with the viewpoint that the fivebrane may be
regarded as
fundamental in its own right, Duff and Lu \duflfb\ then
constructed the
{\it elementary fivebrane} solution by coupling the $7$-form
version of
supergravity to a fivebrane $\sigma$-model source in analogy
 with the
elementary string. This carries an electric charge $\tilde e_6$.
String/fivebrane duality then suggested that by coupling the
$7$-form version
of supergravity to super Yang-Mills (without a $\sigma$-model
source),
one ought to find a nonsingular heterotic string soliton
carrying a
topological magnetic charge $\tilde g_2$. Here one would expect
 an
 eight-dimensional Yang-Mills instanton in the eight directions
 transverse to
the string. This was indeed the case \duflhs, but scaling
arguments required
an unconventional Yang-Mills Lagrangian, quartic in the field
strengths,
which, however, is only to be expected in a fivebrane loop
expansion
\refs{\duflhs,\duflloop}.

Somewhat surprisingly, the elementary fivebrane, as pointed out
by
Callan, Harvey and Strominger \refs{\calhs,\calhsone}, could
 also be
regarded as a soliton when viewed from the dual perspective,
with
$g_6=\tilde e_6$. In other words, it provides a nonsingular
solution of the
source-free $3$-form equations even without the presence of
Yang-Mills fields.
By the same token, when viewed from the dual perspective,
the elementary
string provides a nonsingular solution of the source-free
$7$-form equations
 with $\tilde g_2=e_2$ \dufklsin.

\newsec{Supersymmetric extended objects as solitons:
the brane-scan}

As the $p$-brane moves through spacetime, its trajectory is
 described by the
functions $X^M (\xi)$ where $X^M$ are the spacetime
coordinates ($M = 0, 1,
\ldots, D - 1$) and $\xi^i$ are the worldvolume
coordinates ($i = 0, 1, \ldots,
d - 1$).  It is often convenient to make the so-called
 ``static gauge choice''
by making the $D = d + (D - d)$ split
\eqn\xysplit{X^M (\xi) = (X^{\mu} (\xi), Y^m (\xi)),}
where $\mu = 0, 1, \ldots, d - 1$~and~$m = d, \ldots, D - 1$,
and then setting
\eqn\xeta{X^{\mu} (\xi) = \xi^{\mu}.}
Thus the only physical worldvolume degrees of freedom are given
 by the $(D -
d)~Y^m (\xi)$.  So the number of on-shell bosonic degrees of
freedom is
\eqn\nonsh{N_B = D - d.}

To describe the super $p$-brane we augment the $D$ bosonic
coordinates $X^M
(\xi)$ with anticommuting fermionic coordinates
$\theta^{\alpha} (\xi)$.
Depending on $D$, this spinor could be Dirac, Weyl, Majorana or
Majorana-Weyl.
The fermionic $\kappa$-symmetry means that half of the spinor
degrees of
freedom are redundant and may be eliminated by a physical gauge
choice.  The
net result is that the theory exhibits a {\it $d$-dimensional
 worldvolume
supersymmetry} where the number of fermionic generators is
exactly half of the
generators in the original spacetime supersymmetry.  This
 partial breaking of
supersymmetry is a key idea.  Let $M$ be the number of real
components of the
minimal spinor and $N$ the number of supersymmetries in $D$
spacetime
dimensions and let $m$~and~$n$ be the corresponding quantities
in $d$
worldvolume dimensions.  Let us first consider $d > 2$.  Since
$\kappa$-symmetry always halves the number of fermionic degrees
of freedom and going
on-shell halves it again, the number of on-shell fermionic
degrees of freedom
is
\eqn\nonshf{N_F = {1\over 2}~mn = {1\over 4}~MN.}
Worldvolume supersymmetry demands $N_B = N_F$ and hence
\eqn\dmnone{D - d = {1\over 2}~mn = {1\over 4}~MN.}
A list of dimensions, number of real components of the minimal
spinor and
possible supersymmetries is given in Table 1, from which we see
that there are
only 8 solutions to \dmnone\ all with $N = 1$, as shown in
Fig. 1.  We note in
particular that $D_{{\rm max}} = 11$ since $M \geq 64$ for
$D \geq 12$ and
hence \dmnone\ cannot be satisfied.  Similarly
 $d_{{\rm max}} = 6$ since
$m \geq
16$ for $d \geq 7$.  The case $d = 2$ is special because of the
 ability to
treat left and right moving modes independently.  If we require
 the sum of both
left and right moving bosons and fermions to be equal, then we
again find the
condition \dmnone.  This provides a further 4 solutions all
with $N = 2$,
corresponding to Type II superstrings in $D = 3, 4, 6$~and~$10$
(or 8 solutions
in all if we treat Type IIA and Type IIB separately.  The
 gauge-fixed Type IIB
superstring will display (8, 8) supersymmetry on the worldsheet
 and the Type
IIA will display (16, 0), the opposite
 \refs{\calhs,\calhsone}\ of what one
might naively expect).
If we require only left (or right) matching, then \dmnone\ is
 replaced by
\eqn\dmntwo{D - 2 = n = {1\over 2}~MN,}
which allows another 4 solutions in $D = 3, 4, 6$~and~$10$,
all with $N = 1$.
The gauge-fixed theory will display (8,0) worldsheet
 supersymmetry.  The
heterotic string falls into this category.  The results are
shown in Fig. 1 \achetw.

An equivalent way to arrive at the above conclusions is to list
all scalar
supermultiplets in $d \geq 2$ dimensions and to interpret the
 dimension of the
target space, $D$, by
\eqn\dsca{D - d =~{\rm number~of~scalars}.}
A useful reference is \strath, which provides an exhaustive
classification
of all unitary representations of supersymmetry with maximum
spin 2.  In
particular, we can understand $d_{{\rm max}} = 6$ from this
 point of view since
this is the upper limit for scalar supermultiplets.
In summary, according to
the above classification, Type II $p$-branes do not exist for
$p > 1$. We
shall return to this issue, however, in section 4.
\bigskip

\halign{\indent #&\qquad\hfil# \hfil&\quad\hfil
#\hfil&\quad\hfil # \hfil &
\quad \hfil # \hfil &\quad \hfil # \hfil &\quad #\hfil\cr
&&&Dimension & Minimal Spinor& Supersymmetry&\cr
&&&($D$ or $d$) & ($M$ or $m$) & ($N$ or $n$)&\cr
&&&11 & 32 & 1&\cr
&&&10 & 16 & 2, 1&\cr
&&&9 & 16 & 2, 1&\cr
&&&8 & 16 &2, 1&\cr
&&&7 & 16 & 2, 1&\cr
&&&6 & 8 & 4, 3, 2, 1&\cr
&&&5 & 8 & 4, 3, 2, 1&\cr
&&&4 & 4 & 8, $\ldots$, 1&\cr
&&&3 & 2 & 16, $\ldots$, 1&\cr
&&&2 & 1 & 32, $\ldots$, 1&\cr}

\noindent
Table 1.  Minimal spinor components and supersymmetries.

\newsec{The Montonen-Olive conjecture}
Following Callan, Harvey and Strominger \calhstwo,
we briefly review some general
aspects of solitons.
Two important features of solitons are the following \raj:

1) Most solitons are non-perturbative, i.e. they are solutions
to
nonlinear field equations which cannot be found by perturbation
of the
linearized field equations. Another non-perturbative feature is
the fact
that the mass per unit $p$-volume is inversely proportional to
some power of a
dimensionless coupling constant, so that the weak-coupling
 perturbative
limit corresponds to the strong-coupling limit for the solitons.
 Since
the classical solutions are non-perturbative, the quantum
effects obtained
from soliton interactions are also non-perturbative, and vanish
to all orders in perturbation theory.

2) Most solitons are characterized by a conserved topological
index which
after quantization becomes a conserved quantum number. This in
contrast to
the conserved Noether charge associated with a continuous
symmetry of the
lagrangian.

In addition, soliton solutions typically depend on a finite
number of
parameters called moduli which act as coordinates on the moduli
space of
soliton solutions of fixed topological charge.

The simplest example of a soliton with these properties is
the ``kink''
solution in $1+1$ spacetime dimensions. The solution to the
Lagrangian
\eqn\kinklag{{\cal L} = - \half \partial_\mu  \phi \partial^\mu
\phi -
U(\phi),}
with potential $U(\phi)=\lambda (\phi^2 - m^2/\lambda)^2/4$
and
dimensionless coupling $g \equiv \lambda/m^2$ has conserved
topological charge
\eqn\qkinktwo{Q = { \sqrt{g} \over 2} (\phi(+\infty) -
\phi(-\infty)).}
$Q=\pm 1$ for a kink (anti-kink) in which $\phi$ varies from
the minimum
of $U$ at $\phi = \mp 1/\sqrt{g}$ at $x=-\infty$ to the minimum
 at
$\phi= \pm 1/\sqrt{g}$ at $x=+\infty$.
The energy (rest mass) of the kink is given by
\eqn\kinkenergy{E = \int dx \half (\phi')^2 + U(\phi) =
{2 \sqrt{2} \over 3}
{m \over g},}
so that the kink mass is proportional to $1/g$, and the
solution
is non-perturbative. It is straightforward to use the
collective coordinates
method \raj\ to separate out explicitly the dependence on the
zero modes
and to be left with a well defined perturbation theory for the
non-zero modes.

A supersymmetric version of the kink solution has Lagrangian
\eqn\lgkinkfer{{\cal L} = - \half (\partial_\mu \phi)^2 +
 \half \bar \psi i
\gamma^\mu\partial_\mu \psi - \half V^2(\phi) - \half V'(\phi)
\bar \psi \psi ,}
where $\psi$ is a Majorana fermion and
$V=\lambda(\phi^2-a^2)$. This theory has two chiral
supercharges given by
\eqn\qkinkf{Q_\pm = \int dx (\dot \phi \pm \phi')\psi_\pm \mp
V(\phi) \psi_\mp,}
where $\psi_\pm$ are the left- and right-handed components of
 $\psi$.
The corresponding supersymmetry algebra is given by \wito
\eqn\ssalg{Q_+^2 = P_+, \quad Q_-^2 = P_-,
\quad \{Q_+,Q_-\}= T,}
with $P_\pm = P_0 \pm P_1$, and
where the central term $T$ is esssentially the topological
 charge. The relation
\eqn\ppqq{P_+ + P_- = (Q_+ + Q_-)^2 -T = (Q_+ - Q_-)^2 +T}
implies a Bogomolnyi bound, $M \ge T/2$,  where $M$ is the rest
mass.
This bound is saturated precisely for those states
$|s \rangle$  for which
$(Q_+ \pm Q_-) |s \rangle= 0$, i.e. for states annihilated by
some combination
of the supersymmetry charges. In fact, demanding unbroken
supersymmetry is
equivalent to solving the ``square root'' of the equations of
motion, and the
fact that the kink solution is annihilated by one combination
of the
supercharges implies
saturation of the Bogomolnyi bound. The other combination of
supercharges does
not annihilate the kink state but instead produces a fermion
zero mode in the
kink background. This follows from the fact that a
supersymmetry variation of
a solution to the bosonic equations of motion if it is nonzero
produces a
solution to the fermionic equations of motion in the bosonic
background.

The property that half of the supercharges annihilate the
classical solution
leading to saturation of a Bogomolnyi bound, while the other
half acting on
the soliton produce fermion zero modes in the soliton
background is found in
most known examples of solitons in supersymmetric theories.
As a result,
searching for configurations which preserve
some of the supercharges provides a shortcut to solving the
full equations
of motion since the resulting equations are typically first
 order
as compared to the second order equations of motion. We will
find this to be
the case for the string soliton solutions discussed in this
Report.

We now turn to the magnetic monopole solution in Yang-Mills
 theory.
The action for a Yang-Mills Higgs theory with gauge group
$SU(2)$ and a
Higgs field $\Phi$ in the adjoint representation is given by
\eqn\ymhig{S = \int d^4 x \big( -{1\over 4}Tr F_{\mu \nu}
F^{\mu \nu} - \half
Tr D^\mu \Phi D_\mu \Phi - V(\Phi) \big),}
where $V$ has a minimum at $\langle Tr \Phi^2 \rangle = v^2$
which breaks $SU(2)$ down to $U(1)$.
For a static configuration the energy is given by
\eqn\staten{E = \int d^3 x Tr(B^i \pm D^i \Phi)^2
                 + \int d^3 x V(\Phi) + 4 \pi v |Q_M|,}
where $v$ is the asymptotic vacuum expectation value of $\Phi$.
The energy thus satisfies a Bogomolnyi bound
\eqn\bogbnd{E \ge  4 \pi v |Q_M|,}
where
\eqn\qmb{Q_M = \int {\bf B} \cdot d {\bf S}}
is the magnetic charge and {\bf B}
is the asymptotic value of the gauge invariant $U(1)$ field
strength.
The magnetic charge $Q_M$ is also related through the equations
of motion
to the element of $\pi_2(SU(2)/U(1))=Z$ which labels the
topological charge
carried by the Higgs field configuration. The Bogomolnyi bound
is saturated if
$V(\phi) \equiv 0$ and $B^i = \pm D^i \Phi$, which can be
integrated to give
explicit monopole solutions. The charge one solution was found
in \pras;
multi-monopole solutions are discussed in \corg. The universal
 formula for the
classical mass of the particles of the theory is given by
\eqn\bogmass{M^2 =  (4 \pi)^2 v^2 (Q_E^2 + Q_M^2),}
where $Q_E$ and $Q_M$ are the electric and magnetic charges of
the particle
respectively. For monopoles \bogmass\ is simply the saturation
of the
Bogomolnyi bound. For massive gauge bosons,
it is just the usual relation between the gauge boson mass and
 the Higgs
vacuum expectation value.

The zero potential limit implies
that the static force between two monopoles of like charge or
 between
two gauge bosons of like charge vanishes. This is due to a
cancellation
between a repulsion due to photon (vector) exchange and an
 attraction due
to massless Higgs boson (scalar) exchange \man. This type of
``zero-force
condition" holds for all multi-soliton solutions we consider in
this Report.

Based on ideas of \godno, Montonen and Olive \mono,
conjectured the existence of a ``dual'' formulation of gauge
theory
in which the roles of gauge bosons and magnetic monopoles are
exchanged.
An example of this type of duality in $1+1$ spacetime
dimensions is the
the Thirring model -- sine-Gordon duality, in which
topological and Noether charges are exchanged.

In $3+1$ dimensions, however, monopoles have spin zero while
gauge bosons
have spin one, and the mass formula is probably not exact when
 quantum
corrections are included, as the vanishing of the potential is
not natural
quantum mechanically. Embedding the theory in $N=2$ super
Yang-Mills theory
solves this problem, since then the charges $Q_E$ and $Q_M$
appear as central
charges in the supersymmetry algebra as in the kink solution
and the mass
formula \bogmass\ is exact for supersymmetric states \wito.
However, in $N=2$ the monopole states fill out a matter
supermultiplet consisting of spin zero and spin one-half states.
In order to construct monopoles with spin one it is necessary
to extend the supersymmetry to $N=4$, the maximal allowable
 global
supersymmetry in $3+1$ dimensions. This theory has a number of
remarkable
features. First, the structure of the fermion zero modes is
such that
the monopole supermultiplet now coincides with the gauge
 supermultiplet
and includes states of spin $1$, $1/2$, and $0$ \osb.
Second, the scalar potential has exact flat directions due to
supersymmetry
and again the mass formula is exact. Finally, this theory is
finite
with vanishing beta-function, so that a duality which relates
$g \rightarrow
1/g$ can make sense quantum mechanically at all scales. Thus in
this special
theory all of the simple objections to the existence of the
sort of duality
suggested by Montonen and Olive disappear. Of course this is a
far cry
from showing that such a duality actually holds, but the
evidence is suggestive
enough that the idea is well worth pursuing.

Finally, the
Montonen-Olive conjecture
for more general gauge groups says that the dual gauge group
should have
a weight lattice dual to the weight lattice of the original
group.
It is tempting to speculate \duflrsfd\  that this is related
to ten-dimensional heterotic
string theory, where the gauge groups $SO(32)/Z_2$ and
$E_8 \times E_8$ with
self-dual lattices are singled out by anomaly cancellation.

\chapter{Strings and fivebranes in D=10}

\def\sqr#1#2{{\vbox{\hrule height.#2pt\hbox{\vrule width
.#2pt height#1pt \kern#1pt\vrule width.#2pt}\hrule height.#2pt}}}
\def\Box{\mathchoice\sqr64\sqr64\sqr{4.2}3\sqr33}

\newsec{The elementary string}

We begin by recalling the elementary string solution of
\dabghr.
We want to find a vacuum-like supersymmetric configuration with
$D = 2$
super-Poincare symmetry from the 3-form version of
$D = 10, N = 1$ supergravity
theory. As usual, the fermionic fields should vanish for this
configuration.
We start by making an ansatz for the $D = 10$ metric
 $g_{MN}$, 2-form
$B_{MN}$ and dilaton $\phi$ ($M = 0, 1, \cdots,9$)
corresponding to the
most general eight-two split invariant under
$P_2 \times SO(8)$, where $P_2$ is
the $D = 2$ Poincare group. We split the indices
\eqn\essplit{x^M = (x^\mu, y^m),}
where $\mu = 0,1$ and $m = 2, \cdots, 9$, and write the line
 element as
\eqn\eslinel{ds^2 = e^{2A} \eta_{\mu\nu} dx^\mu dx^\nu + e^{2B}
\delta_{mn} dy^m
dy^n ,}
and the two-form gauge field as
\eqn\btwo{B_{01} = - e^C .}
 All other
components of $B_{MN}$ and all components of the gravitino
$\psi_M$ and
dilatino $\lambda$ are set zero. $P_2$ invariance requires that
the arbitrary
functions $A, B$ and $C$ depend only on $y^m$; $SO(8)$
invariance then requires
that this dependence be only through
$y = \sqrt {\delta_{mn} y^m y^n}$.
Similarly, our ansatz for the dilaton is
\eqn\esdil{\phi = \phi(y).}

As we shall now show, the four arbitrary functions $A, B, C,$
and $\phi$ are
reduced to one by the requirement that the field configurations
 \eslinel,
\btwo\ and \esdil\ preserve some unbroken supersymmetry. In
other words, there
must exist Killing spinors $\varepsilon$ satisfying \dabghr\
\eqn\eskillsp{\delta \psi_M = D_M \varepsilon +
{1\over 96}\,e^{-\phi/
2}\big(\Gamma_M\,^{NPQ} - 9~\delta_M\,^N \Gamma^{PQ}\big)
 H_{NPQ}~
\varepsilon = 0,}
\eqn\eskillsptwo{\delta \lambda = - {1\over2\sqrt{2}}~\Gamma^M
\partial_M \phi
\varepsilon + {1\over 24\sqrt{2}}~e^{-\phi/2}~\Gamma^{MNP}
H_{MNP}~\varepsilon = 0,}
where
\eqn\hthree{H_{MNP} = 3\partial_{\lbrack M} A_{NP\rbrack}.}
Here $\Gamma_A$ are the $D = 10$ Dirac matrices satisfying
\eqn\gamten{\lbrace \Gamma_A, \Gamma_B \rbrace = 2\eta_{AB}.}
$A, B$ refer to the $D = 10$ tangent space,
$\eta_{AB} = (-, +,\cdots, +)$, and
\eqn\gamgam{\Gamma_{AB\cdots C} = \Gamma_{\lbrack A}
\Gamma_{B \cdots}
\Gamma_{C\rbrack},}
thus $\Gamma_{AB} = {1\over 2}~(\Gamma_A \Gamma_B - \Gamma_B
\Gamma_A)$, etc.
The $\Gamma$'s with world-indices $P, Q, R, \cdots$ in
\eskillsp\ and \eskillsptwo\
have been converted using vielbeins $e_M\,^A$. We make an
eight-two split
\eqn\esgamsplit{\Gamma_A = (\gamma_\alpha\otimes 1, \gamma_3
\otimes \Sigma_a),}
where $\gamma_\alpha$ and $\Sigma_a$ are the $D = 2$ and
$D = 8$ Dirac matrices,
 respectively. We also define
\eqn\gamthree{\gamma_3 = \gamma_0 \gamma_1,}
so that $\gamma_3^2 = 1$ and
\eqn\gamnine{\Gamma_9 = \Sigma_2 \Sigma_3 \cdots \Sigma_9 ,}
so that $\Gamma_9^2 = 1$. The most general spinor consistent
with $P_2
\times SO(8)$ invariance takes the form
\eqn\spinsplit{\varepsilon (x, y) = \epsilon \otimes \eta ,}
where $\epsilon$ is a spinor of $SO(1,1)$ which may be further
decomposed into
chiral eigenstates via the projection operators
$(1\pm \gamma_3)$ and $\eta$ is
an $SO(8)$ spinor which may  further be decomposed into chiral
 eigenstates via
the projection operators $(1\pm\Gamma_9)$. The $N=1, D=10$
supersymmetry
parameter is, however, subject to the ten-dimensional chirality
condition
\eqn\chicon{\Gamma_{11}~\varepsilon = \varepsilon ,}
where $\Gamma_{11} = \gamma_3\otimes\Gamma_9$ and so the $D=2$
 and $D=8$
chiralities are correlated.

Substituting the ansatz \eslinel, \btwo\ and \esdil\ into
\eskillsp\ and
\eskillsptwo\ leads to the solution \dabghr
\eqn\spsptwo{\varepsilon = e^{3\phi/8} \epsilon_0\otimes\eta_0,}
where $\epsilon_0$ and $\eta_0$ are constant spinors satisfying
\eqn\epet{(1 - \gamma_3)\epsilon_0 =0,\quad (1 - \Gamma_9)
\eta_0 = 0,}
and where
\eqn\abc{\eqalign{A&= {3\phi\over 4} +c_A ,\cr
B&= -{\phi\over 4} + c_B ,\cr
C&= 2\phi +2 c_A ,\cr}}
where $c_A$ and $c_B$ are constants. If we insist that the
 metric is
asymptotically Minkowskian, then
\eqn\cacb{c_A = -~{3\phi_0\over 4},\quad c_B = ~{\phi_0\over 4},}
where $\phi_0$ is the value of $\phi$ at infinity i.e.
 the dilaton vev
$\phi_0 =~<\phi>$. The condition \epet\ means that one half of
the
supersymmetries are broken.

At this stage the four unknown functions $A$, $B$, $C$ and
$\phi$ have been
reduced to one by supersymmetry. To determine $\phi$, we must
substitute the
ansatz into the field equations which follow from the action
$I_{10}({\rm string}) + S_2$ where $I_{10}({\rm string})$
 is the bosonic sector of the 3-form version of $D =10, N = 1$
supergravity
given by
\eqn\sthreef{I_{10}({\rm string}) = {1\over 2\kappa^2}
\int d^{10}x~\sqrt{-g}~\bigg(R - {1\over 2}
(\partial\phi)^2 - {1\over 2\cdot 3!}~e^{-\phi} H^2\bigg),}
 and $S_2$ is the string
$\sigma$-model action. In \eskillsp, \eskillsptwo\ and
\sthreef\ we have
employed
the canonical choice of metric for which the gravitational
 action is the
conventional Einstein-Hilbert action. This metric is related to
the metric
appearing naturally in the string $\sigma$-model by
\eqn\gstsm{g_{MN}({\rm string}~\sigma{\rm\!-\!model}) =
e^{\phi/2} g_{MN}
({\rm canonical}),}
which will be derived in the section 2.3.
In canonical variables,
therefore, the string $\sigma$-model action is given by
\eqn\stwo{\eqalign{S_2 = - T_2 \int d^2 \xi \bigg(&{1\over 2}
\sqrt{-\gamma}~
\gamma^{ij}\partial_i X^M \partial_j X^N g_{MN}~
e^{\phi/2} - 2~\sqrt{-\gamma}
\cr&+{1\over 2!}~\varepsilon^{ij} \partial_i X^M \partial_j X^N
 B_{MN}\bigg).
\cr}}
We have denoted the string tension by $T_2$. The supergravity
 field equations
are
\eqn\sgfe{\eqalign{R^{MN}&- {1\over 2}~g^{MN} R -
{1\over 2}~\bigg(\partial^M
\phi~\partial^N \phi - {1\over 2}~g^{MN} (\partial \phi)^2 \bigg)
\cr
&-{1\over 2\cdot 2!} \bigg(H^M\,_{PQ} H^{NPQ} - {1\over6}
{}~g^{MN}
H^2\bigg) e^{-\phi}\cr&=\kappa^2 T^{MN}(\rm string),\cr}}
where
\eqn\tmnst{T^{MN}({\rm string}) = -T_2 \int d^2\xi\sqrt{-\gamma}
{}~\gamma^{ij}
\partial_i X^M \partial_j X^N e^{\phi/2}~{\delta^{10} (x - X)
\over \sqrt{-g}},}
\eqn\dhthree{\eqalign{\partial_M (&\sqrt{-g}~e^{-\phi} H^{MNP})
\cr
&=2\kappa^2 T_2 \int d^2 \xi~\varepsilon^{ij} \partial_i X^N
\partial_j X^P
\delta^{10} (x - X),\cr}}
\eqn\phihh{\Box ~\phi + {1\over 2\cdot 3!}~e^{-\phi} H^2 =
{\kappa^2 T_2\over 2}
\int d^6 \xi \sqrt{-\gamma}~\gamma^{ij}\partial_i X^M
\partial_j X^N g_{MN}
e^{\phi/2}{\delta^{10} (x - X)\over\sqrt{-g}}.}
Furthermore, the string field equations are
\eqn\stfe{\eqalign{&\partial_i (\sqrt{-\gamma}~\gamma^{ij}
\partial_j X^N
g_{MN}~e^{\phi/2} ) - {1\over 2}~\sqrt {-\gamma}~\gamma^{ij}
\partial_i X^N
\partial_j X^P \partial_M (g_{NP}~e^{\phi/2} )\cr
&-{1\over 2}~\varepsilon^{ij} \partial_i X^N \partial_j
X^P H_{MNP} = 0, \cr}}
and
\eqn\gamij{\gamma_{ij} = \partial_i X^M \partial_j X^N g_{MN}
e^{\phi/2}.}
To solve these coupled supergravity-string equations we make
the static
gauge choice
\eqn\statg{X^\mu = \xi^\mu, \quad \mu = 0,1}
and the ansatz
\eqn\statan{X^m=Y^m = {\rm constant}, \qquad m=2,...,9.}
As an example, let us now substitute \abc, \statg\ and
 \statan\ into the
2-form equation \dhthree. We find
\eqn\poisdil{\delta^{mn} \partial_m \partial_n e^{-2\phi} =
 - 2\kappa^2 T_2
e^{-\phi_0/2}\delta^8 (y),}
and hence
\eqn\dilsol{e^{-2\phi} = e^{-2\phi_0} \left(1 + {k_2\over y^6}
\right),}
where the constant $k_2$ is given by
\eqn\kcharge{k_2 \equiv {\kappa^2 T_2\over 3\Omega_7}
{}~e^{3\phi_0/2},}
and $\Omega_n$ is the volume of the unit $n$-sphere $S^n$. One
 may verify
that all the field equations \sgfe\ to \gamij\ are reduced to a
 single
equation \poisdil\ by using \cacb, \statg, \statan\ and the
 expressions for the
Ricci tensor $R^{MN}$ and Ricci scalar $R$ in terms of $A$ and
 $B$. For future
reference, we shall compute these for arbitrary worldvolume
 dimension $d$ and
spacetime dimension $D$. The
ansatz is
\eqn\apmet{d s^2 = e^{2A} \eta_{\mu\nu}d x^\mu d x^\nu + e^{2B}
\delta_{mn}
d y^m d y^n ,}
where $\mu = 0, 1, \cdots, d -1; m = d, d+1, \cdots, D-1$ and
$A = A(y), B = B(y)$. For the above
metric, we have
\eqn\aprone{R_{\mu\nu} = -\eta_{\mu\nu} \delta^{mn}e^{2(A - B)}
\bigg(\partial_m
\partial_n A + d \partial_m A \partial_n A  + \tilde {d}
\partial_m A
\partial_n B \bigg),}
\eqn\aprtwo{\eqalign{\quad R_{mn} =&- \tilde{d} \partial_m
\partial_n B -
\delta_{mn} \delta^{kl} \partial_k \partial_l B - d \partial_m
 \partial_n A \cr
&+ d \bigg(\partial_m A \partial_n B + \partial_n A
\partial_m B - \delta_{mn}
\delta^{kl} \partial_k A \partial_l B \bigg) - d \partial_m A
\partial_n B \cr
&+ \tilde {d} \partial_m B \partial_n B - \tilde {d}
\delta_{mn} \delta^{kl}
\partial_k B \partial_l B ,\cr}}
\eqn\aprthree{\eqalign{R = e^{-2B} \bigg [ &- 2d \delta^{mn}
\partial_m
\partial_n A -d(d+1) \delta^{mn} \partial_m A \partial_n A \cr
&-2d\tilde {d} \delta^{mn} \partial_m A\partial_n B\cr
& - 2(\tilde {d} + 1) \delta^{mn}\partial_m \partial_n B
-\tilde {d} (\tilde {d} + 1) \delta^{mn} \partial_m B
\partial_n B \bigg ],\cr}}
where $\tilde {d} \equiv D - d - 2$.

Having established that the supergravity configuration
preserves half the
supersymmetries, we must also verify that the string
configuration \statg\
and \statan\ also preserve these supersymmetries. As discussed
 in \berdps, the
criterion is that in addition to the existence of Killing
spinors satisfying
\eskillsp\ and \eskillsptwo\ we must also have
\eqn\poschi{(1 - \Gamma)\varepsilon = 0,}
where the choice of sign is dictated by the choice of the sign
in the
Wess-Zumino term in \stwo, and where
\eqn\gam{\Gamma \equiv {1\over 2! \sqrt{-\gamma}}
{}~\varepsilon^{ij}
\partial_i X^M \partial_j X^N \Gamma_{MN}.}
Since $\Gamma^2 = 1$ and tr $\Gamma = 0,\, {1\over 2} (1 \pm
\Gamma)$ act as
projection operators. From \statg\ and \statan\ we see that
\eqn\gamtwo{\Gamma = \gamma_3 \otimes 1,}
and hence \poschi\ is satisfied as a consequence of \epet.
Equation \poschi\
explains, from a string point of view, why the solutions we
 have found
preserve just half the supersymmetries. It originates from
the fermionic
$\kappa$-symmetry of the superstring action. The fermionic
zero-modes
on the worldvolume are just the Goldstone fermions associated
with the broken
supersymmetry \refs{\huglp,\tow}.

As shown in \dabghr\ the elementary string solution saturates a
Bogolmol'nyi
bound for the mass per unit length
\eqn\mtwo{{\cal M}_2 = \int d^8 y~\theta_{00},}
where $\theta_{MN}$ is the total energy-momentum pseudotensor
of the
combined gravity-matter system. One finds
\eqn\mtwobog{\kappa {\cal M}_2 \geq {1\over \sqrt{2}} |e_2|
e^{ \phi_0/2},}
where $e_2$ is the Noether ``electric '' charge whose
conservation follows the
equation of
motion of the 2-form \dhthree, namely
\eqn\etwo{e_2 = {1\over {\sqrt{2}\kappa}} \int\limits_{S^7}
e^{-\phi}\,^{\ast}
H,}
where $^{\ast}$ denotes the Hodge dual using the canonical
 metric and the
integral is over an asymptotic
seven-sphere surrounding the string. We find for our solution
that
\eqn\mtwotwo{{\cal M}_2 = e^{\phi_0/2}~T_2,}
and
\eqn\etwotwo{e_2 = \sqrt{2}\kappa~T_2.}
Hence the bound is saturated. This provides another way, in
addition to
unbroken
supersymmetry, to understand the stability of the solution.

A straightforward generalization to exact, stable multi-string
configurations
can be obtained by replacing the single string $\sigma$-model
source by a
superposition of $N$ string sources. The resulting solution is
a linear
superposition of solutions to \poisdil\
\eqn\multist{e^{-2\phi} = e^{-2\phi_0}~\Bigg[1 + \sum_{\ell}
{k_2\over \mid
\vec y - \vec y_{\ell} \mid^6}\Bigg],}
where $\vec y_{\ell}$ corresponds to the position of each
string.
The ability to superpose solutions of this kind is a well-known
 phenomenon
in soliton and instanton physics and goes by the name of the
``no-force
condition". In the present context, it means that the
gravitational-dilatonic
attractive force acting on each of the strings is exactly
 cancelled
by an equal but repulsive force from the 2-form. This effect is
also present in other elementary and solitonic solutions and is
closely related both to
the saturation of the Bogomol'nyi bound and to the existence of
unbroken
supersymmetry. In the supersymmetric context the no-force
condition is
sometimes called ``antigravity". To see this explicitly,
consider the
multi-string configurations \multist\ with, for example, $N$
strings as
sources. In general, we do not have the transverse $SO(8)$
symmetry, but we
still have the $P_2$ symmetry for the configurations \multist.
Let each
string with label $l$ satisfy $X^\mu ( l ) = \xi^\mu$ so that,
in
particular, they all have the same orientation. The lagrangian
for this
$N$-string configuration in the fields of the sources given by
\eslinel\ and
\btwo\ is,
from \stwo,
\eqn\ltwo{{\cal L}_2 = - T_2 \Bigg[ \sqrt {-{\rm det} (e^{2 A +
\phi/2} \eta_{ij}
+ e^{2 B + \phi/2} \partial_i Y^m (l) \partial_j Y_m (l))} -
 e^C \Bigg],}
corresponding to a potential
\eqn\vtwo{V = T_2 ( e^{2A + \phi/2} - e^C).}
But this vanishes by the supersymmetry conditions \abc. On the
 other hand,
if the $N$-strings had the opposite orientation, and hence the
opposite
sign $e_2$, then the sign change in the Wess-Zumino term in
\ltwo\ would
result
in a net attractive force and therefore the corresponding
configurations
cannot be stable.

\newsec{The solitonic fivebrane}

The elementary string discussed above is a solution
 of the
coupled field-string system with action $I_{10}({\rm string})+
S_2$.  As such
it exhibits
$\delta$-function singularities at $y = 0$.  It is
characterized by a
non-vanishing
Noether electric charge $e_2$.  By contrast, we now wish to
find a solitonic
fivebrane, corresponding to a solution of the source free
equations
 resulting
from $I_{10}({\rm string})$ alone and which will be
characterized
by a
non-vanishing topological ``magnetic'' charge $g_6$.

To this end, we now make an ansatz invariant under
$P_6 \times
SO(4)$.  Hence we write \essplit\ and \eslinel\ as before where
now
$\mu = 0, 1 \ldots 5$ and $m = 6, 7, 8, 9$.  The ansatz
for the
antisymmetric tensor, however, will now be made on the field
 strength rather
than on the
potential.  From section 2.1 we recall that a non-vanishing
electric charge
corresponds to
\eqn\elcharge{{1\over \sqrt{2} \kappa} e^{-\phi}{}^{\ast}H
 = e_2 \varepsilon_7/\Omega_7,}
where $\varepsilon_n$ is the volume form on $S^n$. Accordingly,
to obtain a non-vanishing magnetic charge, we make the ansatz
\eqn\hgsix{{1\over \sqrt{2} \kappa} H = g_6
\varepsilon_3/\Omega_3.}
 Since this is an
harmonic
form, $H$ can no longer be written globally as the curl of $B$,
 but it
satisfies the
Bianchi identity.  It is now not difficult to show that all the
 field
equations are
satisfied. The solution is given by
\eqn\dflufbsol{\eqalign{e^{2\phi}&=e^{2\phi_0}\left(1+{k_6\over
 y^2}\right),\cr
 ds^2&=e^{-(\phi-\phi_0)/2}\eta_{\mu\nu}dx^\mu dx^\nu+e^{3(\phi-
\phi_0)/2}
\delta_{mn}dy^mdy^n,\cr
 H&=2k_6 e^{\phi_0/2}\varepsilon_3,\cr}}
where $\mu,\nu=0,1,...,5$, $m,n=6,7,8,9$ and where
\eqn\ksix{k_6={\kappa g_6 \over \sqrt{2}\Omega_3}e^{-\phi_0/2}.}
It follows that the mass per unit 5-volume now saturates a bound
involving the magnetic charge
\eqn\bogomsolfb{{\cal M}_6={1\over \sqrt{2}} \mid g_6 \mid
e^{-\phi_0/2}.}
Note that the $\phi_0$ dependence is such that ${\cal M}_6$ is
large
for small ${\cal M}_2$ and vice-versa.

The electric charge of the elementary solution and the magnetic
charge of
the soliton
solution obey a Dirac quantization rule \refs{\nep,\tei}
\eqn\etwogsix{e_2 g_6 = 2 \pi n, \qquad n = {\rm integer},}
and hence from \etwotwo\
\eqn\gsix{g_6 = 2\pi n/\sqrt{2}
\kappa T_2.}

\newsec{String/fivebrane duality}

The 3-form version of $D = 10, N =1$ supergravity admits the
 elementary string
as a solution \dabghr. In this and the next section we shall
show that the
7-form version admits the elementary fivebrane
as a solution \duflfb. The elementary string can also be
interpreted as
a magnetic-like solution of the 7-form version, as can the
elementary fivebrane
for the 3-form version. This suggests that the heterotic string
and the heterotic fivebrane may be dual to each other in the
sense that they
are equivalent descriptions of the same underlying physical
theory. In the
dual formulation, the metric $g_{MN}$ and the dilaton $\phi$
are the same,
whereas the 7-form field strength $K$ of the fivebrane is dual
 to the 3-form
field strength of the string $H$.  More precisely,
\eqn\hdualk{H = e^\phi {^\ast} K ,}
where $^\ast$ denotes the Hodge dual using the canonical metric,
 so that the
field equation of the 3-form becomes the Bianchi identity of
the 7-form and
vice-versa. Hence
\eqn\sftfb{I_{10}({\rm fivebrane}) = {1\over 2\kappa^2} \int
 d^{10} x
\sqrt{- g}
\big(R - {1\over 2 } (\partial \phi)^2 - {1\over 2\cdot 7!}
 e^\phi K^2 \big),}
where $K$ is the curl of the 6-form $A$:
\eqn\kda{K = d A.}

String/fivebrane duality will tell us that
the fivebrane loop coupling constant ${\rm g}_6$ is given by
the inverse cube
root of the string loop coupling constant ${\rm g}_2$,
\eqn\gtwosix{{\rm g}_6 = {\rm g}_2^{-1/3}=e^{-\phi_0/3}, }
which implies that the strongly coupled heterotic string
corresponds to the
weakly coupled fivebrane, and vice versa. We will now derive
\gtwosix\ and the
following relations between the canonical gravitational metric
and the metrics
which appear naturally in
the string and fivebrane $\sigma$-models:
\eqn\metrels{g_{MN} ({\rm canonical}) = e^{-\phi/2} g_{MN}({\rm
 string}) =
e^{\phi/6} g_{MN}({\rm fivebrane}).}
In general, we have
\eqn\stomcan{g_{MN} ({\rm string}) = \Omega_s (\phi) g_{MN}
({\rm canonical}),}
and
\eqn\fbomcan{g_{MN} ({\rm fivebrane} ) = \Omega_f (\phi) g_{MN}
({\rm canonical}). }
The corresponding lowest-order $\sigma$-model actions
written in terms of the canonical metric are given by
\eqn\Stwo{S_2 = - T_2 \int d^2 \xi ({1\over 2} \sqrt {-\gamma}
\gamma^{ij}
\partial_i X^M \partial_j X^N \Omega_s(\phi) g_{MN} + {1\over 2}
\varepsilon^{ij}\partial_i X^M \partial_j X^N B_{MN} ),}
for the string and
\eqn\Ssix{\eqalign{S_6 = - T_6 \int d^6 \xi \big(&{1\over 2}
\sqrt {-\gamma}
\gamma^{ij}\partial_i X^M \partial_j X^N \Omega_f(\phi) g_{MN}
 - 2\sqrt {-
\gamma}\cr &+{1\over 6!} \varepsilon^{ijklmn} \partial_i X^M
\partial_j X^N
\partial_k X^O\partial_l X^P \partial_m X^Q
\partial_n X^R A_{MNOPQR} \big)\cr}}
for the fivebrane. Note that in this case
\eqn\Gamij{\gamma_{ij} = \partial_i X^M
\partial_j X^N \Omega_f (\phi) g_{MN}.}
To lowest order, the actions $I_{10}({\rm string})$ and
$S_2$ describe the
same string (for
example, the elementary string), and the actions
$I_{10}({\rm fivebrane})$
and $S_6$ describe the
same fivebrane. Moreover,
string/fivebrane duality implies that either
$I_{10}({\rm string})$ or
$I_{10}({\rm fivebrane})$ describes not only the string but
also the
fivebrane by using \hdualk. Therefore, we would expect that
 both
$I_{10}({\rm string})$ and $I_{10}({\rm fivebrane})$
should scale in the same way as both $S_2$ and $S_6$ under the
following constant rescalings:
\eqn\gscale{g_{MN} \rightarrow \alpha g_{MN},}
\eqn\phiscale{e^\phi \rightarrow \beta e^\phi ,}
\eqn\bscale{B_{MN} \rightarrow \lambda^2 B_{MN} ,}
\eqn\ascale{A_{MNOPQR} \rightarrow \sigma^6 A_{MNOPQR}.}
{}From the above requirements, each of the actions
 $I_{10}({\rm string})$,
$I_{10}({\rm fivebrane})$, $S_2$ and $S_6$ should also scale
homogeneously.
It follows from \Stwo\ that
\eqn\stwos{S_2 \rightarrow \lambda^2 S_2 ,}
and
\eqn\omss{\Omega_s (\phi) \rightarrow {\lambda^2 \over \alpha}
\Omega_s (\phi),}
from \Ssix\ that
\eqn\ssixs{S_6 \rightarrow \sigma^6 S_6 ,}
and
\eqn\omfs{\Omega_f (\phi) \rightarrow {\sigma^2 \over \alpha}
\Omega_f (\phi),}
from \sthreef\ that
\eqn\sftsts{ I_{10}({\rm string}) \rightarrow \alpha^4
 I_{10}({\rm string}),}
and
\eqn\labeal{\lambda^4 = \beta \alpha^2,}
and from \sftfb\ that
\eqn\sftfbs{I_{10}({\rm fivebrane}) \rightarrow
\alpha^4 I_{10}({\rm fivebrane}),}
and
\eqn\albesi{\alpha^6 = \beta \sigma^{12}.}
As expected, $I_{10}({\rm string})$ scales the same way as
$I_{10}({\rm fivebrane})$. Combining either
\sftsts\ or \sftfbs\ with \stwos\ and \ssixs, we obtain
\eqn\allasi{\alpha^4 = \lambda^2 \sigma^6.}
Hence, by using \labeal, \albesi\ and \allasi, we obtain
\eqn\allasitwo{\alpha = \lambda^{1/2} \sigma^{3/2},}
and
\eqn\belasi{\beta = \lambda^3 \sigma^{- 3}.}
Substituting \belasi\ into \phiscale, and \allasitwo\ into
\gscale,
\omss\ and \omfs, respectively, we obtain
\eqn\phisc{e^\phi \rightarrow \lambda^3 \sigma^{- 3} e^\phi ,}
\eqn\gsc{g_{MN} \rightarrow \lambda^{1/2} \sigma^{3/2} g_{MN} ,}
\eqn\omssc{\Omega_s (\phi) \rightarrow (\lambda^3
\sigma^{- 3})^{1/2}
\Omega_s
(\phi),}
and
\eqn\omfsc{\Omega_f (\phi) \rightarrow (\lambda^3
\sigma^{- 3})^{- 1/6}
\Omega_f (\phi).}
Comparing \omssc\ and  \omfsc\ with \phisc, we immediately
 have
\eqn\omsomf{\Omega_s (\phi) = e^{\phi/2}, \qquad \qquad
\Omega_f (\phi) =
e^{- \phi/6},}
which implies \metrels. This result agrees with what one
obtains in string
theory by setting the $\beta$-functions of $g_{MN}$, $B_{MN}$
 and $\phi$ to
zero, i.e. by using string worldsheet conformal invariance.

We are now ready to derive \gtwosix. This can be achieved
simply by writing
$I_{10}({\rm string})$ and $I_{10}({\rm fivebrane})$ in terms
of string and fivebrane metrics given in
\metrels, respectively. We obtain for the string action
\eqn\sstring{I_{10}({\rm string}) = {1\over 2\kappa^2}
\int d^{10} x \sqrt {-g}
e^{-2\phi} \big(R + 4 (\partial\phi)^2 - {1\over 2\cdot 3!}
 H^2 \big),}
where the common factor $e^{-2\phi}$ for each term implies
that the string loop counting parameter ${\rm g}_2$ is given
by
\eqn\stloop{{\rm g}_2 = e^{\phi_0}.}
For the fivebrane action,
\eqn\sfbrane{I_{10}({\rm fivebrane}) = {1\over 2 \kappa^2}
 \int d^{10} x
\sqrt {-g}
e^{2\phi/3}\big(R - {1\over 2\cdot 7!} K^2 \big),}
and an analogous situation arises, namely a common factor
$e^{2\phi/3}$ in
front of each term. Although we have yet to construct the
heterotic fivebrane
action and quantize it, string/fivebrane duality tells that if
heterotic string
theory is a sensible theory, then so is heterotic fivebrane
theory. So we would
expect that the action \sfbrane\ would be the field-theory
limit of the
heterotic fivebrane. In analogy with the case of \sstring, the
common factor
of \sfbrane\ suggests that the fivebrane loop counting
parameter ${\rm g}_6$
is
given by
\eqn\fbloop{{\rm g}_6 = e^{-\phi_0/3}.}
Combining \stloop\ and \fbloop\ we obtain \gtwosix.

\newsec{The elementary fivebrane}

In this section, we rederive the fivebrane soliton as an
elementary
electric-like singular solution, the ``elementary fivebrane"
\duflfb\ of the
7-form version of $D =10, N = 1$ supergravity, which is the
analog of a static
electric-like singular ``elementary string" solution \dabghr\
of the 3-form
version.

We want to find a vacuum-like supersymmetric configuration with
 $D = 6$
super-Poincare symmetry from the 7-form version of
$D = 10, N = 1$ supergravity
theory. As usual, the fermionic fields should vanish for this
configuration.
We start by making an ansatz for the $D = 10$ metric $g_{MN}$,
 6-form
$A_{MNOPQR}$ and dilaton $\phi$ ($M = 0, 1, \cdots,9$)
 corresponding to the
most general six-four split invariant under $P_6 \times SO(4)$,
where $P_6$ is
the $D = 6$ Poincare group. We split the indices
\eqn\fbsplit{x^M = (x^\mu, y^m),}
where $\mu = 0, \cdots, 5$ and $m = 6, \cdots, 9$, and write
the line element as
\eqn\fblinel{ds^2 = e^{2A} \eta_{\mu\nu} dx^\mu dx^\nu + e^{2B}
 \delta_{mn}
dy^m dy^n ,}
and the six-form gauge field as
\eqn\asix{A_{\mu\nu\rho\sigma\lambda\tau} = - {1\over ^6 g}
\varepsilon_{\mu\nu\rho\sigma\lambda\tau} e^C ,}
where $^6 g$ is the determinant of
$g_{\mu\nu}$, $\varepsilon_{\mu\nu\rho\sigma
\lambda\tau} \equiv g_{\mu\alpha} g_{\nu\beta} g_{\rho\gamma}
g_{\sigma\delta}
g_{\lambda\epsilon} g_{\tau\xi}
\varepsilon^{\alpha\beta\gamma\delta\epsilon\xi}
$ and $\varepsilon^{012345} = 1$, i.e. $A_{012345} = - e^C$.
 All other
components of $A_{MNOPQR}$ and all components of the gravitino
 $\psi_M$ and
dilatino $\lambda$ are set zero. $P_6$ invariance requires that
the arbitrary
functions $A, B$ and $C$ depend only on $y^m$; $SO(4)$
 invariance then requires
that this dependence be only through $y = \sqrt {\delta_{mn}
y^m y^n}$.
Similarly, our ansatz for the dilaton is
\eqn\fbdil{\phi = \phi(y).}

As we shall now show, the four arbitrary functions $A, B, C,$
and $\phi$ are
reduced to one by the requirement that the field configurations
\fblinel,
\asix\ and \fbdil\ preserve some unbroken supersymmetry. In
other words, there
must exist Killing spinors $\varepsilon$ satisfying
\eqn\killsp{\delta \psi_M = D_M \varepsilon +
{1\over{2 \cdot 8!}}\,e^{\phi/
2}\big(3~\Gamma_M\,^{NOPQRST} - 7~\delta_M\,^N
 \Gamma^{OPQRST}\big)
K_{NOPQRST}~
\varepsilon = 0,}
\eqn\killsptwo{\delta \lambda = - {1\over2\sqrt{2}}~\Gamma^M
\partial_M \phi
\varepsilon - {1\over 2\cdot 2\sqrt{2} \cdot 7!}~e^{\phi/2}
{}~\Gamma^{MNOPQRS}
K_{MNOPQRS}~\varepsilon = 0,}
where
\eqn\kseven{K_{MNOPQRS} = 7\partial_{\lbrack M} A_{NOPQRS
\rbrack}.}
We make a six-four split
\eqn\fbgamsplit{\Gamma_A = (\gamma_\alpha\otimes 1, \gamma_7
\otimes \Sigma_a),}
where $\gamma_\alpha$ and $\Sigma_a$ are the $D = 6$ and
 $D = 4$ Dirac matrices,
respectively. We also define
\eqn\gamseven{\gamma_7 = \gamma_0 \gamma_1 \gamma_2 \gamma_3
\gamma_4
\gamma_5 ,}
so that $\gamma_7^2 = 1$ and
\eqn\gamfive{\Gamma_5 = \Sigma_6 \Sigma_7 \Sigma_8 \Sigma_9 ,}
so that $\Gamma_5^2 = 1$. The most general spinor consistent
 with $P_6
\times SO(4)$ invariance takes the form
\eqn\fbspinsplit{\varepsilon (x, y) = \epsilon \otimes \eta ,}
where $\epsilon$ is a spinor of $SO(1, 5)$ which may be further
 decomposed into
chiral eigenstates via the projection operators
$(1\pm \gamma_7)$ and $\eta$ is
an $SO(4)$ spinor which may  further be decomposed into chiral
eigenstates via
the projection operators $(1\pm\Gamma_5)$. The $N=1, D=10$
 supersymmetry
parameter is, however, subject to the ten-dimensional chirality
 condition
\eqn\fbchicon{\tilde \Gamma_{11}~\varepsilon = \varepsilon ,}
where $\Gamma_{11} = \gamma_7\otimes\Gamma_5$ and so the $D=4$
 and $D=6$
chiralities are correlated.

In our background \fblinel, \asix\ and \fbdil, the
 transformation rules \killsp\
and \killsptwo\ reduce to
\eqn\trone{\partial_\mu ~\varepsilon -
{1\over 8}~\gamma_\mu ~\Sigma^m~
\Bigg[{1\over 2}~e^{\phi/2 -6A} \partial_m e^C - 4~\gamma_7
\partial_m A\Bigg]
\varepsilon =0,}
\eqn\trtwo{\eqalign{\partial_m~\varepsilon&+{1\over 2}
(\Sigma_m \Sigma^n
\partial_n B - \partial_m B)~\varepsilon\cr
&+{1\over 16}~e^{\phi/2 - 6A}~\gamma_7~(3~\Sigma_m
\Sigma^n \partial_n e^C +
4~\partial_m e^C)~\varepsilon = 0,\cr}}
and
\eqn\trthree{\Sigma^m \bigg({1\over 2}~e^{\phi/2 -6A}
 \partial_m e^C +
\gamma_7 \partial_m \phi \bigg)~\varepsilon =0.}
Note that the $\gamma_\mu$ and $\Sigma_m$ carry world indices.
 Hence we find
that \duflfb
\eqn\fbspsptwo{\varepsilon = e^{-\phi/8} \epsilon_0\otimes\eta_0,
}
where $\epsilon_0$ and $\eta_0$ are constant spinors satisfying
\eqn\fbepet{(1 - \gamma_7)\epsilon_0 =0,\quad (1 - \Gamma_5)
\eta_0 = 0,}
and where
\eqn\fbabc{\eqalign{A&= - {\phi\over 4} +c_A ,\cr
B&= {3\phi\over 4} + c_B ,\cr
C&= -2\phi +6 c_A ,\cr}}
where $c_A$ and $c_B$ are constants. If we insist that the
 metric is
asymptotically Minkowskian, then
\eqn\fbcacb{c_A = +~{\phi_0\over 4},\quad c_B =
-~{3\phi_0\over 4}.}
 The condition \fbepet\ means that one half of the
supersymmetries are broken.

At this stage the four unknown functions $A$, $B$, $C$ and
$\phi$ have been
reduced to one by supersymmetry. To determine $\phi$, we must
substitute the
ansatz into the field equations which follow from the action
$I_{10}({\rm fivebrane})+S_6$.
In canonical variables, the supergravity field equations are
\eqn\fbsgfe{\eqalign{R^{MN}&- {1\over 2}~g^{MN} R -
{1\over 2}~\bigg(\partial^M
\phi~\partial^N \phi - {1\over 2}~g^{MN} (\partial \phi)^2
\bigg)\cr
&-{1\over 2\cdot 6!} \bigg(K^M\,_{OPQRST} K^{NOPQRST} -
{1\over14}~g^{MN}
K^2\bigg) e^{\phi}\cr&=\kappa^2 T^{MN}(\rm fivebrane),\cr}}
where
\eqn\tmnfb{T^{MN}({\rm fivebrane}) = -T_6 \int d^6\xi
\sqrt{-\gamma}~\gamma^{ij}
\partial_i X^M \partial_j X^N e^{-\phi/6}~{\delta^{10}
 (x - X)\over \sqrt{-g}},}
\eqn\dkseven{\eqalign{\partial_M (&\sqrt{-g}~e^{\phi}
 K^{MNOPQRS})\cr
&=2\kappa^2 T_6 \int d^6 \xi~\varepsilon^{ijklmn}
\partial_i X^N \partial_j X^O
\partial_k X^P \partial_l X^Q \partial_m X^R \partial_n X^S
\delta^{10} (x - X),
\cr}}
\eqn\phikk{-\Box ~\phi + {1\over 2\cdot 7!}~e^{\phi} K^2 =
{\kappa^2 T_6\over 6}
\int d^6 \xi \sqrt{-\gamma}~\gamma^{ij}\partial_i X^M
\partial_j X^N g_{MN}
e^{-\phi/6}{\delta^{10} (x - X)\over\sqrt{-g}}.}
Furthermore, the fivebrane field equations are
\eqn\fbfe{\eqalign{&\partial_i (\sqrt{-\gamma}~\gamma^{ij}
\partial_j X^N
g_{MN}~e^{-\phi/6} ) - {1\over 2}~\sqrt {-\gamma}~\gamma^{ij}
\partial_i X^N
\partial_j X^P \partial_M (g_{NP}~e^{-\phi/6} )\cr
&-{1\over 6!}~\varepsilon^{ijklmn} \partial_i X^N \partial_j
 X^O \partial_k X^P
\partial_l X^Q \partial_m X^R \partial_n X^S K_{MNOPQRS} = 0,
 \cr}}
and
\eqn\fbgamij{\gamma_{ij} = \partial_i X^M \partial_j X^N
g_{MN}e^{-\phi/6}.}
To solve these coupled supergravity-fivebrane equations we make
the static
gauge choice
\eqn\fbstatg{X^\mu = \xi^\mu, \quad \mu = 0, 1, 2, 3, 4, 5}
and the ansatz
\eqn\fbstatan{X^m=Y^m = {\rm constant}, \qquad m=6,7,8,9.}
As an example, let us now substitute \fbabc, \fbstatg\ and
\fbstatan\ into the
6-form equation \dkseven. We find
\eqn\fbpoisdil{\delta^{mn} \partial_m \partial_n e^{2\phi} = -
2\kappa^2 T_6
e^{3 \phi_0/2}\delta^4 (y),}
and hence
\eqn\fbdilsol{e^{2\phi} = e^{2\phi_0} \left(1 + {k_6\over y^2}
\right),}
where the constant $k_6$ is now given by
\eqn\fbkcharge{k_6 \equiv {\kappa^2 T_6\over \Omega_3}
{}~e^{-\phi_0/2}.}
One may verify
that all the field equations \fbsgfe\ to \fbgamij\ are reduced
to a single
equation \fbpoisdil\ by using \fbcacb, \fbstatg, \fbstatan\ and
the expressions for
the
Ricci tensor $R^{MN}$ and Ricci scalar $R$ in terms of $A$ and
 $B$ given in the
general case above.

Having established that the supergravity configuration
 preserves half the
supersymmetries, we must also verify that the fivebrane
 configuration \fbstatg\
and \fbstatan\ also preserve these supersymmetries. As
 discussed in \berdps,
the
criterion is that in addition to the existence of Killing
spinors satisfying
\killsp\ and \killsptwo\ we must also have
\eqn\fbposchi{(1 - \tilde\Gamma)\varepsilon = 0,}
where the choice of sign is dictated by the choice of the sign
in the
Wess-Zumino term in \Ssix, and where
\eqn\gam{\tilde\Gamma \equiv {1\over 6! \sqrt{-\gamma}}
{}~\varepsilon^{ijklmn}
\partial_i X^M \partial_j X^N \partial_k X^O \partial_l X^P \partial_m X^Q
\partial_n X^R \Gamma_{MNOPQR}.}
Since $\tilde\Gamma^2 = 1$ and
$tr \tilde \Gamma = 0$,\ ${1\over 2} (1 \pm \tilde\Gamma)$ act as
projection operators. From \fbstatg\ and \fbstatan\ we see that
\eqn\fbgamtwo{\tilde\Gamma = \gamma_7 \otimes 1,}
and hence \fbposchi\ is satisfied as a consequence of \fbepet.
\fbposchi\
explains, from a fivebrane point of view, why the solutions we
 have found
preserve just half the supersymmetries. It originates from the
 fermionic
$\kappa$-symmetry of the superfivebrane action. Again the
 fermionic zero-modes
on the worldvolume are just the Goldstone fermions associated
 with the broken
supersymmetry \refs{\huglp,\tow}.

Using the same method as in section 2.1, we may also establish
a bound for the mass per unit area of the
fivebrane
\eqn\msix{{\cal M}_6 = \int d^4 y~\theta_{00},}
where $\theta_{MN}$ is the total energy-momentum pseudotensor of the
combined gravity-matter system. One finds
\eqn\msixbog{\kappa {\cal M}_6 \geq {1\over \sqrt{2}}
 |\tilde e_6| e^{- \phi_0/2},}
where $\tilde e_6$ is the Noether charge whose conservation
follows the equation of
motion of the 6-form \dkseven, namely
\eqn\esix{\tilde e_6 = {1\over \sqrt{2}\kappa} \int\limits_{S^3}
e^{\phi}\,^{\ast} K,}
where $^{\ast}$ denotes the Hodge dual and the integral is over
an asymptotic
three-sphere surrounding the fivebrane. We find for our solution that
\eqn\msixtwo{{\cal M}_6 = e^{-\phi_0/2}~T_6,}
and
\eqn\esixtwo{\tilde e_6 = \sqrt{2}\kappa~T_6.}
Hence the bound is saturated. Once again, this provides another way, in
addition to unbroken supersymmetry, to understand the stability of the
solution.

As in the case of the elementary string, there is a straightforward
generalization to exact,
stable multi-fivebrane configurations by linear superposition of
\eqn\multifb{e^{2\phi} = e^{2\phi_0}~\Bigg[1 + \sum_{\ell}
{k_6\over \mid
\vec y - \vec y_{\ell} \mid^2}\Bigg],}
where $\vec y_{\ell}$ corresponds to the position of each fivebrane.
In this case the no-force
condition results from the cancellation of the gravitational-dilatonic
attractive force with the repulsive force from the 6-form. To see this
explicitly, consider the
multi-fivebrane configurations \multifb\ with, for example, $N$ fivebranes as
sources.
In general, we do not have the transverse $SO(4)$ symmetry, but we
still have the $P_6$ symmetry for the configurations \multifb. Let each
fivebrane with label $l$ satisfy $X^\mu ( l ) = \xi^\mu$ so that, in
particular, they all have the same orientation. The lagrangian for this
$N$-fivebrane in the fields of the sources given by \fblinel\ and \asix\ is,
from \Ssix,
\eqn\lsix{{\cal L}_6 = - T_6 \Bigg[ \sqrt {-{\rm det} (e^{2 A - \phi/6}
\eta_{ij}
+ e^{2 B - \phi/6} \partial_i Y^m (l) \partial_j Y_m (l))} - e^C \Bigg],}
corresponding to a potential
\eqn\vsix{V = T_6 ( e^{6A - \phi/2} - e^C).}
But this vanishes by the supersymmetry conditions \abc. On the other hand,
if the $N$-fivebrane had the opposite orientation, and hence the opposite
$\tilde e_6$, then the sign change in the Wess-Zumino term in
\lsix\ would result
in a net attractive force and therefore the corresponding configurations
cannot be stable.

\newsec{The solitonic string}

Finally, just as the fivebrane was seen as a solitonic solution of
string theory, one can re-interpret the elementary string solution of section
2.1 as a soliton of the fivebrane theory, i.e. as a solution of the source-free
equations resulting from $I_{10}({\rm fivebrane})$. The solution
will be characterized by a non-vanishing topological ``magnetic'' charge $g_2$
\eqn\kgtwo{{1\over \sqrt{2} \kappa} K = \tilde g_2
\varepsilon_7/\Omega_7.}
The solution is given by
\eqn\dabsol{\eqalign{e^{-2\phi}&=e^{-2\phi_0}
\left(1+{k_2\over y^6}\right),\cr
 ds^2&=e^{3(\phi-\phi_0)/2}\eta_{\mu\nu}dx^\mu dx^\nu+e^{-(\phi-\phi_0)/2}
\delta_{mn}dx^mdx^n,\cr
 K&=6k_2 e^{-3\phi_0/2}\varepsilon_7,\cr}}
where $\mu,\nu=0,1$, $m,n=2,3,...,9$ and where
\eqn\ktwo{k_2={\kappa \tilde g_2 \over 3\sqrt{2}\Omega_7}
e^{3\phi_0/2}.}
It follows that the mass per unit length now saturates a bound
involving the magnetic charge
\eqn\bogomsolfb{{\cal M}_2={1\over \sqrt{2}} \mid \tilde g_2 \mid
e^{\phi_0/2}.}
Note that the $\phi_0$ dependence is such that ${\cal M}_2$ is large
for small ${\cal M}_6$ and vice-versa.

The electric charge of the elementary fivebrane and the magnetic charge of
the solitonic string obey a Dirac quantization rule \refs{\nep,\tei}
\eqn\esixgtwo{\tilde e_6 \tilde g_2 = 2 \pi n, \qquad n = {\rm integer},}
and hence from \esixtwo\
\eqn\gtwo{\tilde g_2 = 2\pi n/\sqrt{2}
\kappa T_6.}

The introduction of the elementary fivebrane and the solitonic string allows
us to translate the Dirac quantization rules into a condition on the tensions
of the string and fivebrane. From \hdualk, \etwo\ and \esix\ it follows that
\eqn\egsame{ e_2=\tilde g_2, \qquad \qquad \tilde e_6=g_6.}
It then follows from \etwotwo, \esixtwo, \etwogsix\ and \esixgtwo\ that
\eqn\ttwosix{\kappa^2 T_2 T_6 = |n| \pi, \qquad n = {\rm integer}.}

\newsec{Singular or non-singular?}

Neither the elementary string nor the elementary fivebrane qualify for the
epithet ``soliton''. They were obtained by coupling supergravity to the
corresponding $\sigma$-models and hence display a $\delta$-function
singularity at the location of the source $y=0$. Moreover, in both cases the
curvature calculated from the $\sigma$-model metric blows up at $y=0$. To see
this we
rewrite the string solution \eslinel\ in the string $\sigma$-model metric
(in what follows we set $\phi_0=0$ for simplicity)
\eqn\stline{ds^2=\left(1+{k_2\over y^6}\right)^{-1} \eta_{\mu\nu} dx^\mu dx^\nu
+ \delta_{mn} dx^m dx^n.}
We may verify that this metric exhibits a curvature singularity by computing
the scalar curvature
\eqn\stcurv{R_{\rm string}({\rm string\ \sigma-model})\sim -y^{-2}.}
Similarly, we rewrite the fivebrane solution \fblinel\ in the fivebrane
$\sigma$-model metric
\eqn\fbline{ds^2=\left(1+{k_2\over y^2}\right)^{-1/3}\eta_{\mu\nu} dx^\mu
dx^\nu
+ \left(1+{k_2\over y^2}\right)^{2/3} \delta_{mn} dx^m dx^n.}
We may also verify that this metric exhibits a curvature singularity by
computing the scalar curvature
\eqn\fbcurv{R_{\rm fivebrane}({\rm fivebrane\ \sigma-model})\sim -y^{-2/3}.}
(Incidentally, $R_{\rm string}$ and $R_{\rm fivebrane}$ both blow up as
$y^{-1/2}$ in canonical variables.)

To justify that the {\it solitonic} string and fivebrane are worthy of their
name, therefore, we must demonstrate that they are nonsingular. To show this,
we now rewrite the string metric in fivebrane variables
\eqn\dabfb{ds^2=\left(1+{k_2\over y^6}\right)^{-2/3}\eta_{\mu\nu}
dx^\mu dx^\nu + \left(1+{k_2\over y^6}\right)^{1/3}\left(dy^2+y^2d\Omega_7^2
\right),}
and the fivebrane metric in string variables
\eqn\dflust{ds^2=\eta_{\mu\nu}
dx^\mu dx^\nu + \left(1+{k_6\over y^2}\right)\left(dy^2+y^2d\Omega_3^2\right).}
Remarkably, both are free of curvature singularities,
as may be seen by noting
that, as $y\to 0$, the radius of $S^7$ in \dabfb\ tends to the finite
value $k_2^{1/6}$ and the radius of $S^3$ in \dflust\ tends to the finite
value $k_6^{1/2}$. This is confirmed by a calculation of the scalar
curvatures. We find
\eqn\scurvtwo{R_{\rm string}({\rm fivebrane\ \sigma-model})\sim +k_2^{-1/3}}
\eqn\fcurvtwo{R_{\rm fivebrane}({\rm string\ \sigma-model})\sim +k_6^{-1}.}
Note that throughout this report we have employed the fivebrane $\sigma$-model
metric of \refs{\duflfb,\duflrsfd}\ given in \fbomcan, for
which
$g_{MN}({\rm fivebrane\ \sigma-model})=e^{-2\phi/3}
g_{MN}({\rm string\ \sigma-model})$. Now any metric
$e^{-c\phi}g_{MN}({\rm string\ \sigma-model})$ will yield string
solutions with non-singular curvature and any metric
$e^{+c\phi}g_{MN}({\rm fivebrane\ \sigma-model})$
will yield fivebrane solutions with  non-singular curvature provided
$c\geq 2/3$. Interestingly enough, however, only the unique choice $c=2/3$
yields metrics which are also free of conical singularities
\refs{\gibt,\dufgt}. From the point
of view of metric singularities, therefore, the string is a
 singular solution
 of string theory and a nonsingular solution of fivebrane
theory, whereas the
fivebrane is a singular solution of fivebrane theory and a
nonsingular
solution of string theory. Thus the singularity structure is
entirely
symmetric between strings and fivebranes, in accordance with
string/fivebrane
duality.

We shall now show how to reach the same conclusion from a
 somewhat more
physical test-probe/source approach \dufklsin.
Does a test-probe fall into the source in
a finite proper time, as measured by its own clock, in
which case the singularity is real, or in an infinite proper
time, in
which case the singularity is harmless? We find that the answer
is
probe-dependent. We shall show that when test probe and source
are both
strings or both fivebranes the singularity is real. By contrast, we shall
show
that if one is a string and the other a fivebrane, the
singularity is
harmless.

Let us consider the trajectory of a test string falling
radially into
a source string, oriented along $x^1=\xi^1$. For simplicity,
 let the test
string lie either parallel or antiparallel to the source string. If we
eliminate $\gamma_{ij}$ from \stwo\ and substitute the solution
of section 2.1, we find that
the lagrangian governing the dynamics of the test string is
 given by
\eqn\stradlag{{\cal L}_2=-T_2 e^{2\phi}\left(\sqrt{\dot t^2-\dot y^2
e^{-2\phi}}\mp \dot t\right),}
where the minus (plus) sign corresponds to the parallel (antiparallel)
configuration. The time derivative is with respect to $\xi^0$,
which
we choose to be the proper time $\tau$ measured by a clock at rest in
the frame of the test string. From section 2.1 this is given
by
\eqn\stlinel{d\tau^2=-e^{\phi/2}ds^2=e^{2\phi}dt^2-dy^2.}
Thus the calculation has been reduced to a one-dimensional problem and
the dynamics of \stradlag\ is similar to that of a point particle whose
mass is equal or opposite to its electric charge. Since there is
no explicit time-dependence in ${\cal L}_2$, we have the following constant of
the motion
\eqn\estst{{\partial{\cal L}_2\over \partial \dot t}=-T_2 e^{2\phi}
\left({\dot t\over \sqrt{\dot t^2-\dot y^2 e^{-2\phi}}}\mp 1\right)=
-T_2E.}
$E$ is the constant energy per unit mass of the motion and is determined
from the intial conditions. Note that for the parallel strings case, we
recover the zero static force result by noting that if $\dot y=0$ initially,
then $E=0$ and $\dot y=0$ everywhere. We also recover
the vanishing leading order (in the velocity) dynamic force result found in
\calk.
{}From \estst\ we obtain an expression for the coordinate velocity
\eqn\ssv{\left({dy\over dt}\right)^2={E^2 e^{-2\phi}\pm 2E
\over \left(E e^{-2\phi}\pm 1\right)^2}.}
We now wish to relate the radial position to the proper time.
Combining \ssv\ and \stlinel\ we obtain
\eqn\sspv{\left({dy\over d\tau}\right)^2=e^{-2\phi}\left(E^2 e^{-2\phi}\pm
2E\right)}
for the proper velocity in terms of the radial position. The
acceleration can be obtained by differentiating \sspv\ with respect to
$\tau$ and replacing \sspv\ in the resulting expression. The acceleration
written in terms of the position is independent of the sign of the velocity
and is given by
\eqn\sspa{{d^2y\over d\tau^2}=-{6k_2E^2\over y^7}\left(1+{k_2\over y^6}
\pm E^{-1}\right).}
For parallel strings, the force is always attractive when initially
$\dot y\neq 0$. For antiparallel strings,
the acceleration is always inward, and the test string does
indeed fall towards the source string. We may thus choose the negative
sign for the square roots in \ssv\ and \sspv. To calculate the proper time
taken for the test string to reach the source string, we rewrite \sspv\
and integrate
\eqn\sspt{\tau_0=\int_0^{\tau_0}d\tau=\int_0^{y_0}{dy\over
\sqrt{e^{-2\phi}\left(E^2e^{-2\phi}\pm 2E\right)}}.}
On using the expression for $\phi$ from section 2.1, we
note that $\tau_0$ is finite.
Thus the test string falls into the source string
in a finite amount of time, and the singularity is real. In particular,
let us focus on the case where the test string is antiparallel to the source
string. If $\dot y=0$ at $y=y_0$, then
\eqn\estinit{E=2e^{2\phi(y_0)}={2\over 1+k_2/y_0^6}.}
Let $x\equiv y/y_0$, then $\tau_0$ can be written as
\eqn\sspttwo{\tau_0={e^{-2\phi(y_0)}y_0^4\over 2\sqrt{k_2}}
\int_0^1 {dx x^6\over \sqrt{(x^6+k_2/y_0^6)(1-x^6)}}.}
For large $y_0$, we find that $\tau_0\sim k_2^{-1/2}$. Since the mass per
unit length of the string is given by ${\cal M}_2=T_2$ \dabghr, this means
that
$\tau_0\sim {\cal M}_2^{-1/2}$, which is the same dependence of the time on the
 mass
for an observer falling into a Schwarzschild black hole. Just as for the
black hole case, moreover, it is easy to see from \sspv\ and \sspa\ that
the proper velocity and acceleration both tend to infinity as the
test string approaches the singularity.
To further strengthen the analogy with a black hole-type singularity,
one can calculate the elapsed distant observer time for the fall. In this case
one can easily show that $t_0\to\infty$, $dy/dt\to 0$ and $d^2y/dt^2\to 0$
as the test string approaches the singularity. In other words, the distant
observer never sees the test string reach the singularity. In this case,
the event horizon is at the singularity.

We shall now repeat the above calculation for a test-fivebrane falling
radially into a source fivebrane oriented along $x^a=\xi^a$, ($a=1,...,5$).
Again, we let the test fivebrane lie either parallel or antiparallel
to the source fivebrane, i.e. with the same or opposite orientation. If
we eliminate $\gamma_{ij}$ from \Ssix\ and the fivebrane solution of section
2.4, we find that the
lagrangian governing the dynamics of the test fivebrane is given by
\eqn\fbradlag{{\cal L}_6=-T_6 e^{-2\phi}\left(\sqrt{\dot t^2-\dot y^2
e^{2\phi}}
\mp \dot t\right),}
where the minus (plus) sign corresponds to the parallel (antiparallel)
configuration. The time derivative is with respect to $\xi^0$, which
we choose to be the proper time $\tau$ measured by a clock at rest in
the frame of the test fivebrane. From section 2.4 this is given
by
\eqn\fbline{d\tau^2=-e^{-\phi/6}ds^2=e^{-2\phi/3}\left(dt^2-e^{2\phi}dy^2
\right).}
This time the Euler-Lagrange equations yield the following constant of the
motion
\eqn\efbfb{{\partial{\cal L}_6\over \partial \dot t}=-T_6 e^{-2\phi}
\left({\dot t\over \sqrt{\dot t^2-\dot y^2 e^{2\phi}}}\mp 1\right)=
-T_6E.}
{}From \efbfb\ we obtain an expression for the coordinate velocity
\eqn\ffv{\left({dy\over dt}\right)^2={E^2 e^{2\phi}\pm 2E\over
\left(E e^{2\phi}\pm 1\right)^2}.}
Combining \ffv\ and \fblinel\ we obtain
\eqn\ffpv{\left({dy\over d\tau}\right)^2=e^{2\phi/3}
\left(E^2 e^{2\phi}\pm 2E\right)}
for the proper velocity in terms of the radial position. The
acceleration can be obtained by differentiating \ffpv\ with
respect to
$\tau$ and replacing \ffpv\ in the resulting expression. The
 acceleration
written in terms of the position is independent of the sign of
the velocity
as in the string-string case and is again attractive, so the
test fivebrane
 does
indeed fall towards the source fivebrane. To calculate the
proper time
taken for the test fivebrane to reach the source fivebrane we
 rewrite \ffpv\
and integrate
\eqn\ffpt{\tau_0=\int_0^{\tau_0}d\tau=\int_0^{y_0}{dy\over
\sqrt{e^{2\phi/3}\left(E^2e^{2\phi}\pm 2E\right)}}.}
On using the expression for $\phi$ from section 2.4, we note
again that $\tau_0$ is manifestly finite.
Thus the test fivebrane falls into the source fivebrane
in a finite amount of time, and the singularity is real. In the
antiparallel
case, the dependence of the time on the mass of a source for
the test fivebrane
initially at rest and for large initial separation is again of
the form
$\tau_0\sim k_6^{-1/2}$. Since the mass per unit $5$-volume of
 the fivebrane
is given by ${\cal M}_6=T_6$ \duflfb, this means $\tau_0\sim
{\cal M}_6^{-1/2}$ as for the
string. It is easy to see
that the proper velocity and acceleration both $\to\infty$ as
 the
test fivebrane approaches the singularity
and that $t_0\to\infty$, $dy/dt\to 0$ and $d^2y/dt^2\to 0$.
Once more, the event horizon is located at the singularity.

An entirely different state of affairs holds for a test string
 moving
in the background of a source fivebrane or, by duality, a test
fivebrane
moving in the background of a source string. In this case, the
test probe
takes an infinite amount of proper time to reach the source.
First we consider the trajectory of a test string falling
radially into a
source fivebrane, oriented along $x^a=\xi^a$ ($a=1,2,...,5$).
Let the test
string lie either parallel or antiparallel to one of the
fivebrane directions,
say $x^1$. From section 2.4, the only nonvanishing components
of $K$
are of the form $K_{012345m}$, where the directions $m=6,7,8,9$
are transverse to the
fivebrane. By dualizing, we see that the only nonzero
components of
$H=dB$ are $H_{pqr}(y)$, where again, $p,q,r=6,7,8,9$. It
follows that
the only nonzero components of $B_{MN}$ occur when $M,N=6,7,8,9$.
 It then
follows that the WZW term $\epsilon^{ij}\partial_iX^M
\partial_jX^NB_{MN}$
vanishes. Substituting the fivebrane solution in \stwo, we find that the test
string
lagrangian reduces to
\eqn\stfblag{{\cal L}_2=-T_2\sqrt{\dot t^2-e^{2\phi}\dot y^2}}
for purely radial motion. From \dflufbsol\ and \gstsm, the proper time
is given by
\eqn\stfblinel{d\tau^2=-e^{\phi/2}ds^2=dt^2-e^{2\phi}dy^2.}
Again we have a constant of the motion
\eqn\stfbenergy{{\partial{\cal L}_2\over \partial\dot t}=-T_2{\dot t\over
\sqrt{\dot t^2-e^{2\phi}\dot y^2}}=-T_2 E.}
Note that $E=1$ corresponds to a zero static force.
We invert \stfbenergy\ to obtain the coordinate velocity
\eqn\sfv{\left({dy\over dt}\right)^2=e^{-2\phi}\left(1-1/E^2\right).}
Combining \stfblinel\ and \stfbenergy, we obtain the proper velocity
\eqn\sfpv{\left({dy\over d\tau}\right)^2=\left(E^2-1\right)e^{-2\phi}.}
The acceleration is given by
\eqn\sfpa{{d^2y\over d\tau^2}={k_6(E^2-1)e^{-4\phi}\over y^3}.}
Note that the acceleration is repulsive in this case.
In both the $y\to 0$ and $y\to\infty$ limits, the acceleration vanishes
(an asymptotic freedom of some sort). We assume that the string is directed
towards the fivebrane initially. The time taken for the fall from
an initial position $y_0$
\eqn\sfpt{\tau_0={1\over\sqrt{E^2-1}}\int_0^{y_0} e^\phi dy}
diverges logarithmically with $y$. Therefore it takes the string an
infinite amount of proper time to reach the singularity. In other words,
the string never sees the singularity, and as far as the string is concerned,
the singularity is not real.

An analogous calculation for a test fivebrane falling
towards a source string shows that the string is nonsingular
as a source for fivebranes. For a test fivebrane with one of its
spatial directions parallel to the string, the WZW term again vanishes, as
in the above case. In this case, substituting the string solution into \Ssix,
the lagrangian reduces to
\eqn\fbstlag{{\cal L}_6=-T_6\sqrt{\dot t^2-e^{-2\phi}\dot y^2}}
for purely radial motion. From the \dabsol\ and \fbomcan, the proper time is
given by
\eqn\fbstlinel{d\tau^2=-e^{-\phi/6}ds^2=e^{4\phi/3}\left(dt^2-e^{-2\phi}dy^2
\right).}
Again we have a constant of the motion
\eqn\fbstenergy{{\partial{\cal L}_6\over \partial\dot t}=-T_6{\dot t\over
\sqrt{\dot t^2-e^{2\phi}\dot y^2}}=-T_6 E.}
Again $E=1$ corresponds to a zero static force.
We invert \fbstenergy\ to obtain the coordinate velocity
\eqn\fsv{\left({dy\over dt}\right)^2=e^{2\phi}\left(1-1/E^2\right).}
Combining \fsv\ and \fbstlinel\ we obtain the proper velocity
\eqn\fspv{\left({dy\over d\tau}\right)^2=\left(E^2-1\right)e^{2\phi/3}.}
The acceleration is again found to be repulsive and is given by
\eqn\fspa{{d^2y\over d\tau^2}={k_2(E^2-1)e^{8\phi/3}\over y^7}.}
Again the acceleration vanishes in both the $y\to 0$ and $y\to\infty$ limits.
Now assume that the fivebrane is directed
towards the string initially. The time taken for the fall from
an initial position $y_0$ is given by
\eqn\fspt{\tau_0={1\over\sqrt{E^2-1}}\int_0^{y_0} e^{-\phi/3} dr,}
and again diverges logarithmically with $y$. Therefore it takes the
fivebrane an infinite amount of proper time to reach the singularity. In
other words, the fivebrane never sees the singularity, and as far as the
fivebrane is concerned, the string singularity is not real.

We have seen that, as far as singularities are concerned, the superstring
and the superfivebrane solitons are on an equal footing: the fivebrane is
a singular solution of fivebrane theory but a non-singular solution of
string theory while the string is a singular solution of string theory
but a non-singular solution of fivebrane theory. What is asymmetric,
however, is the state of current technology. One can prove rigorously
that $I_{10}({\rm string})$ describes the field theory limit of string
theory and that the string loop coupling constant is, from \gstsm\
given by $g({\rm string})=e^{\phi_0}$; one has only plausibility arguments
that the dual action $I_{10}({\rm fivebrane})$ describes the field-theory limit
of fivebrane theory and that the fivebrane loop coupling constant is,
from \fbomcan, $g({\rm fivebrane})=e^{-\phi_0/3}$ and hence that the strong
coupling limit of the string corresponds to the weakly coupled fivebrane and
vice-versa \duflrsfd.

\newsec{Strings and fivebranes as interpolating solitons}

   In this section we discuss how in string $\sigma$-model metric, the
solitonic fivebrane
  solution of $D=10$ supergravity interpolates \gibt\ between $D=10$ Minkowski
  spacetime and a supersymmetric $S^3$ compactification to a linear
  dilaton vacuum \duftv. Furthermore, in {\it fivebrane} sigma-model
  metric, the solitonic string solution of $D=10$ supergravity
  interpolates \dufgt\ between $D=10$ Minkowski spacetime and a
  supersymmetric $S^7$ compactification to a
  three-dimensional anti-de Sitter generalization of the linear dilaton
  vacuum, which may be invariantly characterized in terms of conformal
  Killing vectors \duftv.

  Let us first consider the solitonic fivebrane solution of section 2.4 in
string variables rewritten in terms of a new radial coordinate $r^2=y^2+b^2$
where $b^2=k_6$
  \eqn\onetwo{
  \eqalign{
  ds^2 &= -dt^2 + d{\bf x}\cdot d{\bf x} +
  \left(1-{b^2\over r^2}\right)^{-2} dr^2 + r^2 d\Omega_3^2\cr
  e^{-2\phi} &= 1-{b^2\over r^2}\cr
  H &=2b^2\varepsilon_3\ ,\cr}}
    where $d{\bf x}\cdot d{\bf x}$ is the Euclidean 5-metric,
  $d\Omega_3^2$ the standard metric on the unit 3-sphere and
  $\varepsilon_3$ its volume 3-form.
To determine the asymptotic form of the metric near $r=b$ we let
\eqn\rarhofb{r=b\left(1+{e^{2\rho/b}\over 2}\right).}
In the limit $r=b$ \onetwo\ reduces to
\eqn\fbinter{\eqalign{
  ds^2 &\sim (-dt^2 +d{\bf x}\cdot d{\bf x} +
  d\rho^2) + b^2d\Omega_3^2\cr
  \phi &\sim -{1\over b}\rho\cr
  H &\sim 2b^2 \varepsilon_{3}\cr}}
which is $M_7 \times S^3$, with a linear dilaton
  vacuum. That is, in string sigma-model metric the
  solitonic fivebrane interpolates between ten-dimensional
  Minkowski spacetime and the product of $S^3$ with a seven-dimensional
{\it Minkowski} spacetime.

Next we consider the solitonic string solution of section 2.5 in fivebrane
variables written in terms of a new radial coordinate
$r^6=y^6+b^6$ where $b^6=k_2$
  \eqn\onethree{\eqalign{ds^2 &= \left(1-{b^6\over r^6}\right)^{2\over3}
\left(-dt^2
  +d\sigma^2\right) + \left(1-{b^6\over r^6}\right)^{-2} dr^2 + r^2
  d\Omega_7^2 \ \cr
  e^{-2\phi}&= \left(1-{b^6\over r^6}\right)^{-1}\cr
  K&= 6b^6\varepsilon_7. \cr}}
  This time we define
\eqn\rarhost{r=a\left(1+{e^{6\rho/b}\over 6}\right).}
 Near $r=b$ \onethree\ reduces to
\eqn\stinter{\eqalign{
  ds^2 &\sim e^{4\rho/b}\left(-dt^2 +d\sigma^2\right) +
  d\rho^2 + b^2d\Omega_7^2\cr
  \phi &\sim {3\over b}\rho\cr
  K &\sim 6b^6 \varepsilon_{7}.\cr}}
The metric is the standard metric on the product of $S_7$ with
three-dimensional anti-de Sitter space, $adS_3$.
Clearly, the dilaton is linear only for the
  special choice of coordinates used here, and since the spacetime factor
  of the asymptotic metric is not Minkowski (as it was for the fivebrane
  solution) but rather $adS_3$, the geometrical significance of
  linearity is unclear from the above result.
  We verify below that the $S^7$ compactification of D=10 supergravity
  implied by this analysis indeed exists, incidentally elucidating the
  geometrical significance of `linear' in the $adS$ context.

      Let $\{ x^M; \, M=0,1,\dots ,9\}$ be coordinates for the
  ten-dimensional spacetime, and define
  \eqn\twoone{
  \Phi= e^{2\phi/3}.}
  The field equations of \sfbrane\ can now be written as
  \eqn\twotwoa{  R_{MN} = \Phi^{-1}D_MD_N\Phi
  + {1\over 2. 6!}K_{MP_1\dots P_6}K_N{}^{P_1\dots P_6} -
  {1\over 3.7!}g_{MN} K^2}
    \eqn\twotwob{\partial_M (\sqrt{-g}\Phi K^{MN_1\dots N_6})=0 }
   \eqn\twotwoc{\Box\Phi =  {1\over 3.7!}K^2 \Phi .}

  We now split the coordinates $\{x^M\}$ into $\{x^\mu, y^m\}$ with
  $\mu=0,1,2$ and $m=1,\dots, 7$, and seek product metrics of the form
 \eqn\twothree{  g_{\mu\nu}= g_{\mu\nu}(x) \qquad g_{mn} =g_{mn}(y)}
    with $g_{\mu n}=0$. In this case, $\sqrt{-g}=e_3(x) e_7(y)$ where
  $e_3$ and $e_7$ are the scalar density volume factors for the three
  and seven-dimensional spaces, respectively. It follows that \twotwob\ is
  now solved by setting
  \eqn\twofour{
  K^{m_1\dots m_7} = 3m(e_7)^{-1}\varepsilon^{m_1\dots m_7}
  }
  for some constant $m$, with all other components of $K$ vanishing,
  whereupon \twotwoc\ becomes
  \eqn\twofive{
  (\Box -3m^2)\Phi=0
  }
  We now suppose that $\Phi=\Phi(x)$, and further that
  \eqn\twosix{
  D_\mu \partial_\nu \Phi ={1\over 3} g_{\mu\nu} \Box \Phi
  }
  or, equivalently, in view of \twofive,
  \eqn\twoseven{
  D_\mu \partial_\nu \Phi = m^2 g_{\mu\nu}\Phi\
  }
    which implies \twofive\ and therefore supercedes it. Given \twoseven,
\twotwoa\
  reduces to the two equations
  \eqn\twoeight{
  R_{\mu\nu} = -2m^2 g_{\mu\nu} \qquad R_{mn} = {3\over 2} m^2 g_{mn}
  }
  with $R_{m\nu}=0$. These equations are solved by the standard
  invariant metrics on $adS_3$ and $S^7$ respectively. It remains to
  solve \twoseven. We can choose coordinates $(t,\sigma, \rho)$ on $adS_3$
  such that the $adS_3$ metric is
  \eqn\twonine{
  ds^2 =  e^{2m\rho}\left(-dt^2 + d\sigma^2\right) + d\rho^2
  }
  In these coordinates it is straightforward to verify that
  \eqn\twoten{
  \phi = {3m\over 2}\rho
  }
  solves \twoseven. We have therefore found an $S^7$ compactification of
  $D=10$ supergravity to $adS_3$ with a linear dilaton. Setting $m=2/b$
  we recover \stinter, found previously as an asymptotic limit of
  the extreme string solution.

  Observe now that \twosix\ implies that $ g^{\mu\nu}\partial_\mu
  \Phi\,\partial_\nu $ is a {\it conformal Killing vector} of $adS_3$. We
  have shown above that an eigenfunction of the Dalembertian on $adS_3$
  with eigenvalue $3m^2$ is the potential for a conformal Killing vector
  of $adS_3$. A similar observation was made previously in the context of
  an $S^3$ compactification of $D=10$ supergravity \duftv\ to $adS_7$.
  In fact, both the $S^3$ and $S^7$ compactifications exhibited above are
obtained from the
  solutions found in \duftv\ by the
  analytic continuation $m\rightarrow im$. We now see that
  the linear dilaton can be characterized in a coordinate-free way as
  proportional to the logarithm of a conformal Killing potential.

\chapter{(d-1)-branes in diverse dimensions}

\def\dg{\hbox{$^\dagger$}}

\def\half{{1\over2}}

\def\goto#1{\mathrel {\mathop {\longrightarrow} \limits_{#1}}}

\def\a{\alpha}
\def\b{\beta}

\def\p{\partial}
\def\goto#1{\mathrel {\mathop {\longrightarrow} \limits_{#1}}}
\def\lr{\goto{r\to\infty}}

\newsec{Diverse dimensions}

In the previous section, we focused on spacetime dimension $D=10$,
worldvolume dimension $d=2$ and dual worldvolume dimension $\tilde
d=8-2=6$.
The purpose of the present section is to see how far these results
generalize to arbitrary $D$, $d$, and $\tilde d$. We present a
generic Lagrangian, in arbitrary spacetime dimension $D$, describing
the interaction of a dilaton, a graviton and an antisymmetric tensor
of arbitrary rank $d$.  For each $D$~and~$d$, we find black
$\tilde{p}$-brane solutions where $\tilde{p} = \tilde{d} -
1$~and~$\tilde d = D - d - 2$.  These solutions display a spacetime
singularity surrounded by an event horizon, and are characterized by
a mass per unit $\tilde p$-volume ${\cal M}_{\tilde{d}}$, and
topological ``magnetic'' charge $g_{\tilde{d}}$, obeying
$\sqrt{2}\kappa {\cal M}_{\tilde{d}} \geq g_{\tilde{d}}$.
 The theory also
admits
elementary $p$-brane solutions with ``electric'' Noether charge
$e_d$, obeying
the Dirac quantization rule $e_d g_{\tilde{d}} = 2\pi n$, $n
=$~integer.  We
then present the Lagrangian describing the theory dual to the
original theory,
whose antisymmetric tensor has rank $\tilde{d}$ and for which the
roles of
topological and elementary solutions are interchanged. In the extreme
limits
$\sqrt{2}\kappa {\cal M}_{\tilde{d}} = g_{\tilde{d}}$ or
$\sqrt{2}\kappa {\cal M}_d = e_d$, the
singularity and
event horizon coalesce. As discussed in section 4,
for specific values of $D$~and~$d$, these extreme solutions also
exhibit supersymmetry and may be identified with previously
classified
heterotic, Type IIA and Type IIB super $\tilde p$-branes. Curiously
enough,
the results obtained in section 2 on the singularity structure of
strings and fivebranes do not generalize to arbitrary $d$.

In section 3.2, we write down a general
action in $D$ spacetime dimensions describing the interaction of an
antisymmetric tensor potential of
rank $d$ with gravity and a dilaton.  We allow these fields to couple
to an elementary
$d$-dimensional extended object, (a $p$-brane, with $d = p + 1$) and
define an
``electric'' Noether charge associated with it.  This construction
involves, in particular, the identification of the $p$-brane
$\sigma$-model metric in terms of the dilaton and the canonical
metric.

In section 3.3, we show how the combined field equations admit
solutions describing such elementary objects, in much the same way as
the elementary string emerged as a solution of supergravity coupled
to a string $\sigma$-model source in section 2.1. The mass per unit
$p$-volume ${\cal M}_d$ and the Noether charge $e_d$ of these
elementary solutions satisfy the equality
$\sqrt{2}\kappa {\cal M}_d=|e_d|$, which is the same relation found
in  section 2 for the supersymmetric string and fivebrane. One may
also demonstrate a no-force condition by showing that the mutual
gravitational-dilaton attraction of two such $p$-branes of the same
orientation is exactly cancelled by an equal and opposite
contribution from the antisymmetric tensor.  This permits the
construction of stable
multi-$p$-brane solutions.

In addition to the singular elementary $(d - 1)$-brane solutions
carrying
non-zero ``electric'' Noether charge $e_d$, the theory also admits
non-singular soliton $(\tilde{d}- 1$)-brane solutions, where
$\tilde{d} = D - d - 2$.  As described in section 3.4, these
solutions are dual to the elementary solutions and carry a non-zero
``magnetic'' topological charge $g_{\tilde{d}}$, obeying the Dirac
quantization rule \refs{\nep,\tei}
\eqn\diracquan{e_d g_{\tilde{d}} = 2\pi n, \qquad  n = {\rm
integer}.}

In section 3.5 we consider the theory dual to the theory of section
3.2, for which the roles of antisymmetric tensor field equations and
Bianchi identities, and hence electric and magnetic charges, are
interchanged.  This, together with our result for the $\sigma$-model
metric of section 3.2, leads to a relation between the loop expansion
parameter ${\rm g}_d$ of the $(d - 1)$-brane and the loop expansion
parameter ${\rm g}_{\tilde{d}}$ of the $(\tilde{d} - 1)$-brane.  We
find
\eqn\looparam{{\rm g}_d^d = 1/{\rm g}_{\tilde{d}}^{\tilde{d}},}
thus confirming that strongly coupled $(d - 1)$ branes correspond to
weakly
coupled $(\tilde{d} - 1)$-branes and vice versa. The introduction of
the $\sigma$-model metric permits a useful comparison with
Brans-Dicke theory.

$D=6$ is of special interest because in this dimension a string is
dual to another string. Moreover, as discussed in section 3.6, this
permits the construction of a ``dyonic'' \duffkr\ solution carrying
both electric charge
$e_2$ and magnetic charge $g_2$. In the limit $e_2=g_2$, this
solution reduces to the previously derived self-dual superstring
\duflblacks.

Although many of the results of section 2 generalize in an obvious
way to arbitrary $d$ and $D$, this is not the case for the issue of
singularities. We show in section 3.7 that although the curvature of
the $(\tilde d-1)$-brane metric written in the dual $(d-1)$-brane
$\sigma$-model variables is still finite
at $y=0$ for all $d$ and $D$, the proper time for an infalling test
$(d-1)$-brane to reach a source $(\tilde d-1)$-brane is finite only for
particles and strings and their duals.

Black $p$-branes, solutions with ${\cal M}_{\tilde{d}} \geq {1\over
\sqrt{2}} g_{\tilde{d}}$ are discussed in section 3.8.  These
solutions exhibit singularities shielded by an event horizon.  As
special cases, we recover the $D = 10$ black $p$-branes $(p = 0,
\ldots 6)$ of
 \hors, the $D = 11$ black $p$-branes $(p = 2, 5)$ of \guv\ and the
$D = 4$ Kaluza-Klein black hole $(p = 1)$
\refs{\gib,\dobm,\chod,\pol}.

\newsec{The action}

Consider an antisymmetric tensor potential of rank $d,~A_{M_1 M_2
\ldots M_d}$,
in $D$
spacetime dimensions ($M = 0, 1, \ldots (D-1)$) interacting with
gravity,
$g_{MN}$, and
the dilaton, $\phi$, via the action
\eqn\lgeact{I_D (d) = {1\over 2\kappa^2} \int d^D x \sqrt{-g} \Bigg(R
-
{1\over 2} (\partial\phi)^2 -
{1\over 2(d+1)!} e^{-a (d) \phi} F^2_{d+1}\Bigg),}
where the rank ($d+1$) field strength $F_{d+1}$ is given by
\eqn\defF{F_{d+1} = dA_d\hfil,}
and $a (d)$ is an, as yet undetermined, constant.  Special cases
of
this action have
been considered before in the context of classical solutions
\gibm.  Here we
keep both
$D$ and $d$ arbitrary.  We allow these fields to couple to an
elementary
$d$-dimensional
extended object (a ``($d-1$)-brane'') whose trajectory is given by
$X^M (\xi^i)~(i = 0, 1,\ldots (d-1))$, worldvolume metric by
$\gamma_{ij} (\xi)$, and tension by $T_d$,
 via the
action
\eqn\gepbact{\eqalign{S_d = T_d \int d^d\xi \Bigg(&-{1\over 2}
\sqrt{-\gamma}
 \gamma^{ij}
\partial_i X^M \partial_j X^N g_{MN} e^{a(d)\phi/d} +
{(d-2)\over 2}
\sqrt{-\gamma}\cr
&-{1\over d!} \varepsilon^{i_1 i_2 \ldots i_d} \partial_{i_1} X^{M_1}
\partial_{i_2}
X^{M_2} \ldots \partial_{i_d} X^{M_d} A_{M_1 M_2 \ldots
M_d}\Bigg).\cr}}
The $\phi$ dependence is chosen so that under the rescaling
\eqn\resclaw{\eqalign{g_{MN}&\rightarrow \lambda^{2d/(D-2)}
g_{MN},\cr
A_{M_1 M_2 \ldots M_d}&\rightarrow \lambda^d A_{M_1 M_2 \ldots
M_d},\cr
e^{\phi}&\rightarrow \lambda^{2d (D-d-2)/(D-2)a (d)}
e^{\phi},\cr
\gamma_{ij}&\rightarrow \lambda^2 \gamma_{ij},\cr}}
both actions scale the same way
\eqn\actscal{\eqalign{I_D (d)&\rightarrow \lambda^d I_D (d),\cr
S_d&\rightarrow \lambda^d S.\cr}}

The field equations and Bianchi identities of the $A$ field may be
written
\eqn\fieldeq{d^{\ast} (e^{-a (d)\phi} F) = 2\kappa^2
(-)^{d^2}~^{\ast}J,}
\eqn\bianchi{dF \equiv 0,}
where the rank $d$ source $J$ is given by
\eqn\defj{J^{M_1 \ldots M_d} = T_d \int d^d \xi
\varepsilon^{i_1 i_2 \ldots i_d}
\partial_{i_1} X^{M_1} \partial_{i_2} X^{M_2} \ldots \partial_{i_d}
X^{M_d}{\delta^D (x-X)\over \sqrt{-g}}.}
Let us introduce the dual worldvolume dimension, $\tilde{d}$, by
\eqn\dtilde{\tilde{d} \equiv D - d - 2.}
We may now define two conserved charges:  the Noether ``electric''
charge
\eqn\elcharge{e_d = {1\over \sqrt{2}\kappa}
\int\limits_{S^{\tilde{d}+1}}
e^{-a (d)\phi}{}^{\ast}F,}
where $S^{\tilde{d}+1}$ is the ($\tilde{d}+1$)-sphere surrounding the
elementary
 ($d-1$)
brane, and the topological ``magnetic'' charge
\eqn\macharge{g_{\tilde{d}} = {1\over \sqrt{2} \kappa}
\int\limits_{S^{d+1}}
F.}
This latter charge will be non-zero if the action $I_D$ admits a
solitonic
$\tilde{d}$-dimensional extended object (a
``($\tilde{d}-1$)-brane'').  These
charges
obey a Dirac quantization condition \refs{\nep,\tei},
\eqn\charquan{{e_d g_{\tilde{d}}\over 4\pi} = {n\over 2}, \qquad n =
{\rm integer}}
analogous to the ($d=1, D=4$) condition that relates electric and
magnetic
charges.  At
this stage, of course, it is not yet obvious that the system admits
either
elementary or
solitonic extended object solutions, nor if they do, what are the
values of
the
electric
and magnetic charges $e_d$ and $g_{\tilde{d}}$.

Let us first consider the field equations resulting from $I_D + S_d$.
The Einstein
equation is
\eqn\einstein{\eqalign{\sqrt{-g}&\Bigg[R^{MN} - {1\over 2} g^{MN} R -
{1\over
2} (\partial^M \phi\partial^N \phi - {1\over 2} g^{MN}
(\partial\phi)^2)\cr
&~-{1\over 2} {1\over d!} (F^M\,_{M_1 \ldots M_d} F^{NM_1 \ldots M_d}
- {1\over
2(d+1)}
g^{MN} F^2) e^{-a (d)\phi}\Bigg]\cr
&=\kappa^2 \sqrt{-g} T^{MN} ((d-1)-{\rm brane}),\cr}}
where the energy-momentum tensor is given by
\eqn\engmtensor{T^{MN} ((d-1)-{\rm brane}) = -T_d \int d^d \xi
\sqrt{-\gamma}
 \gamma^{ij} \partial_i
X^M \partial_j X^N e^{a\phi/d} {\delta^D (x-X)\over \sqrt{-g}},}
the antisymmetric tensor equation is
\eqn\antieq{\partial_M (\sqrt{-g} e^{-a\phi} F^{M M_1 \ldots
M_d}) =
 2\kappa^2 T_d \int d^d \xi
\varepsilon^{i_1 \ldots i_d} \partial_{i_1} X^{M_1} \ldots
\partial_{i_d}
X^{M_d}
\delta^D (x-X),}
and the dilaton equation is
\eqn\dilaeq{\eqalign{&\partial_M (\sqrt{-g} g^{MN} \partial_N \phi)+
{a (d)\over 2(d+1)!}\sqrt{-g} e^{-a (d)\phi} F^2 \cr
&={a (d) \kappa^2 T_d\over d}\int d^d \xi \sqrt{-\gamma}
\gamma^{ij}
\partial_i X^M \partial_j X^N g_{MN} e^{a (d)\phi/d} \delta^D
(x-X).\cr}}
Furthermore, the ($d-1$)-brane field equations are
\eqn\braneq{\eqalign{\partial_i (\sqrt{-\gamma} \gamma^{ij}
\partial_j X^N
g_{MN}e^{a (d)\phi/d})&-{1\over 2} \sqrt{-\gamma} \gamma^{ij}
\partial_i X^N
\partial_j X^P\partial_M (g_{NP} e^{a (d)\phi/d})\cr
&-{1\over d!} \varepsilon^{i_1 \ldots i_d} \partial_{i_1} X^{M_1}
\ldots
\partial_{i_d} X^{M_d} F_{M M_1 \ldots M_d} = 0,\cr}}
and
\eqn\defgama{\gamma_{ij} = \partial_i X^M \partial_j X^N g_{MN}
e^{a (d)\phi/d}.}
\newsec{The elementary ($d-1$)-brane}

To solve these coupled field-$(d-1)$-brane equations we begin by
making an
ansatz
for the $D$-dimensional metric $g_{MN}$, $d$-form $A_{M_1} \ldots
M_d$,
dilaton
$\phi$
and coordinates $X^M (\xi)$ corresponding to the most general
$d/(D-d)$ split
invariant
under $P_d \times SO (D-d)$ where $P_d$ is the $d$-dimensional
Poincar\'e
group.
  We
split the indices
\eqn\splitx{x^M = (x^{\mu}, y^m),}
where $\mu = 0, 1 \ldots (d-1)$ and $m = d, d+1, \ldots (D-1)$, and
write the
line-element as
\eqn\supmetr{ds^2 = e^{2A} \eta_{\mu\nu} dx^{\mu} dx^{\nu} + e^{2B}
\delta_{mn}
 dy^m dy^n,}
and the $d$-form gauge field as
\eqn\dform{A_{\mu_1 \ldots \mu_d} = - {1\over ^dg} \varepsilon_{\mu_1
\ldots
\mu_d} e^C,}
where $^dg$ is the determinant of $g_{\mu\nu}$, $\varepsilon_{\mu_1
\ldots
\mu_d} \equiv g_{\mu_1 \nu_1} \ldots g_{\mu_d \nu_d}
\varepsilon^{\nu_1 \ldots \nu_d}$ and
$\varepsilon^{012 \ldots (d-1)} = 1$ i.e. $A_{01 \ldots (d-1)} = -
e^C$.  All
other
components of $A_{M_1 \ldots M_d}$ are set to zero.  $P_d$ invariance
requires
that the
arbitrary functions A, B, C depend only on $y^m$; $SO(D-d)$
invariance then
requires that
this dependence be only through $y = \sqrt{\delta_{mn} y^m y^n}$.
Similarly our
 ansatz
for the dilaton is
\eqn\ansaph{\phi = \phi (y).}
In the ($d-1$)-brane sector we also split
\eqn\bransplit{X^M = (X^{\mu}, Y^m),}
and make the static gauge choice
\eqn\staticgauge{X^{\mu} = \xi^{\mu},}
and the ansatz
\eqn\ansatzy{Y_m = {\rm constant}.}
Substituting these ansatz into \defgama\ yields
\eqn\newgama{\gamma_{ij} = e^{2A + a(d)\phi/d} \eta_{ij},}
and the only non-vanishing components of the field strength are
\eqn\fieldstr{F_{m\mu_1 \ldots \mu_d} = - {1\over ^dg}
\varepsilon_{\mu_1
\ldots \mu_d} \partial_m e^C.}
Then the $\mu\nu$ components of the Einstein equation \einstein\
reduce to a
 single equation
\eqn\munucomp{\eqalign{&e^{(d-2) A + {\tilde d} B} \delta^{mn}
\bigg[(d-1)
\partial_m \partial_n A +
{d(d-1)\over 2} \partial_m A \partial_n A + ({\tilde d} + 1)
\partial_m \partial_n B\cr
&+{({\tilde d} + 1) {\tilde d}\over 2} \partial_m B \partial_n B +
\tilde d
(d-1) \partial_m A\partial_n B\cr
&+ {1\over 4} e^{-2dA + 2C - a (d)\phi} \partial_m C \partial_n
C +
{1\over 4} \partial_m \phi \partial_n \phi\bigg]\cr
&= - \kappa^2 T_d e^{(d-2) A + a(d) \phi/2} \delta^{D-d}
(y),\cr}}
and the $mn$ components reduce to
\eqn\mncomp{\eqalign{&e^{dA + (\tilde d - 2) B} \bigg[-\tilde d
\partial^m \partial^n B + \delta^{m n}\tilde d  \delta^{kl}
\partial_k \partial_l B\cr
&-d\partial^m \partial^n A + d \delta^{mn} \delta^{kl} \partial_k
\partial_l A
 - d\partial^m A \partial^n A + {d(d+1)\over 2} \delta^{mn}
\delta^{kl}
\partial_k A\partial_l A\cr
&+d(\partial^m A \partial^n B + \partial^m B \partial^n A + (\tilde d
- 1)
\delta^{mn} \delta^{kl} \partial_k A \partial_l B)\cr
&-{1\over 2} \partial^m \phi \partial^n \phi + {1\over 4} \delta^{mn}
\delta^{kl} \partial_k \phi \partial_l \phi\bigg] \cr
&- {1\over 2} e^{-dA + (\tilde d - 2) B +
2C - a (d)\phi}
\bigg[-\partial^m C \partial^n C + {1\over 2} \delta^{mn} \delta^{kl}
\partial_k C \partial_l C\bigg]\cr &=0.\cr}}
The antisymmetric tensor field equation \antieq\ becomes
\eqn\newanti{\delta^{mn} \partial_m \bigg[e^{-a (d)\phi - dA +
\tilde d B}
\partial_n e^C\bigg] = 2\kappa^2 T_d \delta^{D-d} (y),}
and the dilaton equation \dilaeq\ becomes
\eqn\newdila{\eqalign{&\delta^{mn} \partial_m \bigg(e^{dA + \tilde d
B}
\partial_n \phi\bigg) - {a (d)\over 2}
e^{-dA + \tilde d B + 2C - a (d)\phi} \delta^{mn} \partial_m C
\partial_n C\cr
&=a (d) \kappa^2 T_d e^{dA + a (d)\phi/2} \delta^{(D-d)}
(y).\cr}}
Finally, the ($d-1$)-brane equation \braneq\ becomes
\eqn\newbrane{\partial_m (e^{dA + a (d)\phi/2} - e^C) = 0.}
Hence we have five equations for the four unknown functions
$A$, $B$, $C$,
$\phi$ and
the
unknown parameter $a (d)$.

The unique solution, assuming that $g_{MN}$ tends asymptotically to
$\eta_{MN}$,
 is given
by
\eqn\solution{\eqalign{A&={\tilde{d}\over 2(d+\tilde{d})} (C -
C_0),\cr
B&= - {d\over 2(d+\tilde{d})} (C - C_0),\cr
{a (d)\over 2} \phi&={a^2 (d)\over 4} (C - C_0) + C_0,\cr}}
where $C_0 = a \phi_0/2$ and $\phi_0$ is the dilaton vev.
$C$ is given by
\eqn\cexpre{\eqalign{e^{-C}&=e^{-C_0} + {k_d\over y^{\tilde{d}}},
\qquad
\tilde{d} > 0\cr
&=e^{-C_0} - {\kappa^2 T_d\over \pi} ln~y, \qquad \tilde{d} = 0\cr}}
and
\eqn\defk{k_d = 2\kappa^2 T_d/\tilde{d}~\Omega_{\tilde{d}+1},}
where $\Omega_{\tilde{d}+1}$ is the volume of $S^{\tilde{d}+1}$.  The
parameter
$a (d)$ is
given by
\eqn\aexpre{a^2 (d) = 4 - {2d\tilde{d}\over d + \tilde{d}}.}

Note, incidentally, that for these solutions, the coefficients of the
$\delta$-function in \munucomp\ and \newdila\ vanish at $y = 0$. So
the
Einstein equation and the dilaton equation are essentially
source-free;
only in the antisymmetric tensor equation is there
a $\delta$-function
source.
We shall return to this in section 3.4.

A crucial result of this section is that we have fixed the constant
$a (d)$ as in \aexpre\ by the requirement that our theory
\lgeact\ yield elementary $(d - 1)$-brane solutions.

The mass per unit ($d-1$)-volume of the elementary ($d-1$)-brane is
given by
\eqn\defmass{{\cal M}_d = \int d^{D-d} y \theta_{00},}
where $\theta_{MN}$ is the total energy-momentum pseudotensor of the
combined
gravity-matter system.  We find
\eqn\masexpre{{\cal M}_d = T_d e^{C_0}.}
To compute the electric charge $e_d$ of \elcharge\ it is convenient
to introduce polar coordinates
\eqn\polar{y^m = (y, \theta^i),}
where $i = 1, \ldots, (\tilde{d}+1)$, so that
\eqn\metricy{\delta_{mn} dy^m dy^n = dy^2 + y^2
d\Omega^2_{\tilde{d}+1},}
where $d\Omega^2_{\tilde{d}+1}$ is the metric on the unit
$S^{\tilde{d}+1}$.
Then we note from \fieldstr\ that
\eqn\strexpre{F_{y \mu_1 \ldots \mu_d} = -{1\over ^dg}
\varepsilon_{\mu_1
\ldots \mu_d} \partial_y e^C ,}
The dual of $F,~^{\ast}F$, has non-vanishing components only in the
$\theta^i$ directions
\eqn\dualf{\sqrt{-g}{} ^{\ast}F^{\theta_1 \ldots \theta_{D-d-1}} =
-(-)^{(D-d)
(d+1)} e^{2C} \partial_y e^{-C},}
Hence, using \supmetr,\solution-\defk, we find
\eqn\dualfa{e^{-a\phi}{}^{\ast}F_{\theta_1 \ldots
\theta_{D-d-1}} =
(-)^{(D-d) (d+1)} 2\kappa^2 T_d {\varepsilon_{\theta_1 \ldots
\theta_{D-d-1}}
\over \Omega_{\tilde{d}+1}}.}
It follows from \elcharge\ that
\eqn\elchargea{e_d = \sqrt{2} \kappa T_d (-)^{(D-d) (d+1)},}
and hence
\eqn\massagain{{\cal M}_d = {1\over \sqrt{2}} \mid e_d \mid e^{a
(d)
\phi_0/2}.}
Thus we find that the mass and charge obey the same equality as we
found for the supersymmetric solutions of section 2 even though as
yet no supersymmetry has been assumed. This is in fact a consequence
of assuming that the ratio of coefficients for the kinetic term and
Wess-Zumino term in the $p$-brane $\sigma$-model are as given in
\gepbact.

There is a straightforward generalization to exact, stable
multi-($d-1$)-brane configurations obtained by a linear superposition
of the solutions \cexpre,
\eqn\multisol{e^{-C} = e^{-C_o} + \sum\limits_l {k_d\over \mid
\vec y
 - \vec y_l \mid^2},}
where $\vec y_l$ corresponds to the position of each ($d-1$)-brane. To see
the no-force condition explicitly, consider the multi-($d-1$)-brane
configuration
\multisol\ with, for example, $N$ \  ($d - 1$)-branes as sources. In
general, we do
not have the transverse $SO(D - d)$ symmetry, but we still have the
$P_d$
Poincare symmetry for the configuration \multisol. Let each ($d -
1$)-brane with
label $l$ satisfy $X^\mu (l) = \xi^\mu$ so that, in particular, they
all have
the same orientation. The lagrangian for each of the ($d - 1$)-branes
with
label $l$ in the fields of the sources given by \splitx-\ansaph\ is,
from \gepbact\
\eqn\noforce{{\cal L}_d = -T_d \Bigg[ \sqrt{-det (e^{2A + a
(d)\phi/d}
\eta_{ij} + e^{2B +a (d)\phi/d} \partial_i Y^m (l) \partial_j
Y_m (l)} -
e^C\Bigg]}
corresponding to a potential
\eqn\forcepoten{V = T_d (e^{dA + a (d)\phi/2} - e^C),}
but this vanishes by \newbrane.  This generalizes to arbitrary $d$
and $D$ the
no-force condition for strings and fivebranes discussed in section 2.
Expanding out \noforce\ we find
\eqn\lagexpand{{\cal L} = - {T_d \over2}~e^{(d-2) A + 2B + a
(d)\phi/2}
\eta^{ij} \partial_i Y^m \partial_j Y_m +\ldots ,}
and so the absence of velocity-dependent forces corresponds to
\eqn\absencevdf{(d-2) A + 2B + a (d)\phi/2 = {\rm constant},}
which is indeed satisfied by virtue of \solution\ and we find that
the
constant is just $C_0$.  This generalizes to arbitrary $d$ and $D$,
the absence
 of velocity dependent
forces for strings and fivebranes in $D = 10$ \calk.

\newsec{The solitonic ($\tilde{d}-1$)-brane}

The elementary ($d-1$)-branes we have discussed so far correspond to
solutions
 of the
coupled field-brane system with action $I_D (d) + S_d$.  As such they
exhibit
$\delta$-function singularities at $y = 0$.  They are characterized
by a
non-vanishing
Noether electric charge $e_d$.  By contrast, we now wish to find
solitonic
($\tilde{d}-1$)-brane, corresponding to solutions of the source free
equations
 resulting
from $I_D (d)$ alone, which are regular at $y = 0$, and which
will be
characterized
by a
non-vanishing topological magnetic charge $g_{\tilde{d}}$.  (Recall
that
$\tilde{d} = D-d-2$).

To this end, we now make an ansatz invariant under $P_{\tilde{d}}
\times
SO(D-\tilde{d})$.  Hence we write \splitx\ and \supmetr\ as before
where now
$\mu = 0, 1 \ldots
(\tilde{d}-1)$ and $m = \tilde{d}, \tilde{d} + 1, \ldots (D-1)$.  The
ansatz
for the
antisymmetric tensor, however, will now be made on the field strength
rather
than on the
potential.  From section 3.2 we recall that a non-vanishing electric
charge
corresponds to
\eqn\elchargeb{{1\over \sqrt{2} \kappa}
e^{-a\phi}{}^{\ast}F_{\tilde{d}+1}
 = e_d \varepsilon_{\tilde{d}+1}/\Omega_{\tilde{d}+1},}
where $\varepsilon_{\tilde{d}+1}$ is the volume form on
$S^{\tilde{d}+1}$.
 Accordingly,
to obtain a non-vanishing magnetic charge, we make the ansatz
\eqn\ansazmag{{1\over \sqrt{2} \kappa} F_{d+1} = g_{\tilde{d}}
\varepsilon_{d+1}/\Omega_{d+1},}
where $\varepsilon_{d+1}$ is the volume form on $S^{d+1}$.  Since
this is an
harmonic
form, $F$ can no longer be written globally as the curl of $A$, but
it
satisfies the
Bianchi identities.  It is now not difficult to show that all the
field
equations of $I_D(d)$ are
satisfied simply by making the replacement
$d \rightarrow \tilde{d}$,
and hence
$a(d) \rightarrow a (\tilde{d}) = - a (d)$
in \einstein-\dilaeq\ with the source terms set to zero.
For future reference we write the explicit solution
in the case $\phi_0 = 0$
\eqn\solisolution{\eqalign{ds^2 &= \bigg(1 + {k_{\tilde d}\over
y^d}\bigg)^{
-d/(d + {\tilde d})} dx^\mu dx_\mu + \bigg(1 + {k_{\tilde d}\over
y^d}\bigg)^{
\tilde d /(d + {\tilde d})} dy_m dy_m,\cr
e^{ 2\phi} & = \bigg(1 + {k_{\tilde d}\over
y^d}\bigg)^{a(d)},\cr
F_{d + 1} &= \sqrt {2} \kappa g_{\tilde d} \varepsilon_{d + 1}
/ \Omega_{d + 1}.\cr}}
Note that by this device, we have found solutions everywhere
including
$y = 0$, since the $\delta$-functions were already absent in the
Einstein
and dilaton equations.

It follows that the mass per unit ($\tilde{d}-1$)-volume now
satisfies
\eqn\magbogom{\eqalign{{\cal M}_{\tilde{d}}&={1\over \sqrt{2}} \mid
g_{
\tilde{d}} \mid e^{a (\tilde{d}) \phi_0/2}\cr
&={1\over \sqrt{2}} \mid g_{\tilde{d}} \mid e^{-a (d)
\phi_0/2}.\cr}}
Note that the $\phi_0$ dependence is such that ${\cal M}_{\tilde{d}}$
is large
for small ${\cal M}_d$ and vice-versa.

The electric charge of the elementary solution and the magnetic
charge of
the soliton
solution obey a Dirac quantization rule \refs{\nep,\tei}
\eqn\pbdirac{e_d g_{\tilde{d}} = 2 \pi n, \qquad n = {\rm integer},}
and hence from \elchargea\
\eqn\machargea{(-)^{(D-d)(d+1)} {g}_{\tilde{d}} = 2\pi n/\sqrt{2}
\kappa T_d,}

\newsec{Duality}

We now wish to consider the theory ``dual'' to \lgeact\ for which the
roles of
 field
equations \fieldeq\ and Bianchi identities \bianchi\ are
interchanged.  To this
end let us write the action
\eqn\dualact{\tilde{I}_D (\tilde{d}) = {1\over 2 \kappa^2} \int d^Dx
\sqrt{-g}
 \Bigg(R - {1\over 2}
(\partial\phi)^2 - {1\over 2(\tilde{d}+1)!} e^{a (d)\phi}
\tilde{F}_{\tilde{d}+1}^2\Bigg),}
where the rank ($\tilde{d}+1$) field strength $\tilde{F}$ is given by
\eqn\dualstr{\tilde{F}_{\tilde{d}+1} = d \tilde{A}_{\tilde{d}},}
$a (d)$ is the same constant as appearing in \lgeact\ but
appears with
opposite sign, i.e
\eqn\dualpha{a(\tilde d ) = - a (d).}
  Allow
these fields to couple to an elementary $\tilde{d}$-dimensional
extended object
(a``($\tilde{d}-1$)-brane'') with action
\eqn\dualpbact{\eqalign{\tilde{S}_{\tilde{d}} = T_{\tilde{d}} \int
d^{\tilde{d}
} \xi \Bigg(&-{1\over 2}
\sqrt{-\gamma} \gamma^{ij} \partial_i X^M \partial_j X^N g_{MN}
 e^{-a (d)\phi/
\tilde{d}} + {(\tilde{d}-2)\over 2} \sqrt{-\gamma}\cr
&-{1\over \tilde{d}!} \varepsilon^{i_1 i_2 \ldots i_{\tilde{d}}}
\partial_i X^{
M_1}\partial_{i_2} X^{M_2} \ldots \partial_{i_d} X^{M_{\tilde{d}}}
\tilde{A}_{
M_1 M_2 \ldots M_{\tilde{d}}}\bigg).\cr}}
The $\phi$ dependence is such that under the rescaling
\eqn\dualscaling{\eqalign{g_{MN}&\rightarrow
\tilde{\lambda}^{2\tilde{d}/(D-2)}
 g_{MN},\cr
\tilde{A}_{M_1 \ldots M_{\tilde{d}}}&\rightarrow
\tilde{\lambda}^{\tilde d}
\tilde{A}_{
M_1 \ldots M_{\tilde{d}}},\cr
e^{\phi}&\rightarrow \tilde{\lambda}^{-2\tilde{d}
(D-\tilde{d}-2)/(D-2)
a (d)} e^{\phi},\cr
\gamma_{ij}&\rightarrow \tilde{\lambda}^2 \gamma_{ij},\cr}}
both actions scale the same way
\eqn\dualactscal{\eqalign{\tilde{I}_D (\tilde{d})&\rightarrow
\tilde{\lambda}^{
\tilde{d}} I_D (d),\cr
\tilde{S}_{\tilde{d}}&\rightarrow \tilde{\lambda}^{\tilde{d}}
\tilde{S}_{\tilde{d}}.\cr}}
The field equations and Bianchi identities of the $\tilde{A}$ field
may be
written
\eqn\dufieldeq{d^{\ast} (e^{a (d)\phi} \tilde{F})=
2\kappa^2 (-)^{\tilde{d}^2}{}^{\ast}\tilde{J},}
\eqn\dubianchi{d\tilde{F} = 0.}
It should be clear that the system described by $\tilde{I}_D
(\tilde{d}) +
\tilde{S}_{\tilde{d}}$ admit the same elementary solutions as that
described by
$I_D (d)+ S_d$ and that $\tilde{I}_D (\tilde{d})$ alone admits the
same
solitonic solutions as
$I_D (d)$ alone, provided we everywhere make the replacement $d
\rightarrow
\tilde{d}$
and hence $a (d) \rightarrow a (\tilde{d}) = - a (d)$.
 In
particular the
Noether electric charge is given by
\eqn\duelcharge{\tilde{e}_{\tilde{d}} = {1\over \sqrt{2}\kappa}
\int\limits_{
S^{d+1}} e^{a\phi}{}^{\ast}\tilde{F}_{d+1},}
and the topological magnetic charge by
\eqn\dumacharge{\tilde{g}_d = {1\over \sqrt{2} \kappa}
\int\limits_{S^{
\tilde{d}+1}}\tilde{F}_{\tilde{d}+1},}
and they obey the condition
\eqn\dualdirac{{\tilde e}_{\tilde{d}} {\tilde g}_d = 2\pi n.}

So far we have discovered that the equations of $I_D (d)$ admit an
elementary
($d-1$)-brane solution and a solitonic ($\tilde{d}-1$)-brane
solution.
Conversely, the
equations of $\tilde{I}_D (\tilde{d})$ admit an elementary
($\tilde{d}-1$)-brane
 solution
and a solitonic ($d-1$)-brane solution.  We now wish to go a step
further and
assert that
the ($d-1$)-brane is ``dual'' to the ($\tilde{d}-1$)-brane.  In its
strongest
sense this
means that the two theories are equivalent descriptions of the same
physics.
In the
present context, however, we simply make the assumption that the $I_D
(d)$ and
$\tilde{I}_D (\tilde{d})$ are equivalent, i.e we assume that the
metric $g_{MN}$
and
dilaton $\phi$ are the same and that the ($\tilde{d}+1$)-form field
strength
$\tilde{F}_{\tilde{d}+1}$ is dual to the ($d+1$)-form field strength
$F_{d+1}$.
 More
precisely,
\eqn\duality{\tilde{F}_{\tilde{d}+1} = e^{-a
(d)\phi}{}^{\ast}F_{d+1},}
so that the (source-free) field equations and Bianchi identities of
$I_D (d)$,
\fieldeq\ and \bianchi, become the Bianchi identities and
(source-free) field
equations of $\tilde{I}_D (\tilde{d})$, \dubianchi\ and \dufieldeq.
This leads
 immediately to
\eqn\charelation{\eqalign{e_d&=\tilde{g}_d,\cr
g_{\tilde{d}}&=\tilde{e}_{\tilde{d}},\cr}}
and hence
\eqn\tensiondirac{\kappa^2 T_d T_{\tilde{d}} =  |n| \pi.}

The duality assumption also leads to a relation between the
dimensionless loop
expansion
parameters of the ($d-1$)-brane and the ($\tilde{d}-1$)-brane.  To
see this we
note that
metrics appearing naturally in ($d-1$)-brane and
($\tilde{d}-1$)-brane
$\sigma$-models \gepbact\ and \dualpbact\ are
\eqn\sigmacano{g_{MN} (d)=e^{a (d)\phi/d} g_{MN} ({\rm
canonical}),}
\eqn\dusigmac{g_{MN} (\tilde{d})=e^{-a (d)\phi/\tilde{d}} g_{MN}
({\rm
canonical}).}
If we rewrite $I_D (d)$ and $\tilde{I}_D (\tilde{d})$ in these
variables we
find
\eqn\sigmact{\eqalign{I_D(d)&={1\over 2\kappa^2} \int d^Dx \sqrt{-g}
e^{-(D-2)
a (d)\phi/2d} \Bigg[R\cr
&\quad -{1\over 2} \bigg(1 - {a^2 (D-1) (D-2)\over 2d^2}\bigg)
(\partial
\phi)^2 - {1\over 2 \cdot (d+1)!} F^2_{d+1}\Bigg],\cr}}
and
\eqn\dusigmact{\eqalign{\tilde{I}_D (\tilde{d})&={1\over 2\kappa^2}
\int d^Dx
\sqrt{-g} e^{(D-2)a (d)\phi/2\tilde{d}} \Bigg[R\cr
&\quad -{1\over 2} \bigg(1 - {a^2 (D-1) (D-2)\over
2\tilde{d}^2}\bigg)
(\partial\phi)^2 -
{1\over 2 (\tilde{d}+1)!} \tilde{F}^2_{\tilde{d}+1}\Bigg].\cr}}
Note that in both cases a common dilaton-dependent factor appears.
This
reveals that the
($d-1$)-brane loop counting parameter is
\eqn\looparam{{\rm g}_d = e^{(D-2) a (d) \phi_0/4d},}
and the ($\tilde{d}-1$)-brane loop counting parameter is
\eqn\dulooparam{{\rm g}_{\tilde{d}} = e^{-(D-2) a
(d)\phi_0/4\tilde{d}}.}
Hence
\eqn\loopduality{{\rm g}_d^d = 1/{\rm g}_{\tilde{d}}^{\tilde{d}},}
and strongly coupled ($d-1$) branes correspond to weakly coupled
($\tilde{d}-1$)
 branes
and vice-versa.

Finally we note that, in the case of $d = 2$, the following field
redefinition
\eqn\fieldredef{(D-2) a (2) \phi = 8 \Phi}
yields from \sigmact\ an $I_D (d)$ which is $D$-independent, namely
\eqn\stringact{I_D (2) = {1\over 2\kappa^2} \int d^Dx \sqrt{-g}
e^{-2\Phi}
\Bigg[R + 4(\partial\Phi)^2 - {1\over 2.3!} F_3^2\Bigg].}
This is a well-known result in string theory.  Curiously, there is no
field
redefinition which renders the integrand of
$I_D (d)$ independent of $D$ for $d \not=2$.
However, we may
dualize \stringact\ to obtain
\eqn\dustringact{\tilde{I}_D (D-4) = {1\over 2\kappa^2} \int d^Dx
\sqrt{-g}~e^{
4\Phi/D-4} \Bigg[R
- {4 (D-10)\over  (D-4)^2} (\partial\Phi)^2 - {1\over 2(D-3)!}
\tilde{F}^2_{D-3}\Bigg].}
In these string variables the metric of the elementary string is
given by
\eqn\stringmetric{ds^2 = \bigg(1 + {k_2 e^{C_0}\over
y^{D-4}}\bigg)^{-1} \eta_{
\mu\nu} dx^{\mu} dx^{\nu} + \delta_{mn} dy^m dy^n}
with $\mu = 0,1$ and $m = 1 \ldots D-2$.  Also
\eqn\stringalpha{a (2) = \sqrt{{8\over D-2}},}
so
\eqn\redefphi{\Phi = {1\over 2} (C - C_0) + {D-2\over 4} C_0 ,}
where
\eqn\stringc{\eqalign{e^{-C}&=e^{-C_0} + {k_2\over y^{D-4}}, \qquad D
> 4 \cr
&=e^{-C_0} - {\kappa^2 T_2\over \pi}~ln~y \qquad D = 4 .\cr}}

On the other hand the solitonic ($D-5$)-brane is given by
\eqn\dustringmetr{ds^2 = \eta_{\mu\nu} dx^{\mu} dx^{\nu} + \bigg(1 +
{k_{D-4}
\over y^2} e^{C_0}\bigg)\delta_{mn} dy^m dy^n ,}
where $\mu = 0 \ldots D-5$ and $m = D - 4, \ldots, D - 1$.  Also
\eqn\dualpha{a (D-4) = - \sqrt{{8\over D-2}},}
so
\eqn\dualphi{\Phi = - {1\over 2}~(C - C_0) - {(D-2)\over 4} C_0 ,}
where
\eqn\dualc{e^{-C} = e^{-C_0} + {k_{D-4}\over y^2}.}
We note that in these string $\sigma$-model variables the transverse
part of
 the
 metric
in \stringmetric\ is flat and the spacetime part of the metric in
\dustringmetr\
 is flat.  These
 are
therefore free field theories from the point of view of conformal
field theory.

Having constructed the action in $\sigma$-model variables \sigmact,
it is instructive to compare it with Brans-Dicke theory.
The action for Brans-Dicke gravity (generalized from $4$ to $D$
dimensions) may
be written in
terms of a scalar field $\eta$ and some metric $g_{MN}(BD)$
\eqn\appenone{I ({\rm Brans-Dicke}) = {1\over 2\kappa^2} \int d^D x
\sqrt{-g}
\bigg[\eta R -{\omega\over \eta} (\partial\eta)^2\bigg] +
\int d^D x {\cal L} ({\rm matter},g),}
where $\omega$ is a free parameter and where, by construction,
${\cal L}$ (matter, $g$) is
independent of $\eta$.  In comparing this to our general action $I_D
(d)$ we
have to decide
what is meant by ${\cal L}$ (matter, $g$).  Let us first suppose that
this
refers not to
the antisymmetric tensor action of \lgeact\ but to the ($d -
1$)-brane action
 $S_d$ of \gepbact.
Then we must make the identification
\eqn\appentwo{g_{MN} (BD) = g_{MN} (d),}
where $g_{MN} (d)$ is the ($d - 1$)-brane $\sigma$-model metric of
\sigmacano.
 Comparison with
\sigmact\ then yields the identifications
\eqn\appenthree{\eta = e^{-(D - 2) a (d) \phi/2d},}
\eqn\appenfour{\omega = {2d^2\over (D - 2)^2 a^2 (d)} -
{D - 1\over D - 2} = - {(D - 1) (d - 2) -d^2\over (D - 2) (d - 2) -
d^2},}
where we have used \aa.  It is interesting to note, for
example, that
in $D = 10$
strings ($d = 2$) correspond to $\omega = - 1$, fivebranes ($d = 6$)
to
$\omega = 0$ and
threebranes ($d = 4$) to $\omega = \infty$.

\newsec{The self-dual string and dyonic string in $D=6$}

The case $D=6$ is special because in this case the theory dual to the
superstring is itself a superstring. The two strings are related by
the strong/weak coupling replacement $\phi \rightarrow - \phi$.
Compare \stringact\ with \dustringact, which are the bosonic sectors
of $D=6$ supergravity and its dual. This permits the construction of
a dyonic string which carries both electric and magnetic charges,
which we shall shortly discuss.

However, there is another supersymmetric solitonic string in $D = 6$:
 the
{\it self-dual superstring} which falls outside our
previous discussions and requires a special treatment.  This is the
$D = 6$ counterpart of the self-dual superthreebrane in $D = 10$ of
section 4. Our starting point is the $N = 2$, $D = 6$ self-dual
supergravity \refs{\rom,\sala} which, in common with the Type IIB
superstring
in $D =10$ discussed in section 4, admits
covariant field equations, but no manifestly covariant field
equations.  It
describes a
graviton $e_M\,^A$, two left-handed gravitini $\psi_{Ma}$ and one
tensor field
 $B_{MN}$ with
self-dual field strength $G_{MNP}$.  The gravitini transformation
rules are
(in our notation)
\eqn\gtinitransf{\delta \psi_M = \nabla_M \varepsilon - {1\over 8}
G_{MNP} \Gamma^{NP}\varepsilon.}
So if we make a two/four split as in section 3.2 with
\eqn\splitgama{\eqalign{\Gamma_A&= (\gamma_{\alpha} \otimes 1,
\gamma^3
\otimes \Sigma_m), \qquad \Gamma^7 = \gamma^3 \otimes \Gamma^5,\cr
\gamma^3&=\gamma^0 \gamma^1, \qquad \Gamma^5 = \Sigma^2 \Sigma^3
\Sigma^4
\Sigma^5,\cr}}
the criterion for unbroken supersymmetry, $\delta \psi_M = 0$,
reduces to
\eqn\criterion{\eqalign{\partial_{\mu} \varepsilon - {1\over 2}
\gamma^3
\gamma_\mu \otimes \Sigma^n (\partial_n A
+ {1\over 2} e^{-2A} \partial_n e^C \gamma^3) \varepsilon&=0,\cr
\partial_m \varepsilon + {1\over 2} \partial_m B \varepsilon -
{1\over 2}
(\delta^n\,_m +
\Sigma^n\,_m) (\partial_n B - {1\over 2} e^{-2A} \partial_m e^C
\gamma^3)
\varepsilon&=0,\cr}}
and hence supersymmetry requires
\eqn\susygiving{C = 2A, \qquad B = - A, \qquad \varepsilon = e^{-B/2}
\varepsilon_0,}
where $\varepsilon_0$ obeys $\gamma^3 \varepsilon_0 = -
\varepsilon_0$, and
one half of the
supersymmetries are broken.

The bosonic equations of motion are
\eqn\einsteineq{R_{MN} - {1\over 2} g_{MN} R = {1\over 4}
G_M\,^{PQ} G_{NPQ}}
\eqn\selfdueq{G_{MNP} =- \tilde{G}_{MNP},}
and substituting \susygiving\ yields
\eqn\munucomp{e^{6A} \delta^{mn}\partial_m
\partial_n e^{-2A} = 0}
for the $\mu\nu$ components of the Einstein equation and
\eqn\mncomp{e^{2A} \delta^{mn} \partial_m \partial_n e^{-2A} = 0}
for the $mn$ components, so that
\eqn\solutiontwo{e^{-2A} = 1 + {k_2\over y^2}.}
All the properties of the dyonic self-dual threebrane \duflselft\ to
be discussed in section 4  apply, mutatis mutandis, to the
dyonic self-dual string, including Dirac quantization rules and the
saturation
of the Bogolmol'nyi bound.

The effective bosonic equations of motion of this string are
\eqn\effecact{\eqalign{&\partial_i (\sqrt{-\gamma} \gamma^{ij}
\partial_j X^N g_{MN}) - {1\over 2}
\sqrt{-\gamma} \gamma^{ij} \partial_i X^N \partial_j X^P \partial_M
g_{NP}\cr
&={1\over 2} G_{MNP} \partial_i X^N \partial_j X^P
\varepsilon^{ij},\cr}}
but, since $G_{MNP} = - \tilde{G}_{MNP}$, there is no manifestly
covariant
worldsheet action.  It would be interesting to include the fermionic
degrees of freedom and construct the spacetime supersymmetric,
$\kappa$-symmetric,  Green-Schwarz string equations, but this has not
yet been done.

We shall now discuss the dyonic string solution \duffkr\ of the $D=6$
string action \stringact\ and show that in the limit $e_2=g_2$ it
reduces to the self-dual string configuration. The solution is
\eqn\dyst{\eqalign{e^{-2\Phi_E}&=1+{k_2\over y^2},\qquad
e^{2\Phi_M}=1+{\tilde k_2\over y^2},\cr
\Phi&=\Phi_E + \Phi_M,\cr
ds^2&=e^{2\Phi_E} (-d\tau^2 + d\sigma^2) + e^{2\Phi_M}(dy^2 + y^2
d\Omega^2),\cr
G&=2\tilde k_2\epsilon_3,\qquad
\tilde G= e^{-2\Phi}{} ^\ast G=2k_2\epsilon_3,\cr}}
where $k_2=\kappa e_2/\sqrt{2}\Omega_3$ and $\tilde k_2=\kappa
g_2/\sqrt{2}\Omega_3$. For $e_2=g_2$ \dyst\ reduces to the self-dual
string
above. However, since the self-dual string
 was shown above to break $1/2$ of the
supersymmetries of the self-dual theory, it breaks $3/4$ of the
supersymmetries of the non-self-dual theory, a result that can be
shown directly from the supersymmetry transformation rules of the
non-self-dual theory even when $e_2 \neq g_2$ \duffkr. One recovers
$1/2$ of the supersymmetries when either $e_2$ or $g_2$ vanishes.

\newsec{Singularity structure and interpolation}

In section 2.6, we had shown using test-probe computations and
by examining curvature singularities that while the
string is a singular solution of string theory and
the fivebrane is a singular solution of fivebrane theory, the string
and fivebrane are mutually nonsingular in the
sense that each could be viewed as a nonsingular soliton of
the other theory. This symmetry in the singularity structure was used
to support the string/fivebrane duality conjecture. One might naively
have expected that this state of affairs would generalize to
arbitrary $p$-branes and their duals, but in this section we shall
show that this is not the case.
We find that only particles and strings are mutually nonsingular with
their duals from the point of view of the test-probe/source approach,
even though
the absence of curvature singularities at $y=0$ persists for all $p$.

In probing the singularity structure, we consider the radial
trajectory of the $(d-1)$-brane infalling into the dual $(\tilde
d-1)$-brane background. We assume that the
$(d-1)$-brane and $(\tilde d-1)$-brane are nonintersecting,
as is generically the case for $D=d+\tilde d + 2$. As before,
the contribution of the
antisymmetric field strength to the worldvolume action vanishes.
The effective action reduces to
\eqn\act{{\cal L}=\sqrt{-\gamma},}
where $\gamma_{ij}=g_{MN}\partial_i X^M\partial_j X^N$ is the
worldvolume metric and where $g_{MN}$ is the
$(\tilde d - 1)$-brane metric in the
$(d-1)$-brane variables and is given by
\eqn\met{ds^2=\Delta(d)^{1-2/d} \eta_{mu\nu} dx^\mu dx^\nu +
\Delta(d)^{-2/d}\delta_{mn}dy^m dy^n,}
where $\Delta(d)=(1+(b/y)^d)^{-1}$, $\mu,\nu=0,1,...,\tilde d-1$
are
$(\tilde d-1)$-brane indices, $m,n=\tilde d,...,D-1$ are
transverse space indices and $y$ is the radial coordinate in the
transverse space. A $(d-1)$-brane propagating in this
background will
in general have $n$ of its $d-1$ spatial directions parallel to the
$(\tilde d-1)$-brane and $d-n-1$ directions perpendicular to the
$(\tilde d-1)$-brane, where $n\leq n_0=min(d-1,\tilde d-1)$.

Our criterion for the singularity of the $(\tilde d-1)$-brane is as
follows: if for some $n$ a $(d-1)$-brane test-probe views the
singularity at the
$(\tilde d-1)$-brane source (i.e. the test-probe falls into the source
in finite proper time) then the $(\tilde d-1)$-brane is viewed as a
singular solution of $(d-1)$-brane theory. Otherwise, the dual solitons
are mutually nonsingular, as was demonstrated for strings and
fivebranes in particular.

Replacing \met\ in \act\ for a given choice of $n$, the
lagrangian
reduces in the limit of radial motion to
\eqn\lagone{{\cal L}=\sqrt{\Delta^{n-1}\dot t^2 -
\Delta^{n-2}\dot
y^2}.}
It follows that
\eqn\econ{E={\Delta^{n-1}\dot t \over
\sqrt{\Delta^{n-1}\dot t^2 - \Delta^{n-2}\dot y^2}}}
is a constant of the motion.
We combine \econ\ with the geodesic condition $\gamma_{00}=-1$
for massive test-probes,
which follows from setting $X^0=\tau$ on the worldsheet and
compute the proper time for the radial fall of the
test-probe into the dual
source to be
\eqn\time{\tau=\int_0^{y_0} {dy \Delta^{-1/d}\over
\sqrt{\Delta^{-n+1}E^2-1}},}
where $y_0$ is the initial radial separation.
Note that for $n=0$ there is a turning point at $\Delta(y_T)=E^2$, so
that no singularity can be observed (the force in this case being
repulsive). So without loss of generality, we let $n\geq 1$ below.
The question of singularity now reduces to whether or not $\tau$ in
\time\ converges or not near $y=0$. Convergence of the integral
implies a finite proper time and a singular dual
object while divergence implies infinite proper time and mutual
nonsingularity of probe and source. Since $\Delta\to (y/b)^d$ as
$y\to 0$, it is straightforward to show that finiteness of $\tau$ is
equivalent to finiteness of
\eqn\integ{I(d,n)=\int_0^{y_0} dy y^{d(n-1)/2 -1}.}
$I$ is finite provided $d(n-1)/2>0$, or $n >1$, since $d>0$.
Since $n\leq n_0=min(d-1,\tilde d-1)$, it follows that $I$
cannot be finite when either of the dual objects is a particle
($d=1$) or a string ($d=2$).
This means that strings and particles are always mutually nonsingular
with their duals, in whatever spacetime dimension they are embedded.
If, however,
$n_0\geq 2$, or in other words both $p$-branes are membranes or
higher dimensional objects, then the choice $n=n_0$ yields finite $I$
and the dual objects are mutually singular. Note that this analysis
does not depend on the spacetime dimension $D$ at all. Also note that
in our earlier analysis for strings and fivebranes we set
$n=n_0$, which in view of \integ\ and our
singularity criterion is a consistent choice in the general case
(since $n=n_0$ is the case in which $I$ is ``most likely'' to
converge, and so that if there is a singularity it would certainly be
observed in this case).

Next we turn to the question of vacuum interpolation \dufgt.
  To determine the asymptotic form of the
  metric of near $y=0$ we introduce the new radial coordinate
$r^d=y^d + a^d$ and let
  \eqn\drrho{
  r= b\left( 1+{e^{d\rho/b}\over d}\right).}
  Near $r=b$ we get
  \eqn\rneara{
  \eqalign{
  ds^2 &\sim e^{(d-2)\rho/b}(-dt^2 +d{\bf x}\cdot d{\bf x}) +
  d\rho^2 + b^2d\Omega_{d+1}^2\cr
  \phi &\sim -{da\over 2b}\rho\cr
  F_{d+1} &\sim db^d \varepsilon_{d+1} .\cr}}

  If $d\ne 2$ the asymptotic spacetime is $(adS)_{\tilde
  d+1}\times S^{d+1}$, and there is an event horizon at $r=b$.
If
$d=2$ the
  dual $(\tilde d -1)$-brane interpolates between
 $D$-dimensional
  Minkowski spacetime and the product of $S^3$ with a
 $(\tilde d
  +1)$-dimensional {\it Minkowski} spacetime, generalizing the
 case
of
  the $D=10$ fivebrane.
  For these cases the asymptotic behaviour of the dilaton near
 the
  $p$-brane core can be invariantly characterized as linear in
an
ignorable
  coordinate associated with a space-translation Killing vector.

\newsec{Blackbranes}

Finally we turn to the two-parameter solitonic solutions of the
theory which
 display event horizons: the ``blackbranes''.

  Using the canonical metric, the $(\tilde{d} - 1)$-brane
 black
soliton
 solution  may be written for all $\tilde{d} \geq 1$ as
\eqn\blacksolution{\eqalign{ds^2=& - \Delta_+
\Delta_-^{-\tilde{d}/(d +
 \tilde{d})} dt^2 \cr
 & + \Delta_+^{-1} \Delta_-^{{a^2\over 2d} - 1} dr^2 \cr
 & + r^2 \Delta_-^{{a^2\over 2d}}  d \Omega^2_{d + 1} \cr
 & + \Delta_-^{{d\over d + \tilde{d}}}  dx^i dx_i, \qquad i = 1
 \ldots \tilde{d} - 1, \cr
e^{-2\phi}=& \Delta_-^a, \cr
\Delta_{\pm}=& \bigg[1 - \bigg({r_{\pm}\over r}\bigg)^d\bigg],
\cr
F_{d + 1}=& d(r_+ r_-)^{d/2} \varepsilon_{d + 1}, \cr}}
where the magnetic charge $g_{\tilde{d}}$ and the mass per unit
$(\tilde{d} - 1)$-volume ${\cal M}_{\tilde{d}}$ are related to
$r_{\pm}$ by
 \lu\
\eqn\bmcharge{g_{\tilde d} = {\Omega_{d + 1}\over \sqrt{2} \kappa}
 d (r_+ r_-)^{d/2},}
and
\eqn\bmass{{\cal M}_{\tilde d} = {\Omega_{d + 1}\over 2 \kappa^2}
 [(d + 1)
 r_+^d - r_-^d].}
 The solutions possess an $R \times SO (d + 2) \times E (\tilde{d} - 1)$
 symmetry  where  $E (n)$ denotes the $n$-dimensional Euclidean
group.  The
 solutions exhibit an event  horizon at $r= r_+$ and an inner horizon
at
$r = r_-$. The absence of naked singularities, $r_+\geq r_-$,
translates into
the same Bogomol'nyi bound
$\sqrt{2} \kappa {\cal M}_{\tilde d} \geq |\tilde g_{\tilde d}|$.
In the special case  $D = 11$, $\tilde{d} = 3, 6$ they reduce to the
black
 membrane and black fivebrane of \guv.   In the special case $D = 10$
i.e
 $\tilde d = 8 - d$, they reduce to the black $p$-brane solutions of
\hors.
  In the special case $D = 4$, $\tilde d = 1$ they reduce to  the
  Kaluza-Klein black hole solution of \refs{\gib,\dobm,\chod,\pol}.
In the
 limit of zero charge, $r_-=0$, the dilaton and antisymmetric tensor
are
 trivial and the metric reduces to
\eqn\trivialmetric{ds^2=-\Delta_+ dt^2 + \Delta_+^{-1} dr^2 + r^2
  d\Omega_{d + 1}^2 + dx^i dx_i.}
In \grel\ it was argued that these solutions
 are
 classically unstable.  More interesting is the extreme mass $=$
charge
limit $r_+ = r_-$ where the metric component $g_{00}$ becomes equal
to the one
multiplying  $dx^i dx^i$ and the symmetry is enlarged to
$SO (d + 2) \times P(\tilde{d})$.  It is convenient to  introduce the change of
variables $y^d = r^d - r_-^d$, then \blacksolution\ becomes
 \eqn\extreme{\eqalign{ds^2 &=\Delta_-^{d/(d + \tilde{d})}
dx^\mu
 dx_\mu +\Delta_-^{-\tilde{d}/(d + \tilde{d})}  (dy^2 + y^2
d\Omega^2_{d
 +1}),\cr
e^{-2\phi}&={\Delta_{-}}^{a},\cr
\Delta_-&=\bigg[1 + \bigg({r_-\over y}\bigg)^d\bigg]^{-1},\cr
F_{d + 1}&=d r_-^d \varepsilon_{d + 1}.\cr}}
But in the case $1 \leq d \leq 7$, $1 \leq \tilde{d} \leq 7$, these
are
  precisely the super $p$-branes, so $r_+ = r_-$ also corresponds to
the
 appearance of supersymmetry.   It is also possible to find {\it
elementary}
 black $(d - 1)$-branes with  parameters ${\cal M}_d$~and~$e_d$
obeying the
 bound  $\sqrt{2}\kappa {\cal M}_d \geq  |e_d|$, by including a source
term on
 the right hand side of the equations.  In this case however,  it
would
be necessary to relax the equality of the kinetic and WZW term
coefficients in
  \gepbact\ to allow for mass $\neq$ charge.  (This equality is
forced on us
in the supersymmetric  case, by virtue of $\kappa$-symmetry \berst).

For specific values of $d$ and $D$, the extreme solutions also
exhibit
 supersymmetry and hence stability is guaranteed.
Some may be identified with the previously
classified
heterotic, Type IIA and Type IIB super $\tilde p$-branes. It is
to
this subject
 that we now turn.

\chapter{The brane-scan revisited}

\def\dg{\hbox{$^\dagger$}}

\def\lam{\hbox{$\lambda\kern-6pt^{\_\_}$}}

\def\r{\vbox{\hbox{\raise1.5mm\hbox{$>$}}
\kern-18pt\hbox{\lower1.5mm\hbox{$\sim$}}}}
\def\l{\vbox{\hbox{\raise1.5mm\hbox{$<$}}
\kern-18pt\hbox{\lower1.5mm\hbox{$\sim$}}}}
\def\one{\phantom{0}}
\def\two{\phantom{00}}
\def\a{\hbox{$^{a]}$}}

\def\leaderfill{\leaders\hbox to 1em{\hss.\hss}\hfill}
\def\rs{\vbox{\hbox{\raise1.1mm\hbox{$>$}}
\kern-18pt\hbox{\lower1.1mm\hbox{$\sim$}}}}
\def\ls{\vbox{\hbox{\raise1.1mm\hbox{$<$}}
\kern-18pt\hbox{\lower1.1mm\hbox{$\sim$}}}}

\newsec{Bose-Fermi matching: a necessary condition}

As pointed out in \dabghr, the string configuration of chapter
 2 also solves both the Type IIA and Type IIB supergravity
equations. Once again, each breaks half of the spacetime
supersymmetries. As discussed in section 4.3, the Type IIA
solution may be shown to follow by simultaneous dimensional
reduction of the $D=11$ supermembrane of section 4.2 The $N=1$
solution of section 2.1 then follows by truncation. Together
 with the $N=1$ fivebrane of section 2.4, all these solutions
correspond to known points on the brane-scan
of supersymmetric extended objects classified in \achetw\ and
discussed in the Introduction. According to this classification,
 no Type II fivebranes (or indeed any Type II $p$-branes with
$p > 1$) could exist. However, it was pointed out in \calhs\
that the fivebrane configuration of chapter 2 also solves both
the Type IIA and Type IIB
supergravity equations and hence that Type II superfivebranes
exist after all.
Moreover, the Type IIB theory also admits a self-dual
superthreebrane \duflselft.
The no-go theorem is circumvented because in addition to the
superspace coordinates $X^M$ and $\Theta^\alpha$ there are also
higher spin fields on the worldvolume: vectors or antisymmetric
tensors. This raises the question: are there other
 super $p$-branes and if so, for what $p$ and $D$?

We begin by asking what new points on the brane-scan are
permitted by bose-fermi matching alone.  There are surprisingly
few:  $p=5$ in $D=11$; $p = 3,
4, \ldots 9$ in $D = 10$; $p = 3, 4, 5$ in $D = 6$ and $p = 3$
in $D = 4$.  The
much harder task is to narrow down these possibilities to
 objects that actually
exist.  One obvious handicap is that, unlike the $p$-branes
discussed in \achetw\ and in the Introduction, no-one has yet
 succeeded in writing down the action for
these new Type II $p$-branes.  The existence of the
 $p = 3$~and~$p = 5$ objects
mentioned above was established indirectly:  by showing that
 they emerge as
soliton solutions of either Type IIA or Type IIB supergravity.
 The nature of
the worldvolume fields is then established by studying the zero
 modes of the
soliton.  In particular, a super $p$-brane requires that the
 soliton solution
preserves some unbroken supersymmetry and hence that the zero
 modes form a
supermultiplet.  Although we know of no general proof that all
supersymmetric
extended objects correspond to a soliton, this is true of all
 those on the old
brane-scan and thus seems a good guide to constructing the new
one.  Following
this route we shall conclude that of all the possible $D = 10$
Type II super
$p$-branes permitted by bose-fermi matching alone, only those
 with $p = 0$
(Type IIA), $p = 1$ (Type IIA and IIB), $p = 3$ (Type IIB),
 $p = 4$ (Type IIA)
$p = 5$ (Type IIA and IIB) and $p = 6$ (Type IIA) actually exist.
  [The reader
may wonder why there seems to be a gap at $p = 2$.  Indeed,
 duality would seem
to demand that in $D = 10$ a Type IIA superfourbrane should
imply a Type IIA
supermembrane.  This object does indeed exist but it should not
be counted as a
new theory since vectors are dual to scalars in $d = 3$ and so its worldvolume
action is simply obtained by dualizing one of the 11 $X^M$ of
the $D = 11$
supermembrane.]  Our results thus confirm the conjecture of
\hors\ that super Type II $p$-brane solitons in $D = 10$ exist for all
$0 \leq p \leq 6$.

Given that the gauge-fixed theories display worldvolume
supersymmetry, and
given that we now wish to include the possibility of vector
(and/or
antisymmetric tensor) fields, it is a relatively
straightforward exercise to
repeat the bose-fermi matching conditions of the Introduction
 for vector (and/or antisymmetric
tensor) supermultiplets.  Once again, we may proceed in one of
 two ways.
First, given that a worldvolume vector has ($d - 2$) degrees of
freedom, the
scalar multiplet condition \dmnone\ gets replaced by
\eqn\dmnthree{D - 2 = {1\over 2}~mn = {1\over 4}~MN .}
Alternatively, we may simply list all the supermultiplets in the
classification of \strath\ and once again interpret $D$ via \dsca.
The results are shown in Fig. 2.

Several comments are now in order:

\item{1)} Vector supermultiplets exist only for $4 \leq d
\leq 10$ \strath.  In $d = 3$ vectors have only 1 degree of
freedom and are dual to scalars.  So these multiplets will
 already have been included as scalar multiplets in section 1.
 In $d = 2$, vectors have no degrees of freedom.

\item{2)} The number of scalars in a vector supermultiplet is
 such that, from
\dsca, $D = 4, 6$ or $10$ only, in accordance with \dmnthree.

\item{3)} One must now repeat the analysis for antisymmetric
tensors to see
if
any new points are introduced on the scan.  For example in
$d = 6$ there is a
chiral $(2,0)$
tensor supermultiplet, with a second rank tensor whose field
strength is
self-dual:  $(B_{\mu\nu}^-, \lambda^I, \phi^{[IJ]})$, $I = 1,
\ldots, 4$,
corresponding to the Type IIA fivebrane and a non-chiral
$(1,1)$ vector
multiplet $(B_{\mu}, \chi^I, A^J\,_I, \xi)$, $I = 1, 2$,
 corresponding to the
Type IIB fivebrane \refs{\calhs,\calhsone}.  However, both
 occupy the same $(d = 6, D = 10)$ slot in
Fig. 2. Nevertheless, there is a new point on the scan,
($d=6$, $D=11$),
namely the $D=11$ superfivebrane, the corresponding $d=6$ supermultiplet
being identical to that of the Type IIA fivebrane. This last observation
corrects an omission in \dufliib.\footnote{$^\ast$}{We are grateful to
Paul Townsend for pointing this out.}

\item{4)} We emphasize that Fig. 2 merely tells us what is
 allowed by
bose/fermi matching.  We must now try to establish which of
these possibilities
actually exists.

All of the circles on the brane-scan are known to correspond to
soliton
solutions of an underlying supersymmetric field theory
\refs{\huglp,\tow,\dabghr,\str,\dufs,\duflfb,\duflhs}.  As for
the crosses, supersymmetric soliton solutions of both Type IIA
and
Type IIB supergravity have been found for the case
$(d = 6, D = 10)$ \calhs\ and of Type IIB for
$(d = 4, D = 10)$ \duflselft\
and of $N=1$ supergravity for the case
$(d=6, D=11)$ \guv. What about the others?
In the next section we shall show that in $D=10$ they exist for
 worldvolume dimensions $d=1$ (Type IIA), $d=2$ (Types IIA and
IIB), $d=3$ (Type IIA), $d=4$ (Type IIB), $d=5$ (Type IIA),
$d=6$ (Types IIA and IIB) and $d=7$ (Type IIA).

\halign{\indent #&\qquad\hfil#\hfil&\quad\hfil # \hfil &
\quad \hfil # \hfil &\quad \hfil # \hfil &\quad \hfil# \hfil &
\quad \hfil # \hfil &\quad \hfil # \hfil&\quad #\hfil \cr
&&$d$ & $\tilde{d}$ & $\alpha (d)$ & $A$ & $B$ & $\phi$&\cr
&&1 & 7 & $-3/2$ & $7C/16$ & $-C/16$ & $-3C/4$&\cr
&&2 & 6 & 1 & $3C/8$ & $-C/8$ & $C/2$&\cr
&&3 & 5 & $-1/2$ & $5C/16$ & $-3C/16$ & $-C/4$&\cr
&&4 & 4 & 0 & $C/4$ & $-C/4$ & 0&\cr
&&5 & 3 & $1/2$ & $3C/16$ & $-5C/16$ & $C/4$&\cr
&&6 & 2 & $-1$ & $C/8$ & $-3C/8$ & $-C/2$&\cr
&&7 & 1 & $3/2$ & $C/16$ & $-7/16$ & $3C/4$&\cr}

\noindent
Table 2.  The functions $A$, $B$ and $\phi$ in terms of $C$ as demanded
by supersymmetry.
\medskip

\newsec{Type II $(d-1)$-branes}

Let us begin in $D = 10$ with Type IIA supergravity, whose bosonic action is
given by
\eqn\actiia{\eqalign{I_{10} (IIA)&={1\over 2\kappa^2}~\int d^{10} x \sqrt{-g}
\Bigg[R - {1\over 2}~(\partial\phi)^2 - {1\over 2.3!}~e^{-\phi} F_3\,^2\cr
&-{1\over 2.2!}~e^{3 \phi/2} F_2\,^2 - {1\over 2.4!}~e^{\phi/2}
F'_4\,^2\Bigg]\cr
&- {1\over 4\kappa^2} \int F_4 \wedge F_4 \wedge A_2 ,\cr}}
where
\eqn\ffour{F'_4 = dA_3 + A_1 \wedge F_3 .}
{}From \aexpre\ we see that the kinetic terms for gravity, dilaton and
antisymmetric
tensors are also correctly described by the generic action $I_{10} (d)$ of
\lgeact\ with $d
= 1, 2, 3$ (i.e $\tilde{d} = 7, 6, 5$).  Both the elementary string ($d = 2$)
and fivebrane $(d = 6)$ solutions of $N = 1$ supergravity described above
continue to provide solutions to Type IIA supergravity, as may be seen by
setting $F_2 = F_4 = 0$.  [This observation is not as obvious as it may seem in
the case of the elementary fivebranes or solitonic strings, however, since it
assumes that one may dualize $F_3$.  Now the Type IIA action follows by
dimensional reduction from the action of $D = 11$ supergravity which contains
$F_4$.  There exists no dual of this action in which $F_4$ is replaced by $F_7$
essentially because $A_3$ appears explicitly in the Chern-Simons term $F_4
\wedge F_4 \wedge A_3$ \nictv.  Since $F_4$~and~$F_3$ in $D = 10$ originate
from
$F_4$ in $D = 11$, this means that we cannot {\it simultaneously} dualize
$F_3$~and~$F_4$ but one may do either {\it separately}.\footnote\dg{We are
grateful to H. Nishino for this observation.}  By partial integration one may
choose to have no explicit $A_3$ dependence in the Chern-Simons term of (4.18)
or no explicity $A_2$ dependence, but not both.]  Furthermore, by setting $F_2
= F_3 = 0$ we find elementary membrane $(d = 3)$ and solitonic fourbrane
$(\tilde{d} = 5)$ solutions, and then by dualizing $F_4$, elementary fourbrane
$(d = 5)$ and solitonic membrane $(\tilde{d} = 3)$ solutions.
Finally, by
setting $F_3 = F_4 = 0$, we find elementary particle $(d = 1)$ and solitonic
sixbrane $(\tilde{d} = 7)$ solutions and then by dualizing $F_2$, elementary
sixbrane $(d = 7)$ and solitonic particle $(\tilde{d} = 1)$ solutions.

Next we consider Type IIB supergravity in $D = 10$ whose bosonic sector
consists of the graviton $g_{MN}$, a complex scalar $\phi$, a
complex 2-form
$A_2$ (i.e with $d = 2$ or, by duality $d = 6$) and a real
4-form $A_4$ (i.e
with $d = 4$ which in $D = 10$ is self-dual).  Because of this
self-duality of
the 5-form field strength $F_5$, there exists no covariant action principle of
the kind \lgeact\ and, strictly speaking, our previous analysis
 ceases to apply.
Nevertheless we can apply the same logic to the equations of
motion and we find
that the solution again falls into the generic category
\solution-\aexpre.  First of all,
by truncation it is easy to see that the same string $(d = 2)$
and fivebrane
$(d = 6)$ solutions of $N = 1$ supergravity continue to solve
the field
equations of Type IIB.  On the other hand, if we set to zero
 $F_3$ and solve
the self-duality condition $F_5 = - ^{\ast}\!F_5$ then we find
the special case
of \solution\ with $d = \tilde{d} = 4$ and hence
$\alpha = 0$~and~$\phi = 0$.  This is the self-dual
 superthreebrane \duflselft\ discussed in more detail below.

All of the above elementary and solitonic solutions satisfy the
mass = charge conditions \massagain\ and \magbogom.
Our next task is to check for supersymmetry.
We begin by making the same ansatz as in section 3.3, namely
\splitx-\dform\ but this
time substitute into the supersymmetry transformation rules
rather than the
field equations, and demand unbroken supersymmetry.  This
 reduces the four
unknown functions $A$, $B$, $C$ and $\phi$ to one.  We then
 compare the results
with the known solutions.

For Type IIA supergravity with vanishing fermion background,
the gravitino
transformation rule is
\eqn\gravitino{\eqalign{\delta\psi_M&=D_m \varepsilon +
{1\over 64}~e^{3\phi/4}
(\Gamma_M\,^{M_1M_2} - 14 \delta_M\,^{M_1} \Gamma^{M_2})
\Gamma^{11}
\varepsilon F_{M_1M_2}\cr
&\qquad +{1\over 96}~e^{-\phi/2} (\Gamma_M\,^{M_1M_2M_3} - 9
\delta_M\,^{M_1}
\Gamma^{M_2M_3}) \Gamma^{11} \varepsilon F_{M_1M_2M_3}\cr
&\qquad +{i\over 256}~e^{\phi/4} (\Gamma_M\,^{M_1M_2M_3M_4} -
{20\over
3}~\delta_M\,^{M_1} \Gamma^{M_2M_3M_4}) \varepsilon
F_{M_1M_2M_3M_4},\cr}}
and the dilatino rule is
\eqn\dilatino{\eqalign{\delta\lambda&={1\over 4}~\sqrt{2}~D_M
\phi \Gamma^M \Gamma^{11}
\varepsilon + {3\over 16}~{1\over \sqrt{2}}~e^{3\phi/4} \Gamma^{M_1M_2}
\varepsilon F_{M_1M_2}\cr
&\qquad \qquad \qquad + {1\over 24}~{i\over \sqrt{2}}~e^{-\phi/2}
\Gamma^{M_1M_2M_3} \varepsilon F_{M_1M_2M_3}\cr
&\qquad \qquad \qquad - {1\over 192}~{i\over \sqrt{2}}~e^{\phi/4}
\Gamma^{M_1M_2M_3M_4} \varepsilon F_{M_1M_2M_3M_4},\cr}}
where $\Gamma^M$ are the $D = 10$ Dirac matrices, where the covariant
derivative is given by
\eqn\covdir{D_M = \partial_M + {1\over 4}~\omega_{MAB} \Gamma^{AB}}
with $\omega_{MAB}$ the Lorentz spin connection, where
\eqn\gammai{\Gamma^{M_1M_2\ldots M_n} = \Gamma^{[M_1} \Gamma^{M_2} \ldots
\Gamma^{M_n]}}
and where
\eqn\gammaii{\Gamma^{11} = i \Gamma^0 \Gamma^1 \ldots \Gamma^9.}
Similarly the Type IIB rules are
\eqn\iibruls{\eqalign{\delta \psi_M = D_M \varepsilon&+{i\over 4 \times
480}~\Gamma^{M_1M_2M_3M_4} \Gamma_M \varepsilon F_{M_1M_2M_3M_4}\cr
&+{1\over 96}~(\Gamma_M\,^{M_1M_2M_3} - 9 \delta_M\,^{M_1} \Gamma^{M_2M_3})
\varepsilon^{\ast} F_{M_1M_2M_3}\cr}}
and
\eqn\iiblam{\delta\lambda = i \Gamma^M \varepsilon^{\ast} P_M - {1\over 24}~i
\Gamma^{M_1M_2M_3} \varepsilon F_{M_1M_2M_3},}
where
\eqn\pmphi{P_M = \partial_M \phi/(1 - \phi^{\ast} \phi).}
In the Type IIB case, $\varepsilon$ is chiral
\eqn\epchiral{\Gamma_{11} \varepsilon = \varepsilon.}

The requirement of unbroken supersymmetry is that there exist
Killing spinors
$\varepsilon$ for which both $\delta\psi_M$~and~$\delta\lambda$
 vanish.
Substituting our ansatze into the transformation rules we find
that for every
$1 \leq d \leq 7$ there exist field configurations which break
exactly half the
supersymmetries.  This is just what one expects for
 supersymmetric extended
object solutions \refs{\huglp,\tow,\dabghr,\str,\dufs,\duflfb}\
and is intimately related to the
$\kappa$-symmetry discussed in the Introduction and the
Bogomoln'yi bounds.
The corresponding values of $A$, $B$ and $\phi$ in terms of $C$
 are given in
Table 2.  The important observation, from \solution, is that
the values required by
supersymmetry also solve the field equations.  Thus in addition
 to the $D = 10$
super $(d - 1)$ branes already known to exist for $d = 2$
(Heterotic, Type IIA
and Type IIB), $d = 4$ (Type IIB only) and $d = 6$ (Heterotic,
 Type IIA and
Type IIB), we have established the existence of a Type IIA
 superparticle $(d =
1)$, a Type IIA supermembrane $(d = 3)$, a Type IIA
superfourbrane $(d = 5)$
and a Type IIA supersixbrane $(d = 7)$.

It is perhaps worth saying a few more words about the self-dual
superthreebrane. By virtue of the (anti) self-duality condition
 $F_5=-^\ast F_5$, the electric Noether
charge coincides with the topological magnetic charge
\eqn\egfour{e_4 = -g_4 .}
(Note that such a condition is possible only in theories
allowing a real
self-duality condition i.e. in 2 mod 4 dimensions, assuming
 Minkowski
signature. The self-dual string of section 3.6 is another
example.) The usual Dirac quantization rule for ($d-1$) branes
\pbdirac\ may,
following \refs{\dir,\zwa,\schwing}, be generalized to dyons
 carrying both
electric and magnetic charges. In $D$ spacetime dimensions, a
 ($d-1$)-brane
with charges ($e_{1d}$, $g_{1\tilde{d}}$) is related to another
with charge
($e_{2d}$, $g_{2\tilde{d}}$) by
\eqn\egd{e_{1d} g_{2\tilde{d}} - e_{2d} g_{1\tilde{d}} = 2\pi n .
}
Note, however, that this by itself says nothing about the
quantization of
the product $e_{1d} g_{1\tilde {d}}$. (Witten \witone\ has
 provided such a
dyon
quantization rule, but it requires either $T$ invariance,
 which is violated
by the self-duality condition, or else an action principle, which is
also absent for Type IIB supergravity.) So we cannot combine \egfour\ and
\egd\ to obtain $e_4$ and $g_4$ as pure numbers.

Finally we count bosonic and fermionic zero modes. We know
that one half of the supersymmetries are broken, hence we have 16 fermionic
zero modes.  Regrouping these 16 fermionic zero modes, we
get four Majorana spinors in $d = 4$. Hence the $d = 4$ worldvolume
supersymmetry is $N = 4$. Worldvolume supersymmetry implies that the
number of fermionic and bosonic on-shell degrees of freedom must be equal,
so we need a total of eight bosonic zero modes. There are the usual six
bosonic translation zero modes, but we are still short of two. The two
extra zero modes come from the excitation of the complex antisymmetric
field strength $G_{MNP}$. The equation of motion for small fluctuations of
the two-form potential $b$ in the soliton background is
\eqn\smfl{D^P G_{MNP} = - {i\over 6}~F_{MNPQR} G^{PQR} .}
This is solved by
\eqn\bzsol{b = e^{ik \cdot x} E \wedge de^{2A} ,}
\eqn\gdb{G = db + i \ast (db) ,}
where $k$ is a null vector in the two Lorentzian dimensions tangent to the
worldvolume. $E$ is a constant polarization vector orthogonal to $k$ but
tangent to the worldvolume and $\ast$ the Hodge dual in the worldvolume
directions. Although $G$ is a complex tensor, the zero-modes
solution
gives only one real vector field on the worldvolume which
 provides the
other two zero modes. These two zero modes turn out to be pure
gauge at
zero worldvolume momentum. Together with the other zero modes,
these
fields make up the $d = 4, N = 4$ matter supermultiplet
$(A_{\mu}, \lambda^I,\phi^{[IJ]})$.

Note that in the self-dual case the condition \loopduality\
simply becomes
\eqn\gddsd{{\rm g}^d_d = 1/{\rm g}^d_d = 1.}
We note as in \hors\
that, by virtue of its dimensionality, this self-dual
threebrane comes closest
to realizing the old idea of ``spacetime as a membrane".

We have classified all supersymmetric extended objects in
$D=10$ that correspond to
solitons of a Poincar\'e supersymmetric field theory in the
usual spacetime
signature which break half the spacetime supersymmetries.
We cannot at the present time rigorously rule out the existence
 of other super
$p$-branes which do
not correspond to solitons.  However, we regard their existence
as unlikely.
  Further progress would require that
we construct the spacetime Green-Schwarz supersymmetric and
$\kappa$-symmetric
actions for these new Type II $p$-branes and, to date, this has
 not been done.
All we know is that, in a physical gauge, the worldvolume
theory corresponding
to the zero modes of the soliton is described by vector or antisymmetric tensor
supermultiplet as in Table 3.
\bigskip

\halign{\indent #&\hfil # \hfil &\quad \hfil # \hfil &\quad \hfil # \hfil &
\quad \hfil# \hfil &\quad \hfil # \hfil&\quad #\hfil \cr
&$d = 7$ & Type IIA & $(A_{\mu}, \lambda, 3\phi)$ &  & $n = 1$&\cr
&$d = 6$ & Type IIA & $(B_{\mu\nu}^-, \lambda^I, \phi^{[IJ]})$ & $I = 1,
\ldots, 4$ & $(n_+, n_-) = (2,0)$&\cr
&   & Type IIB & $(B_{\mu}, \chi^I, A^I\,_J, \xi)$ & $I = 1, 2$ & $(n_+, n_-)
= (1,1)$&\cr
&$d = 5$ & Type IIA & $(A_{\mu}, \lambda^I, \phi^{[IJ]\mid})$ & $I = 1, \ldots,
4$ & $n = 2$&\cr
&$d = 4$ & Type IIB & $(B_{\mu}, \chi^I, \phi^{[IJ]})$ & $I = 1, \ldots, 4$ &
$n = 4$&\cr
&$d = 3$ & Type IIA & $(\chi^I, \phi^I)$ & $I = 1, \ldots, 8$ & $n = 8$&\cr
&$d = 2$ & Type IIA & $(\lambda_L\,^I, \phi_L\,^I)$ & $I = 1, \ldots, 16$ &
$(n_+, n_-) = (16,0)$&\cr
&      & Type IIB & $(\chi^I_L, \phi_L\,^I), (\chi^I_R, \phi^I_R)$ & $I = 1,
\ldots, 8$ & $(n_+, n_-) = (8,8)$&\cr}

\itemitem{Table 3:} Gauge-fixed theories on the worldvolume,
corresponding to
the zero modes of the soliton, are described by the above
 supermultiplets.
\medskip

In our classification, we have also omitted supersymmetric
solitons which break
{\it more than half} the supersymmetries since these solutions
 presumably admit
no $\kappa$-symmetric Green-Schwarz action (at least, not of
 the kind
presently known).  Examples of this are provided by the $D=10$
 double-instanton string of \khubifb\ (which breaks $3/4$),
the $D = 10$ octonionic string of \hars\ (which breaks $15/16$),
 and the $D = 11$ extreme black fourbrane of \guven\ and
extreme black sixbrane of \guv\ (which break $3/4$ and $7/8$,
 respectively).

Finally, we ask what are the implications of our results for
the idea of
duality, in the sense that one theory is simply providing a
 dual
description of the same physics of another theory with the
 weak-coupling
regime of one being the strong-coupling of the other? At the
classical
level discussed in this section, we see that in $D=10$
supersymmetry has
narrowed down the possibilities to just four, namely
 particle/sixbrane duality
(Type IIA only), string/fivebrane duality (Heterotic, Type IIA
 or Type IIB),
membrane/fourbrane duality (Type IIA only) and threebrane
self-duality (Type
IIB only).

Although strictly speaking $D=11$ lies outside the realm of
 superstring theory, the $D=11$ supersymmetric extended objects
 serve the purpose of completing the brane-scan and of
 illustrating the utility of simultaneous dimensional
 reduction.
Before turning our attention to $N = 1$, $D = 11$ supergravity, however,
it is convenient to make the replacement \solution\ in \sigmact\ so that
\eqn\sigmactc{\eqalign{I_D (d) =&{1\over 2\kappa^2} \int d^Dx \sqrt{-g}
e^{-(D-2) a^2C/4d}\bigg[R\cr
&-{a^2\over 8} \bigg(1 - {a^2 (D-1) (D-2)\over 2d^2}\bigg)
(\partial C)^2 -
{1\over 2 \cdot (d+1)!} F_{d+1}^2\bigg],\cr}}
where we have set $C_0 = 0$ for simplicity.  If we now focus on the case
($D = 11$, $d= 3$) we find from \aexpre\ that
\eqn\elevenalpha{a (3) = 0,}
and hence
\eqn\elevenact{I_{11} (3) = {1\over 2\kappa^2} \int d^{11}x \sqrt{-g}
\bigg[R - {1\over 2.4!}F_4^2\bigg].}
This is to be compared with the bosonic sector of $D = 11$ supergravity
\eqn\realeleven{I (D = 11 SUGRA) = {1\over 2\kappa^2} \int d^{11}x \sqrt{-g}
\bigg[R - {1\over 2.4!}F_4^2\bigg] +
{1\over 12\kappa^2}\int F_4 \wedge F_4 \wedge A_3.}
As discussed above, there is no dualized form of this action
since $A_3$ enters explicitly.  We can however find an
 elementary membrane solution.
Once again, this is just the $d= 3$, $\tilde{d} = 6$,
$a (3) = 0$ special case of our general solutions
\solution-\aexpre.
This is the solution of \dufs\ which breaks half the
supersymmetries and
corresponds to the eleven-dimensional supermembrane of \berst,
 for which there is a covariant $\kappa$-symmetric
 Green-Schwarz action. This theory also exhibits a soliton
solution which is the $\tilde d=6, d=3$ superfivebrane of
\guv, for which no such Green-Schwarz action is known.

\newsec{Simultaneous dimensional reduction}

Simple dimensional reduction allows us to derive the actions
$I_D (d)$~and~$S_d$
 for a
($d - 1$)-brane moving in a $D$-dimensional spacetime from the actions
$I_{D + 1}(d)$~and~$S_d$ corresponding to a ($d - 1$)-brane in a
$(D + 1)$-dimensional spacetime.  This
corresponds to
\eqn\simpred{\eqalign{D + 1&\rightarrow D,\cr
d&\rightarrow d,\cr
\tilde{d} + 1&\rightarrow \tilde{d},\cr}}
and takes us vertically on the brane-scan.  Double dimensional reduction
\dufhis, on the other
hand, allows us to derive the actions $I_D (d)$~and~$S_d$ for a
($d - 1$)-brane moving
in $D$-dimensional spacetime from the actions
$I_{D + 1} (d + 1)$~and~$S_{d + 1}$.  This
corresponds to
\eqn\doubred{\eqalign{D + 1&\rightarrow D,\cr
d + 1&\rightarrow d,\cr
\tilde{d}&\rightarrow \tilde{d},\cr}}
and takes us diagonally on the brane-scan.  The first example
 of this was to
rederive the Type IIA superstring in $D = 10$ from the
 supermembrane in $D = 11$ \dufhis. This process
 thus allows us, for example, to rederive the elementary string
of section 2.1 in
$D = 10$ from the $D = 11$ supermembrane discussed above.

To see how it works in general, let us denote all ($D + 1, d + 1$)-dimensional
 quantities
by a hat and all ($D, d$) dimensional quantities without.  Then with
\eqn\split{\eqalign{\hat{X}^{\hat{M}}&= (X^M, X^d), \qquad M = 0, 1, \ldots,
 (d - 1),  (d +1), \ldots, D - 1\cr
\hat{\xi}^{\hat{\mu}}&= (\xi^i, \xi^d),\cr}}
double dimensional reduction consists in setting
\eqn\set{\xi^d = X^d,}
taking $X^d$ to be the coordinate on a circle of radius $R$, and discarding
all but the
zero modes.  In practice, this means taking the background fields $\hat{\phi}$,
$\hat{g}_{\hat{M}\hat{N}}$~and~$\hat{A}_{\hat{M}\hat{N} \ldots \hat{M}_d}$ to
be
independent of $X^d$.  To recover $S_d$, with only background fields $\phi$,
$g_{MN}$~and~$A_{M_1 M_2 \ldots M_{d-1}}$, a further truncation is necessary.
Specifically we write
\eqn\splitfield{\hat{g}_{\hat{M}\hat{N}} (\sigma-{\rm model})=
e^{-2 \beta\phi/d + 1}
\bigg(\matrix{g_{MN} (\sigma-{\rm model})&0\cr
 0&e^{2\beta\phi}\cr}\bigg),}
where $\beta$ is a, for the moment, arbitrary constant and
\eqn\reduten{\hat{A}_{0 1 2 \ldots d + 1} = A_{0 1 2 \ldots d},}
with other components set to zero.  The condition \splitfield\ ensures from
\defgama\ that
\eqn\relgama{\sqrt{-\hat{\gamma}} = \sqrt{- \gamma}}
and hence, together with condition \reduten, we recover the correct
$\sigma$-model
action for $S_{d - 1}$ starting from $\hat{S}_d$ provided
\eqn\redutension{ 2\pi R \hat{T}_{d + 1} =  T_d.}
We fix $\beta$ and the relation between $\hat \phi$ and $\phi$ by requiring
that we obtain the correct background field action
 $I_D (d)$
starting from $I_{D + 1} (d + 1)$. So from \sigmact\
\eqn\reduricci{\eqalign{&e^{- (D - 1) \hat{a} \hat{\phi}/2 (d + 1)}
\sqrt{-\hat{g}} \bigg[\hat{R} - {1\over 2}\bigg(1 - {{\hat a}^2 ( D
(D - 1) \over 2 (d + 1)^2}\bigg)(\partial \hat {\phi})^2 \bigg] \cr
&= e^{-(D - 2) a \phi/2d} \sqrt{-g} \bigg[R - {1 \over 2} \bigg(1 -
{a^2 (D - 1)(D - 2)\over 2 d^2}\bigg) (\partial \phi)^2\bigg],\cr}}
which gives
\eqn\or{\eqalign{\hat{\phi} &= \delta \phi,\cr
 {(D - 1) \hat {a}  \over 2 (d + 1)} \delta &= {(D - 2)a
\over 2 d} - {\tilde d \beta \over d + 1},\cr
1 - {a^2 (D - 1)(D - 2)\over 2 d^2}
&= \delta^2 \bigg(1 - {{\hat a}^2  D (D - 1) \over 2 (d + 1)^2}
\bigg) \cr
&\quad - 4\beta {\tilde d + 1\over d + 1} {(D - 2) a \over 2d}\cr
&\quad + 2 \beta^2 {D(D - 1) - 2(d + 1)(\tilde d + 1)\over (d + 1)^2},\cr}}
and hence
\eqn\betaexp{\beta = {2\over d a },}
\eqn\gammaexp{\delta = {{\hat a}\over a}}
from solving eqs. \or.
We also require
\eqn\redukappa{\hat{\kappa}^2 = 2\pi R \kappa^2.}
Note that the Dirac quantization rule \tensiondirac\ involving
$\kappa^2$~and~$T$ follows from that
involving $\hat{\kappa}^2$~and~$\hat{T}$ on using \redutension\ and
\redukappa.  In canonical variables,
we have
\eqn\redumetric{\eqalign{\hat{g}_{MN} ({\rm canonical})&=
e^{-2\tilde{d}\phi/a(d) (d + \tilde{d}) (d +1 + \tilde{d})}
g_{MN} ({\rm canonical}),\cr
\hat{g}_{dd} ({\rm canonical})&=
e^{{2\tilde d} \phi/(d + 1 +\tilde{d})a (d)}.\cr}}

As an application of simultaneous dimensional reduction, we may derive the
elementary string solution in $D = 10$ from the $D = 11$
membrane solution of \dufs.
The $D = 10$ fields $g_{MN}$, $A_{MN}$~and~$\phi$ are given by
\eqn\duffstelle{\eqalign{\hat{g}_{MN}&= e^{-\phi/6} g_{MN} ({\rm canonical}),
\cr
\hat{g}_{22}&= e^{4\phi/3},\cr
\hat{A}_{012}&= A_{01}.\cr}}
[Curiously, the metric $\hat{g}_{MN}$ in \duffstelle\ bears the
same relation
to $g_{MN}$ (canonical) as does the fivebrane $\sigma$-model
 metric in
\dusigmac\ since $a (d = 2) = 1$~and~$\tilde{d} = 6$.  This
phenomenon
happens in general whenever $\hat{a} = 0$
i.e~~for ($d + 1 = 3, \tilde{d} = 6$), ($d + 1 = 4,
\tilde{d} = 4$) and
($d + 1 = 6,\tilde{d} = 3$)].  Similarly starting from the
sixbrane in $D = 10$ we
 may proceed diagonally
down the brane-scan to a particle in $d = 4$.  It is not
 difficult to show that
the
solutions so obtained will continue to preserve exactly one
half of the
supersymmetries.
Starting from the $d \leq 7$ solutions in $D = 10$ we can thus
 fill out the
triangle of
supersymmetric extended objects shown in Fig. 3. At first
sight, this seems to contradict Fig. 2 since solutions appear
where no
supermultiplet is allowed.  The resolution is simply that only
 the cases $d =
1, 3, 4, 5, 6$~and~$7$ in $D = 10$ are fundamental.  All the
others are
obtained by simply dimensional reduction of these or the
$D = 11$
supermembrane, and are thus described by the same gauge-fixed
action.

Finally, it is shown in \gibt\ that in the special cases $\tilde d=3,d=6$;
$\tilde d=6,d=3$ and $\tilde d=4,d=4$ (all of which, as remarked above, have
$a=0$) the endpoints of the interpolation, namely Minkowski space
$M_D$ at $y=\infty$ and $(AdS)_{\tilde d +1} \times S^{d+1}$ at $y=0$, are
{\it maximally} supersymmetric.

\chapter{Heterotic strings and fivebranes in D=10}

\def\wbg{\overline{\beta}^G_{MN}}
\def\wbb{\overline{\beta}^B_{MN}}
\def\wbd{\overline{\beta}^\phi}
\def\rbg{\beta^G_{MN}}
\def\rbb{\beta^B_{MN}}
\def\rbd{\beta^\Phi}
\def\sqr#1#2{{\vbox{\hrule height.#2pt\hbox{\vrule width
.#2pt height#1pt \kern#1pt\vrule width.#2pt}\hrule height.#2pt}}}
\def\Box{\mathchoice\sqr64\sqr64\sqr{4.2}3\sqr33}
\def\RIJKL{R^I{}_{JKL}}
\def\GRIJKL{\hat R^I{}_{JKL}}

\def\grijkl{\hat R^i{}_{jkl}}

\def\met {g_{mn}}

\def\b#1{\vec\beta_{#1}}
\def\x#1{\vec X_{#1}}

\newsec{Inclusion of Yang-Mills fields}

In this section we  focus on soliton solutions of the heterotic
 string \grohmr. Of particular importance is the heterotic
fivebrane of \str, which did much to lend credence to the
conjecture of string/fivebrane duality {\duf,\str}. The  field
theory limit of the heterotic string is the $2$-form version of
$D=10$ supergravity coupled to either $E_8\times E_8$ or
 $SO(32)$ super Yang-Mills. Including the Lorentz Chern-Simons
term, which follows from stringy corrections
\canhsw,
the bosonic sector of the action in string $\sigma$-model
 variables is
\refs{\berrwv,\cham}
\eqn\sbhs{I_{10} ({\rm heterotic}) = {1\over 2 \kappa^2}
\int d^{10} x \sqrt {-g} e^{-2\phi}\left( R +
4(\partial \phi )^2 - {1\over 2 \cdot 3!} H^2 +
{\alpha' \over 8}\left( tr \hat R^2 - trF^2\right)\right),}
where $\hat R$, the curvature generated by the generalized
connection
\eqn\omab{{\Omega_\pm}_M\,^{AB} = \omega_M\,^{AB} \pm {1\over 2} H_M\,^{AB},}
where $\omega_M\,^{AB}$ is the usual spin-connection, is
explicitly given by
\eqn\gcurv{\GRIJKL=\RIJKL\mp {1\over
2}\left(\nabla_LH^I{}_{JK}-\nabla_KH^I{}_{JL}\right)
+{1\over 4}\left(H^M{}_{JK}H^I{}_{LM}-H^M{}_{JL}H^I{}_{KM}\right).}
$dH$ now obeys the Bianchi identity modified by Chern-Simons
 terms
\footnote{$^\dagger$}{The normalization used below follows that
 of \refs{\dixdp,\towspan}, and  differs from that of \str\ by
 a factor of $8$. This discrepancy was first brought to our
 attention by Paul Townsend.}
\eqn\dhrf{dH = {\alpha'\over 4} \big( tr\hat R^2 - trF^2\big).}
 The trace is in the fundamental representation in the case of
$SO(32)$ and is defined to be $1/30$ times the trace in the
adjoint representation in the case of $E_8\times E_8$.

We begin in section 5.2 with a review of the 't Hooft ansatz
 for the Yang-Mills
instanton \refs{\thoone\wil\corf{--}\jacnr}\ and then turn to
the analogous axionic instanton solution discussed in
\refs{\rey,\khuinst}. The generalized curvature of this and all
other ``fivebrane ansatz" solutions possesses a (anti)
 self-dual structure similar to that of the 't Hooft ansatz,
 albeit in the gravitational sector of the string.

In section 5.3 we write down the perturbative ``gauge''
heterotic soliton solution of \str\ and the ``symmetric''
solution of \refs{\calhs,\calhsone}, which generalizes the
axionic instanton and the gauge solution to a heterotic
multi-fivebrane solution with a YM instanton in the gauge
sector and an axionic instanton in the gravitational sector in
the
four-dimensional space transverse to the fivebrane
We also discuss the zero modes of these solutions.

In section 5.4 we write down the string soliton solution of
\duflhs, which
extends the heterotic string soliton solution of chapter 2 to
incorporate a Yang-Mills field with eight-dimensional
instanton structure. In section 5.5 we summarize the octonionic
 string soliton of \hars\ and the double-instanton string
solution of \khubifb.

\newsec{'t Hooft ansatz and the axionic instanton}

Consider the four-dimensional Euclidean action
\eqn\ymact{S=-{1\over 2g^2}\int d^4y {} tr F_{mn}F^{mn},
\qquad\qquad m,n =1,2,3,4.}
For gauge group $SU(2)$, the fields may be written as $A_m=(g/2i)
\sigma^a A_m^a$ and $F_{mn}=(g/2i)\sigma^a F_{mn}^a$\ \
(where $\sigma^a$, $a=1,2,3$ are the $2\times 2$ Pauli matrices).
The equation of motion derived from this action is solved by the
't Hooft ansatz \refs{\thoone\wil\corf{--}\jacnr}
\eqn\hfanstz{A_{mn}=i \overline{\Sigma}_{mn}\partial_n \ln f,}
where $\overline{\Sigma}_{mn}=\overline{\eta}^{imn}(\sigma^i/2)$
for $i=1,2,3$, where
\eqn\hfeta{\eqalign{\overline{\eta}^{imn}=-\overline{\eta}^{inm}
&=\epsilon^{imn},\qquad\qquad m,n=1,2,3,\cr
&=-\delta^{im},\qquad\qquad n=4 \cr}}
and where $f^{-1}\Box\ f=0$. The above solution obeys the self-duality
condition
\eqn\Finst{F_{mn} = \tilde F_{mn} = {1\over 2} \epsilon_{mn}\,^{pq} F_{pq}.}
The ansatz for the anti-self-dual solution
$F_{mn} = - \tilde F_{mn}$
is similar, with the $\delta$-term in \hfeta\ changing sign.
To obtain a multi-instanton solution, one solves for $f$ in the
four-dimensional space to obtain
\eqn\finst{f=1+\sum_{i=1}^k{\rho_i^2\over |\vec y - \vec a_i|^2},}
where $\rho_i$ is the instanton scale size, $\vec a_i$ the location in
four-space of the $i$th instanton and
\eqn\inwind{k = {1\over 16 \pi^2} \int_{M^4} tr F^2}
is the instanton number. Note that this solution has $5k$ parameters,
while the most general (anti) self-dual solution has $8k$ parameters, or
$8k-3$ if one excludes the $3$ zero modes associated with global $SU(2)$
rotations.
We do not exclude these modes, however, because as explained
 below, they belong to the same supermultiplet as the dilatational zero mode.

Now consider the ansatz
\eqn\sansatz{\eqalign{\met&=e^{2\phi}\delta_{mn}\qquad m,n=6,7,8,9,\cr
g_{\mu\nu}&=\eta_{\mu\nu}\qquad\quad   \mu,\nu=0,1,2,3,4,5,\cr
H_{mnp}&=\pm 2\epsilon_{mnpk}\partial^k\phi
\qquad m,n,p,k=6,7,8,9.\cr}}
Then provided $e^{-2\phi}\Box\ e^{2\phi}=0$, \sansatz\ is a solution to the
low-energy string effective action \stringact\ written in terms of string
$\sigma$-model variables (where here $F_3=H$), and which breaks $1/2$ the
spacetime supersymmetries.
In particular, for
\eqn\efinst{e^{2\phi}=e^{2\phi_0}\left(1+\sum_{i=1}^N{\rho_i^2\over
|\vec y - \vec a_i|^2}\right),}
we recover the multi-fivebrane solution of section 2.4. The
ansatz \sansatz\ in fact possesses a (anti) self-dual structure
in the transverse space $(6789)$, which can be seen by
expressing the generalized curvature in covariant form in terms
of the dilaton field as \khuinst\
\eqn\gcurvphi{\grijkl=\delta_{il}\nabla_k\nabla_j\phi
-\delta_{ik}\nabla_l\nabla_j\phi+\delta_{jk}\nabla_l\nabla_i\phi
-\delta_{jl}\nabla_k\nabla_i\phi\pm\epsilon_{ijkm}\nabla_l\nabla_m\phi
\mp\epsilon_{ijlm}\nabla_k\nabla_m\phi.}
It easily follows that \khumonin\
\eqn\axin{\grijkl=\mp {1\over 2} \epsilon_{kl}{}^{mn}\hat R^i_{jmn}.}
So the (anti) self-duality appears in the gravitational sector of the string
in terms of its generalized curvature thus justifying the name ``axionic
instanton" for the four-dimensional solution first found in \rey.

\newsec{The heterotic fivebrane as a soliton: ``gauge'' versus ``symmetric''}

In terms of string $\sigma$-model variables, \eskillsp\ and \eskillsptwo\
are rewritten as
\eqn\ssrone{\delta \lambda = - {1\over 2{\sqrt 2}} \Gamma^M \partial_M \phi
\varepsilon + {1\over 2\cdot 2{\sqrt 2} \cdot 3!} \Gamma^{MNP} H_{MNP}
\varepsilon = 0,}
\eqn\ssrtwo{\delta \psi_M = \partial_M \varepsilon + {1\over 4}
{\Omega_-}_M\,^{AB}\Gamma_{AB} \varepsilon = 0.}
 For nontrivial Yang-Mills
field, the gaugino $\chi$ supersymmetry transformation for zero
 Fermi fields is given by
\eqn\ssthree{\delta \chi = F_{MN} \Gamma^{MN} \varepsilon=0,}
which is solved \str\ by setting $F_{\mu\nu}=F_{\mu m}= 0$ and
keeping just an $SU(2)$ subgroup of the gauge group and
identifying the
corresponding gauge field in the transverse directions with the
instanton
configuration \Finst\ for
\eqn\epeta{\varepsilon = \epsilon_0 \otimes \eta_0 ,}
where
\eqn\gams{(1 - \gamma_7) \epsilon_0 = 0, \qquad (1 - \Gamma_5)
\eta_0 = 0.}
This can be seen explicitly from
\eqn\insol{\eqalign{\delta \chi =& F_{MN} \Gamma^{MN}
 \varepsilon \cr
                      =& F_{mn} \Sigma_{mn} \varepsilon \cr
                      =& {1\over 2}F_{mn} \Sigma^{mn} (1 -
\Gamma_5) \varepsilon
\cr
                      =& 0, \cr}}
where we have used the identity
${1\over 2}\epsilon_{mn}\,^{pq} \Sigma_{pq} = - \Gamma_5
\Sigma_{mn}$.
In a similar manner, the equations \ssrone\ and \ssrtwo\ for
the ansatz \sansatz\ and constant chiral spinor \epeta\ reduce
to
\eqn\omgam{{\Omega_-}_M\,^{AB} \Gamma_{AB} \varepsilon = 0,}
which is solved precisely because of the self-duality relation
\eqn\sdom{{\Omega_-}_M\,^{mn} = {1\over 2} \epsilon^{mn}\,_{pq}
 {\Omega_-}_M\,
^{pq}}
of the generalized connection. The condition \gams\ once again
 means that one half of the supersymmetries are broken. The
 generalized
connection ${\Omega_-}_M\,^{ab}$ lies in the same $SU(2)$
subgroup of $SO(4)$
as $F_{mn}$ does. Note that we have still not specified the
 dilaton field, which will be determined from the Bianchi
identity \dhrf.

It is at this stage that we may proceed in one of two
directions, depending on whether we consider the pure
supergravity theory or whether we include the
stringy Lorentz Chern-Simons correction. The first route
(which we shall shortly justify) leads to the `gauge'' solution
of \str, while the second route leads to the ``symmetric''
solution of \refs{\calhs,\calhsone}. In the first case, using
$Tr_{E_8} T^a T^b$ ( adjoint) $= 15 Tr_{SU(2)}T^a T^b$ (adjoint)
as given in \bercgw\ and substituting a single instanton
\Finst\ with scale size $\rho$, the Bianchi identity \dhrf\
 reduces for the ansatz \sansatz\ to
\eqn\Dileq{\Box e^{2\phi} = - \alpha' {24 \rho^4 \over (r^2
+ \rho^2)^4} e^{3\phi_0/2},}
which gives
\eqn\Dilsol{e^{2\phi} = e^{2\phi_0} +  \alpha'  {r^2
+ 2\rho^2 \over (r^2 + \rho^2)^2} e^{3\phi_0/2}.}
In analogy with the
comparison between the t'Hooft-Polyakov monopole and Dirac's
magnetic monopole,
the large-distance behavior of the gauge heterotic fivebrane is
 the same as that of
the elementary fivebrane. Therefore, we expect that the
 Bogomol'nyi bound
is also saturated since the mass and the charge depend only on
 the large
distance behavior of the solution. This is indeed the case
\str. For a general instanton solution, the magnetic charge is
given by
\eqn\topwind{g_6={1\over \sqrt{2}\kappa} \int _{S^3}
H={\sqrt{2}\alpha' k \Omega_3\over \kappa},}
where $k$ is defined in \inwind.
Using $\alpha'=1/2\pi T_2$ and comparing \topwind\ with \gsix,
 we find that the integer in the Dirac quantization rule is
 given by $n=k$. This implies in particular that the
single-instanton solution yields a heterotic fivebrane with a
tension equal to that of the fundamental fivebrane.

If one keeps the $tr \hat R^2$ term, the so-called
 ``symmetric'' solution arises \refs{\calhs,\calhsone}. This
can be seen by equating the generalized connection
 $\Omega_{\pm M}$ to the gauge
connection $A_M$ \chad\ so that the corresponding
curvature $\hat R(\Omega_{\pm})$ cancels against the
Yang-Mills field strength $F$
and \dhrf\ reduces to $dH=0$, which then implies
 $e^{-2\phi}\Box\ e^{2\phi}=0$ and \gcurvphi\ and \axin.
It then follows that the solution satisfies
\eqn\exsol{F_{pq}{}^{mn}=\hat R_{pq}{}^{mn},}
where both $F$ and $R$ are (anti)self-dual.
We shall argue in section 8 that this solution is exact
since
$A_M=\Omega_{\pm M}$ implies that all the higher order
 corrections vanish.
The solution is now given by \sansatz\ with
 $e^{2\phi}=e^{2\phi_0} f$, where
$f$ is given by \finst, but where from the Dirac quantization
condition it
follows that
$\rho_i^2=e^{-\phi_0/2} n_i \alpha'$.
For this solution, then, the curvature $\hat R$ is of the
same order in $\alpha'$ as $F$ and cannot be dropped in a
perturbative
approximation. By contrast, had we included the $tr \hat R^2$
 term when
solving for the gauge fivebrane, we would have found that $tr R^2$ vanishes
 and hence that $tr \hat R^2$ is higher order in $\alpha'$ \str.

We finally come to the bosonic and fermionic zero modes
associated with
the above heterotic fivebranes. Since the core of the fivebrane
is essentially a four-dimensional Yang-Mills instanton dressed
up with axion and dilaton fields, the zero modes that we are
going to count are actually those of the instanton. For
example, the bosonic (fermionic) zero modes arising from
translation (supertranslation) invariance are the same in both
the gauge and the gravitational sectors, therefore they need
only be counted once. Hence counting the zero modes associated
with the instanton is sufficient. There are four translational
zero modes (the location of the instanton center) and one
dilatational zero mode (the instanton size). Global $SU(2)$ rotations of the
instanton add three more zero modes. In addition there are 112 zero modes
associated with the minimal embedding of $SU(2)$ in $E_8$ or $SO(32)$. These
are actually related to the generators of the coset $E_8/E_7 \times SU(2)$ or
$SO(32)/ SO(28)\times SO(4)$ which do not leave the $SU(2)$ subgroup invariant.
So we have a total of 120 bosonic zero modes. The fermionic zero modes can be
determined from the Atiyah-Singer index theorem
\eqn\atsi{n_{-} - n_{+} = {1\over 8 \pi^2} \int Tr F^2,}
which from \inwind\ and $k=1$ gives 60 anti-chiral fermionic
zero modes. Since
each Weyl spinor in Euclidean four-dimensional space gives $4$
off-shell or
2 on-shell degrees of freedom, 60 fermionic zero modes give
$240$ off-shell
or 120 on-shell degrees of freedom. The dynamics of these zero
 modes can also
be described by 60 six-dimensional sympletic Majorana-Weyl
spinors transforming
covariantly under $SO(1, 5)$. This can be achieved through the
(super)
collective coordinate expansion discussed in some detail in
\refs{\str,\calhsone}. As expected,
there are equal on-shell bosonic and fermionic degrees of
 freedom and the gauge-fixed theory on the worldvolume is given
by a $d=6,(2,0)$ supersymmetric $\sigma$-model on a hyperkahler
 manifold. This suggests that there might exist a
Green-Schwarz-like formulation with worldvolume
$\kappa$-symmetry for the heterotic fivebrane. For the
symmetric solution, since the scale size is quantized in
units of $\alpha'$, the dilatational zero mode, and all its
 superpartners, are absent.

\newsec{Heterotic string solitons}

In the previous section, we saw that the field theory limit of
the heterotic string admits a gauge heterotic fivebrane as a
 soliton. In this section, we show the converse, a result which
 lends further support to the idea of string/fivebrane duality.
 After constructing the solution, we examine its zero-modes and
suggest that they might correspond to those of the fundamental
heterotic string written in a physical gauge.

Our first task is to construct the fivebrane analog of \sbhs,
i.e. to generalize \sfbrane\ to include the Yang-Mills fields.
 We claim that the
result is
\eqn\sbfb{\eqalign{\tilde I_{10}({\rm heterotic}) =
{1\over 2\kappa^2}~\int d^{10}x &\sqrt{-g}~e^{2\phi/3}
\bigg(R -{1\over 2\cdot 7!} K^2 +{\beta'\over 24} t^{LMNOPQRS}
trF_{LM} F_{NO} F_{PQ} F_{RS}\cr
& - {1\over 8} trF_{LM} F_{NO} tr\hat R_{PQ} \hat R_{RS}
+ {1\over 32} tr\hat R_{LM} \hat R_{NO} tr\hat R_{PQ} \hat R_{RS}\cr
& + {1\over 8} tr\hat R_{LM} \hat R_{NO} \hat R_{PQ} \hat R_{RS} ) + \cdots
\bigg),\cr}}
where the tensor $t^{IJKLMNPQ}$ is given in \gresw\ by
\eqn\tijklmnpq{\eqalign{t^{IJKLMNPQ}=&-{1\over 2} (g^{IK} g^{JL} - g^{IL}
g^{JK}) (g^{MP} g^{NQ} - g^{MQ} g^{NP})\cr
&-{1\over 2} (g^{KM} g^{LN} - g^{KN} g^{LM}) (g^{PI} g^{QJ} -
 g^{PJ} g^{QI})\cr
&-{1\over 2} (g^{IM} g^{JN} -g^{IN} g^{JM}) (g^{KP} g^{LQ} -
 g^{KQ} g^{LP})\cr
&+{1\over 2} (g^{JK} g^{LM} g^{PN} g^{QI} + g^{JM} g^{NK} g^{LP} g^{QI}\cr
&\qquad + g^{JM} g^{NP} g^{KQ} g^{LI} + \,{\rm permutations}).
\cr}}
$dK$ now obeys the Bianchi identity modified by Chern-Simons
corrections
\eqn\fdomfr{dK ={\beta'\over 24}\Biggl(trF^4 -
{1\over 8}trF^2{}tr\hat R^2  +{1\over 32} (tr\hat R^2)^2 +
{1\over 8}tr\hat R^4\Biggr).}
The fivebrane tension $T_6$ is given by $1/\beta' = (2\pi)^3 T_6$. This
unconventional quartic action, and the corresponding quartic
Chern-Simons terms require some justification. This will
necessarily be
indirect since, although the super fivebrane
\refs{\achetw,\berst}
$\sigma$-model is
well-known, the heterotic fivebrane $\sigma$-model has yet to
be constructed.
Even if we knew it, the quantization of fivebranes is still in
 its infancy,
and it is doubtful that \sbfb\ could yet be derived as
rigorously as \sbhs.
As in section 2.3, the point of view we adopt is that the
fivebrane action is
obtained by dualizing the string action. In particular, the
Bianchi identity for $K$ follows from the field equation for $H$.
 However, this process does not respect the loop expansion, and
 what is a tree-level effect in string perturbation theory
maybe a one-loop effect in fivebrane perturbation theory and
vice-versa. To understand this, we recall the relationship
\gtwosix\ between the string loop coupling constant ${\rm g}_2$,
  the fivebrane loop coupling constant
${\rm g}_6$ and $\phi_0$. In string variables, each term in the
string
tree-level action $I_{10}({\rm heterotic})$ is proportional
to
$e^{-2\phi}$. Similarly, in fivebrane variables each term in
 the fivebrane tree-level action
$\tilde I_{10}({\rm heterotic})$ is proportional to
$e^{2\phi/3}$. Thus the Green-Schwarz \gres\ anomaly
cancellation term
\eqn\banom{B\wedge\Biggl(trF^4 -{1\over 8}trF^2{}tr\hat R^2  +
{1\over 32} (tr\hat R^2)^2 + {1\over 8}tr\hat R^4\Biggr),}
which provides a correction to the $H$ field equation, has no
$\phi$-dependence in string variables and is therefore seen to
be $1$-loop in string perturbation theory. Similar remarks
apply to the quartic
Yang-Mills and gravitational terms which appear in $1$-loop
corrections to the effective action \elljm.

On the other hand, both these terms are {\it tree-level} in
fivebrane
perturbation theory, because they both behave like $e^{2\phi/3}$ in
fivebrane variables. The $1$-loop Green-Schwarz corrections to
the $H$ field equation now become the tree-level Chern-Simons
 corrections to the $K$ Bianchi identity \fdomfr. Similarly,
 the quartic terms must now be included in the fivebrane
tree-level action $\tilde I_{10}({\rm heterotic})$. By the same
 token, the quadratic Yang-Mills term in
$I_{10}({\rm heterotic})$ and the Chern-Simons term in \dhrf\
corresponding to
$(tr F^2-tr \hat R^2) \wedge A$ are
$1$ loop in fivebrane perturbation theory since they are
independent of $\phi$ when written in fivebrane variables. We
 therefore omit them from
$\tilde I_{10}({\rm heterotic})$. In arriving at \sbfb\ and
\fdomfr, we have also employed the equation \ttwosix\ with
 $n =1$, $2\kappa^2 = (2\pi)^5\alpha'\beta'$  which relates
 the two fundamental tensions. To further justify \sbfb\ and
\fdomfr, we note the following. First, $\alpha'$ has dimension
$-2$ and $\beta'$ has dimension $-6$, so purely on dimensional
 grounds, we would expect a quartic Yang-Mills action.  This
causes no problems with unitarity.  We
emphasize that the {\it exact}
string and fivebrane actions are equivalent; it is merely the
 division into
``classical" plus ``quantum" which is different in the two
cases.
 Secondly, the Yang-Mills Chern-Simons corrections \fdomfr\ can
 be derived directly from the coupling of Yang-Mills fields to
 a fundamental fivebrane \dixds, and a case can be made for the
 gravitational corrections as well \refs{\dixdp,\dufm}.
Thirdly,
 under the two-parameter rescalings of the background fields
discussed in
section 2.3, $g_{MN}\rightarrow \lambda^{1/2} \sigma^{3/2}
g_{MN}$,
$B_{MN} \rightarrow \lambda^2 B_{MN}$, $A_{MNOPQR} \rightarrow
\sigma^6
A_{MNOPQR}, e^{\phi}
\rightarrow \lambda^3 \sigma^{-3} e^{\phi}$, the elementary
fivebrane
$\sigma$-model action $S_6$ and the elementary string
$\sigma$-model action
$S_2$ scale like $S_6 \rightarrow \sigma^6 S_6$ and $S_2
\rightarrow \lambda^2
S_2$. In order that $I_{10}({\rm heterotic})$ admits a
fivebrane as a soliton \str, it must scale the same way under
the $\sigma$ symmetry i.e. $I_{10}({\rm heterotic})
\rightarrow \sigma^6 S I_{10}({\rm heterotic})$. This is indeed
 the case. Similarly, we are encouraged
in our search for a string soliton solution \duflhs\ of $\tilde
I_{10}({\rm heterotic})$ by
noting that it scales in the right way under the $\lambda$
symmetry i.e. $\tilde I_{10}({\rm heterotic}) \rightarrow
\lambda^2  \tilde I_{10}({\rm heterotic})$. Without further
apology, we
now quote the solution which is the analog of the gauge
 fivebrane rather than
the symmetric fivebrane. We therefore ignore the gravitational
 Chern-Simons
terms.
The Yang-Mills field is given by the eight-dimensional
instanton \refs{\gros,\grokstwo}
\eqn\feightin{F_{mn} = f (r)~{i\over2}\Sigma_{mn}
(1 - \Gamma_9)/2, \quad f = 4
\rho^2/(r^2 + \rho^2)^2 }
embedded in an $SO(8)$ subgroup of $SO(32)$ or $E_8$, where
 $\rho$ is the instanton size. The instanton winding number is
\eqn\winks{\tilde k = {1\over 384 \pi^4} \int_{M^8} tr F^4.}
Curiously enough, we find $\tilde k=1$ in the case of $SO(32)$
but $\tilde k=0$
in the case of $E_8$ because the group $E_8$ has no independent
 fourth-order Casimir. We refer the reader to
\refs{\duflhs,\jxthesis}\ for further details.
The supergravity fields are given by
\eqn\hsansatz{\eqalign{ds^2&=e^{4\phi/3}\eta_{\mu\nu} dx^\mu
dx^\nu +
e^{-2\phi/3} \delta_{mn} dy^m dy^n, \cr
B_{01}&=-e^{2\phi}\cr}}
and
\eqn\hssol{\eqalign{e^{-2\phi} &= 1 + k_2 {r^6 + 6 r^4
\rho^2
+ 15 r^2 \rho^4 + 20 \rho^6\over (r^2 + \rho^2)^6},\qquad
{\rm for}\,
SO(32) \cr &= 1 + {|n|k_2\over r^6},
                         \qquad \qquad \quad {\rm for}\,E_8
\cr}}
where for convenience we have set $\phi_0=0$ and where the
constant $k$ is
given by $k_2 = \pi/(3 T_6\Omega_7)$ as in \ktwo, $n$ is an
integer and $\Omega_7$ is the volume of the unit seven-sphere.
In analogy with the
comparison between the t'Hooft-Polyakov monopole and Dirac's
 magnetic monopole,
the large-distance behavior of the gauge heterotic string is
the same as that of the elementary string. Therefore, we expect
 that the Bogomol'nyi bound
is also saturated since the mass and the charge depend only on
the large
distance behavior of the solution. This is indeed the case
\duflhs. This implies in particular that the single-instanton
solution yields a gauge heterotic string with a tension equal
 to that of the fundamental string. We emphasize that the
solitonic string solution
has been established only to tree level in the fivebrane field
theory and are
{\it a priori} valid only for  $e^{-2\phi/3} << 1$. The question of
whether it survives loop corrections remains a topic for future
 research.

Although we omit the string Chern-Simons terms corresponding to
$(tr F^2-tr \hat R^2)$ from our classical fivebrane
considerations, they play an
important role as the fivebrane analogue of the Green-Schwarz
 anomaly
cancellation terms \refs{\dixdp,\dufm}. In the case of strings,
 the combined gravitational and
Yang-Mills anomalies for $N = 1$ supergravity coupled to a
Yang-Mills
supermultiplet (with $n$ left-hand Majorana-Weyl spinors in the adjoint
representation) can be characterized by a certain 12-form, $I_{12}$. As
discussed in \gres, the anomaly can be cancelled only if $I_{12}$
factorizes into an expression of the form $I_{12} = dH{\wedge} X_8$ where $X_8$
is an 8--form. The necessary and sufficient conditions are
\eqn\ifact{\eqalign{n&={\rm dim}~G = 496 ,\cr
Tr F^6&={1\over 48}~Tr F^4 Tr F^2 - {1\over 14,400}~(Tr F^2)^2.
\cr}}
There are only two solutions for $G$: $SO(32)$
 and $E_8 \times E_8$. The anomaly is
then cancelled by the addition of a term in the action
 $B{\wedge} X_8$. In
the case of the fivebrane, we would require that $I_{12}$
factorizes into an
expression of the form $I_{12} = X_4{\wedge} dK$ where $X_4$ is
 a 4-form.
Assuming that the same $I_{12}$ governs both strings and
 fivebranes, we
discover from \gres\ that the necessary and sufficient
conditions for this
to happen are exactly the same as those given in \ifact.
Hence
we find $SO(32)$
and $E_8 \times E_8$ once more. The anomaly is then cancelled
by the term
$X_4{\wedge} A$ \refs{\sala,\gatn}. Thus $I_{12}$ takes on the
string/fivebrane symmetrical form
$I_{12} = dH{\wedge} dK$.

A different extended soliton string solution, the
``octonionic'' string, was constructed in \hars. It differs
 from the gauge heterotic string discussed above by solving
 $I_{10}({\rm heterotic})$ rather than
$\tilde I_{10}({\rm heterotic})$. Once again, an
eight-dimensional instanton makes its appearance in the
transverse space but this time it is the octonionic instanton
of \refs{\fain,\fubn}\ which preserves $SO(7)$ rather than
$SO(8)$.
This solution breaks $15/16$ of the supersymmetries. Because
the Yang-Mills
lagrangian is here quadratic rather than quartic, however, the
scaling
arguments given above for the gauge heterotic string no longer
 apply. As a
consequence, the octonionic string has infinite mass per unit
length.

Another example of a string solution of
$I_{10}({\rm heterotic})$ and which therefore also has
infinite mass per unit length is the double-instanton string
 solution of \khubifb. The ansatz
\eqn\bifbsol{\eqalign{\phi&=\phi_1 + \phi_2,\cr
\met&=e^{2\phi_1}\delta_{mn}\qquad m,n=2,3,4,5,\cr
g_{ij}&=e^{2\phi_2}\delta_{ij}\qquad i,j=6,7,8,9,\cr
g_{\mu\nu}&=\eta_{\mu\nu}\qquad\quad   \mu,\nu=0,1,\cr
H_{mnp}&=\pm 2\epsilon_{mnpq}\partial^q\phi
\qquad m,n,p,q=2,3,4,5,\cr
H_{ijk}&=\pm 2\epsilon_{ijkl}\partial^k\phi
\qquad i,j,k,l=6,7,8,9\cr}}
with constant chiral spinors
$\epsilon_\pm=\epsilon_2 \otimes \eta_4 \otimes \eta_4'$ solves
the
supersymmetry equations \ssrone, \ssrtwo\ and \ssthree\ with
 zero background fermi fields provided the YM gauge
field satisfies the instanton (anti) self-duality condition
\eqn\ymin{\eqalign{F_{mn}&=\pm {1\over 2}\epsilon_{mn}{}^{pq}
F_{pq},\qquad m,n,p,q=2,3,4,5\cr
F_{ij}&=\pm {1\over 2}\epsilon_{ij}{}^{kl} F_{kl},\qquad
 i,j,k,l=6,7,8,9.\cr}}
The chiralities of the spinors $\epsilon_2$, $\eta_4$ and
 $\eta_4'$
are correlated by
\eqn\chicor{(1 \mp \gamma_3)\epsilon_2 = (1 \mp \gamma_5)\eta_4
= (1 \mp \gamma_5)\eta_4'=0,}
so that three-quarters of the spacetime supersymmetries are
broken. An exact
solution is now obtained as follows. Embed the generalized
$\Omega_{\pm M}$ in an $SU(2) \times SU(2)$ subgroup of the
 gauge group,
and equate it
to the gauge connection $A_M$ \chad\ for $M=2,3,4,5,6,7,8,9$ so
 that $dH=0$ and
the corresponding curvature $R(\Omega_{\pm})$ cancels against
 the Yang-Mills
field strength $F$ in both subspaces $(2345)$ and $(6789)$.
For $e^{-2\phi_1}\Box\ e^{2\phi_1}=e^{-2\phi_2}\Box\
 e^{2\phi_2}=0$ it follows that both $F$ and $R$ are (anti)
self-dual in both four-dimensional subspaces.  The explicit
 solution for $\phi_1$ and $\phi_2$ in \bifbsol\ is given by
\eqn\multin{\eqalign{e^{2\phi_1}&=e^{2\phi_{1_0}}\left(1+
\sum_{i=1}^N{\rho_i^2\over |\vec x - \vec a_i|^2}\right),\cr
e^{2\phi_2}&=e^{2\phi_{2_0}}\left(1+
\sum_{j=1}^{M}{\lambda_j^2\over |\vec y - \vec b_j|^2}\right),
\cr}}
where $\vec x$ and $\vec a_i$ are four-vectors and $\rho_i$ instanton scale
sizes in the space $(2345)$, and $\vec y$ and $\vec b_j$ are four-vectors
and $\lambda_j$ instanton scale sizes in the space $(6789)$. Axion charge
quantization then requires that $\rho_i^2=e^{-2\phi_{1_0}}n_i\alpha'$ and
$\lambda_j^2=e^{-2\phi_{2_0}}m_j\alpha'$, where $n_i$ and $m_j$ are integers.
Note that for $N=0$ or $M=0$ we recover the symmetric fivebrane solution
of section 5.2.
It is interesting to note that both the charge $Q_2=-1/2 \int_{S^7}{}^\ast H$
and the mass per unit length ${\cal M}_2$ of the infinite string diverge.
By contrast, all classes of fivebrane solutions have finite charge and mass
per unit length as a result of the preservation of half the spacetime
supersymmetries and the saturation of a Bogomol'nyi bound. The fact that
three-quarters of the spacetime supersymmetries are broken for this solution
means that the saturation of the Bogomol'nyi bound is no longer guaranteed,
but it is unclear as to whether this would necessarily imply infinite mass
per unit length for the string. The divergence of the ADM mass and the
topological charge in fact follows from the $1/r^2$ falloff of the fields,
and is an infrared phenomenon, as in the case of axion strings in four
dimensions. For this reason, the divergence of the energy density and
topological charge should not prevent the exsitence of a finite effective
action describing this type of string soliton at a scale larger than the
core size \hars. It would therefore seem likely that
finite mass per unit length analogs of this solution exist, possibly in the
context of fundamental fivebrane theory.
Another interesting point is that the $D=8$ instanton number
 $N_8$ for this
string solution is in general nonzero for gauge group $E_8\times E_8$
($N_8=NM$, where $N$
and $M$ are the $D=4$ instanton numbers in the $(2345)$ and $(6789)$ spaces
respectively), since in this case $({\rm Tr F}^2)^2$ is
 nonvanishing. This is
to be contrasted with the zero $D=8$ instanton number found
 for the gauge string soliton above.

\chapter{String solitons in D=4}

\def\wh{\widehat}
\def\al{{( a )}}

\def\bet{{( b )}}
\def\ten{{(10)}}

\def\wt{\widetilde}

\def\B{{\cal B}}
\def\C{{\cal C}}
\def\D{{\cal D}}
\def\E{{\cal E}}

\def\K{{\cal K}}

\def\LL{{\cal L}}
\def\M{{\cal M}}
\def\p{{\partial}}

\def\z=\zeta
\def\h{\eta}

\def\b{\beta}
\def\a{\alpha}
\def\l{\lambda}

\def\x{\xi}

\def\r{\rho}

\def\sqr#1#2{{\vbox{\hrule height.#2pt\hbox{\vrule width
.#2pt height#1pt \kern#1pt\vrule width.#2pt}\hrule height.#2pt}}}
\def\Box{\mathchoice\sqr64\sqr64\sqr{4.2}3\sqr33}

\def\met {g_{\mu\nu}}

\newsec{String compactification to four dimensions}

It was pointed out in \str\ that after toroidal
compactification to four dimensions, the fivebrane would appear
 as either a
0-brane, a 1-brane or a 2-brane, depending on how it wraps
around
the compactified directions \refs{\dufhis,\dufipss,\fujku}.
Thus it ought to be possible to find soliton solutions directly
from the
four-dimensional string corresponding to monopoles (0-branes),
 strings
(1-branes) and domain walls (2-branes). It is these
 four-dimensional solitons
that form the subject of this section.
Such fivebrane-inspired supersymmetric monopoles
\footnote{$^\dagger$}{The bosonic version of the monopole
solution we discuss
 was previously discussed in \refs{\rohw,\banddf,\khuinst}.}
were found in
\refs{\khumono,\khumonex,\jjj}\ while the string and domain wall solutions
were found in \dufkexst.

To find these multi-monopole, multi-string and domain wall
solutions we shall follow
the procedure outlined in \refs{\khumono,\khumonex}, where it
was argued that
monopoles solutions of the heterotic string could be obtained by
modifying the 't Hooft ansatz for the Yang-Mills instanton. We
 shall present
them from both the $D=10$ and $D=4$ points of view \dufkexst. As with the
fivebrane, there are three types of string and domain wall solutions: neutral
(i.e. zero Yang-Mills field), gauge and symmetric.  In common with the
symmetric fivebrane of section 5.3, the symmetric solutions are arguably exact
 to all orders in $\alpha'$, as discussed in section 8. Of particular interest
is the solitonic string, since its couplings to the
background fields of supergravity compared to those of the fundamental
string are such that the dilaton/axion field $S$ is replaced by the modulus
field $T$. It thus belongs to an $O(6,22;Z)$ family of dual strings just as
there is an $SL(2,Z)$ family of fundamental strings \sentwo. This accords with
the suggestion of \refs{\schsen,\bin}\ that string/fivebrane duality
interchanges the roles of strong/weak coupling duality and target space
duality, which is the subject of section 6.6. Moreover, we shall argue that
these $H$-monopoles play the role of winding states for the dual string.

We consider heterotic string theory compactified on a six
dimensional torus. The simplest way to derive the low energy
effective action for this theory is to start with the $N=1$
supergravity theory coupled to $N=1$ super Yang-Mills theory in ten
dimensions, and dimensionally reduce the theory from ten to four
dimensions \refs{\ferkp,\ter,\hass,\mahs}.
Since at a generic point in the moduli space only the abelian
gauge fields give rise to massless fields in four dimensions, it
is enough to restrict to the U(1)$^{16}$ part of the
ten-dimensional gauge group. We shall follow the procedure outlined
in
\senrev\
and confine ourselves to the bosonic sector.
The ten-dimensional action is given by,
\eqn\newone{\eqalign{
I_{10}({\rm string})={1\over 32\pi} \int d^{10}z \sqrt{ - G^{(10)}}\,
e^{-\Phi^{(10)}}&\Big(R^{(10)} +G^{\ten MN}\p_M\Phi^\ten \p_N\Phi^\ten
\cr
- &{1\over 12} H^{(10)}_{MNP}
H^{(10)MNP} - {1\over 4} F^{(10)I}_{MN} F^{(10)IMN}\Big),\cr}}
where $G^{(10)}_{MN}$, $B^{(10)}_{MN}$, $A^{(10)I}_M$, and
$\Phi^{(10)}$ are ten-dimensional metric, anti-symmetric tensor
field, $U(1)$ gauge fields and the scalar dilaton field
respectively ($0\le M, N \le 9$, $1\le I\le 16$), and,
\eqn\newtwo{\eqalign{
F^{(10)I}_{MN} &= \p_M A^{\ten I}_N - \p_N A^{\ten I}_M \cr
H^{(10)}_{MNP} &= (\p_M B^\ten_{NP} -{1\over 2} A_M^{\ten I}
F^{\ten I}_{NP}) + \hbox{cyclic permutations in $M$, $N$, $P$}.\cr}}
We have set $\kappa^2=16\pi$ for later convenience.

For dimensional reduction, it is convenient to introduce the
``four-dimensional fields''  $\wh G_{mn}$, $\wh B_{mn}$, $\wh
A^I_m$, $\Phi$, $A_\mu^\al$, $G_{\mu\nu}$ and $B_{\mu\nu}$ ($1\le
m\le 6$, $0\le \mu\le 3$, $1\le a \le 28$) through the
relations \refs{\mahs,\senone,\schtwo}\
\eqn\newthree{\eqalign{
& \wh G_{mn}  = G^\ten_{m+3,n+3}, \quad  \wh B_{mn}  =
B^\ten_{m+3, n+3}, \quad  \wh A^I_m  = A^{\ten I}_{m+3},
\cr
&{1\over 2}A^{(m)}_\mu  = {1\over 2}\wh G^{mn} G^\ten_{n+3,\mu}, \quad
{1\over 2}A^{(I+12)}_\mu = -({1\over 2} A^{\ten I}_\mu - \wh A^I_n
A^{(n)}_\mu), \cr
& {1\over 2} A^{(m+6)}_\mu = {1\over 2}
B^\ten_{(m+3)\mu} - \wh B_{mn} A^{(n)}_\mu + {1\over 2}\wh A^I_m
A^{(I+12)}_\mu, \cr
& G_{\mu\nu} = G^\ten_{\mu\nu} - G^\ten_{(m+3)\mu} G^\ten_{(n+3)\nu} \wh
G^{mn}, \cr
& B_{\mu\nu} = B^\ten_{\mu\nu} - \wh B_{mn} A^{(m)}_\mu
A^{(n)}_\nu - {1\over 2}(A^{(m)}_\mu A^{(m+6)}_\nu - A^{(m)}_\nu
A^{(m+6)}_\mu),
\cr
& \Phi = \Phi^\ten - {1\over 2} \ln\det \wh G, \quad \quad
\quad 1\le m, n \le 6, \quad
0\le \mu, \nu \le 3, \quad 1\le I \le 16.\cr}}
Here $\wh G^{mn}$ denotes the inverse of the matrix $\wh G_{mn}$.
We now combine the scalar fields $\wh G_{mn}$, $\wh B_{mn}$, and
$\wh A_m^I$ into an  $O(6,22)$ matrix valued scalar field $M$.
For this we regard $\wh G_{mn}$, $\wh B_{mn}$ and $\wh A^I_m$ as
$6\times 6$, $6\times 6$, and $6\times 16$ matrices
respectively, and $\wh C_{mn} = {1\over 2} \wh A^I_m
\wh A^I_n$ as a $6\times 6$ matrix, and define $M$ to be the
$28\times 28$ dimensional matrix
\eqn\newfour{
M = \pmatrix{ \wh G^{-1} & \wh G^{-1} (\wh B + \wh
C) & \wh G^{-1}\wh A \cr (-\wh B + \wh C) \wh G^{-1} & (\wh G
- \wh B +
\wh C) \wh G^{-1} (\wh G + \wh B + \wh C) & (\wh G -\wh B +\wh
C)\wh G^{-1} \wh A \cr  \wh A^T \wh G^{-1} &  \wh A \wh G^{-1}
(\wh G + \wh B +\wh
C) & I_{16} + \wh A^T \wh G^{-1} \wh A \cr }}
satisfying
\eqn\newfive{
M L M^T = L, \quad \quad  M^T=M, \quad \quad L =\pmatrix{0 & I_6
& 0  \cr I_6 & 0 & 0 \cr 0 & 0 & -I_{16}},}
where  $I_n$ denotes the $n\times n$ identity matrix.

The effective action that governs the dynamics of the massless
fields in the four-dimensional theory is obtained by
substituting the expressions for the ten-dimensional fields in
terms of the four-dimensional fields in \newone, and taking all
field configurations to be independent of the internal coordinates. The
result is
\eqn\newsix{\eqalign{
I_4({\rm string}) &= {1\over 32\pi} \int d^4 x \sqrt{-  G} \, e^{-\Phi}
\big[ R_G +
G^{\mu\nu}
\p_\mu \Phi \p_\nu\Phi -{1\over 12} G^{\mu\mu'} G^{\nu\nu'}
G^{\rho\rho'} H_{\mu\nu\rho} H_{\mu'\nu'\rho'} \cr
&\quad\quad  - {1\over 4} G^{\mu\mu'} G^{\nu\nu'} F^\al_{\mu\nu} (LML)_{ab}
F^\bet_{\mu'\nu'} + {1\over 8} G^{\mu\nu} Tr (\p_\mu M L \p_\nu
M L) \big],\cr}}
where
\eqn\newseven{\eqalign{
F^\al_{\mu\nu} &= \p_\mu A^\al_\nu - \p_\nu A^\al_\mu ,\cr
H_{\mu\nu\rho} &= (\p_\mu B_{\nu\rho} + 2 A^\al_\mu
L_{ a  b } F^\bet_{\nu\rho}) + \hbox{cyclic permutations of
$\mu$, $\nu$, $\rho$},\cr}}
and $R_G$ is the scalar curvature associated with the
four-dimensional metric $G_{\mu\nu}$. In deriving this result we have taken
$\int d^6 y=1$, where $y^m$ ($1\le m\le 6$) denote the coordinates
labeling the six-dimensional torus.
 Here $\Phi$ is the $D=4$ dilaton, $R_G$ is the
scalar curvature formed from the string metric $G_{\mu\nu}$, related to the
canonical metric $g_{\mu\nu}$ by $G_{\mu\nu}\equiv e^{\Phi}g_{\mu\nu}$.
$B_{\mu\nu}$ is the 2-form which couples to the string worldsheet and
$A_{\mu}{}^a$ ($a=1,...,28$) are the abelian gauge fields. $M$ is a
symmetric $28\times28$ dimensional matrix of scalar fields satisfying
$MLM=L$ where $L$ is the invariant metric on $O(6,22)$:
\eqn\neweight{
L=\pmatrix{0&I_6&0\cr I_6&0&0 \cr 0&0&-I_{16}}.}

\newsec{$T$ duality and $S$ duality}
The action is invariant under the $O(6,22)$ transformations
$M\rightarrow\Omega M\Omega^T$,
$A_{\mu}{}^a\rightarrow\Omega^{a}{}_{b}A_{\mu}{}^b$, $G_{\mu\nu}\rightarrow
G_{\mu\nu}$, $B_{\mu\nu}\rightarrow B_{\mu\nu}$, $\Phi\rightarrow\Phi$,
where $\Omega$ is an $O(6,22)$ matrix satisfying $\Omega^TL\Omega=L$.
$T$-duality corresponds to the $O(6,22;Z)$ subgroup and is known to be an
exact symmetry of the full string theory \tduality. The
 equations of motion, though
not the action, are also invariant under the $SL(2,R)$
transformations:
${\cal M}\rightarrow \omega{\cal M}\omega^T,
{\cal F}_{\mu\nu}{}^{a\alpha}
  \rightarrow  \omega^{\alpha}{}_{\beta}{\cal F}_{\mu\nu}{}^{a\beta},
 g_{\mu\nu}\rightarrow g_{\mu\nu},\,\,M\rightarrow M$
where $\a=1,2$ with ${\cal F}_{\mu\nu}{}^{a1}=F_{\mu\nu}{}^{a}$ and ${\cal
F}_{\mu\nu}{}^{a2}=\left(\lambda_2(ML)^a{}_{b}\tilde
F_{\mu\nu}{}^{b}+
\lambda_1 F_{\mu\nu}{}^{a}\right)$, where $\omega$ is an
$SL(2,R)$ matrix
satisfying  $\omega^T{\cal L}\omega={\cal L}$ and where
\eqn\drlafour{
{\cal M}={1\over \lambda_2}\pmatrix{
1&\lambda_1\cr
\lambda_1&|\lambda|^2 \cr}, {}
{\cal L}=\pmatrix{
0&1\cr
-1&0\cr}.}
$\l$ is given by $\lambda=\Psi+ie^{-\Phi}\equiv\lambda_1+
i\lambda_2$. The
axion $\Psi$ is defined through the relation
$\sqrt{-g}H^{\mu\nu\rho}=-e^{2\Phi}
\epsilon^{\mu\nu\rho\sigma}\partial_{\sigma}\Psi$.
 $S$-duality
corresponds
to the $SL(2,Z)$ subgroup and there is now a good deal of
evidence \refs{\fonilq,\reyone,\senone,\senph,\senrev,\senmod,
\schsentwo,\schtwo,\dufkexst,\dufr}\
in favor of its also being an exact symmetry of the full
string theory. See also \dufldr. For the restricted class of
configurations obtained by
setting to zero the $16$ gauge fields $F^{13\rightarrow28}$
originating from
the ten-dimensional gauge fields, it is possible to define a dual action
\schtwo\ which has manifest $SL(2,R)$ symmetry. The field strengths
$F^{1\rightarrow6}$, whose origin resides in the $D=10$ metric, remain the
same but the  $F^{7\rightarrow12}$, whose origin resides in the $D=10$
$2$-form, are replaced by their duals. The equations of motion are also
invariant under $O(6,6)$; the action is not except for the $SL(6,R)$
subgroup which acts trivially. This action is precisely the one obtained by
dimensional reduction from the dual ($6$-form) version of $D=10$
supergravity which couples to the worldvolume of the fivebrane
\refs{\duf,\str}\
and for which the axion is just the $6$-form component lying in the extra
$6$ dimensions.
\newsec{Fivebrane compactification to four dimensions}

Once again we shall follow the procedure of \senrev. For
metric $\wt G^{\ten}_{MN}$, a six-form field
$\wt B^\ten_{M_1\ldots M_6}$, and dilaton field
$\wt\Phi^\ten$ , we can  rewrite \sfbrane\ as
\eqn\srone{\eqalign{
I_{10}({\rm fivebrane}) = {1\over 32\pi} \int d^{10} z &\sqrt{- \wt G^\ten}\,
e^{\wt\Phi^\ten/3}
\Big( \wt R^\ten \cr
- {1\over 2 \times 7!} &\wt G^{\ten M_1 N_1} \cdots \wt
G^{\ten M_7 N_7}
\wt H^\ten_{M_1\ldots M_7} \wt H^\ten_{N_1 \ldots N_7}\Big),\cr}}
where
\eqn\srtwo{
\wt H^\ten_{M_1 \ldots M_7} = \p_{[M_1} \wt B^\ten_{M_2 \ldots M_7]}.}
We showed in section 2 that the equations of motion and the Bianchi identities
derived from this
action can be shown to be identical to those derived from the action
\newone\ provided we make the identifications
\eqn\srthree{\eqalign{
& \wt\Phi^\ten =\Phi^\ten, \quad \quad \wt G^\ten_{MN} = e^{-\Phi^\ten/3}
G^\ten_{MN}, \cr
& \sqrt{- \wt G^\ten}\, e^{\wt \Phi^\ten/3} \wt G^{\ten M_1 N_1} \cdots
\wt G^{\ten M_7 N_7} \wt H^\ten_{N_1 \ldots N_7}
= - {1\over 3!} \epsilon^{M_1 \ldots M_{10}} H_{M_8 M_9 M_{10}}.\cr}}

In order to carry out the dimensional reduction of this theory
from ten to four dimensions, it is convenient to introduce the
``four-dimensional fields'' $\lambda$, ${\cal C}_\mu^m$, ${\cal D}_\mu^m$,
$\wh G_{mn}$, ${\cal B}_{\mu\nu}^{mn}$, ${\cal E}_{\mu\nu\rho}^{mnp}$ and
$g_{\mu\nu}$ through the relations \schsen:
\eqn\srfour{\eqalign{
\wh G_{mn} =& e^{\wt \Phi^\ten/3} \wt G^\ten_{m+3,n+3}, \quad \quad
\lambda_1 = {1\over 6!}
\wt B^\ten_{m_1+3, \ldots m_6+3} \epsilon^{m_1\ldots m_6},
\quad \quad \lambda_2 = \sqrt{\det\wh G}\,  e^{-\wt \Phi^\ten},
\cr
\C^m_\mu =& e^{\wt \Phi^\ten/3} \wh G^{mn} \wt G^\ten_{(n+3)\mu},
\quad \quad
\D^{m_1}_\mu = {1\over 5!} \epsilon^{m_1\ldots m_6} \wt B^\ten_{
\mu (m_2+3)\ldots (m_6+3)} - \lambda_1 \C^{m_1}_\mu
\cr
\B^{m_1 m_2}_{\mu\nu} =& {1\over 4!} \epsilon^{m_1\ldots m_6}
\wt B^\ten_{\mu\nu (m_3+3) \ldots (m_6+3)} \cr
&  - [(\lambda_1 \C^{m_1}_\mu
\C^{m_2}_\nu +{1\over 2} \D^{m_1}_\mu \C^{m_2}_\nu -{1\over 2}
\D^{m_1}_\nu \C^{m_2}_\mu) - (m_1\leftrightarrow m_2)]\cr
\E^{m_1 m_2 m_3}_{\mu\nu\rho} =& {1\over 3!} \epsilon^{m_1 \ldots m_6}
\wt B^\ten_{\mu\nu\rho (m_4+3) \ldots (m_6+3)}, \cr
g_{\mu\nu} =& (\lambda_2)^{2/3} (\det\wh G)^{{1\over 6}}
(\wt G^\ten_{\mu\nu} - \wt G^\ten_{(m+3)(n+3)} \C^m_\mu \C^n_\nu),\cr}}
and the corresponding field strengths,
\eqn\srfive{\eqalign{
F^{(\C)m}_{\mu\nu} =& \p_\mu \C^m_\nu - \p_\nu \C^m_\mu, \quad \quad
F^{(\D)m}_{\mu\nu} = \p_\mu \D^m_\nu - \p_\nu \D^m_\mu \cr
K^{mn}_{\mu\nu\rho} =&  \Big(\big[ \p_\mu \B^{mn}_{\nu\rho}
-{1\over 2} \big\{ (\C^n_\rho F^{(\D)m}_{\mu\nu} + \D^n_\rho F^{(\C)m}_{
\mu\nu}) - (m\leftrightarrow n)\big\}\big] \cr
& + \hbox{ cyclic permutations of } \mu, \nu, \rho\Big)
\cr
\K^{mnp}_{\mu\nu\rho\sigma} =& \big[ \p_\mu \E^{mnp}_{\nu\rho\sigma}
+(-1)^P \cdot \hbox{ cyclic permutations of } \mu, \nu, \rho, \sigma\big]
\cr
& - \big[ (\C^p_\sigma K^{mn}_{\mu\nu\rho} +  \hbox{  cyclic
permutations of } m, n, p)\cr
&  +(-1)^P \cdot \hbox {cyclic permutations of } \mu, \nu,
\rho, \sigma\big]\cr & -\big[ \big\{ \C^p_\sigma
\C^n_\rho (F^{(\D)m}_{\mu\nu} + \lambda_1 F^{(\C)m}_{\mu\nu}) +
(-1)^P \cdot  \hbox{ all permutations of } m, n, p\big\}
\cr
&  +(-1)^P \cdot  \hbox{ inequivalent permutations
of } \mu, \nu, \rho, \sigma \big] \cr
&  -\big[ (
\C^p_\sigma \C^n_\rho \C^m_\nu \p_\mu \lambda_1 + (-1)^P \cdot
 \hbox{  all permutations of } m, n, p) \cr
& +(-1)^P \cdot  \hbox{  cyclic permutations of } \mu, \nu,
\rho, \sigma\big].\cr}}
Using the relationship between the fields in the two
formulations of the ten-dimensional $N=1$ supergravity theory
given in \srthree, and the definition of the fields
$\lambda_1$, $\lambda_2$, $\wh G_{mn}$ and $g_{\mu\nu}$ in the
two formulations, one can easily verify that the two sets of
definitions lead to identical $\lambda$, $\wh G_{mn}$ and
$g_{\mu\nu}$.

The action \srone, expressed in terms of these
``four-dimensional fields'', is given by,
\eqn\srsix{\eqalign{
S =& {1\over 32\pi} \int d^4 x \sqrt{-g} \Big[ R - {1\over 2
(\lambda_2)^2}
g^{\mu\nu} \p_\mu \bar\lambda \p_\nu\lambda +{1\over 4} g^{\mu\nu}
\hbox{Tr}(\p_\mu\wh G \p_\nu \wh G^{-1}) \cr
& -{1\over 4} \wh G_{mn} g^{\mu\rho} g^{\nu\sigma}
\pmatrix{F^{(\C)m}_{\mu\nu
} & -F^{(\D)m}_{\mu\nu}} \LL^T\M \LL
\pmatrix{F^{(\C)n}_{\rho\sigma} \cr
- F^{(\D)n}_{\rho\sigma}}\cr
& - {1\over 2\times 2! \times 3!} \wh G_{m_1 n_1} \wh G_{m_2 n_2}
g^{\mu_1\nu_1}\cdots g^{\mu_3\nu_3} K^{m_1m_2}_{\mu_1\mu_2\mu_3}
K^{n_1n_2}_{\nu_1\nu_2\nu_3}\cr
& - {\lambda_2 \over 2 \times 3! \times 4!} \wh G_{m_1n_1}\cdots \wh
G_{m_3n_3}
g^{\mu_1\nu_1}\cdots g^{\mu_4\nu_4} \K^{m_1\ldots m_3}_{\mu_1\ldots \mu_4}
\K^{n_1\ldots n_3}_{\nu_1\ldots \nu_4}\Big],\cr}}
where $\M$ has been defined in \drlafour, and Tr denotes trace over
the indices $m,n$ ($1\le m,n\le 6$).
The equation of motion for $\E^{m_1m_2m_3}_{\mu_1\mu_2\mu_3}$ gives
\eqn\srseven{
\p_{\nu_1}\big[\lambda_2\sqrt{-g}\, \wh G_{m_1n_1} \ldots \wh G_{m_3n_3}
g^{\mu_1\nu_1}\ldots g^{\mu_4\nu_4} \K^{n_1\ldots n_3}_{\nu_1\ldots \nu_4}
\big]=0.}
Since $\K^{n_1\ldots n_3}_{\nu_1\ldots \nu_4}$ is antisymmetric in
$\nu_1, \ldots \nu_4$, we may write
\eqn\sreight{
\lambda_2\sqrt{-g} \, \wh G_{m_1n_1} \ldots \wh G_{m_3n_3}
g^{\mu_1\nu_1}\ldots g^{\mu_4\nu_4} \K^{n_1\ldots n_3}_{\nu_1\ldots \nu_4}
=\epsilon^{\mu_1\ldots \mu_4} H_{m_1m_2m_3}}
for some $H_{mnp}$. The
equation \srseven\ then takes the form:
\eqn\srnine{
\p_\nu H_{m_1m_2m_3} = 0,}
showing that $H_{mnp}$ is a constant. Comparison with the original
formulation of the theory shows that the $H_{mnp}$ are proportional to the
internal components of the three form field strength $H^{(10)}_{MNP}$.
During the dimensional reduction of the original ten-dimensional $N=1$
supergravity theory, we had set these constants to zero. Hence, if we want
to recover the same theory, we must set them to zero here too. This gives
\eqn\srten{
\K^{m_1\ldots m_3}_{\mu_1\ldots \mu_4} =0.}
The action \srsix\ now reduces to
\eqn\sreleven{\eqalign{
S =& {1\over 32\pi} \int d^4 x \sqrt{-g} \Big[ R - {1\over 4}
g^{\mu\nu} tr(\p_\mu \M \LL \p_\nu\M \LL)
+{1\over 4} g^{\mu\nu}
\hbox{Tr}(\p_\mu\wh G \p_\nu \wh G^{-1}) \cr
& -{1\over 4} \wh G_{mn} g^{\mu\rho} g^{\nu\sigma}
\pmatrix{F^{(\C)m}_{\mu\nu
} & -F^{(\D)m}_{\mu\nu}} \LL^T\M \LL
\pmatrix{F^{(\C)n}_{\rho\sigma} \cr
- F^{(\D)n}_{\rho\sigma}}\cr
& - {1\over 2\times 2! \times 3!} \wh G_{m_1 n_1} \wh G_{m_2 n_2}
g^{\mu_1\nu_1}\cdots g^{\mu_3\nu_3} K^{m_1m_2}_{\mu_1\mu_2\mu_3}
K^{n_1n_2}_{\nu_1\nu_2\nu_3}\Big],\cr}}
and has manifest SL(2,R) invariance
\eqn\srtwelve{
\M\to \omega \M \omega^T, \quad \quad \pmatrix{\C^m_\mu \cr -\D^m_\mu}
\to \omega \pmatrix{\C^m_\mu \cr -\D^m_\mu},}
with all other fields remaining invariant under the SL(2,R)
transformation.

This shows that the $SL(2,R)$ symmetry arises naturally in the
four-dimensional theory obtained from the dimensional reduction
of the dual formulation of the $N=1$ supergravity theory in ten
dimensions, just as the $O(6,6)$ or $O(6,22)$ symmetry arises
naturally in the dimensional reduction of the usual $N=1$
supergravity theory from ten to four dimensions.
This result provides another reason for believing that the roles
of $S$ and $T$ duality are interchanged in going from string to fivebrane,
and is entirely consistent with an earlier observation that the dual theory
interchanges the worldsheet and spacetime loop expansions \duflloop. In
this light, the need to treat the above $16$ gauge fields on a different
footing is only to be expected since in the dual formulation their kinetic
terms are $1$-loop effects.

\newsec{Montonen-Olive revisited}

Following \refs{\fonilq,\reyone}\ (see also \dufldr), and
 generalizing an earlier idea of
Montonen and Olive \refs{\mono,\wito,\osb}, Schwarz and Sen
 have
conjectured
\refs{\schsentwo,\senrev}\ on the basis of string/fivebrane
duality that,
when the solitonic excitations are included, the full string spectrum is
invariant not only under the target space $O(6,22;Z)$ ($T$-duality) but also
under the strong/weak coupling $SL(2,Z)$ ($S$-duality). They have
constructed a manifestly $S$ and $T$ duality invariant mass spectrum.
$T$-duality transforms electrically charged winding states into electrically
charged Kaluza-Klein states, but $S$-duality transforms elementary
electrically charged string states into solitonic monopole and dyon states.

We now turn to the electric and magnetic charge spectrum. Schwarz
and Sen \refs{\schsentwo,\senrev}\  present an $O(6,22;Z)$ and $SL(2,Z)$
invariant expression for the mass of particles saturating the strong
Bogomol'nyi bound $m=|Z_1|=|Z_2|$:
\eqn\drlafive{
m^2={1\over 16}(\alpha^a~~\beta^a){\cal M}^{0}(M^{0}+L)_{ab}
    \pmatrix{
\alpha^b\cr
\beta^b\cr},}
where a superscript $0$ denotes the constant asymptotic values of the
fields. Here $\alpha^a$ and $\beta^a$ ($a=1,...,28$) each belong to an even
self-dual Lorentzian lattice $\Lambda$ with metric given by $L$ and are
related to the electric and magnetic charge vectors $(Q^a,P^a)$ by
$(Q^a,P^a)=\left(M_{ab}{}^0(\alpha^b +
\lambda_{1}{}^0\beta^b)/\lambda_2{}^0,
L_{ab}\beta^b\right)$.

Eq. \drlafive\ suggests that it is natural to combine the
vectors $\alpha^a$
and $\beta^a$ into a single $56$-dimensional vector
$\xi=(\alpha^a,\beta^a)$
 which now belongs to a $56$-dimensional lattice $\Gamma$. The
new lattice
$\Gamma$ is self-dual not only with respect to the metric
$\hat {\cal L}={\cal L} \otimes L$. The latter condition says
 that, for any two vectors
$\xi=(\alpha^a,\beta^a)$ and $\xi'=(\alpha'^a,\beta'^a)$
 belonging to the
lattice $\Gamma$
\eqn\dszw{\xi^T \hat {\cal L}\eta'=\alpha^a L_{ab}\beta'^b -
\alpha'^a L_{ab}\beta^b={\rm integer}.}
This is just the Dirac-Schwinger-Zwanziger-Witten
 quantization
condition for the magnetic charge
 \refs{\dir,\schwing,\zwa,\witone}. We shall return to this
Sen-Schwarz spectrum in section 6.8.


\newsec{Monopoles, strings and domain walls}

In this section we now turn to the explicit solutions for the monopole, string
and domain wall obtained by wrapping the fivebrane around the extra dimensions.
We shall see that the neutral monopoles may be identified with the monopole
states in the Sen-Schwarz spectrum. Moreover, we shall argue that they are the
winding states of the dual string soliton regarded as a fundamental theory in
its own right.
{}From the 't Hooft ansatz for the Yang-Mills instanton (see section
5), depending on how many of the four coordinates $f$ is
allowed to depend and depending on whether we compactify, we shall obtain
$D=10$ multi-fivebrane and $D=4$ multi-monopole,
multi-string and domain wall solutions. This follows from the
observation that the arguments of chapter 5 do not depend on the precise
form of $f$ or the dilaton function \dufkexst.

We rewrite the ansatz \sansatz\ with the following notation
\eqn\nsansatz{\eqalign{\met&=e^{2\phi}\delta_{mn}\qquad m,n=1,2,3,4,\cr
g_{\mu\nu}&=\eta_{\mu\nu}\qquad\quad   \mu,\nu=0,5,6,7,8,9,\cr
H_{mnp}&=\pm 2\epsilon_{mnpk}\partial^k\phi
\qquad m,n,p,k=1,2,3,4,\cr}}
with $e^{-2\phi}\Box e^{2\phi}=0$.
Let us single out a direction in the transverse four-space (say $x^4$)
and assume all fields are independent of this coordinate. Since all fields
are already independent of $x^5,x^6,x^7,x^8,x^9$, we may consistently assume
the $x^4,x^5,x^6,x^7,x^8,x^9$ are compactified on a six-dimensional torus,
where we shall take the $x^4$ circle to have circumference $2\pi R$
and the rest to have circumference $2\pi\sqrt{\alpha'}$,
so that $(2\pi)^6 R_1 \alpha'^{5/2}\kappa_4^2=\kappa_{10}^2$.
Going back to the 't Hooft ansatz \hfanstz, the solution
for $f$ satisfying the \ $f^{-1} \Box\ f=0$ has the form
\eqn\fmono{f_M=1+\sum_{i=1}^N{m_i\over |\vec x - \vec a_i|},}
where $m_i$ is proportional to the charge and $\vec a_i$ the location in the
three-space $(123)$ of the $i$th instanton string.
If we make the identification $\Phi\equiv A_4$ then the
lagrangian density may be rewritten as
\eqn\relag{F_{\mu\nu}^a F_{\mu\nu}^a =F_{jk}^a F_{jk}^a + 2F_{k4}^a F_{k4}^a
=F_{jk}^a F_{jk}^a + 2D_k \Phi^a D_k \Phi^a,}
where $j,k=1,2,3$.
We now go to $3+1$ space $(0123)$ with the Lagrangian density
\eqn\lagden{{\cal L}=-{1\over 4}G_{\alpha\beta}^a G^{\alpha\beta a} -{1\over 2}
D_\alpha \Phi^a D^\alpha \Phi^a,}
where $\alpha,\beta=0,1,2,3$.
It follows that the above ansatz is a static solution with
$A_0^a=0$ and all time derivatives vanish. The solution in $3+1$ dimensions
has the form
\eqn\monsol{\eqalign{\Phi^a&=\mp{1\over g}\delta^{aj}\partial_j \ln f_M,\cr
A_k^a&={1\over g}\epsilon^{akj}\partial_j \ln f_M,\cr}}
where $j,k=1,2,3$ and $g$ is the YM coupling constant. This solution
represents a multi-instanton string
configuration with sources at $\vec a_i, i=1,2...N$ \refs{\khumono,\khumonex}.
\footnote{$^*$}{This modified 't Hooft ansatz does not represent a true
multi-monopole solution of the pure Yang-Mills field theory but nevertheless
possesses some properties analogous to those of a multi-monopole solution
\refs{\gropy,\khumono,\khumonex,\jjj}.}
For $e^{2\phi} = e^{2\phi_0} f_M$, we obtain a neutral
solution and a symmetric solution \refs{\khumono,\khumonex}\ respectively
depending on whether we set the gauge field equal to zero or to the
generalized connection. In both cases, the magnetic charge is given by
$\tilde g_1=\sqrt{2}\kappa_4 \widetilde{T_1}$,
where $\widetilde{T_1}=\widetilde{T_6} (2\pi\sqrt{\alpha'})^5$
obeys, from \ttwosix, the
quantization condition
\eqn\stmondual{2\pi R_1\kappa_4^2 T_2 \widetilde{T_1}=n\pi.}
This implies $m_i=n_i\alpha'/2R_1$. Similarly the ``electric''
charge of the fundamental string is $e_1=\sqrt{2} \kappa_4 T_1$, where
$T_1=T_2 2\pi R_1$, and hence
\eqn\dirac{e_1\tilde g_1=2\pi n}
as expected. A noteworthy feature of this solution is that the divergences
from both gauge and gravitational sectors cancel in the classical action.

It is straightforward to reduce the above solution to an explicit
solution in the four-dimensional space $(0123)$. The gauge field
reduction is exactly as above, i.e. we replace $A_4$ with the scalar
field $\Phi$. In the gravitational sector, the reduction from ten to
five dimensions is trivial, as the metric is flat in the subspace
$(56789)$. In going from five to four dimensions, one follows the usual
Kaluza-Klein procedure of replacing $g_{44}$ with a scalar field
$e^{-2\sigma_1}$. The tree-level effective action reduces in four dimensions to
\eqn\redmon{I_4={1\over 2\kappa_4^2}\int d^4 x \sqrt{-g} e^{-2\phi - \sigma_1}
\left( R + 4(\partial\phi)^2 + 4\partial\sigma\cdot\partial\phi -
e^{2\sigma_1} {M_{\alpha\beta}M^{\alpha\beta}\over 4} \right),}
where $\alpha,\beta=0,1,2,3$ and
where $M_{\alpha\beta}=H_{\alpha\beta 4}=\partial_\alpha B_{\beta 4} -
\partial_\beta B_{\alpha 4}$. The four-dimensional monopole solution for this
reduced action is then given by
\eqn\redmsol{\eqalign{e^{2\phi}=e^{-2\sigma_1}&=e^{2\phi_0}\left(
1+\sum_{i=1}^N{m_i\over |\vec x - \vec a_i|}\right),\cr
ds^2&=-dt^2 + e^{2\phi}\left(dx_1^2 + dx_2^2 + dx_3^2\right),\cr
M_{ij}&=\pm\epsilon_{ijk}\partial_k e^{2\phi},\qquad i,j,k=1,2,3.\cr}}
For a single monopole, in particular, we have
\eqn\hmono{M_{\theta\phi}=\pm m \sin\theta,}
which is the magnetic field strength of a Dirac monopole. Note, however, that
this monopole did not arise from the Yang-Mills field strength $F_{MN}$ but
from the compactified three-form $H$, and arises in all versions of this
solution. In particular, one may obtain a multi-magnetic monopole solution
of purely bosonic string theory \khuinst.

We now modify the solution of the 't Hooft ansatz even further and choose
two directions in the four-space $(1234)$ (say $x^3$ and $x^4$) and assume all
fields are independent of both of these coordinates. We may now consistently
assume that $x^3,x^4,x^6,x^7,x^8,x^9$ are compactified on a six-dimensional
torus, where we shall take the $x^3$ and $x^4$ circles to have
circumference
$2\pi R_2$ and the remainder to have circumference
$2\pi\sqrt{\alpha'}$, so that
$(2\pi)^6 R_2^2\alpha'^2\kappa_4^2=\kappa_{10}^2$. Then the
solution
for $f$ satisfying $f^{-1} \Box\ f=0$ has multi-string
structure
\eqn\fstring{f_S=1-\sum_{i=1}^N \lambda_i \ln |\vec x - \vec a_i|,}
where $\lambda_i$ is the charge per unit length and $\vec a_i$
 the location in
the two-space $(12)$ of the $i$th string. If we make the
identification
$\Phi\equiv A_4$ and $\Psi\equiv A_3$
then the lagrangian density for the above ansatz can be
 rewritten as
\eqn\slag{F_{\mu\nu}^a F_{\mu\nu}^a = F_{jk}^a F_{jk}^a +
2D_k \Phi^a D_k \Phi^a + 2D_k \Psi^a D_k \Psi^a ,}
where $j,k=1,2$. We now go to the $3+1$ space $(0125)$ with the
 lagrangian
density
\eqn\lagden{{\cal L}=-{1\over 4}G_{\rho\sigma}^a
G^{\rho\sigma a} -{1\over 2}
D_\rho \Phi^a D^\rho \Phi^a -{1\over 2} D_\rho \Psi^a D^\rho
\Psi^a,}
where $\rho,\sigma=0,1,2,5$.
It follows that the multi-string ansatz is a static solution
with
$A_0^a=0$ and all time derivatives vanish. The solution in
 $3+1$ dimensions
has the form
\eqn\stsol{\eqalign{\Phi^a&=\mp{1\over g}\delta^{aj}\partial_j
 \ln f_S,\cr
\Psi^k&={1\over g}\epsilon^{kj}\partial_j \ln f_S,\cr
A_k^a&=-\delta^{a3}{1\over g}\epsilon^{kj}\partial_j
\ln f_S,\cr}}
where $j,k=1,2$. This solution represents a
multi-string configuration with sources at $\vec a_i, i=1,2...N$.
 By setting
$e^{2\phi} = e^{2\phi_0}f_S$, we obtain from the fivebrane
 ansatz a neutral
multi-string solution and an exact heterotic multi-string
 solution. The
neutral single-string solution coincides with that of
\gresvy\
in the far-field limit. The solitonic string  tension
$\widetilde{T_2}$ is given by
$\widetilde{T_6} (2\pi)^4\alpha'^2$ and from \ttwosix\ is
related to the fundamental string tension $T_2$ by
\eqn\ststdual{(2\pi R_2)^2\kappa_4^2 T_2 \widetilde{T_2}=n\pi.}
This implies $\lambda_i=n_i\alpha'/2\pi R_2^2$.
Like the monopole, the
lagrangian per unit length for the string solution is finite
 as
 a result of the
cancellation of divergences between the gauge and gravitational
 sectors.

As in the multi-monopole case, it is straightforward to reduce
the multi-string solution to
a solution in the four-dimensional space $(0125)$.
The gauge field
reduction is done in \stsol. In the gravitational sector, the
 reduction from
ten to six dimensions is trivial, as the metric is flat in the
 subspace
$(6789)$. In going from six to four dimensions, we compactify
 the $x_3$ and
$x_4$
directions and again follow the Kaluza-Klein procedure by
replacing $g_{33}$
and $g_{44}$ with a scalar field $e^{-2\sigma_2}$. The
 tree-level effective
action reduces in four dimensions to
\eqn\redst{S_4={1\over 2\kappa_4^2}\int d^4 x \sqrt{-g} e^{-2\phi - 2\sigma_2}
\left( R + 4(\partial\phi)^2 + 8\partial\sigma_2\cdot\partial\phi +
2(\partial\sigma_2)^2 - e^{4\sigma_2} {N_\rho N^\rho\over 2} \right),}
where $\rho=0,1,2,5$, where $N_\rho=H_{\rho 34}=\partial_\rho B$, and where
$B=B_{34}$. The four-dimensional
string soliton solution for this reduced action is then given by
\eqn\redssol{\eqalign{e^{2\phi}=e^{-2\sigma_2}&=e^{2\phi_0}\left(
1-\sum_{i=1}^N \lambda_i \ln |\vec x - \vec a_i|\right),\cr
ds^2&=-dt^2 + dx_5^2 + e^{2\phi}\left(dx_1^2 + dx_2^2\right),\cr
N_i&=\pm \epsilon_{ij} \partial_j e^{2\phi}.\cr}}

We complete the family of solitons that can be obtained from the solutions
of the 't Hooft ansatz by demanding that $f$ depend on only one coordinate,
say $x^1$. We may now consistently assume that $x^2,x^3,x^4,x^7,x^8,x^9$ are
compactified on a six-dimensional torus, where we shall take the $x^2$,
$x^3$ and $x^4$ circles to have circumference $2\pi R_3$ and the rest to
have circumference $2\pi\sqrt{\alpha'}$, so that
$(2\pi)^6R_3^3\alpha'^{3/2}\kappa_4^2=\kappa_{10}^2$.
Then the solution
of $f^{-1} \Box\ f=0$ has domain wall structure with the ``confining potential"
\eqn\fdom{f_D=1+ \Lambda x_1,}
where $\Lambda$ is a constant. By setting
$e^{2\phi} = e^{2\phi_0} f_D$, we obtain from the fivebrane ansatz
a neutral domain wall solution and an exact heterotic domain wall
solution. The solitonic domain wall
tension $\widetilde{T_3}$ is given by $\widetilde{T_6}
(2\pi \sqrt{\alpha'}^3$
and from \ttwosix\ is related to the fundamental string tension $T_2$ by
\eqn\stdwdual{(2\pi R_3)^3\kappa_4^2 T_2 \widetilde{T_3}=n\pi.}
This implies $\Lambda=n\alpha'/2\pi R_3^3$. Like the monopole
and string we cannot identify an explicit coset conformal field theory near
each source. Again the reduction to $D=4$ is straightforward. In the gauge
sector, the action reduces to YM + three scalar fields $\Phi$, $\Psi$ and
$\Pi$. For the spacetime $(0156)$ the solution for the fields is given by
\eqn\domsol{\eqalign{\Phi^1&=\mp {1\over g}\partial_1 \ln f_D,\cr
\Psi^3&={1 \over g}\partial_1 \ln f_D,\cr
\Pi^2&=-{1 \over g}\partial_1 \ln f_D,\cr
A_\mu &=0,\cr}}
where $\mu=0,1,5,6$. In the gravitational sector the tree-level effective
action in $D=4$ has the form
\eqn\reddom{S_4={1\over 2\kappa^2}\int d^4 x \sqrt{-g} e^{-2\phi - 3\sigma_3}
\left( R + 4(\partial\phi)^2 + 12\partial\sigma_3\cdot\partial\phi +
6(\partial\sigma_3)^2 - e^{6\sigma_3} {P^2\over 2} \right),}
where $P$ is a cosmological constant. Note that \reddom\ is not obtained by
a simple reduction of the ten-dimensional action owing to the nonvanishing
of $H_{234}$. The four-dimensional domain wall
solution for this reduced action is then given by
\eqn\reddomsol{\eqalign{e^{2\phi}=e^{-2\sigma_3}&=e^{2\phi_0}
\left(1+\Lambda x_1\right),\cr
ds^2&=-dt^2 + dx_5^2 + dx_6^2 + e^{2\phi} dx_1^2,\cr
P&=\Lambda.\cr}}
A trivial change of coordinates reveals that the spacetime is, in fact, flat.
Dilatonic domain walls with a flat spacetime have been discussed in a somewhat
different context in \refs{\cvet,\cvegs,\la}.

As for the fivebrane in $D=10$, the mass of the monopole, the mass per unit
length of the string and the mass per unit area of the domain wall saturate
a Bogomol'nyi bound with the topological charge. (In the case of the string
and domain, wall, however, we must follow \dabghr\ and extrapolate the
meaning of the ADM mass to non-asymptotically flat spacetimes.)

\newsec{String/string duality and $S \leftrightarrow T$}

Let us focus on the solitonic string configuration \redssol\ in the case of
a single source. In terms of the complex field
\eqn\tfield{\eqalign{T&=T_1+iT_2\cr &=B_{34}+ie^{-2\sigma} \cr
&=B_{34}+i\sqrt{{\rm det} g^S_{mn}}
\qquad m,n=3,4,\cr}}
where $g^S_{MN}$ is the string $\sigma$-model metric, the
solution for $n=1$ and $R_2=\sqrt{\alpha'}$ takes the
form (with $z=x_1+x_2$)
\eqn\tsol{\eqalign{T&={1\over 2\pi i}\ln {z\over r_0},\cr
ds^2&=-dt^2 + dx_5^2 - {1\over 2\pi} \ln{r\over r_0} dz d\overline z,\cr}}
whereas both the four-dimensional (shifted) dilaton $\eta=\phi + \sigma$
and the four-dimensional two-form $B_{\mu\nu}$ are zero. In terms of the
canonical metric $g_{\mu\nu}$, $T_1$ and $T_2$, the relevant part of the
action takes the form
\eqn\sgtt{S_4={1\over 2\kappa_4^2}\int d^4 x\sqrt{-g}
\left( R - {1\over 2T_2^2}g^{\mu\nu}\partial_\mu T \partial_\nu \overline T
\right)}
and is invariant under the $SL(2,R)$ transformation
\eqn\sltwor{T\to {aT+b\over cT+d},\qquad ad-bc=1.}
The discrete subgroup $SL(2,Z)$, for which $a$, $b$, $c$ and $d$ are
integers, is just a subgroup of the $O(6,22;Z)$ {\it target space duality},
which can be shown to be an exact symmetry of the compactified string theory
at each order of the string loop perturbation expansion.

This $SL(2,Z)$ is to be contrasted with the $SL(2,Z)$ symmetry of the
elementary four-dimensional solution of \dabghr\ and of
section 3.3.
In the latter solution $T_1$ and $T_2$ are zero, but $\eta$ and $B_{\mu\nu}$
are non-zero. The relevant part of the action is
\eqn\dabactt{S_4={1\over 2\kappa_4^2}\int d^4 x\sqrt{-g}
\left( R - 2g^{\mu\nu}\partial_\mu \eta \partial_\nu \eta
-{1\over 12} e^{-4\eta} H_{\mu\nu\rho} H^{\mu\nu\rho}
\right).}
The equations of motion of this theory also display an $SL(2,R)$ symmetry,
but this becomes manifest only after dualizing and introducing the axion
field $a$ via
\eqn\axfield{\sqrt{-g}g^{\mu\nu}\partial_\nu a=
{1\over 3!}\epsilon^{\mu\nu\rho\sigma} H_{\nu\rho\sigma} e^{-4\eta}.}
Then in terms of the complex field
\eqn\comps{\eqalign{S&=S_1 + iS_2 \cr
&=a + ie^{-2\eta} \cr}}
the Dabholkar {\it et al.} fundamental string solution may be written
\eqn\dabs{\eqalign{S&={1\over 2\pi i} \ln {z\over r_0},\cr
ds^2&=-dt^2 + dx_5^2 - {1\over 2\pi} \ln {r\over r_0} dz d\overline z.\cr}}
Thus \tsol\ and \dabs\ are the same with the replacement $T\leftrightarrow
S$. It has been conjectured that this second $SL(2,Z)$ symmetry may also be a
symmetry of string theory \refs{\fonilq,\senone,\schtwo}, but this is far from
obvious order by order in the string loop expansion since it involves a
strong/weak coupling duality $\eta\to - \eta$. What interpretation
are we to give to these two $SL(2,Z)$ symmetries: one an obvious symmetry of
the fundamental string and the other an obscure symmetry of the fundamental
string?

Related issues are brought up in the recent
interesting papers by Sen \sentwo, Schwarz and Sen \schsen\ and Bin\'etruy
\bin. In particular, Sen draws attention to the Dabholkar {\it et al.} string
solution \dabs\ and its associated $SL(2,Z)$ symmetry as supporting evidence
in favor of the conjecture that $SL(2,Z)$ invariance may indeed be an exact
symmetry of string theory. He also notes
that the spectrum of electric and magnetic charges is consistent with the
proposed $SL(2,Z)$ symmetry \sentwo.\footnote{$^\dagger$}
{Sen also discusses the concept of a ``dual string", but for him this is
obtained from the fundamental string by an $SL(2,Z)$ transform. For us, a
dual string is obtained by the replacement $S\leftrightarrow T$.}

All of these observations fall into place if one accepts the proposal of
Schwarz and Sen \schsen: {\it under string/fivebrane duality the roles of
the target-space duality and the strong/weak coupling duality are
interchanged !} This proposal is entirely consistent with an earlier one that
under string/fivebrane duality the roles of the $\sigma$-model loop
expansion and the string loop expansion are interchanged \duflloop. In this
light, the two $SL(2,Z)$ symmetries discussed above are just what one
expects. From the string point of view, the $T$-field $SL(2,Z)$ is an
obvious target space symmetry, manifest order by order in string loops
whereas the $S$-field $SL(2,Z)$ is an obscure strong/weak coupling symmetry.
{}From the fivebrane point of view, it is the $T$-field $SL(2,Z)$ which is
obscure while the $S$-field $SL(2,Z)$ is an ``obvious" target space
symmetry.
(This has not yet been proved except at the level of the low-energy field
theory, however. It would be interesting to have a proof starting from the
worldvolume of the fivebrane.)
This {\it duality of dualities} is illustrated in Table 4.
\vfill\eject

\halign{\indent #&\hfil # \hfil &\quad \hfil # \hfil
 &\quad \hfil # \hfil&\quad #\hfil \cr
& &{\it Fundamental string}&{\it Dual string} &\cr
& & & & \cr
&{\it Moduli}& $T=b+ie^{-\sigma}$ & $S=a+ie^{-\eta}$ &\cr
&{\it Worldsheet coupling}&$<e^{\sigma}>=\alpha'/R^2$
&$<e^{\eta}>=g^2$&
\cr
&{\it Large/small radius} &$R\rightarrow \alpha'/R$
&$g\rightarrow 1/g$
&\cr
&{\it T-duality}&$O(6,22;Z)$&$SL(2,Z)$&\cr
&{\it Axion/dilaton}&$S=a+ie^{-\eta}$&$T=b+ie^{-\sigma}$ &\cr
&{\it Spacetime coupling}&$<e^{\eta}>=g^2$
&$<e^{\sigma}>=\alpha'/R^2$
& \cr
&{\it Strong/weak coupling}&$g\rightarrow 1/g$
&$R\rightarrow \alpha'/R$
& \cr
&{\it S-duality}&$SL(2,Z)$&$O(6,22;Z)$ & \cr}

\itemitem{Table 4:}
Duality of dualities.
\medskip

String theory requires two kinds of loop expansion: classical
$(\alpha')$ worldsheet loops with expansion parameter
$<e^{\sigma}>$ where
$\sigma$ is a modulus
field, and quantum $(\hbar)$ spacetime loops with expansion
parameter
$<e^{\eta}>$ where $\eta$ is the
dilaton field.  Introducing the axion field $a$ and another
pseudoscalar
modulus field $b$,
four-dimensional string/string duality interchanges the roles of
$S=a+ie^{-\eta}$ and
$T=b+ie^{-\sigma}$, and hence interchanges classical and quantum. Thus this
duality of dualities
exhibited by four-dimensional strings is entirely consistent with the earlier
result that
ten-dimensional string/fivebrane duality interchanges the spacetime and
worldsheet loop expansions
\duflloop, and is entirely consistent with the Dirac
quantization rule \ststdual\ that follows
from the string/fivebrane rule \second:
\eqn\glasone{8GR^2=n\alpha' \tilde \alpha' \qquad\qquad
n=integer,}
where $2\pi\tilde \alpha'$ is the inverse tension of the dual
string.
 Thus, for $n=1$, we have
\eqn\glastwo{\eqalign{
<e^{\eta}>&=g^2=8G/\alpha'=\tilde \alpha'/R^2 \cr
<e^{\sigma}>&=\tilde g^2=8G/\tilde\alpha'=\alpha'/R^2\cr}}
where $\tilde g$ is the dual string spacetime loop expansion
parameter. Invariance of the Dirac quantization rule requires
that an $S$ ($T$) transformation of the fundamental string be
accompanied by a $T$ ($S$) transformation of the dual string.

Group theoretically, these dualities
are given by $O(6,22;Z)$ in the case of $T$-duality and
$SL(2,Z)$ in the case
of $S$-duality. It has
been suggested \refs{\dufldr,\dufr,\duffkk,\glasgow}
 that these two kinds of duality should be
united into a bigger group
$O(8,24;Z)$ which contains both as subgroups (see section 6.8).
  This would have the bizarre
effect of
eliminating the distinction between classical and quantum.

This interchange in the roles of the $S$ and $T$ field in
going from the string to the fivebrane has also been noted by Bin\'etruy
\bin. It is made more explicit when $S$ is expressed in terms of the
variables appearing naturally in the fivebrane version
\eqn\fbvers{\eqalign{S&=S_1 + iS_2 \cr
&=A_{034789} + ie^{-2\eta},\cr
&=A_{034789} + i\sqrt{{\rm det} g^F_{mn}}, \qquad m,n=3,4,6,7,8,9, \cr}}
where $g^F_{MN}=e^{-\phi/3}g^S_{MN}$ is the fivebrane $\sigma$-model metric
\duflrsfd\ and
$A_{MNPQRS}$ is the 6-form which couples to the 6-dimensional worldvolume of
the fivebrane, in complete analogy with \tfield.

Note, however, that unlike the Dabholkar {\it et al.} solution and the
Greene {\it et al.} solution, the
symmetric solution \stsol\ also involves the non-abelian gauge fields
$A_\rho,
\Phi,\Psi$ whose interactions appear to destroy the $SL(2,Z)$. This remains
a puzzle. (A generalization of the $D=4$ Dabholkar {\it et al.} solution
involving gauge fields may also be possible by obtaining it as a soliton of
the fivebrane theory. This would involve a $D=4$ analogue of the $D=10$
solution discussed in section 5.4.)

It may at first sight seem strange that a string can be dual to another
string in $D=4$. After all, the usual formula relating the dimension of an
extended object, $d$, to that of the dual object, $\tilde d$, is
$\tilde d=D-d-2$. So one might expect string/string duality only in $D=6$
\duflloop. However, when we compactify $n$ dimensions and allow
the dual object to wrap around $m\leq \tilde d-1$ of the compactified
directions
we find $\tilde d_{{\rm effective}}=\tilde d -m=D_{{\rm effective}}-d-2
+(n-m)$, where $D_{{\rm effective}}=D-n$. In particular for $D_{{\rm
effective}}=4$, $d=2$, $n=6$ and $m=4$, we find $\tilde d_{{\rm
effective}}=2$.

Thus the whole string/fivebrane duality conjecture is put in a different
light when viewed from four dimensions. After all, our understanding of the
quantum theory of fivebranes in $D=10$ is rather poor, whereas the quantum
theory of strings in $D=4$ is comparatively well-understood (although we
still have to worry about the monopoles and domain walls). In particular,
the dual string will presumably exhibit the normal kind of mass spectrum
with linearly rising Regge trajectories, since the classical
($\hbar$-independent) string expression $\widetilde T_6 L^4 \times
({\rm angular\ momentum})$ has dimensions of $({\rm mass})^2$, whereas
the analogous classical expression for an uncompactified fivebrane is
$(\widetilde T_6)^{1/5} \times ({\rm angular\ momentum})$ which has
dimensions $({\rm mass})^{6/5}$ \duf. Indeed, together with the observation
that the $SL(2,Z)$ strong/weak coupling duality appears only after
compactifying at least $6$ dimensions, it is tempting to revive the earlier
conjecture \refs{\duf,\fujku}\ that the internal consistency of the
fivebrane may actually {\it require} compactification.

\newsec{Kaluza-Klein black holes in string theory}

 In this section,
we extend the three solutions of section 6.5 to two-parameter solutions of
 the low-energy
equations of the four-dimensional heterotic string, characterized by a mass
per unit $p$-volume, ${\cal M}_{p+1}$,  and magnetic charge,
$g_{p+1}$, where $p = 0, 1$ or 2. We show them to be special cases of the
generic black $p$-branes discussed in section 3.  The neutral solitons
discussed in section 6.5 are recovered
in the extremal limit, $ \sqrt{2} \kappa {\cal M}_{p+1} =  g_{p+1}$ and are
non-singular in the sense that the curvature
singularity disappears when
expressed in terms of the dual $\sigma$-model metric. The
two-parameter solution extending the supersymmetric monopole corresponds to
a magnetically charged black hole, while the solution extending the
supersymmetric domain wall corresponds to a black membrane. By contrast, the
two-parameter string solution does not possess a finite horizon and
corresponds to a naked singularity.

All three solutions involve both the
dilaton and the modulus fields, and are thus to be contrasted with pure dilaton
solutions. In particular, the effective scalar coupling to the Maxwell field,
$e^{-a\phi} F_{\mu\nu} F^{\mu\nu}$, gives rise to a new string black hole
with $a = \sqrt{3}$, in contrast to the pure dilaton solution of the
heterotic string which has $a = 1$
\refs{\gib,\gibm,\ghs,\shatw,\klopv,\kal,\ko,\hor,\senprl,\gidps}.
It thus resembles the black hole
previously studied in the context of Kaluza-Klein theories
\refs{\dobm,\chod,\pol,\grop,\sor,\gibp,\gib,\hor} which also
has $a = \sqrt{3}$, and which reduces to the Pollard-Gross-Perry-Sorkin
\refs{\pol,\grop,\sor} magnetic monopole in the extremal limit.
As we shall see, this is no coincidence, but a
consequence of the target space duality that interchanges
winding modes and Kaluza-Klein modes.
In this connection, we recall \hw, according to which
$a > 1$ black holes behave like elementary particles!

The fact that the heterotic string admits $a = \sqrt{3}$ black holes also
has implications for string/fivebrane duality \refs{\duf\str{--}\duflrsfd}.
We shall show that electric/magnetic duality in $D=4$ may be seen as a
consequence of string/fivebrane
duality in $D=10$.

We begin with the two-parameter black hole.
The solution of the action \redmon\ is given by
\eqn\bmonrs{\eqalign{e^{-2\Phi}&=e^{2\sigma_{1}}=\left(1 - {r_-\over r}
\right),\cr
ds^2&=-\left(1-{r_+\over r}\right)\left(1-{r_-\over r}\right)^{-1}dt^2 +
\left(1-{r_+\over r}\right)^{-1}dr^2 +
r^2\left(1-{r_-\over r}\right)d\Omega_2^2,\cr
F_{\theta\varphi}&=\sqrt{r_+r_-} \sin\theta ,\cr}}
where here, and throughout this section, we set the dilaton vev $\Phi_0$ equal
to zero. This represents a magnetically charged black hole with
event horizon at $r=r_+$ and inner horizon at $r=r_-$.
The magnetic charge and mass of the black hole are given by
\eqn\massch{\eqalign{g_{1}&={4\pi\over {\sqrt{2}\kappa}}(r_+r_-)^{{1}\over{2}},
   \cr
{\cal M}_1&={{2 \pi}\over {{\kappa}^2}}(2r_+ - r_-) .\cr}}
Changing coordinates via $y=r-r_-$ and taking the extremal limit $r_+=r_-$
yields:
\eqn\monex{\eqalign{e^{2\Phi}&=e^{-2\sigma_{1}}=\left(1 + {r_-\over y}
\right),\cr
ds^2&=-dt^2 + e^{2\Phi}\left(dy^2 + y^2d\Omega_2^2\right),\cr
F_{\theta\varphi}&=r_- \sin\theta,\cr}}
which is just the neutral (i.e. no Yang-Mills) version of the supersymmetric
monopole solution of section 6.5 which saturates
the Bogomol'nyi bound $\sqrt{2} \kappa{\cal M}_1\ge g_1$.

Next we derive a two-parameter string solution which, however,
does not possess a finite event horizon and consequently cannot
be
interpreted as a black string. A two-parameter family of
solutions of the
action \redst\ is now
given by
\eqn\stsol{\eqalign{e^{2\Phi}&=e^{-2\sigma_{2}}=
(1+k/2-\lambda \ln y),\cr
ds^2&=-(1+k)dt^2 + (1+k)^{-1}(1+k/2-\lambda \ln y)dy^2 +
y^2(1+k/2-\lambda \ln y)d\theta^2 + dx_3^2,\cr
F_\theta&=\lambda \sqrt{1+k}.\cr}}
Note that for $k\neq 0$ this describes a string with non-zero
deficit angle
whereas for $k=0$ we recover the supersymmetric string soliton
solution of
section 6.5 which, as shown in section 6.6, is dual to the
 elementary string
solution of Dabholkar {\it et al.} The solution shown in
\stsol\ in fact represents a
naked singularity,
since the event horizon is pushed out to $r_+=\infty$, which
agrees with
the Horowitz-Strominger ``no-$4D$-black-string'' theorem \hors.

Finally, we consider the two-parameter black membrane solution.
The two-parameter black membrane solution of the action
 \reddom\ is then
\eqn\bmemrs{\eqalign{e^{-2\Phi}&=e^{2\sigma_{3}}=
\left(1-{r\over r_-}
\right),\cr
ds^2&=-\left(1-{r\over r_+}\right)\left(1-{r\over r_-}\right)^{-1}dt^2 +
\left(1-{r\over r_+}\right)^{-1}\left(1-{r\over r_-}
\right)^{-4}dr^2 +
dx_2^2 + dx_3^2,\cr
F&=-(r_+r_-)^{-1/2}.\cr}}
This solution represents a black membrane with event horizon
 at
$r=r_+$ and inner horizon at $r=r_-$.
Changing coordinates via $y^{-1}=r^{-1}-r_{-}^{-1}$ and taking the extremal
limit yields
\eqn\domex{\eqalign{e^{2\Phi}&=e^{-2\sigma_3}=\left(1+{y\over r_-}
\right),\cr
ds^2&=-dt^2 + dx_2^2 + dx_3^2 + e^{2\Phi} dy^2,\cr
F&=-{1\over r_-},\cr}}
which is just the supersymmetric domain wall solution of section 6.5.

 The black hole, string and domain wall
solutions of the heterotic string shown above are nothing but
the $({\tilde d}=1, d=1, a=\sqrt{3}, \omega=-{4/3})$,
$({\tilde d}=2, d=0, a=2, \omega=-{3/ 2})$
and $({\tilde d}=3, d=-1, a=\sqrt{7}, \omega=-{10/ 7})$
special cases of the solutions of section 3.8.
To confirm this, it is sufficient to
show that the three actions $S_1$, $S_2$ and $S_3$ may be cast
into the form \lgeact. This is achieved by transforming the metric
$g_{\mu \nu}$ and scalars $\Phi$ and $\sigma_{i}, (i=1, 2, 3),$ to the
canonical metric
$g_{\mu \nu}(\rm can)$ and scalars $\phi$ and $\lambda$ via the following field
redefinitions:

{\it monopole}
\eqn\one{\eqalign{g_{\mu \nu}&=
e^{{1 \over {\sqrt{3}}}({\sqrt{2}}{\lambda}-\phi)}g_{\mu \nu}({\rm can}),
\cr
\Phi&={1 \over {2 \sqrt{3}} } \left({\lambda \over \sqrt{2}}-2\phi \right), \cr
\sigma_{1}&={1 \over \sqrt{3} } \left({\lambda \over \sqrt{2}}+\phi\right) .}}

{\it string}
\eqn\two{\eqalign{g_{\mu \nu}&=
e^{\lambda}g_{\mu \nu}({\rm can}), \cr
\Phi&={1\over 2} (\lambda-\phi), \cr
\sigma_{2}&={1 \over 2}\phi .}}

{\it membrane}
\eqn\three{\eqalign{g_{\mu \nu}&=
e^{{1\over \sqrt{7}}({\sqrt{6}}{\lambda}+\phi)}g_{\mu \nu}({\rm can}),
 \cr
\Phi&=
{1\over {2\sqrt{7}}} \left({3\sqrt{3}\over \sqrt{2}}{\lambda}-2\phi \right),
\cr
\sigma_{3}&={1\over \sqrt{7}} \left(-{\lambda\over \sqrt{6}}+\phi \right) .}}
Having done this, we can then set $\lambda=0$ to obtain the desired result.
Note that by analytically continuing the general solution to the cases $d=0$
and $d=-1$, we are extrapolating the meaning of the ADM mass and topological
charge to non-asymptotically flat spacetimes.

We note that the black hole solution corresponds to a
Maxwell-scalar coupling $e^{-a \phi} F_{\mu \nu} F^{\mu \nu}$ with
$a=\sqrt{3}$. This is to be contrasted with
the pure dilaton black hole solutions of the heterotic string
that have attracted much attention
recently
\refs{\gib,\gibm,\ghs,\shatw,\klopv,\kal,\ko,\hor,\senprl,\gidps}
and have $a=1$
\footnote{$^\ast$}{Contrary to some claims in the
literature, the pure Reissner-Nordstr\"om
black hole with $a=0$ is also a solution of the low energy heterotic string
equations. This may be seen by noting that it provides a solution to $(N=2,
\, D=4)$ supergravity which is a consistent truncation of toroidally
compactified $N=1, \, D=10$ supergravity \dupo.}. The case $a=\sqrt{3}$
also occurs when
the Maxwell field and the scalar field $\phi$ arise
from a Kaluza-Klein reduction of
pure gravity from $D=5$ to $D=4$:
\eqn\redmet {\hat{g}_{MN} = e^{\phi\over{\sqrt{3}}}
   \pmatrix { g_{\mu\nu}+e^{-\sqrt{3} \phi} A_{\mu}A_{\nu} &
                    e^{-\sqrt{3} \phi} A_{\mu} \cr
          e^{-\sqrt{3} \phi} A_{\nu} &   e^{-\sqrt{3} \phi} \cr},}
where $\hat{g}_{MN} \, (M,N=0,1,2,3,4)$ and $g_{\mu\nu} \,
(\mu,\nu=0,1,2,3)$ are the canonical metrics in 5 and 4 dimensions
respectively. The resulting action is given by
\eqn\redact{S={1\over2 \kappa^2} \int d^4 x \sqrt{-g} \left[ R -{1\over 2}
   (\partial \phi)^2 -{1\over 4} e^{-\sqrt{3} \phi} F_{\mu \nu} F^{\mu \nu}
   \right]}
and it admits as an ``elementary'' solution the $a=\sqrt{3}$ black hole
metric \bmonrs,  but with the scalar field
\eqn\die{e^{-2 \phi}= \Delta_{-}^{\sqrt{3}}}
and the electric field
\eqn\fse{{1\over \sqrt{2} \kappa}e^{-\sqrt{3} \phi \, \ast} F_{\theta\varphi}=
         {e\over 4 \pi} \sin\theta}
corresponding to an electric monopole with Noether charge $e$.
This system also admits the topological magnetic solution with
\eqn\dim{e^{-2 \phi}= \Delta_{-}^{-\sqrt{3}}}
and the magnetic field
\eqn\fsm{{{1}\over{\sqrt{2}\kappa}} F_{\theta\varphi}={g\over4 \pi}\sin\theta}
corresponding to a magnetic monopole with  topological magnetic
charge $g$ obeying the Dirac quantization rule
\eqn\diqua{eg=2 \pi n, \, \, \, n={\rm integer.}}
In effect, it was for this reason that the $a=\sqrt{3}$ black hole
was identified as a solution of the Type II string in \duflblacks, the fields
$A_\mu$ and $\phi$ being just the abelian gauge field and the dilaton of
($N=2,\, D=10$) supergravity which arises from Kaluza-Klein compactification
of ($N=1,\, D=11$) supergravity.

Some time ago, it was pointed out in \gibp\ that $N=8$ supergravity,
compactified from $D=5$ to $D=4$, admits an infinite tower of
elementary states
and that these elementary states fall into $N=8$ supermultiplets. They also
pointed out that this theory admits an infinite tower of solitonic states
which also fall into the same $N=8$ supermultiplets. The authors of \gibp\
 conjectured,
$\acute{\rm a}$ la Olive-Montonen \mono, that there should
exist a dual formulation of the theory for which the roles of electric
elementary states and magnetic solitonic states are interchanged.
It was argued in \duflblacks\ that this electric/magnetic duality conjecture
in $D=4$ could be reinterpreted as a particle/sixbrane duality conjecture
in $D=10$.

To see this, consider the action dual to $S$, with $a=-\sqrt{3}$, for
which the roles of Maxwell field equations and Bianchi identities are
interchanged:
\eqn\duact{\tilde{S}=
        {1\over2 \kappa^2} \int d^4 x \sqrt{-g} \left[ R -{1\over 2}
   (\partial \phi)^2 -{1\over 4} e^{\sqrt{3} \phi}
      \tilde{F}_{\mu \nu} \tilde{F}^{\mu \nu}
   \right],}
where $\tilde{F}_{\mu\nu} =e^{-\sqrt{3} \phi \, \ast} F_{\mu\nu} .$
This is precisely the action obtained by double dimensional reduction of
a dual formulation of ($D=10, \, N=2$) supergravity in which the two-form
$F_{MN}$ ($M,N=0,...,9$) is swapped for an 8-form
$\tilde{F}_{M_1..M_8}$, where $\tilde{F}_{\mu\nu}=
\tilde{F}_{\mu\nu456789}$.
This dual action also admits both electric and magnetic
monopole solutions
but because the roles of field equations and Bianchi identities
are
interchanged, so are the roles of electric and magnetic. Since
the 1-form
and 7-form potentials, which give rise to these 2-form and
8-form
field strengths, are those that couple naturally to the
worldline of
a point particle or the worldvolume of a 6-brane, we see that
the
Gibbons-Perry ($N=8,\, D=4$) electric/magnetic duality conjecture may be
re-expressed as a (Type II, $D=10$) particle/sixbrane duality conjecture.
Indeed, the $D=10$ black sixbrane of \hors\ is simply obtained
by adding 6 flat dimensions to the $D=4,\, a=\sqrt{3}$ magnetic
black hole.

In general, the $N=8$ theory will admit black holes with
mass $M$ and magnetic charge $g$ saturating the Bogomol'nyi
bounds
\eqn\bogo{4\pi G(M^2+\Sigma^2)=4\pi GM^2(1+a^2)=g^2}
where $\Sigma=a M$ is the scalar charge.

The solutions presented in this section now allow us to discuss
the
$a=\sqrt{3}$
electric/magnetic duality from a totally different perspective
from above.
For concreteness, let us focus on  generic toroidal
compactification of the
heterotic string. Instead of $N=8$ supergravity, the
four-dimensional theory is now
$N=4$ supergravity coupled to 22  $N=4$ vector
multiplets.
The same dual Lagrangians \redact\ and \duact\ still emerge
but
 with
completely different origins. The Maxwell field $F_{\mu\nu}$
(or $\tilde{F}_
{\mu\nu}$) and the scalar field $\phi$ do not come from the
$D=10$ \ $2$-form (or
$8$-form) and dilaton of the Type II particle (or sixbrane),
but rather from
the $D=10$ \ $3$-form (or $7$-form) and dilaton plus modulus
field of the
heterotic string (or heterotic fivebrane). Thus, the $D=4$
 electric/magnetic
duality can now be re-interpreted as a $D=10$ string/fivebrane
duality! (Note that in addition to the $6$ Maxwell fields
coming from the $D=10$ $3$-form, there will be another $6$
coming from
the $D-10$ metric. Moreover, these are interchanged under
an $O(6,6;Z)$ subgroup of the target-space duality discussed
below. A truncation of a single $U(1)$ similar to \redmon\
also yields $a\sqrt{3}$; a result well-known from Kaluza-Klein
theory\footnote{$^\dagger$}
{Gibbons discusses both the $a=1$ black hole of pure $N=4$
supergravity and the $a=\sqrt{3}$ Kaluza-Klein black hole in the same paper
\gib, as does Horowitz \hor.
Moreover, black
holes in pure $N=4$ supergravity are treated by Kallosh {\it et al.}
\refs{\klopv\kal{--}\ko}.
The reader may therefore wonder why the $a=\sqrt{3}$ $N=4$ black hole
discussed here was overlooked. The reason is that pure $N=4$
supergravity does not admit the $a=\sqrt{3}$ solution; it is crucial that
we include the $N=4$ vector multiplets in order to introduce the modulus
fields.}. That both kinds of black-hole solutions have
$a=\sqrt{3}$ can thus be understood as a consequence of
duality.)

Because of the non-vanishing modulus field  $g_{44}=e^{-2\sigma}$, however,
the $D=10$ black fivebrane solution is not obtained by adding 6
flat
dimensions to the $D=4$ black hole. Rather the two are connected by wrapping
the fivebrane around 5 of the 6 extra dimensions \dufkexst.

Of course, we have established only that these two-parameter
configurations are solutions of the field theory limit of the
heterotic string.
Although the extreme one-parameter solutions are expected to be
exact to all
orders in $\alpha'$, the same reasoning does not carry over to
the new
two-parameter solutions. It would be interesting to pursue
conformal field
theory arguments (see section 8),
perhaps along the lines recently suggested in \gidps.

It would be also interesting to see whether the generalization
of the
one-parameter solutions of section 6.5 to the two-parameter
solutions of this
section can be carried out when we include the Yang-Mills
coupling.
 This would necessarily involve giving up
the self-duality condition on the Yang-Mills field strength,
 however, since the
self-duality condition is tied to the extreme, $\sqrt{2}
\kappa {\cal M}_{p+1}
=  g_{p+1}$, supersymmetric solutions.

Finally, there is the question of whether these solutions are
 peculiar
to the toroidal compactification or whether they survive in
more realistic
orbifold or Calabi-Yau models \dauria. Although the actions $S_{1}$, $S_{2}$
and $S_{3}$ were originally derived in the context of the torus \dufkexst, they
also appear in a large class of $N=1$ supergravity theories \sgone.

\newsec{Massive string states as extreme black holes}

The idea that elementary particles might behave like black holes is not a
new one \salam. Intuitively, one might expect that
a pointlike object whose mass exceeds the Planck mass, and whose Compton
wavelength is therefore less than its Schwarzschild radius, would exhibit
an event horizon. In the absence of a consistent quantum theory of gravity,
however, such notions would always remain rather vague. Superstring theory,
on the other hand, not only predicts such massive states but may provide us
with a consistent framework in which to discuss them. In this
section we confirm the claims of the previous section
that
certain massive excitations of four-dimensional superstrings
are indeed
black holes.  Of course, non-extreme black holes would be
unstable due to
the Hawking effect. To describe stable elementary particles,
therefore, we
must focus on extreme black holes whose masses saturate a
 Bogomol'nyi bound
\footnote{$^*$}{The relationship between extremal black holes
and the
gravitational field around some of the elementary string states
has also been
discussed in \refs{\fabjr,\senone}.}. We
therefore remain agnostic concerning the stronger claims
\refs{\ellmn,\susskind}\
that {\it all} black holes are single string states or,
conversely, that all
massive string states are black holes.

We have just seen that this theory exhibits both electrically
 and magnetically charged
black hole solutions corresponding to scalar-Maxwell parameter
$a=0,1,\sqrt3$. In other words, by choosing appropriate
combinations of
dilaton and moduli fields to be the scalar field $\phi$ and
appropriate
combinations of the field strengths and their duals to be the
Maxwell field
$F$, the field equations can be consistently truncated to a form given by
the Lagrangian
\eqn\mbone{
{\cal L}= {1\over 32\pi}\sqrt{-g}\left [R-{1\over 2}(\partial
\phi)^2-{1\over 4}e^{-a \phi}F^2 \right]}
for these three values of $a$. (A {\it consistent} truncation
is defined to be
one for which all
solutions of the truncated theory are solutions of the original
theory). In the case of zero angular momentum, the bound
between the black hole ADM mass
$m$, and the electric charge $Q=\int e^{-a\phi}{}{\tilde F}/8\pi$, where a
tilda denotes the dual, is given by
\eqn\mbtwo{
m^2\geq Q^2/4(1+a^2),}
where, for simplicity, we have set the asymptotic value of
$\phi$ to zero.
The $a=0$ case yields the Reissner-Nordstrom solution which,
notwithstanding
contrary claims in the literature, does solve the low-energy
string
equations. The $a=1$ case yields the dilaton black hole
\refs{\gib,\ghs,\gidps}. The $a=\sqrt{3}$ case corresponds to
the
Kaluza-Klein black hole and the ``winding" black hole
\dufkmr\ which are
related to each other
by $T$-duality. The Kaluza-Klein solution has been known for
some time \gib\ but only
recently recognized \dufkmr\ as a heterotic string solution.

Let us denote by $N_L$ and $N_R$
the number of left and right oscillators respectively.
We shall consider the Schwarz-Sen \refs{\schsentwo,\senrev}\
$O(6,22;Z)$ invariant
spectrum of elementary electrically charged massive $N_R=1/2$
states of this
four-dimensional heterotic string, and show that the spin zero
states
correspond to extreme limits of  black hole solutions which
preserve
$1/2$ of the spacetime supersymmetries. By supersymmetry, the black hole
interpretation then applies to all members of the $N=4$ supermultiplet
\refs{\gibbons,\aichelburg},
which has $s_{max}=1$. For a subset of states the low-energy string action can
be truncated
to \mbone.  The scalar-Maxwell parameter is given by $a=\sqrt 3$ for
$N_L=1$ and $a=1$ for $N_L>1$ (and vanishing
left-moving internal momenta). The other states with $N_L>1$
are extreme
black holes too, but are not
described by a single scalar truncation of the type \mbone.
The $N=4$ supersymmetry algebra
possesses two central charges $Z_1$ and $Z_2$.  The $N_R=1/2$
states correspond
to that subset of
the full spectrum that belong to the $16$ complex dimensional
($s_{max}\geq1$)
representation of
the $N=4$ supersymmetry algebra, are annihilated by half of the
supersymmetry
generators and
saturate the strong Bogomol'nyi bound $m=|Z_1|=|Z_2|$. As
discussed in
\refs{\wito,\osb,\schsentwo,\senrev}, the
reasons for focussing on this $N=4$ theory, aside from its
simplicity, is that
one expects that the
allowed spectrum of electric and magnetic charges is not
renormalized by
quantum corrections, and
that the allowed mass spectrum of particles saturating the
Bogomol'nyi bound is
not renormalized
either.
As discussed in \refs{\schsentwo,\senrev}\ only a subset of the
conjectured spectrum corresponds to elementary string states.
 First of all
these states will be only electrically charged, i.e. $\b =0$,
but there
will be restrictions on $\a$ too. Without loss of generality
let us focus on
a compactification with $M{}^0=I$ and $\lambda_{2}{}^{0}=1$.
Any other toroidal
compactifications can be brought into this form by $O(6,22)$
transformations
and a constant shift of the dilaton.  The mass formula \drlafive\
now becomes
\eqn\mbseven{
m^2={1\over 16}
 \alpha^a (I+L)_{ab} \alpha^b=
 {1\over 8}
 (\alpha_R)^2}
with $\alpha_R = {1\over 2}(I+L) \alpha$ and
$\alpha_L = {1\over 2} (I-L) \alpha$. In the string
language $\a_{R(L)}$ are the right(left)-moving internal
momenta.
The mass of a generic string state in the Neveu-Schwarz sector
(which is
degenerate with the Ramond sector) is given by
\eqn\mbeight{
m^2={1\over 8} \left\{
 \left(\alpha_R \right)^2 +2 N_R -1 \right\}=
 {1\over 8} \left\{
 \left(\alpha_L \right)^2 +2 N_L -2 \right\}.}
A comparison of \mbseven\ and
\mbeight\ shows that the string states
satisfying the Bogomol'nyi bound all have $N_R=1/2$. One then
finds
\eqn\mbnine{
 N_L-1={1\over 2}\left(\left( { \alpha}_R \right)^2
 -\left( { \alpha}_L \right)^2\right)
 ={1\over 2} { \alpha}^{T} L { \alpha},}
leading to ${ \alpha}^{T} L { \alpha}\geq -2$.  We shall now
show that
extreme black holes with $a=\sqrt{3}$ are string states with
${
\alpha}^{T} L { \alpha}$ null ($N_L=1$) and those with $a=1$ are string
states with ${ \alpha}^{T} L { \alpha}$ spacelike $(N_L>1)$. We
have
been unable to identify solutions of the low-energy field
equations of \newsix\
corresponding to states with ${ \alpha}^{T} L { \alpha}$
timelike
$(N_L<1)$. \footnote{$^\dagger$}{In the {\it non-abelian}
theory Sen
\senrev\
identifies these states with the electric analogues of BPS
monopoles.}

 Let us first focus on the $a=\sqrt{3}$ black hole. To identify
it as a state
in the spectrum we have to find the corresponding charge vector
 ${\bf \a}$
and to verify that the masses calculated by the formulas
\mbtwo\ and
\drlafive\ are identical. The action \newsix\ can be
consistently truncated
by keeping the metric $g_{\mu\nu}$, just one field strength
($F=F^1$, say), and
one scalar field $\phi$ via the ansatz $\Phi=\phi/\sqrt{3}$ and
$M_{11}=e^{2\phi/\sqrt{3}}=M_{77}^{-1}$. All other diagonal
components of
$M$ are set equal to unity and all non-diagonal components to
zero. Now
\newsix\
reduces to \mbone\ with $a=\sqrt{3}$. (This yields the electric
 and magnetic
Kaluza-Klein (or ``$F$") monopoles. This is
not quite the truncation chosen in \dufkmr, where just $F^7$ was
retained and $M_{11}=e^{-2\phi/\sqrt{3}}=M_{77}^{-1}$. This yields the
electric and
magnetic winding (or ``$H$") monopoles.  However, the two are related by
$T$-duality).  We shall
restrict ourselves to the purely electrically charged solution
 with charge
$Q=1$, since this one is
expected to correspond to an elementary string excitation. The
charge vector
${\bf \a}$ for this
solution is obviously given by $\alpha^a=\delta^{a,1}$ with
 ${\bf \alpha}^{T} L
{\bf \alpha}=0$.
Applying \drlafive\ for the mass of the state we find
$m^2=1/16=Q^2/16$, which
coincides with
\mbtwo\ in the extreme limit.  This agreement confirms the claim that this
extreme $a=\sqrt{3}$
black hole is a state in the Sen-Schwarz spectrum and preserves $2$
supersymmetries.

Next we turn to the $a=1$ black hole. The theory is consistently truncated
by keeping the metric, $F=F^1=F^7$ and setting $M=I$. The only
non-vanishing
scalar is the dilaton $\Phi\equiv\phi$.  Now \newsix\ reduces to
\mbone\
with $a=1$ but $Q^2=2$. An extreme $a=1$ black hole with
electric charge $Q$
is then represented by the charge configuration
$\alpha^a=\delta^{a,1}+\delta^{a,7}$. Applying \drlafive\ we
 find
$m^2=1/4=Q^2/8$ which coincides with \mbtwo\ in the extreme
limit.
Therefore the $a=1$ extreme solution is also in the spectrum,
and has ${\bf
\a}^{T} L {\bf \a}=2$ or $N_L=2$.

Although physically very different, we can see with hindsight
 that both the
$a=\sqrt3$ and $a=1$
black holes permit a uniform mathematical treatment by noting
that both may be
obtained from the
Schwarzschild solution by performing an $[O(6,1)
 \times O(22,1)]/[O(6) \times
O(22)$ transformation
\senone. The $28$ parameters of this transformation correspond
to the $28$
$U(1)$ charges.  If
$\gamma$ and ${\bf u}$ correspond to the boost angle  and a
$22$ dimensional unit
vector
respectively, associated with $O(22,1)/O(22)$  transformations,
$\delta$ and
${\bf v}$ denote the
boost angle and the $6$  dimensional unit vector respectively,
associated with
the $O(6,1)/O(6)$
transformations, and $m_0$ is the mass of the original
Schwarzschild
black hole, then the mass and charges of the new black hole
solution
are given by \senone:
\eqn\mbten{\eqalign{
m &={1\over 2} m_0 (1+\cosh\gamma \cosh\delta),\cr
\alpha_L &={\sqrt 2} m_0 \cosh\delta \sinh\gamma \, {\bf u},\cr
 \alpha_R &={\sqrt 2} m_0 \cosh\gamma \sinh\delta \, {\bf v}
.\cr}}
(Note that the convention about $R$ and $L$ of \senone\ is
opposite to the
one used in this section). Black holes with
$\alpha^T L\alpha=0$ are generated by setting
$\gamma=\delta$,
whereas black holes with $\alpha^T L\alpha>0$ are generated by
setting
$\gamma<\delta$. The
Bogomol'nyi bound given in \mbseven\  corresponds to
$m^2=( \alpha_R)^2/8$.
This bound is saturated by taking the limit where the mass
$m_0$ of the
original Schwarzschild black hole approaches $0$ and the
parameter
$\delta$ approaches $\infty$, keeping
the product $m_0\sinh\delta$ fixed. As  discussed in \senone,
this
is precisely the extremal limit. Thus we see that extremal
 black holes
satisfy the Bogomol'nyi relation, both for $\alpha^T L
\alpha=0$ and
$\alpha^T L\alpha>0$.

{}From the above $a=\sqrt 3$
solution we can generate the whole set of supersymmetric black
hole solutions
with ${\bf
\alpha}^{T} L {\bf \alpha}=0$ in the following way: first we
 note that we are
interested in
constructing black hole solutions with different charges but
 with fixed
asymptotic values of $M$
(which here has been set to the identity).  Thus we are not
 allowed to make
$O(6,22)$
transformations that change the asymptotic value of $M$. This
leaves
us with only an $O(6) \times O(22)$ group of transformations.
 The effect
of these transformations acting on the parameters given in
\mbten\ above is
to
transform the vectors ${\bf u}$ and ${\bf v}$ by $O(22)$ and
$O(6)$ transformations
respectively without changing the parameters $\gamma$ and
$\delta$.  Now, the
original $a=\sqrt 3$
solution corresponds
to a choice of parameters $\gamma=\delta$,
${\bf u}^m=\delta_{m1}$ and ${\bf v}^m
=\delta_{k1}$. It is clear that an $O(6) \times O(22)$
transformation can rotate
${\bf u}$ and ${\bf v}$ to arbitrary $22$ and $6$ dimensional
unit vectors
respectively, wothout changing $\gamma$ and $\delta$. Since
 this corresponds to the
most general charge vector satisfying $\alpha^T L\alpha=0$,
we see that the $O(6) \times O(22)$ transformation can indeed
generate an
arbitrary black hole solution with $\alpha^TL\alpha=0$ starting
 from the
original $a=\sqrt 3$ solution.
This clearly leaves the mass
invariant, but the new charge vector ${\bf \a}'$ will in
general not be located
on the lattice. To
find a state in the allowed charge spectrum we have to rescale
${\bf \a}'$ by a
constant $k$ so
that ${\bf \a}''=k{\bf \a}'$ is a lattice vector. Clearly the
 masses calculated
by \mbtwo\ and
\drlafive\ still agree (this is obvious by reversing the steps
of rotation and
rescaling), leading
to the conclusion that all states obtained in this way preserve
$1/2$ of the
supersymmetries.
Therefore all states in the spectrum belonging to $s_{max}=1$
supermultiplets
for which
$N_R=1/2,N_L=1$ are extreme $a=\sqrt{3}$ black holes.

Let us now turn to the case of the $a=1$ solution. In this case
the
original solution corresponds to the choice of parameters
$\gamma=0$,
${\bf v}^m=\delta_{m1}$. (For $\gamma=0$, the parameter
${\bf u}$ is
irrelevant).
An $O(6) \times O(22)$ transformation can rotate
${\bf v}$ to any other $6$ dimensional unit vector, but it
cannot change the
parameters $\delta$ and $\gamma$. As a result, the final
solution will
continue to have $\gamma=0$ and hence $\alpha_L=0$.  Since this
does
not represent the most general charge vector $\alpha$, with
$\alpha^T L
\alpha>0$, we see that the most general black hole representing
states
with $\alpha^T L\alpha>0$ is not obtained in this way even after rescaling.
The missing  states with $\a_L\neq 0$  are constructed by choosing
$\gamma$  so that $\tanh ^2 \gamma={\a_L^2}/{\a_R^2}$, and
${\bf u}, \ {\bf v}$ as for the $a=\sqrt{3}$ case, followed by a suitable
$O(6) \times O(22)$ rotation.
Clearly, those solutions  are extreme black
holes too.
However, for these solutions  a truncation to an effective action of the form
\mbone\ is not possible.  The following picture arises: for a fixed
value of $\a_R^2$, $\a_L^2$ can
vary in the range $\a_R^2\geq \a_L^2\geq 0$. The boundary states are
described by the well-known $a=\sqrt{3}$ ($\a_R^2=\a_L^2$) and $a=1$
($\a_L^2=0$) black holes, whereas the states in between
 cannot be related to a single scalar-Maxwell parameter $a$. But all
solutions preserve $1/2$ of the supersymmetries.

 It should also be clear that the purely magnetic extreme black hole
solutions \dufkmr\ obtained from the above by the replacements $
\phi\rightarrow -\phi, \alpha\rightarrow\beta$ will also belong to the
Schwarz-Sen spectrum of solitonic states.  Starting from either the purely
electric or purely magnetic solutions, dyonic states in the spectrum which
involve non-vanishing axion field $\Psi$ can then be obtained by $SL(2,Z)$
transformations. Specifically, a black hole with charge vector $({\bf
\alpha}, 0)$ will be mapped into ones with charges $(a{\bf \alpha}, c{\bf
\alpha})$ with the integers $a$ and $c$ relatively prime
\refs{\schsentwo,\senrev}.

Not all black hole solutions of \newsix\ belong to the Sen-Schwarz
spectrum, however. Let us first consider the Reissner-Nordstrom solution.
Since this black hole solves the equations of $N=2$ supergravity, whose
bosonic sector is pure Einstein-Maxwell, it solves \newsix\ as well.  The
required consistent truncation is obtained by keeping $g_{\mu\nu}$,
$F=F^1=F^7=\tilde F^2=\tilde F^8$ and setting $\Phi=0$, $M=1$. Now
\newsix\ effectively reduces to \mbone\ with $a=0$ but $Q^2=4$. On the
other hand, if it were in the Schwarz-Sen spectrum its charge vectors would
be given by $\alpha^a=\delta^{a,1}+\delta^{a,7}$ with ${\bf \alpha}^{T}L
{\bf \alpha}=2$ and  $\beta^a=\delta^{a,2}+\delta^{a,8}$ with ${\bf
\beta}^{T}L{\bf \beta}=2$.  Applying \drlafive\ for the mass of the state we
find $m^2=1/2$, which disagrees with the result $m^2=1$ obtained from the
extreme limit of \mbtwo\. So the test fails and the $a=0$ black hole does
not belong to the Schwarz-Sen spectrum.  This was only to be expected since
it breaks $3/4$ of the supersymmetries and hence saturates the weaker
Bogomol'nyi bound $m=|Z_1|,|Z_2| =0$ \klopv.  Such black holes
belong to the $32$ complex dimensional ($s_{max}=3/2$) supermultiplet. We
see no reason to exclude these states from the full string spectrum, however.
Another example of a black hole solution not in the Schwarz-Sen spectrum is
the $a=1$ dilaton black hole of \refs{\ghs,\gidps} where the only
non-vanishing gauge field is $F^{13}$. This has mass $m^2=Q^2/8$ but
according to \drlafive\ its mass would vanish.  Again, this contradiction is
only to be expected since this solution breaks all the supersymmetries, in
contrast with the $F=F^1=F^7$ embedding discussed above. We
do not know whether such black holes saturating no
Bogomol'nyi bound ($m>|Z_1|,|Z_2|)$, which include the neutral Schwarzschild
black holes ($Z_1=Z_2=0$), are also in the string spectrum. States with
these quantum numbers would belong to the $256$ dimensional ($s_{max}\geq 2$)
supermultiplets. According to \gibbons, however, black holes breaking
all the supersymmetries do not themselves form supermultiplets. This would
appear to contradict the claim that {\it all} black holes are string states.

In the supersymmetric case, all values of $a$ lead to extreme black holes with
zero entropy but their temperature is zero, finite or infinite according as
$a<1$, $a=1$ or $a>1$, and so in \hw\ the question was posed: can
only
$a>1$ scalar black
holes describe elementary particles? We have not definitively answered
this question but a tentative response would be as follows.  First we note
that the masses and charge vectors are such that the lightest $a=0$ black
hole may be regarded as a bound state (with zero binding energy) of two
lightest $a=1$
black holes
which in turn can each be regarded as bound states (again with zero binding
energy) of two
lightest
$a=\sqrt{3}$ black holes. Thus if by elementary particle one means an object
which cannot be
regarded as a bound state, then indeed extreme scalar black holes with $a>1$
are the only
possibilty, but if one merely means a state in the string spectrum then
$a\leq1$ extreme scalar
black holes are also permitted.

We have limited ourselves to $N_R=1/2$ supermultiplets with
$s_{min}=0$.
Having
established that the $s=0$ member of the multiplet is an extreme black hole,
one
may then use the fermionic zero modes to perform supersymmetry
transformations
to generate the
whole
supermultiplet of black holes \refs{\gibbons,\aichelburg}\ with
 the same mass and
charges. Of
course,
there are $N_R=1/2$ multiplets with $s_{min}>0$ coming from
oscillators with
higher spin and our
arguments have nothing to say about whether these are also
extreme black holes.
They could be
naked
singularities. Indeed, although in this paper we have focussed
primarily on
identifying
certain massive heterotic string states with extreme black holes,
 perhaps
equally remarkable is
that these elementary string states can be described at all by
solutions of the
supergravity
theory.  In a {\it field} theory, as opposed to a {\it string}
theory, one is
used to having as
elementary massive states only the Kaluza-Klein modes with
$s_{max}=2$.
However, as we have
already seen, the  winding states (usually thought of as
intrinsically stringy)
are on the same
footing as Kaluza-Klein states as far as solutions are
concerned, so perhaps
the same is true
for the $s>2$ states.

None of the
spinning $N_R=1/2$ states is described by extreme {\it rotating} black hole
metrics because they
obey the same Bogomol'nyi bound as the $s_{min}=0$ states,
whereas the mass
formula for an extreme
{\it rotating} black hole depends on the angular momentum $J$.
Moreover, it is
the fermion fields
which carry the spin in the $s_{min}=0$ supermultiplet.
(For the $a=0$ black
hole, they yield a
gyromagnetic ratio $g=2$ \aichelburg; the $a=\sqrt{3}$ and $a=1$
superpartner $g$-factors
are unknown to us).  It may be that there are states in the
string spectrum
described by the
extreme rotating black hole metrics but if so they will belong
to the $N_R\neq
1/2$
sector\footnote{$^{**}$}{The gyromagnetic and gyroelectric
ratios of the states in the
heterotic string
spectrum would then have to agree with those of charged
rotating black hole
solutions of the
heterotic string. This is indeed the case: the $N_L=1$ states
\hosoya\ and
the rotating
$a=\sqrt{3}$ black holes \gibw\ both have $g=1$ whereas the
$N_L>1$
states \susskind\
and the rotating $a=1$ \senprl\ (and $a=0$ \schild) black holes
both
have $g=2$. In fact,
it was the observation that the Regge formula $J\sim m^2$ also
describes the
mass/angular momentum
relation of an extreme rotating black hole which first led
Salam \salam\
to imagine that
elementary particles might behave like black holes!}.  Since,
whether rotating
or not, the black
hole solutions are still independent of the azimuthal angle and
independent of
time, the
supergravity theory is effectively {\it two-dimensional} and
therefore possibly
integrable. This
suggests that the spectrum should be invariant under the larger
duality
$O(8,24;Z)$ \dufldr,
which combines $S$ and $T$. The corresponding Kac-Moody
extension would then
play the role of the
spectrum generating symmetry \geroch.

\chapter{Dynamics of string solitons}

\def\b#1{\vec\beta_{#1}}
\def\x#1{\vec X_{#1}}
\def\x{{(R^2-2)v^2\over 1-v^2}}

\newsec{Metric on moduli space}

All the soliton solutions we have discussed so far have the
property, like BPS
magnetic monopoles, that they exert zero static force on
each other and can be superposed to form multi-soliton solutions with
arbitrarily variable collective coordinates. Since these static
properties are also possessed by fundamental strings winding
around an
infinitely large compactified dimension, it was conjectured in
\dabh\ that the
elementary string is actually the exterior solution for an
 infinitely long
fundamental string. In this section we show that, in
contradistinction to the BPS case, the velocity-dependent forces
between these string solitons also vanish,
at least to order $\beta^2$ (where $\beta$ is the velocity).
We also argue that this phenomenon provides further,
dynamical evidence for the identification of the elementary
string solution
with the underlying fundamental string by comparing the
scattering of the
elementary
solutions with expectations from a Veneziano amplitude
computation for
macroscopic fundamental strings \khuscat.

As shown in \absencevdf, the static ansatz leads to a vanishing
velocity-dependent force for a test ($d-1$)-brane propagating
in the background
of an elementary ($d-1$)-brane. By duality, this result also
holds for test
($\tilde d-1$)-branes propagating in the background of a
solitonic
($\tilde d-1$)-brane.
As this is a rather surprising result, we would like to compute
the metric on
moduli space for these solitons.
The geodesics of this metric represent the motion of
quasi-static solutions
in the static solution manifold and in the absence of a full
time dependent
solution provide a good approximation to the low-energy
dynamics of the
solitons. In all cases the metric is found to be flat in
agreement with the
test-soliton approximation, which again implies vanishing
dynamical force in
the low-velocity limit. Here we summarize the computation for
the metric on
moduli space for monopoles discussed in
\refs{\khumonscat,\khumonex}.

Manton's prescription \mantwo\ for the study of soliton
scattering may be
summarized as follows. We first invert the constraint equations
of the system.
The resultant time dependent field configuration does not in
general
satisfy the full time dependent field equations, but provides
an initial data point for the fields and their time derivatives.
Another way of saying this is that the initial motion is tangent to the
set of exact static solutions.  The kinetic action obtained
by replacing the solution to the constraints into the action
defines a
metric on the parameter space of static solutions. This metric defines
geodesic motion on the moduli space \mantwo.

A calculation of the metric on moduli space for the scattering of BPS
monopoles and a description of its geodesics was worked out by Atiyah
and Hitchin \atihone. Several interesting properties of monopole
scattering were found, such as the conversion of monopoles into dyons
and the right angle scattering of two monopoles on a direct collision
course \refs{\atihone,\atihtwo}. The configuration space is found to
be a four-dimensional manifold $M_2$ with a self-dual Einstein metric.

Here we adapt Manton's prescription to study the dynamics of the heterotic
 string monopoles discussed in section 6.5. We follow essentially
the same steps that Manton outlined for monopole scattering, but take
into account the peculiar nature of the string effective action. Since
we work in the low-velocity limit, our kinematic analysis is nonrelativistic.

We first solve the constraint equations for the monopoles.
These equations are simply the $(0j)$
components of the tree-level equations of motion
\eqn\constraints{\eqalign{R_{0j}-{1\over 4}H^2_{0j}+2\nabla_0\nabla_j\phi&=0,
\cr
-{1\over 2}\nabla_kH^k{}_{0j}+H_{0j}{}^k\partial_k\phi&=0.\cr}}
We wish to find an $O(\beta)$ solution to the above equations which
represents a quasi-static version of the neutral multi-monopole solution (i.e.
 a multi-monopole solution with time dependent $\vec a_i$). We consider the
neutral case as it was argued in \khustab\ that the the
cancellation of gauge
and gravitational corrections in the static action for the
symmetric case also
manifests itself in the dynamics, at least to $O(\alpha')$.
 Hence the scattering of symmetric string
 monopoles is expected to be similar to that of neutral string monopoles.
Also, we use the solution \nsansatz\ with
$e^{2\phi}=e^{2\phi_0}f_M$, with $f_M$ given in \fmono\ and
compactify only the directions $x^5,x^6,x^7,x^8,x^9$.
 In other words, we do not make the
replacement $g_{44}=e^{-2\sigma}$ to obtain the
completely compactified version \redmsol, although the results in both cases
are
identical, in order to more easily keep track of the terms in the former case.
We give each monopole an arbitrary transverse velocity
$\vec\beta_n$ in the $(123)$ subspace of the four-dimensional transverse space
and see what corrections to the fields are required by the
constraints. The vector $\vec a_n$ representing the position of the
$n$th monopole in the three-space $(123)$ is given by
\eqn\aunty{\vec a_n(t)=\vec A_n + \vec\beta_nt,}
where $\vec A_n$ is the initial position of the $n$th monopole. Note that at
$t=0$ we recover the exact static multi-monopole solution. Our solution to
the constraints will adjust our quasi-static approximation so that the
initial motion in the parameter space is tangent to the initial
exact solution at $t=0$.
The $O(\beta)$ solution to the constraints is given by
\eqn\orderbeta{\eqalign{e^{2\phi(\vec x,t)}&=1+\sum_{n=1}^N{m_n\over
|\vec x - \vec a_n(t)|},\cr g_{00}&=-1,\qquad g^{00}=-1,\qquad
g_{ij}=e^{2\phi}\delta_{ij},\qquad g^{ij}=e^{-2\phi}\delta_{ij},\cr
g_{0i}&=-\sum_{n=1}^N{m_n\vec\beta_n\cdot \hat x_i\over |\vec x - \vec
a_n(t)|},\qquad g^{0i}=e^{-2\phi}g_{0i},\cr
H_{ijk}&=\epsilon_{ijkm}\partial_m e^{2\phi},\cr
H_{0ij}&=\epsilon_{ijkm}\partial_m g_{0k}=\epsilon_{ijkm}\partial_k
\sum_{n=1}^N{m_n\vec\beta_n\cdot \hat x_m\over |\vec x - \vec a_n(t)|},\cr}}
where $i,j,k,m=1,2,3,4$, all other metric components are flat, all other
components of $H$ vanish, the $\vec a_n(t)$ are given by \aunty\ and we use a
flat space $\epsilon$-tensor. Note that $g_{00}$, $g_{ij}$ and $H_{ijk}$ are
unaffected to order $\beta$. Also note that we can interpret the monopoles
as either strings in the space $(01234)$ or point
objects in the three-dimensional subspace $(0123)$.

The kinetic Lagrangian is obtained by replacing the expressions for the
fields in \orderbeta\ into the string $\sigma$-model action
\stringact\ for $D=5$ \footnote{$^*$}{Strictly speaking one must add to
\stringact\ a surface term to cancel the double derivative terms in the action
\refs{\gibhp,\gibhpone,\brih,\khumant,\khumonscat}\ however the addition of
this term introduces only flat kinetic terms and thus presents no nontrivial
contribution to the metric on moduli space.}. Since \orderbeta\ is a solution
to
order $\beta$, the leading order terms in the action (after the
quasi-static part) are of order $\beta^2$. The
 $O(\beta)$ terms in the solution give $O(\beta^2)$ terms when replaced in the
kinetic action. Collecting all $O(\beta^2)$ terms in \stringact\ we
get the following kinetic Lagrangian density for the volume term:
\eqn\kinlag{\eqalign{{\cal L}_{kin}=-{1\over 2\kappa^2}\Biggl(
&4\dot \phi\vec M\cdot\vec \nabla\phi
-e^{-2\phi}\partial_iM_j\partial_iM_j
-e^{-2\phi}M_k\partial_j\phi\left(\partial_jM_k-\partial_kM_j\right)\cr
&+4M^2e^{-2\phi}(\vec \nabla\phi)^2
+2\partial_t^2e^{2\phi}-4\partial_t(\vec M\cdot\vec \nabla\phi)-4\vec
\nabla\cdot(\dot\phi\vec M)\Biggr),\cr}}
where $\vec M\equiv -\sum_{n=1}^N{m_n\vec \beta_n\over |\vec x - \vec
a_n(t)|}$. Henceforth let $\vec X_n\equiv \vec x - \vec a_n(t)$.
The last three terms in \kinlag\ are time-surface or space-surface terms
which vanish when integrated.
The kinetic Lagrangian
$L_{kin}=\int d^3x{\cal L}_{kin}$ for monopole scattering converges
everywhere. This can be seen simply by studying the limiting behaviour
of $L_{kin}$ near each monopole. For a single monopole at $r=0$ with magnetic
charge $m$ and velocity $\beta$, we collect the logarithmically divergent
pieces
and find that they cancel:
\eqn\logdiv{{m\beta^2\over 2}\int r^2 drd\theta \sin\theta d\phi
\left(-{1\over r^3} + {3\cos^2\theta\over r^3}\right)=0.}

We now specialize to the case of two identical monopoles of magnetic
charge $m_1=m_2=m$ and velocities $\vec\beta_1$ and $\vec\beta_2$.
Let the monopoles be located at $\vec a_1$ and $\vec a_2$.
Our moduli space consists of the configuration space of the relative
separation vector $\vec a\equiv \vec a_2 - \vec a_1$.
The most general kinetic Lagrangian can be written as
\eqn\genkinlag{\eqalign{L_{kin}=&h(a)(\b1\cdot\b1+\b2\cdot\b2)+p(a)\left(
(\b1\cdot\hat a)^2 + (\b2\cdot\hat a)^2\right)\cr
&+2f(a)\b1\cdot\b2 + 2g(a)(\b1\cdot\hat a)(\b2\cdot\hat a).\cr}}
Now suppose $\b1 = \b2 =\vec\beta$, so that \genkinlag\ reduces to
\eqn\boostlag{L_{kin}=(2h+2f)\beta^2+(2p+2g)(\vec\beta\cdot\hat a)^2.}
This configuration, however, represents the boosted solution of the
two-monopole static solution. The kinetic energy should therefore be
simply
\eqn\cmke{L_{kin}={M_T\over 2}\beta^2,}
where $M_T=M_1+M_2=2M$ is the total mass of
the two-monopole solution. It then follows that the anisotropic part of
\boostlag\ vanishes and we have
\eqn\hfpg{\eqalign{g+p&=0,\cr 2(h+f)&={M_T\over 2}.\cr}}
It is therefore sufficient to compute $h$ and $p$. This can be done by
setting $\vec\beta_1=\vec\beta$ and $\vec\beta_2=0$.
The kinetic Lagrangian then reduces to
\eqn\rdkinlag{L_{kin}=h(a)\beta^2 + p(a)(\vec\beta\cdot\hat a)^2.}
Suppose for simplicity
also that $\vec a_1=0$ and $\vec a_2=\vec a$ at $t=0$.
The Lagrangian density of the volume term in this case is given by
\eqn\voltm{\eqalign{{\cal L}_{kin}&={-1\over 2\kappa^2}\Biggl(
{3m^3 e^{-4\phi}\over 2r^4}(\vec\beta\cdot\vec x)\left[
{\vec\beta\cdot\vec x\over r^3} + {\vec\beta\cdot(\vec x-\vec a)
\over |\vec x-\vec a|^3}\right] - {e^{-2\phi}m^2\beta^2\over r^4}\cr
&-{e^{-4\phi}m^3\beta^2\over 2r^4}\left( {1\over r} +
{\vec x\cdot(\vec x-\vec a)\over |\vec x-\vec a|^3}\right) +
{e^{-6\phi}m^4\beta^2\over r^2}\left( {1\over r^4} + {1\over |\vec x-\vec a|^4}
+ {2\vec x\cdot(\vec x-\vec a)\over r^3|\vec x-\vec a|^3}\right)\Biggr).\cr}}

The integration of the kinetic Lagrangian density in \voltm\ over three-space
yields the kinetic Lagrangian from which the metric on moduli space can be
read off. For large $a$, the nontrivial leading order  behaviour of the
components of the metric, and hence for the functions $h(a)$ and $p(a)$, is
generically of order $1/a$. In fact, for Manton scattering of BPS monopoles,
the leading order scattering angle is $2/b$ \mantwo, where $b$ is the impact
parameter. Here we restrict our computation to the leading order
metric in moduli space. A tedious but straightforward collection of $1/a$
terms in the Lagrangian yields
\eqn\leadi{{-1\over 2\kappa^2}{1\over a}\int d^3x\left[ -{3m^4e^{-6\phi_1}
\over r^7}(\vec\beta\cdot\vec x)^2 + {m^3e^{-4\phi_1}\over r^4}\beta^2+
{m^4e^{-6\phi_1}\over r^5}\beta^2 - {3m^5e^{-8\phi_1}\over r^6}\beta^2
\right],}
where $e^{2\phi_1}\equiv 1+m/r$.
The first and third terms clearly cancel after integration over three-space.
The second and fourth terms are spherically symmetric. A simple integration
yields
\eqn\leadii{\int_0^\infty r^2dr \left( {e^{-4\phi_1}\over r^4} -
{3m^2e^{-8\phi_1}\over r^6}\right)
=\int_0^\infty {dr\over (r+m)^2} - 3m^2\int_0^\infty {dr\over (r+m)^4}=0.}
The $1/a$ terms therefore cancel, and the leading order metric on moduli
space is flat. This implies that to leading order the dynamical
 force is zero and the scattering is trivial, in agreement with
 the test-soliton result.
In other words, there is no deviation from the initial
trajectories to
leading order in the impact parameter. Analogous computations
for elementary
strings in
$D=4$ \khugeo\ and fivebranes in $D=10$ \refs{\khumant,\fels}\ lead to the
same result of a flat metric. From $S \leftrightarrow T$ duality (see section
6.6) it follows
that the metric on moduli space for solitonic strings in $D=4$ is also flat.

\newsec{Veneziano amplitude for elementary strings}

We address the scattering problem in this section from the string
theoretic point of view. In particular, we calculate the string
four-point amplitude for the scattering of macroscopic winding state
strings in the infinite winding radius limit. In this scenario, we can
best approximate the soliton scattering problem considered in the previous
section but in the case of elementary strings in $D=4$.
We find that the Veneziano amplitude obtained also indicates trivial
scattering in the large winding radius limit, thus providing evidence
for the identification of the elementary strings with infinitely
long macroscopic fundamental strings. The fivebrane analog of this computation
awaits the construction of a fundamental fivebrane theory. However, a vertex
operator representation of fivebrane solitons (and also of  string monopoles)
should in principle be possible.
The computation of the fivebrane Veneziano amplitude would then represent a
dynamical test for string/fivebrane duality.

The scattering problem is set up in four dimensions, as the kinematics
correspond essentially to a four dimensional scattering problem, and
strings in higher dimensions generically miss each other anyway \polc.
The precise compactification scheme is irrelevant to our purposes.

The winding state strings then reside in four spacetime dimensions
$(0123)$, with one of the dimensions, say $x_3$, taken to be periodic with
period $R$, called the winding radius. The winding number $n$ describes
the number of times the string wraps around the winding dimension:
\eqn\wind{x_3\equiv x_3 + 2\pi Rn}
and the length of the string is given by $L=nR$. The integer $m$, called
the momentum number of the winding configuration, labels the allowed
momentum eigenvalues. The momentum in the winding direction is thus
given by
\eqn\pthree{p^3={m\over R}.}
The number $m$ is restricted to be an integer so that the quantum wave function
$e^{ip\cdot x}$ is single valued.
The total momentum of  each string can be written as the sum of a
right momentum and a left momentum \eqn\tmoment{p^\mu=p^\mu_R+p^\mu_L,}
where $p^\mu_{R,L}=(E,E\vec v,{m\over 2R}\pm nR)$,
$\vec v$ is the transverse velocity and $R$ is the winding radius.
The mode expansion of the general
configuration $X(\sigma,\tau)$ in the winding direction
satisfying the two-dimensional wave equation
and the closed string boundary conditions can be written as the sum of
right moving pieces and left moving pieces, each with the mode expansion
of an open string \gresw~
\eqn\movers{\eqalign{X(\sigma,\tau)&=X_R(\tau -\sigma) + X_L(\tau +\sigma),\cr
X_R(\tau -\sigma)&=x_R + p_R(\tau -\sigma) + {i\over 2}
\sum_{n=0} {1\over n}\alpha_n e^{-2in(\tau - \sigma)},\cr
X_L(\tau +\sigma)&=x_L + p_L(\tau +\sigma) + {i\over 2}
\sum_{n=0} {1\over n}\tilde\alpha_n e^{-2in(\tau + \sigma)}.\cr}}
The right moving and left moving components are then essentially
independent parts with corresponding vertex operators, number operators
and Virasoro conditions.

The winding configuration described by $X(\sigma,\tau)$ describes a
soliton string state. It is therefore a natural choice for us to compare
the dynamics of these states with the soliton-like solutions of
the previous
sections (including the elementary solutions) in order to
 determine whether we can identify the elementary string
solutions of
the supergravity field equations with infinitely long
 fundamental
strings. Accordingly, we study the scattering of the winding states in
the limit of large winding radius.

Our kinematic setup is as follows. We consider the scattering of
two straight macroscopic strings in the center-of-mass (CM)
 frame with
winding number $n$ and momentum number $\pm m$ \refs{\gresw,\polc}.
The incoming momenta in the CM frame are given by
\eqn\imoment{\eqalign{p^\mu_{1R,L}&=(E,E\vec v,{m\over 2R}\pm nR),\cr
p^\mu_{2R,L}&=(E,-E\vec v,-{m\over 2R}\pm nR).\cr}}
Let $\pm m'$ be the outgoing momentum number.
For the case of $m=m'$, the outgoing momenta are given by
\eqn\omoment{\eqalign{-p^\mu_{3R,L}&=(E,E\vec w,{m\over 2R}\pm nR),\cr
-p^\mu_{4R,L}&=(E,-E\vec w,-{m\over 2R}\pm nR),\cr}}
where conservation of momentum and winding number have been used and
where $\pm\vec v$ and $\pm\vec w$ are the incoming and outgoing velocities of
the strings in the transverse $x-y$ plane. The outgoing momenta
winding numbers are not {\it a priori} equal to the initial winding
numbers, but must add up to $2n$. Conservation of energy for
sufficiently large $R$ then results in the above answer. This is also in
keeping with the soliton scattering nature of the problem (i.e. the
solitons do not change ``shape" during a collision).

For now we have assumed no longitudinal excitation ($m=m'$).
We will later relax this condition to allow for such excitation, but
show that our answer for the scattering is unaffected by this
possibility. It follows  from this condition that
$v^2=w^2$. For simplicity we take $\vec v=v\hat x$ and
$\vec w=v(\cos\theta\hat x+\sin\theta\hat y)$, and thus reduce the
problem to a two-dimensional scattering problem.

As usual, the Virasoro conditions $L_0=\widetilde{L}_0=1$ must hold, where
\eqn\vops{\eqalign{L_0&=N+\half (p^\mu_R)^2,\cr \widetilde{L}_0&=\widetilde
{N}+\half (p^\mu_L)^2 \cr}}
are the Virasoro operators \gresw\ and where $N$ and $\widetilde{N}$ are the
number operators for the right- and left-moving modes respectively:
\eqn\numbs{\eqalign{N&=\sum \alpha^\mu_{-n}\alpha_{n\mu},\cr
\widetilde{N}&=\sum \tilde\alpha^\mu_{-n}\tilde\alpha_{n\mu},\cr}}
where we sum over all dimensions, including the compactified ones.
It follows from the Virasoro conditions that
\eqn\evr{\eqalign{\widetilde{N}-N&=mn,\cr
	E^2(1-v^2)&=2N-2+{({m\over 2R}+nR)}^2.\cr}}

In the following we set $n=1$ and consider for simplicity the scattering
of tachyonic winding states. For our purposes, the nature of the string
winding states considered is irrelevant. A similar calculation for
massless bosonic strings or heterotic strings, for example, will be
slightly more complicated, but will nevertheless exhibit the same essential
behaviour. For tachyonic winding states we have $N=\widetilde{N}=m=0$.
Equation \evr\ reduces to
\eqn\tevr{E^2(1-v^2)=R^2-2.}
The Mandelstam variables $(s,t,u)$ are identical for right and left
movers and are given by
\eqn\mandlestam{\eqalign{s&=4\left[\x-2\right],\cr
t&=-2\left[\x\right](1+\cos\theta),\cr
u&=-2\left[\x\right](1-\cos\theta).\cr}}
It is easy to see that
$p_{iR}\cdot p_{jR}=p_{iL}\cdot p_{jL}$ holds
for this configuration so that the tree level 4-point function
reduces to the usual Veneziano amplitude for closed tachyonic strings \polc
\eqn\veneziano{\eqalign{A_4&={\kappa^2\over 4} B(-1-s/2,-1-t/2,-1-u/2)\cr
&=({\kappa^2\over 4}) {\Gamma(-1-s/2)\Gamma(-1-t/2)\Gamma(-1-u/2)\over
\Gamma(2+s/2)\Gamma(2+t/2)\Gamma(2+u/2)}.\cr}}
This can be seen as follows. In the standard computation of the four
point function for closed string tachyons, we rely on the independence
of the right and left moving open strings. For the tachyonic winding
state, we also separate the right and left movers with vertex operators
given by $V_R=e^{ip_R\cdot x_R}$ and $V_L=e^{ip_L\cdot x_L}$ respectively
to arrive at the following expression for the amplitude
\eqn\afour{A_4={\kappa^2\over 4}\int d\mu_4(z)\prod_{i<j}
|z_i-z_j|^{p_{iR}\cdot p_{jR}} |z_i-z_j|^{p_{iL}\cdot p_{jL}}.}
{}From $p_{iR}\cdot p_{jR}=p_{iL}\cdot p_{jL}$, \afour\ reduces to the
expression for the four-point amplitude of a nonwinding closed tachyonic
string, from which the standard Veneziano amplitude in \veneziano\ results.

To compare the implications of $A_4$ with the results of
section 7.1 we take $R\to\infty$. It is
convenient to define $x\equiv\x=s/4+2$, since the Mandelstam variables can
be expressed solely in terms of $x$ and $\theta$. We now have
$A_4=A_4(x,\theta)$, which can be explicitly written as
\eqn\ampone{A_4=({\kappa^2\over 4})
{\Gamma(3-2x)\Gamma(-1+x(1+\cos\theta))\Gamma(-1+x(1-\cos\theta))\over
\Gamma(-2+2x)\Gamma(2-x(1+\cos\theta))\Gamma(2-x(1-\cos\theta))}.}
The problem reduces to studying $A_4$ in the limit $x\to\infty$.
We now use the identity $\Gamma(1-a)\Gamma(a)\sin\pi a=\pi$ to rewrite
$A_4$ as
\eqn\amptwo{\eqalign{A_4=({\kappa^2\over 4\pi})&
\left[{\Gamma(-1+x(1+\cos\theta))
\Gamma(-1+x(1-\cos\theta))\over \Gamma(-2+2x)}\right]^2\cr
&\times\left({\sin(\pi x(1+\cos\theta))\sin(\pi x(1-\cos\theta))\over\sin
2\pi x}\right).\cr}}
{}From the Stirling approximation $\Gamma(u)\sim\sqrt{2\pi}u^{u-1/2}e^{-u}$
for large $u$, we obtain in the limit $x\to\infty$
\eqn\ampthree{\eqalign{A_4\sim&\left[{\left(x(1+\cos\theta)\right)
^{x(1+\cos\theta)}
\left(x(1-\cos\theta)\right)^{x(1-\cos\theta)}\over (2x)^{2x}}\right]^2\cr
&\times\left({\sin(\pi x(1+\cos\theta))\sin(\pi x(1-\cos\theta))\over\sin
2\pi x}\right).\cr}}
Note that the exponential terms cancel automatically. From \ampthree\ we
notice that the powers of $x$ in the first factor also cancel. $A_4$
then reduces in the limit $x\to\infty$ to
\eqn\amp{\eqalign{A_4\sim\left({1+\cos\theta\over 2}\right)^{2x(1+\cos\theta)}
&\left({1-\cos\theta\over 2}\right)^{2x(1-\cos\theta)}\cr
&\times\left({\sin(\pi x(1+\cos\theta))\sin(\pi x(1-\cos\theta))\over\sin
2\pi x}\right).\cr}}
The poles in the third factor in \amp\ are just the usual $s$-channel poles.
It follows from \amp\ that for $\theta\neq 0,\pi$~
$A_4 \to e^{-f(\theta)x}$ as $x\to\infty$,
where $f$ is some positive definite function of $\theta$.
Hence the 4-point function vanishes exponentially with the winding radius
away from the poles.

In general, for finite $R$ and fixed $v$ the strings may scatter into
longitudinally excited final states, {\it i.e.} states not satisfying
the above assumption that $m'=m$. The $4$-point amplitude for each
transition still vanishes exponentially with $R$.  A simple counting
argument shows that the total number of possible final states for a
given $R$ is bounded by a polynomial function of $R$. This counting
argument proceeds as follows:

Without loss of generality, we may assume that our incoming states have
$N=\widetilde{N}=m=0$ with fixed $R$ and $v$. We relax the assumption of
no logitudinal excitation to obtain outgoing states with nonzero $m$.
We still consider $n=1$ winding states for simplicity. Our scattering
configuration can now be described by the incoming momenta
\eqn\imom{\eqalign{p^\mu_{1R,L}&=(E,E\vec v,\pm R),\cr
p^\mu_{2R,L}&=(E,-E\vec v,\pm R).\cr}}
and the outgoing momenta
\eqn\omom{\eqalign{-p^\mu_{3R,L}&=(E_1,E_1\vec w_1,{m\over 2R}\pm R),\cr
-p^\mu_{4R,L}&=(E_2,-E_2\vec w_2,-{m\over 2R}\pm R).\cr}}
Note that in general $E_1$ and $E_2$ are not equal to $E$. Without loss
of generality, we take $m$ to be positive. From
conservation of momentum, however, we have
\eqn\conserv{\eqalign{E_1+E_2&=2E,\cr E_1\vec w_1&=E_2\vec w_2.\cr}}
It follows from the energy momentum relations for the ingoing and
outgoing momenta that
\eqn\enmom{\eqalign{E^2(1-v^2)&=R^2-2,\cr
E_1^2(1-w_1^2)&=2N_1-2+\left({m\over 2R}+R\right)^2,\cr
E_2^2(1-w_2^2)&=2N_2-2+\left(-{m\over 2R}+R\right)^2,\cr}}
where $N_1$ and $N_2$ are the number operators for the the right movers
of the outgoing states.
Subtracting the third equation in \enmom\ from the second equation and
using \conserv\ we obtain the relation
\eqn\nme{N_1-N_2+m=(E_1-E_2)E.}
{}From the first equation in \enmom\ it follows that $E$ is bounded by
some multiple of $R$ for fixed $v$. It then follows from the first
equation in \conserv\ that both $E_1$ and $E_2$ are bounded by a
multiple of $R$. So from \nme\ we see that $N_1-N_2+m$ is bounded by
some quadratic polynomial in $R$. We now add the last two equations in
\enmom\ to obtain
\eqn\eenn{E_1^2(1-w_1^2)+E_2^2(1-w_2^2)=2N_1+2N_2+2R^2+{m^2\over
2R^2}-4.}
The left hand side of \eenn\ is clearly bounded by a quadratic
polynomial in $R$. It follows that $N_1+N_2$ is also bounded by a
quadratic polynomial, and that so are $N_1$ and $N_2$ and also, then,
$N_1-N_2$. From the boundedness of $N_1-N_2+m$ it therefore follows that
$m$ is bounded by a polynomial in $R$. Therefore
the total number of possible distinct excited states (numbered by $m$)
is bounded by a polynomial in $R$. The above argument also goes through
for the case of a nonzero initial momentum number. For each transition,
however, one can show that the Veneziano amplitude is dominated by an
exponentially vanishing function of $R$, from a calculation entirely
analogous to the zero-longitudinal excitation case worked out above.
Hence the total square amplitude of the scattering (obtained by
summing the square amplitudes of all possible transitions) is still
dominated by a factor which vanishes exponentially with the radius,
except at the poles at $\theta=0,\pi$ corresponding to forward and
backward scattering, which are physically equivalent for identical bosonic
strings. This is in agreement with the trivial
scattering found in section 7.1 and provides further evidence for the
identification of the elementary string with the fundamental string.

The above argument can be repeated for any other type of string, including
the heterotic string \grohmr. The kinematics differ slightly from the
tachyonic case but the $4$-point function is still dominated by an
exponentially vanishing factor in the large radius limit. Hence the
scattering is trivial, again in agreement with the result found in
section 7.1.

The Veneziano amplitude result in fact holds for arbitrary incoming winding
states. A considerably more tedious calculation for the general case
shows that in the large winding radius limit the outgoing strings always
scatter trivially and with no change in their individual winding numbers.
In this limit, then, these states scatter as true solitons \khuwind. It
would be interesting to see if this result holds for the full quantum string
loop expansion.

\chapter{Exactness of solutions}

\newsec{Axionic instanton: bosonic solution}

A classical solution of string theory is considered ``exact'' if it satisfies
the string equations of motion to all orders in
the classical string parameter $\alpha'$ (or equivalently if the
Weyl anomaly coefficients of the $\sigma$-model vanish to all orders
in $\alpha'$). Alternatively,
an exact classical solution can be demonstrated by the construction of an exact
$\sigma$-model for the worldsheet of the string or by writing down the
corresponding coset conformal field theory. In this section we follow the
discussion of \khuinst\ to obtain an exact extension of the purely bosonic
version of the axionic instanton solution of section 5.2 in the limit
 in which this solution reduces to a linear dilaton
wormhole \refs{\antben,\antbenone} (see section 2.7).
Exactness is shown by combining
the metric and antisymmetric tensor in a generalized curvature, which
is written covariantly in terms of the tree-level dilaton field, and rescaling
the dilaton order by order in the parameter $\alpha'$. The corresponding
conformal field theory is written down.

 For this purpose we use the theorem of equivalence of the
massless string field equations to the sigma-model Weyl invariance
conditions (demonstrated to two-loop order by Metsaev and
Tseytlin \refs{\mett,\mettone}), which require the Weyl
anomaly coefficients $\wbg$, $\wbb$ and $\wbd$ to vanish identically to
the appropriate order in the parameter $\alpha'$.
The two-loop solution obtained by this method suggests a
representation of the sigma model as the product of a WZW \gepw\ model
and a one-dimensional CFT (a Feigin-Fuchs Coulomb gas) \rey.
This representation allows us to obtain an exact solution.

The bosonic sigma model action can be written as \lov
\eqn\sigmod{I={1\over 4\pi\alpha'}\int d^2x\sqrt{-\eta}
\left(\eta^{ab}
\partial_aX^M\partial_bX^N g_{MN}+i\epsilon^{ab}\partial_aX^M\partial_b
X^N B_{MN}+\alpha'R^{(2)}\phi\right),}
where $g_{MN}$ is the sigma model metric, $\phi$ is the dilaton and $B_{MN}$
is the antisymmetric tensor and where $\eta_{ab}$ is the worldsheet metric
and $R^{(2)}$ the two-dimensional curvature.
The Weyl anomaly coefficients are given by \refs{\mett,\mettone}
\eqn\weyl{\eqalign{\wbg&=\rbg+2\alpha'\nabla_M\nabla_N\phi+\nabla_{(M}
W_{N)},\cr\wbb&=\rbb+\alpha' {H_{MN}}^P\partial_P\phi
+{1\over 2}{H_{MN}}^P W_P,\cr
\wbd&=\rbd+\alpha'(\partial\phi)^2+{1\over 2}W^M\partial_M\phi
,\cr}}
where $\rbg$, $\rbb$ and $\rbd$ are the RG $\beta$ functions and where
$H_{MNP}=\partial_{[M}B_{NP]}$ and
$W_M=-(\alpha'^2/24)\nabla_M H^2$.

We follow Metsaev and Tseytlin's computation of the renormalization
group beta functions for the general sigma-model and combine dimensional
regularization and the minimal subtraction scheme with the following
generalized prescription for contraction of $\epsilon^{ab}$ tensors \mett:
\eqn\reg{\epsilon^{ab}\epsilon^{cd}=f(d)\left(\delta^{ac}\delta^{bd}
-\delta^{ad}\delta^{bc}\right),}
where $f(d)=1-f_1\epsilon+O(\epsilon^2)$ and $\epsilon=d-2$.
We note that the precise form of the renormalization group beta functions at
two-loop order is not scheme-independent but depends on the choice of
$f_1$. Here we set $f_1=-1$, for which Metsaev and Tseytlin obtain the
following two-loop expressions for the Weyl anomaly
coefficients \refs{\mett,\mettone}:
\eqn\weycos{\eqalign{\wbg&=\alpha'(\hat R_{(MN)}+2\nabla_M
\nabla_N\phi)\cr&+{\alpha'^2\over
2}\left(\hat R^{ABC}{}_{(M}\hat R_{N)ABC}
-{1\over 2}\hat R^{BCA}{}_{(M}\hat
R_{N)ABC}+{1\over 2}\hat R_{A(MN)B}
(H^2)^{AB}\right)+\nabla_{(M}W_{N)},\cr
\wbb&=\alpha'(\hat R_{[MN]}+H_{MN}{}^P\partial_P\phi)
\cr&+{\alpha'^2\over
2}\left(\hat R^{ABC}{}_{[M}\hat R_{N])ABC}
-{1\over 2}\hat R^{BCA}{}_{[M}\hat
R_{N]ABC}+{1\over 2}\hat R_{A[MN]B}
(H^2)^{AB}\right)+{1\over 2}H_{MN}{}^P W_P
,\cr
\wbd&={D\over 6}-{\alpha'\over 2}\left(\nabla^2\phi-2(\partial\phi)^2+{1\over
12}H^2\right)\cr&+{\alpha'^2\over 16}\left(2(H^2)^{MN}\nabla_M
\nabla_N\phi+R^2_{PMNK}-{11\over 2}RHH
+{5\over 24}H^4+{11\over 8}(H^2_{MN})^2+{4\over 3}\nabla H\cdot
\nabla H\right)\cr&+{1\over 2}W^M\partial_M\phi,\cr}}
where $\nabla H\cdot\nabla H\equiv\nabla_M H_{NPQ}
\nabla^M H^{NPQ}.$
Unless otherwise indicated, all expressions are written to two loop
order in the beta-functions, which corresponds to $O(\alpha')$ in the effective
action \stringact.
Also, the only nontrivial coefficients occur when all indices are in the
 transverse curved four space, as it is clear that the
flat dimensions do not contribute, so henceforth we restrict ourselves to this
space.

It follows from section 5.2 and the abovementioned theorem of
equivalence that any dilaton function satisfying
$e^{-2\phi}\Box\ e^{2\phi}=0$ with
\eqn\bsansatz{\eqalign{\met&=e^{2\phi}\delta_{mn}\qquad m,n=1,2,3,4,\cr
g_{\mu\nu}&=\eta_{\mu\nu}\qquad\quad   \mu,\nu=5,...,26,\cr
H_{mnp}&=\pm 2\epsilon_{mnpk}\partial^k\phi
\qquad m,n,p,k=1,2,3,4\cr}}
is an $O(\alpha')$
 solution to \weycos. Note that this is essentially the fivebrane ansatz,
except that in bosonic string theory we have $22$ flat directions instead of
$6$. Eqs. \gcurvphi\ and \axin\ follow immediately as in the supersymmetric
case of section 5.2.

We now specialize to the spherically symmetric case of $e^{2\phi}={Q/r^2}$ in
\bsansatz\ and determine the $O(\alpha')$ corrections to the massless fields in
\bsansatz\ so that the Weyl anomaly coefficients vanish to $O(\alpha'^2)$.
For this solution we notice \khuinst
\eqn\ddphi{\nabla_m\nabla_n\phi=0,}
and therefore from \gcurvphi
\eqn\gcurvzero{\grijkl=0,}
and we have what is called a ``parallelizable" space \refs{\mett,\mettone}.
To maintain a parallelizable space to $O(\alpha')$ we keep $\met$
and $H_{\alpha\beta\gamma}$ in their lowest order form and assume
that any corrections to \bsansatz\ appear in the dilaton:
\eqn\tlgansatz{\eqalign{\phi&=\bar\phi+\alpha'\phi_1+...\cr
e^{2\bar\phi}&={Q\over r^2},\cr
\met&=e^{2\bar\phi}\delta_{mn},\cr
H_{mnp}&=\pm 2\epsilon_{mnpk}\partial^k\bar\phi .\cr}}
It follows from \tlgansatz\ that $H^2=24(\partial\bar\phi)^2=24/Q$ and
thus $W_m=0$. It follows from \gcurvzero\ that $\wbg$ and $\wbb$
vanish identically to two loop order and that
\eqn\dweylone{\eqalign{\wbd={D\over 6}+&\alpha'\left((\partial\phi)^2
-{1\over Q} \right)\cr
&+{\alpha'^2\over 16}\left(R^2_{kmnp}-{11\over 2}RHH
+{5\over 24}H^4+{11\over 8}(H^2_{mn})^2+{4\over 3}\nabla H\cdot\nabla H
\right).\cr}}
We use the relations in equation (34) in \mett\ for parallelizable spaces
and the observation that $(H^2_{mn})^2=2H^4=192/Q^2$ for our solution to
get the identities
\eqn\parallel{\eqalign{R^2_{kmnp}&={1\over 8}H^4,\cr
RHH&={1\over 2}H^4,\cr\nabla H\cdot\nabla H&=0 .\cr}}
\dweylone\ then simplifies further to
\eqn\dweyltwo{\wbd={D\over 6}+\alpha'\left((\partial\phi)^2-{1\over Q}\right)
+2{\alpha'^2\over Q^2}.}
The lowest order term in $\wbd$ is proportional to the central charge
and the $O(\alpha')$ terms vanish identically. With the choice
$\overrightarrow\nabla\phi_1=-(1/Q)\overrightarrow\nabla\phi_0$, the
$O(\alpha'^2)$ terms also vanish identically.
The two-loop solution is then given by
\eqn\tlsansatz{\eqalign{e^{2\phi}&={Q\over r^{2(1-{\alpha'\over Q})}},\cr
\met&={Q\over r^2}\delta_{mn},\cr
H_{mnp}&=\pm 2\epsilon_{mnpk}\partial^k\phi_0,\cr}}
which corresponds to a simple rescaling of the dilaton.
A quick check shows that this solution has finite action near the
singularity.

We now rewrite $\wbd$ in \dweyltwo\ in the following suggestive form:
\eqn\dweylsplit{\eqalign{6\wbd&=\left(1+6\alpha'(\partial\phi)^2\right)
+\left(3-6{\alpha'\over Q}+12({\alpha'\over Q})^2\right)\cr&=4.\cr}}
The above splitting of the central charge $c=6\wbd$ suggests
the decomposition of the corresponding sigma model into the
product of a one-dimensional CFT (a Feigin-Fuchs Coulomb gas)
and a three-dimensional WZW model with an $SU(2)$ group manifold
\refs{\mett,\mettone}. This can be seen as follows.
Setting $u=\ln r$, we can rewrite \sigmod\ for our solution \reyone\
in the form $I=I_1+I_3$, where
\eqn\onecft{I_1={1\over 4\pi\alpha'}\int d^2x\left(Q(\partial u)^2
+\alpha' R^{(2)}\phi\right)}
is the action for a Feigin-Fuchs Coulomb gas, which is a one-dimensional
CFT with central charge given by
$c_1=1+6\alpha'(\partial\phi)^2$ \gin. The imaginary charge of the
Feigin-Fuchs Coulomb gas describes the dilaton background growing
linearly in imaginary time and
$I_3$ is the Wess--Zumino--Witten \gepw\ action on an $SU(2)$ group manifold
with central charge
\eqn\threecharge{c_3={3k\over k+2}\simeq 3-{6\over k}+{12\over k^2}+...}
where $k=Q/\alpha'$, called the ``level" of the WZW model, is an
integer. This can be seen from
the quantization condition on the Wess-Zumino term \gepw
\eqn\iwzw{\eqalign{I_{WZ}&={i\over 4\pi\alpha'}\int_{\partial S^{\pm}_3}
d^2x\epsilon^{ab}\partial_ax^m\partial_bx^n B_{mn}\cr
&={i\over 12\pi\alpha'}\int_{S^{\pm}_3}d^3x\epsilon^{abc}
\partial_ax^m \partial_bx^n\partial_cx^p H_{mnp}\cr
&=2\pi i\left({Q\over\alpha'}\right).\cr}}
Thus $Q$ is not arbitrary, but is quantized in units of $\alpha'$. This
quantization condition is equivalent to
\etwogsix\ once we set $Q=k_6$ from \ksix.

We use this splitting to obtain exact expressions
for the fields by fixing the metric and antisymmetric tensor field
in their lowest order form and rescaling the dilaton order by order in
$\alpha'$. The resulting expression for the dilaton is
\eqn\alldilaton{e^{2\phi}={Q\over r^{\sqrt{{4\over 1+{2\alpha'\over Q}}}}}.}

\newsec{Symmetric fivebrane}

The symmetric fivebrane of section 5.3 was obtained
by equating the curvature of the Yang-Mills gauge field with the generalized
curvature of the axionic instanton.
This solution thus represents a supersymmetric extension of
the bosonic solution of section 8.1. Unlike the bosonic solution, however,
the symmetric fivebrane can be argued to be an exact classical solution in the
multi-instanton case, and
not merely in the wormhole limit \refs{\calhs,\calhsone}. In that particular
limit, however, one can
demonstrate exactness of the symmetric fivebrane without
modification to the dilaton
from both the $\beta$-function and CFT points of view.
Furthermore, these exactness arguments can be supplemented by
superconformal worldsheet $\sigma$-model arguments and
nonrenormalization  theorems due to higher worldsheet supersymmetry.

{}From the $\beta$-function (or equations of motion)
point of view, one would guess that
the tree-level symmetric fivebrane solution is exact since
$A_M=\Omega_{\pm M}$ suggests that all the
higher order corrections vanish.
This can be seen by noting that the higher order corrections to the
effective action (up to at least $O(\alpha'^3)$) are
all at least linear in the tensor
 \refs{\berd,\berdone})
\eqn\tabcd
{T_{ABCD}\equiv tr\hat R_{[AB}\hat R_{CD]}-trF_{[AB}F_{CD]},}
which clearly vanishes since the two curvatures are identical.

The fact that the single instanton solution in the
heterotic case carries through even to $O(\alpha')$ without correction to the
dilaton seems to contradict the bosonic solution in the wormhole limit
by suggesting that
the expansion for the Weyl anomaly coefficient $\wbd$ terminates at one loop.
This seeming contradiction is resolved by noting that in the heterotic case
the gauge rotation which decouples the fermions is anomalous \calhstwo, and the
effect of this anomaly is to replace $k$ in \threecharge\ by $k'=k-2$ \alljj.
The bosonic contribution to this part of the central charge is then given by
\eqn\termcharge{c_3={3k'\over k'+2}=3-{6\over k}=3-{6\alpha'\over Q},}
which indeed terminates at one loop order. The exactness of the splitting
then requires that $c_1$ not get any corrections from $(\partial\phi)^2$ so
that $c_1+c_3=4$ is exact for the tree-level value of the dilaton \calhs.
In \refs{\calhs,\calhstwo}, arguments based on $N=4$ worldsheet
supersymmetry were presented to support the claim of exactness of the
heterotic fivebrane. We summarize these arguments below.

To show that a solution is an exact solution in string theory one must in
principle show that the solution derives form a superconformal worldsheet
sigma-model. For the heterotic fivebrane, one can show that the
corresponding $\sigma$-model possesses $(4,4)$ worldsheet supersymmetry,
in which case nonrenormalization theorems show that the tree-level solution
is exact.

For generic fivebrane solutions annihilated by $D=6$, $N=1$ spacetime
supersymmetries, the compactification from $10$ to $6$ dimensions maintains
$N=1$ spacetime supersymmetry. In compactification from $10$ to $4$
dimensions with $N=1$ spacetime supersymmetry, the compactified $6$
dimensions must possess at least $(2,0)$ supersymmetry. Similarly, in
compactifying
to $D=4$, $N=2$, the six-dimensional compactification sigma-model must possess
at least
$(4,0)$ supersymmetry. Since $N=1$, $D=6$ spacetime supersymmetry is
equivalent to $N=2$, $D=4$ spacetime supersymmetry, a generic fivebrane
solution must possess at least $(4,0)$ spacetime supersymmetry.

In writing down a generic heterotic fivebrane sigma-model action
with $(1,0)$ worldsheet supersymmetry, the
only worldsheet fermions that couple nontrivially are four right movers
which couple to the generalized connection $\Omega_-$ and four
left movers
which couple to the gauge field, which must be an instanton to
satisfy
the tree-level supersymmetry equations (see section 5.3). Now
if we demand that the dilaton
satisfy $e^{-2\phi} \Box e^{2\phi} =0$ and consequently the
exactness
condition $A=\Omega_+$, the theory becomes left-right symmetric
 and
thus must possess at least $(1,1)$ worldsheet supersymmetry.
This can also
be seen by observing that for $dH=0$
\eqn\gclr{R(\Omega_+)_{mnpq}=R(\Omega_-)_{pqmn}}
follows immediately from \gcurvphi.  The symmetric fivebrane therefore
possesses
$(4,4)$ worldsheet supersymmetry. For the explicit construction of the
$(4,4)$ worldsheet $\sigma$-model, we refer the reader to
\calhstwo, in which it is argued that the existence of more than one pair
of complex structures elevates the worldsheet supersymmetry to $(4,4)$.
As a consequence of $(4,4)$ supersymmetry, the theory is finite and
from nonrenormalization theorems the tree-level solution is exact.
In a similar manner, it can be argued that
all symmetric solutions correspond to $(4,4)$
supersymmetry on the worldsheet of the fundamental string and are thus
presumably exact to all orders in $\alpha'$. In particular, this includes the
$D=4$ monopole, string and domain wall solutions of section 6.5.

{}From algebraic conformal field theory constructions, $(4,4)$ worldsheet
supersymmetry manifests itself in the wormhole limit
in the form of two $SU(2)$ Kac-Moody
symmetries, one from the $N=4$ superalgebra and the other from the WZW
wormhole throat \refs{\sev,\schout,\calhstwo}.

The double-instanton string solution of section 5.4
can also be argued to be exact using similar arguments as above.
Both the $\beta$-function arguments and the $(4,4)$ worldsheet
supersymmetry arguments above can be used in essentially
the same manner to demonstrate exactness of the string solution.
The explicit construction in this case can be found in \koun.

\newsec{Exact elementary string}

Finally, we summarize the arguments of \refs{\hort,\tseytlin}\ regarding the
exactness of the elementary string.
Consider the family of backgrounds with
metric and antisymmetric tensor
characterized by a single function $F(x)$ and dilaton $\phi(x)$:
\eqn\ffff{ ds^2 = F(x) \eta_{\mu\nu} dx^\mu dx^\nu + dx_i dx^i\ ,
 \qquad B_{01} = {1\over 2}F(x) \ . }
The two functions $F$ and $\phi$ depend only
on the transverse coordinates
$x^i$. The leading order string  equations of \sigmod\
then reduce to  \klim\
\eqn\fmod{ \partial^2 F^{-1} = 2b^i \partial_i F^{-1} , \ \ \
  \phi = \phi_0 + b_i x^i  + {1\over 2}  \ln F (x) \ , \ }
where $b_i$ is a  constant vector.
It was
argued in \refs{\klim,\hort}\ that these solutions should not
receive non-trivial higher order corrections, and that
one can show  \hort\ that there
exists  a scheme in which all of these solutions are exact
and receive no $\alpha'$ corrections. Since the equation for $F^{-1}$ is
linear,
linear combinations of these solutions yield new exact solutions.
For $b_i=0$ one recovers the elementary string solution of \dabghr:
\eqn\fss{ F\inv ={ 1 + {M \over r^{D-4}} }   \ , \ \ D>4 \ ; \ \ \
 F\inv ={ 1 - {M \  \ln \ r } } \ , \ \ \  D=4\ , \ \  \ r^2 = x_i x^i\ ,  }
where $D$ is the number of spacetime dimensions. The case $D=10$ was discussed
in section 2.1. The exactness arguments proceed as follows.

For a more general model with curved transverse space, defined  as
\hort
\eqn\mof{  L_F=F(x) \eta_{\mu\nu}
\partial x^\mu \bar\partial x^\nu +  (G_{ij} + B_{ij})(x)\ \partial x^i \bar
\partial x^j
+ \alpha'{\cal R}\phi(x),  }
the all-order conformal invariance conditions
are satisfied \hort\ provided one is given a
conformal ``transverse'' theory  $(G', B', \phi')$  and
\eqn\summm{  G_{ij} =G'_{ij}  +  {1\over 2} \alpha' \  \partial_{i } {\ln } F \
\partial_{j}
{\ln} F \ , \ \ \  \phi= \phi'   + {1\over 2} \ln F \ , \
\ \ B_{ij}= B'_{ij} \ ,   }
with  $F$ satisfying
\eqn\taccc{ -   \omega' F^{-1} +  \partial^i \phi' \partial_i F^{-1}
=- {1\over 2} \nabla'^2 F^{-1}    +  O(\alpha')  +  \partial^i \phi'
\partial_i F^{-1} =0\ . }
Here $\omega'$  is the anomalous dimension operator depending on $G'$
 \refs{\hort,\tseytlin}.
When   $(G',B,\phi')$ correspond to  a  known CFT
this equation
can be written down explicitly to all orders in $\alpha'$.
Since the relation between $G$ and $G'$ is local  and
the transverse theory  is,  in general,  defined modulo local coupling
redefinitions,  one can argue \hort\ that there
exists a (``leading-order'') scheme in which  the exact solution is represented
by
the conformal transverse theory $(G,B,\phi')$  and  $F$ satisfying \taccc.

The   simplest  example of the  conformal  transverse model is flat space
with  a linear dilaton, in which the  corresponding model  is  represented
(in the leading-order   scheme) by
\eqn\exaav{  G_{ij} =\delta_{ij}  \  , \ \ \
\ \  \phi= \phi_0  + b_i x^i   + {1\over 2}\  \ln \  F \ . }
In this  scheme, the exact  form of the equation for the function $F$ is
simply
\eqn\fff{ - {1\over 2}  \partial^2 F^{-1}  +  b^i \del_i F^{-1} =0 \ , }
i.e. the  model
\eqn\fef{  L_{F}=F(x) \eta_{\mu\nu}\partial x^\mu \bar \partial x^\nu +
\partial x^i \bar \partial x_i    + \alpha'{\cal R}
 (\phi_0 +b_ix^i +    {1\over 2} \ln F ) \  , }
with $F$ satisfying \fff\ is  conformally  invariant to all orders, i.e.
gives an exact string solution.

In this scheme the leading-order duality is exact since the
leading-order  dual to \fef\ is the model
\eqn\feff{  L_{K}=\eta_{\mu\nu} \partial x^\mu \bar\partial x^\nu +
 F^{-1} (x) \partial x^i \bar \partial x_i    + \alpha'{\cal R}
 (\phi_0 +b_ix^i) \  ,   }
which represents an exact  solution \hort\ if $F$ solves \fff.
In particular, one concludes that there exists a scheme in which the elementary
solution \fss\ is a classical string solution to all orders in $\alpha'$.

\chapter{Recent developments}

This report summarizes the status of string solitons up
to the summer of '94. Since the subject is developing so
rapidly it is difficult to know when to draw the line. Here
we record some of the most interesting developments:

1) Seiberg and Witten \seiw, and
Vafa and Witten \vafw\ have provided further evidence for the
$S$-duality of section 6.2 in global supersymmetric Yang-Mills
theories. An interesting attempt to
show evidence for $S$-duality in string theory can be found
in the recent paper of Gauntlett and Harvey \gauh   .

2) Frampton and Kephart \frak\ have suggested that the
identification of Bogomol'nyi string states with extreme
black holes of section 6.8 be generalized and claim that
{\it all} massive string states are black holes. As discussed
in section 6.8, we remain agnostic on this point.

3) The entropy of a scalar/Maxwell parameter $a$ is given
by \hw
\eqn\bhent{S=\pi r_+^2 \left( {r_+ - r_-\over r_+}
\right)^{2a^2\over 1+a^2}}
and so the claim of section 6.8 that the extreme $(r_+=r_-)$
black holes have zero entropy, while valid for $a\neq 0$, is
ambiguous for $a=0$. This ambiguity is resolved in recent
papers by Hawking, Horowitz and Ross \hawhr\ and
Gibbons and Kallosh \gibkal.

4) Both the $a=\sqrt{3}$ and $a=1$ supersymmetric black holes,
shown to be solutions of the heterotic string in
\refs{\dufkmr,\ghs}\ and discussed in sections 6.7
and 6.8 have recently been shown to be exact to all orders
in $\alpha'$ by Horowitz and Tseytlin \hortsey\ (at least
in the context of the bosonic string). Further discussions of
black holes, supersymmetry and duality may be found in the
papers by Cveti\v c and Youm \cvey\ and by Bergshoeff, Kallosh
and Ortin \bko\ (for an earlier reference, see Kalara and
Nanopoulos \kaln).

5) The conjecture that $S$ and $T$ duality be united into
$O(8,24;Z)$, as discussed in section 6.8, has recently been
 taken up by Sen \sentd\ in
the context of the $D=3$ heterotic string. He shows that the
fundamental string is related to the stringy cosmic string
\gresvy\ by an $O(8,24;Z)$ transformation. In \duffkr,
Duff, Ferrara, Khuri and Rahmfeld find
an $O(8,24;Z)$ transformation relating the fundamental string
to the dual string \dufkexst, and make various connections
between the $T$-duality group and spacetime supersymmetry.
See also the paper by Maharana \maha.

6) On the issue of $p$-brane singularities,
Gibbons, Horowitz and Townsend \ght\ have considered further
the higher-dimensional resolution of dilaton black hole
singularities \dufgt\ discussed in sections 3.7 and 6.8. They
conclude that an $a=1/\sqrt{3}$ black hole in $D=4$,
the self-dual string in $D=6$ \duflblacks, the
self-dual threebrane in $D=10$ \refs{\hors,\duflselft}\
and the fivebrane in $D=11$ \guv\ are completely nonsingular.

7) Hull and Townsend \hult\ have generalized the arguments
of Duff and Rahmfeld \dufr\ and of section 6.8 concerning
string states as black holes
to the Type II superstring, where the four-dimensional
supersymmetry is $N=8$ and where the duality group
$E_{7(7)}(Z)$ contains $O(6,6;Z)$ and $SL(2,Z)$ as subgroups
\dufldr. This leads to a new and interesting interplay between
elementary and solitonic string states not found in the
$D=4$ heterotic string. They also make the interesting
observation that the flatness of the moduli space for solitons
that break half the supersymmetry is protected by supersymmetry
for $N=8$ supergravity but not for $N=4$. Thus there is no
apparent reason to expect the moduli space metric of extreme
black hole solitons of the exact heterotic string theory
(to all orders in $\alpha'$  and $g$) to be flat. They even
make the bolder suggestion that when solitons are taken into
account there is no distinction between the field theory and
the string theory.

There is now a consensus that the really interesting questions
of superstring theory will never be answered within the
framework of a weak coupling perturbation expansion. It is
thus refreshing to see this new burst of activity in
non-perturbative string theory.

\extra{Acknowledgements}
We have enjoyed useful conversations with Atish Dabholkar,
John Dixon, Sergio Ferrara,
Gary Gibbons, Chris Hull, Jim Liu, Nick Manton, Ruben Minasian,
Chris Pope, Joachim Rahmfeld,
John Schwarz, Ashoke Sen, Ergin Sezgin and Paul Townsend.
MJD is grateful to the Theory Group, Rutherford-Appleton
Laboratory; the CERN Theory Division; the Director and Staff
of the Isaac Newton Institute, and the organizers of the
{\it Topological Defects} programme, for their hospitality.
RRK is grateful to the Princeton University Physics Department
for its hospitality.
\vfill\supereject
\listrefs
\bye